\documentclass[11pt]{article}
\usepackage{jheppub} 
\usepackage{slashed}
\usepackage{array,multirow,colortbl,longtable,booktabs}
\usepackage{amsmath,amsfonts,amssymb,bbm}
\usepackage[normalem]{ulem}
\usepackage[T1]{fontenc}
\usepackage{mathrsfs,wasysym,cancel}
\usepackage{units}
\usepackage{hyperref}
\usepackage{lscape}
\usepackage{graphicx}
\usepackage{feynmp}
\usepackage[dvipsnames,usenames]{xcolor}
\usepackage{listings}
\usepackage{fontawesome}

\definecolor{logolightgreen}{rgb}{0.686, 0.914, 0.867}

\definecolor{code_frame}{RGB}{207, 207, 207}
\definecolor{code_bg}{RGB}{247, 247, 247}
\colorlet{tab_bg}{code_bg!80!logolightgreen}

\colorlet{tablesColor}{logolightgreen!80}
\def\nn{\nonumber}

\graphicspath{{./Figures/}{./}}
\renewcommand{\O}{\mathcal{O}}
\renewcommand{\P}{\mathcal{P}}
\renewcommand{\a}{\alpha}
\renewcommand{\b}{\beta}
\renewcommand{\d}{\delta}
\newcommand{\e}{\varepsilon}
\newcommand{\g}{\gamma}
\newcommand{\s}{\sigma}

\renewcommand{\dag}{\dagger}
\newcommand{\de}{\partial}

\newcommand{\sw}{s_{\theta}}
\newcommand{\cw}{c_{\theta}}
\newcommand{\hsw}{s_{\hat \theta}}
\newcommand{\hcw}{c_{\hat \theta}}
\newcommand{\hsdw}{s_{2\hat \theta}}
\newcommand{\hcdw}{c_{2\hat \theta}}

\newcommand{\hv}{\hat{v}}
\newcommand{\aem}{\alpha_{\rm em}}
\newcommand{\aew}{\aem}
\newcommand{\Lag}{\mathcal{L}}
\renewcommand{\to}{\rightarrow}
\newcommand{\hc}{\text{h.c.}}

\newcommand{\C}{\bar{C}}

\newcommand{\ascheme}{$\{\aem,m_Z,G_F\}$}
\newcommand{\mwscheme}{$\{m_W,m_Z,G_F\}$}

\newcommand{\bmat}{\begin{pmatrix}}
\newcommand{\emat}{\end{pmatrix}}

\DeclareMathOperator{\re}{Re}
\DeclareMathOperator{\im}{Im}
\DeclareMathOperator{\diag}{diag}
\DeclareMathOperator{\Br}{Br}

\newcommand{\smeftsim}{\texttt{SMEFTsim}}
\newcommand{\dimsixtop}{\texttt{dim6top}}
\newcommand{\smeftatnlo}{\texttt{SMEFT@NLO}}
\newcommand{\feynrules}{\texttt{FeynRules}}
\newcommand{\ufo}{\texttt{UFO}}
\newcommand{\mathematica}{\texttt{Mathematica}}
\newcommand{\madgraph}{\texttt{MadGraph5\_aMC@NLO}}
\newcommand{\general}{\texttt{general}}
\newcommand{\Utf}{\texttt{U35}}
\newcommand{\MFV}{\texttt{MFV}}
\renewcommand{\top}{\texttt{top}}
\newcommand{\topsl}{\texttt{topU3l}}

\newcommand{\tw}{\textwidth}

\lstdefinelanguage{MG}{
    sensitive=true,
    keepspaces=true,
    showspaces=false,
    showstringspaces=false,
    rulecolor=\color{code_frame},
    frame=single,
    frameround={t}{t}{t}{t},
    framexleftmargin=5mm,
    backgroundcolor=\color{code_bg!80!logolightgreen},
    basicstyle=\ttfamily,
    aboveskip=1.2em,
    belowskip=1.2em
}
\lstdefinestyle{MG_wide}{
   language=MG,
    xleftmargin=-10mm,
    xrightmargin=-10mm
}

\subheader{\includegraphics[width=4.5cm]{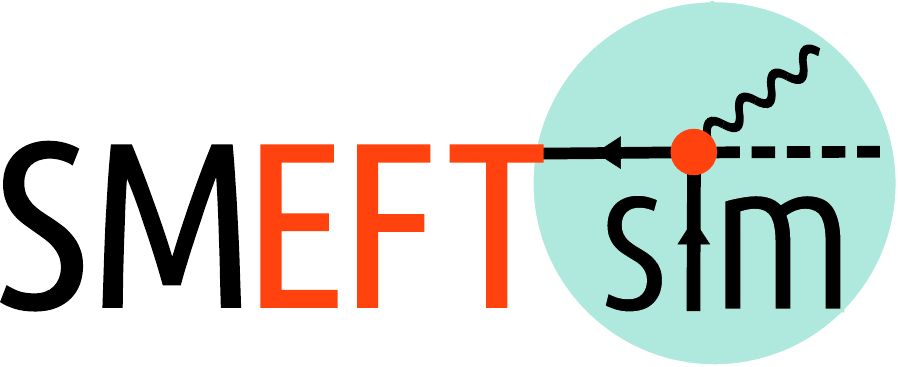}}

\title{SMEFTsim~3.0 --  a practical guide}
\author{Ilaria Brivio}
\affiliation{Institut f\"ur Theoretische Physik, Universit\"at Heidelberg.\\ Philosophenweg 16, 69120 Heidelberg (Germany)}
\emailAdd{brivio@thphys.uni-heidelberg.de}

\abstract{The \smeftsim\ package~\cite{Brivio:2017btx} is designed to enable automated computations in the Standard Model Effective Field Theory (SMEFT),  where the SM Lagrangian is extended with a complete basis of dimension six operators. It contains a set of models written in \feynrules\ and pre-exported to the \ufo\ format, for usage within Monte Carlo event generators. The models differ in the flavor assumptions and in the input parameters chosen for the electroweak sector. 
The present document provides a self-contained, pedagogical reference that collects all the theoretical and technical aspects relevant to the use of \smeftsim\ and it documents the release of version~3.0. Compared to the previous release, the description of Higgs production via gluon-fusion in the SM has been significantly improved,  two flavor assumptions for studies in the top quark sector have been added, and a new feature has been implemented, that enables the treatment of linearized SMEFT corrections to the propagators of unstable particles.

\vspace*{1em}
\noindent
\smeftsim~3.0 is available on the Github~\faGithub\ website~\href{https://SMEFTsim.github.io}{\tt https://SMEFTsim.github.io}\\
and on the \feynrules\ database~\href{http://feynrules.irmp.ucl.ac.be/wiki/SMEFT}{\tt http://feynrules.irmp.ucl.ac.be/wiki/SMEFT}.

}

\begin{document}
\maketitle

\section{Introduction}\label{sec.SMEFT}
LHC physics is about to enter a precision era that will span over the next two decades.
During this time, new opportunities to hunt for new physics will arise: 
direct searches of new particles will be complemented by indirect searches, 
that target possible deviations from the predictions of the Standard Model (SM).
While the isolation of this kind of signatures is not without challenges, indirect searches present some very attractive features. Most notably,  they do not rely on specific assumptions about the nature of the new physics under scrutiny and, at the same time, their sensitivity in terms of new physics scales can potentially extend beyond the energy reach of the collider.

The Standard Model Effective Field Theory (SMEFT) is the best established theory framework to describe such effects. Its formulation employs the degrees of freedom and gauge symmetries of the SM and it is structured as an infinite series of operators sorted by canonical dimension. At the observables level, it reproduces a series expansion in $(E/\Lambda)$, being $E$ the typical energy exchanged in a process and $\Lambda$ the mass scale that characterizes the beyond-SM (BSM) dynamics. The condition $(E/\Lambda)\ll1$, indicating the near decoupling of the new physics sector, is necessarily assumed. 

The SMEFT has been developed extensively in the past ten years, laying the ground for
a systematic program for indirect searches~\cite{Brivio:2017vri,Passarino:2019yjx,David:2020pzt}. The ultimate goal is to measure as many EFT parameters as possible, in a manner that enables the extraction of unbiased information about the underlying physics. The crucial aspect of this program is its transversality: the SMEFT contains a large number of parameters, each typically entering the description of several processes. Combining measurements of different observables is then mandatory in order to preserve the model-independence of the analysis. To date, this principle has been applied within individual sectors as well as across Higgs, electroweak (EW) and top quark measurements, see Refs.~\cite{Almeida:2018cld,Biekotter:2018rhp,Hartland:2019bjb,vanBeek:2019evb,Brivio:2019ius,Falkowski:2019hvp,Bissmann:2019qcd,Dawson:2020oco,Ellis:2020unq,ATLAS:2020naq,Ethier:2021ydt} for recent examples. 
The incorporation of data from flavor observables (including non-LHC experiments) would be very valuable in this context, as most of the SMEFT parameter space is ``flavorful''. 
First steps in this direction were taken in~\cite{Bissmann:2019gfc,Aoude:2020dwv,Bissmann:2020mfi,Bruggisser:2021duo}.

The theory developments have been accompanied by the publication of a number of computing tools that automate most stages of a SMEFT study~\cite{Brivio:2019irc}. These include the definition of non-redundant operator bases and the translation between them~\cite{Falkowski:2015wza,Aebischer:2017ugx,Gripaios:2018zrz,Criado:2019ugp}, the matching to concrete BSM models or to the low-energy EFT and the renormalization group running~\cite{Celis:2017hod,Criado:2017khh,Aebischer:2018bkb,Bakshi:2018ics,Fuentes-Martin:2020zaz,Cohen:2020qvb,Fuentes-Martin:2020udw}, the extraction of the Feynman rules in $R_\xi$ gauges~\cite{Dedes:2017zog,Dedes:2019uzs} and in the background field gauge~\cite{Corbett:2020bqv}, Monte Carlo simulations~\cite{Alloul:2013naa,Brivio:2017btx,AguilarSaavedra:2018nen,Degrande:2020evl} and global  analyses~\cite{Castro:2016jjv,Aebischer:2018iyb,Straub:2018kue,deBlas:2019okz,Ellis:2020unq}.

The \smeftsim\ package~\cite{Brivio:2017btx} was designed in order to enable the Monte Carlo simulation of arbitrary processes in the effective theory, in the spirit of providing a unified, general-purpose tool for SMEFT physics at the LHC. 
It provides complete tree level, unitary gauge predictions
at $\O(\Lambda^{-2})$, including \emph{all} the dimension six operators in the so-called Warsaw basis~\cite{Grzadkowski:2010es}. The field and parameter redefinitions that are required in order to compute physical observables in the SMEFT are conveniently performed internally.
The package contains \feynrules~\cite{Christensen:2008py,Alloul:2013bka} source files and a set of models pre-exported to the \ufo\ format~\cite{Degrande:2011ua}. 
Although the latter are in principle compatible with most Monte Carlo generators, they have been optimized  for the use in \madgraph~\cite{Alwall:2014hca}, that is illustrated in Sec.~\ref{sec.MG_use}. Potential issues due, for instance, to the event generation entering regions where the EFT validity or the unitarity of the $S$-matrix are violated are not addressed within \smeftsim\ itself, but can be generally treated with tools offered by the Monte Carlo generators or with theoretical assessments \emph{a posteriori}.

The \ufo\ models differ in the flavor assumptions and in the choice of the input parameters for the EW sector.\footnote{The original release contained two fully equivalent implementations, that were called model sets A and B. Both were provided for debugging and cross-validation. Set B is not supported anymore starting from version~3.0, which is based on set A. } The original release implemented three alternative flavor scenarios: a general one, a $U(3)^5$-symmetric case and a linear minimal flavor violation (MFV) option where BSM CP-violating phases are forbidden. For each setup, it offered two EW input sets: \ascheme\ or \mwscheme. 

This work documents the release of \smeftsim\ version~3.0, that introduces a number of improvements summarized below. It is also meant as a pedagogical and self-contained reference for its usage, where all the relevant theoretical aspects are reviewed in detail.

\vskip 1em
The present document is structured as follows: Sections~\ref{sec.SMEFT} -- \ref{sec.inputs} review  the theoretical background while
Sections~\ref{sec.Hloops} -- \ref{sec.MG_use} describe technical aspects of the \feynrules\ and \ufo\ implementations and provide recommendations for their use.

The notation is fixed in Section~\ref{sec.notation}.
Section~\ref{sec.redefinitions} focuses on the bosonic sector and it reviews the field and parameter redefinitions required to ensure a canonical parameterization of the kinetic terms and scalar potential. Section~\ref{sec.flavor} is devoted to the flavor structure of the SMEFT and it defines the five scenarios implemented in \smeftsim. Some significant changes have been made compared to version 2, that are documented in detail, and two new flavor options have been introduced (\top, \topsl) that comply with the recommendations for studies of top quark observables~\cite{AguilarSaavedra:2018nen}.
Section~\ref{sec.inputs} provides a general discussion of how the extraction of numerical values for the SM parameters is affected in the presence of higher-dimensional operators, and illustrates the treatment of these effects in \smeftsim.

Section~\ref{sec.Hloops} documents the implementation of Higgs interactions that are purely loop-generated in the SM, namely $h\g\g,\, hZ\g, hgg$: as \smeftsim\ only supports tree-level interactions, these are treated as effective vertices in the large top mass limit, which is a good approximation for Higgs production and decay processes. 
Compared to version 2, the description of Higgs-gluon vertices has been substantially improved, such that it can now model one-loop SM interactions with up to 4 gluons. Section~\ref{sec.propagators} focuses on SMEFT effects in the propagators of unstable particles, that arise due to modifications of their pole masses or decay widths. A new feature has been introduced in version~3.0, that enables the inclusion of such effects, linearized in the EFT parameters, in Monte Carlo simulations. To our knowledge \smeftsim\ is the first publicly available \ufo\ model to implement such a  tool.
Sections~\ref{sec.Mathematica_use},~\ref{sec.MG_use} provide recommendations for the use of \smeftsim\ in \mathematica\ and in \madgraph\ respectively, and in  Section~\ref{sec.conclusion} we conclude.

Additional useful material is provided in the Appendices: analytic expressions of the decay widths implemented in the propagator corrections (App.~\ref{app.dwidthExpressions}), a list of changes made in version~3.0 (App.~\ref{app.new}), tables to facilitate the conversion between flavor assumptions (App.~\ref{app.flavor_comparison}), between theory and code notation (App.~\ref{app.parameters}) and between \smeftsim\ and \dimsixtop\ or \smeftatnlo\ (App.~\ref{app.UFO_comparison}). Finally, App.~\ref{app.validation} documents the validation of the \ufo\ models, that followed the procedure recommended in~\cite{Durieux:2019lnv}.

\subsection{Basics and notation}\label{sec.notation}
We consider the SMEFT Lagrangian truncated at the dimension-6 level:
\begin{equation}
\Lag_{\rm SMEFT} = \Lag_{\rm SM} + \Lag_6\,.
\end{equation} 
We neglect all lepton- and baryon-number violating terms, which includes the dimension-5 Weinberg operator that generates a Majorana mass term for neutrinos.
For future convenience, the SM Lagrangian is split into four terms:
\begin{align}
\Lag_{\rm SM} &= \Lag_{\rm gauge} + \Lag_{\rm fermions} + \Lag_{\rm Yukawa} + \Lag_{\rm Higgs}\,,
\end{align}
where
\begin{align}
\label{eq.LGauge}
\Lag_{\rm gauge} &= -\frac14 B_{\mu\nu} B^{\mu\nu} -\frac14 W^i_{\mu\nu} W^{i\mu\nu}-\frac14 G^a_{\mu\nu}G^{a\mu\nu}
\,,\\[1mm]
\label{eq.LFermions}
\Lag_{\rm fermions} &= \bar q i \slashed{D} q + \bar u i \slashed{D} u + \bar d i \slashed{D} d + \bar l i \slashed{D} l + \bar e i \slashed{D} e
\,,\\[1mm]
\label{eq.LYukawa}
\Lag_{\rm Yukawa} &= - \bar d Y_{d} H^\dag q - \bar u Y_{u} \tilde H^\dag q - \bar e Y_{l} H^\dag l +\hc
\,,\\[1mm]
\label{eq.LHiggs}
\Lag_{\rm Higgs} &= D_\mu H^\dag D^\mu H  + m^2 (H^\dag H) - \lambda (H^\dag H)^2\,.
\end{align}
$q,l$ represent the left-handed quark and lepton doublets respectively, and $u,d,e$ the right-handed quarks and leptons. $H$ is the Higgs doublet and  $\tilde H = i\s^2 H^*$, where $\s^i$, $i=\{1,2,3\}$ are the Pauli matrices.
$Y_d$, $Y_u$, $Y_l$ are the $3\times3$ Yukawa matrices of the down and up quarks and of the charged leptons. 
Covariant derivatives are defined with a plus sign, i.e.\footnote{The covariant derivative sign is handled automatically by \feynrules. The convention chosen here also implies that gauge field strenghts have the form $W_{\mu\nu}^i = \de_\mu W^i_\nu-\de_\nu W^i_\mu - g_W \varepsilon^{ijk} W^j_\mu W^k_\nu$, etc.}
\begin{equation}
D_\mu q = \left[\de_\mu + i g_s T^a G_\mu^a + i \frac{g_W}{2}\s^i W^i_\mu + i {\bf y}_q g_1 B_\mu\right] q\,.
\end{equation} 
$T^a \equiv \lambda^a/2$, $a=\{1,\dots,8\}$ are the $SU(3)_c$ generators, with $\lambda^a$ the Gell-Mann matrices. $\mathbf{y}_q=1/6$ denotes the hypercharge of the $q$ field and $g_s,g_W,g_1$ are the $SU(3)_c\times SU(2)_L\times U(1)_Y$ coupling constants.
As a general rule, color indices are denoted by $a,b,c,d$, $SU(2)_L$ indices by $i,j,k$ and flavor indices by $p,r,s,t$. Summation over identical indices is always understood, unless otherwise specified.

The Lagrangian $\Lag_6$ contains a complete and non-redundant basis of dimension-6 operators $Q_\a$ constructed with the SM fields and invariant under the $SU(3)_c\times U(2)_L\times U(1)_Y$ gauge symmetry. \smeftsim\ implements the Warsaw basis~\cite{Grzadkowski:2010es}, whose operators are collected in 8 groups, following the classification of Ref.~\cite{Alonso:2013hga}. Class 8 is further split into 4 subgroups:\footnote{Note that $\Lag_6^{(7)}$ implicitly contains $(Q_{Hud}+\hc)$, as this operator is not Hermitian.}
\begin{align}\label{eq.L6_splitting}
\Lag_6 &= \Lag_6^{(1)} + \Lag_6^{(2)} + \Lag_6^{(3)} + \Lag_6^{(4)} 
+ \Lag_6^{(7)} + \Lag_6^{(8)}
+ \left[\Lag_6^{(5)} + \Lag_6^{(6)} +\hc\right]\,,
\\
\Lag_6^{(8)} &= \Lag_6^{(8a)} +\Lag_6^{(8b)} + \Lag_6^{(8c)} + \left[\Lag_6^{(8d)}+\hc\right]\,.
\label{eq.L8_splitting}
\end{align}
Each sub-Lagrangian has the form
\begin{equation}\label{eq.L6_classes}
\Lag_6^{(n)} = \frac{1}{\Lambda^2}\sum_\a C_\a Q_\a\,,
\end{equation}
with the sum running over the class-$n$ operators $\{Q_\a\}$ defined in Table~\ref{tab.Warsaw_basis} and $C_\a$ denoting the associated Wilson coefficients. 
Both $Q_\a$ and $C_\a$ generally carry flavor indices, that are implicitly contracted in Eq.~\eqref{eq.L6_classes}. 
In this basis, explicit CP violation is carried by the real coefficients $C_{\widetilde G},\, C_{\widetilde W}, \, C_{H\widetilde G},\, C_{H\widetilde W}, \, C_{H\widetilde B},$ \, $C_{H\widetilde WB}$ and by the imaginary parts of the Wilson coefficients associated to non-Hermitian fermionic operators, namely those in $\Lag_6^{(5),(6),(8d)}$ and $\O_{Hud}$. Baryon-number violating operators are omitted.

The operators definitions use the following notation:
\begin{align}
\widetilde X^{\mu\nu} &= \frac12 \e^{\mu\nu\rho\s}X_{\rho\s}\,,
&
H^\dag i\overleftrightarrow{D}_\mu H  &= H^\dag (iD_\mu H) - (i D_\mu H^\dag)H\,,
\\
\s^{\mu\nu} &= \frac{i}{2}[\g^\mu,\g^\nu]\,,
&
H^\dag i\overleftrightarrow{D}^i_\mu H  &= H^\dag \s^i(i D_\mu H) - (i D_\mu H^\dag)\s^i H\,.
\end{align}

\begin{table}[h!]
\begin{center}
\small
\hspace*{-1.5cm}
 \renewcommand{\arraystretch}{1.7}
 \begin{tabular}{|*3{>{$}c<{$}|>{$}p{4cm}<{$}|}}
 \hline
 \rowcolor{tablesColor}
 \multicolumn{2}{|c|}{$\Lag_6^{(1)}$ -- $X^3$} & 
  \multicolumn{2}{c|}{$\Lag_6^{(6)}$ -- $\psi^2 XH$}&
 \multicolumn{2}{c|}{$\Lag_6^{(8b)}$ -- $(\bar RR)(\bar RR)$}
 \\
\hline
Q_G & 
f^{abc} G_\mu^{a\nu} G_\nu^{b\rho} G_\rho^{c\mu}  &
Q_{eW} & 
(\bar l_p \sigma^{\mu\nu} e_r) \s^i H W_{\mu\nu}^i &
Q_{ee} & 
(\bar e_p \gamma_\mu e_r)(\bar e_s \gamma^\mu e_t) 
\\
Q_{\widetilde G} & 
f^{abc} \widetilde G_\mu^{a\nu} G_\nu^{b\rho} G_\rho^{c\mu}  &
Q_{eB} & 
(\bar l_p \sigma^{\mu\nu} e_r) H B_{\mu\nu} &
Q_{uu} & 
(\bar u_p \gamma_\mu u_r)(\bar u_s \gamma^\mu u_t) 
\\
Q_W & 
\e^{ijk} W_\mu^{i\nu} W_\nu^{j\rho} W_\rho^{k\mu} &
Q_{uG} & 
(\bar q_p \sigma^{\mu\nu} T^a u_r) \widetilde H \, G_{\mu\nu}^a &
Q_{dd} & 
(\bar d_p \gamma_\mu d_r)(\bar d_s \gamma^\mu d_t) 
\\
Q_{\widetilde W}& 
\e^{ijk} \widetilde W_\mu^{i\nu} W_\nu^{j\rho} W_\rho^{k\mu} &
Q_{uW} & 
(\bar q_p \sigma^{\mu\nu} u_r) \s^i \widetilde H \, W_{\mu\nu}^i &
Q_{eu} & 
(\bar e_p \gamma_\mu e_r)(\bar u_s \gamma^\mu u_t) 
\\\cline{1-2}
 \multicolumn{2}{|c|}{\cellcolor{tablesColor}$\Lag_6^{(2)}$ -- $H^6$} &
Q_{uB} & 
(\bar q_p \sigma^{\mu\nu} u_r) \widetilde H \, B_{\mu\nu} &
Q_{ed} & 
(\bar e_p \gamma_\mu e_r)(\bar d_s\gamma^\mu d_t) 
\\\cline{1-2}
Q_H & 
(H^\dag H)^3 &
Q_{dG} & 
(\bar q_p \sigma^{\mu\nu} T^a d_r) H\, G_{\mu\nu}^a &
Q_{ud}^{(1)} & 
(\bar u_p \gamma_\mu u_r)(\bar d_s \gamma^\mu d_t) 
\\
\cline{1-2}
\multicolumn{2}{|c|}{\cellcolor{tablesColor} $\Lag_6^{(3)}$ -- $H^4 D^2$} &
Q_{dW} & 
(\bar q_p \sigma^{\mu\nu} d_r) \s^i H\, W_{\mu\nu}^i &
Q_{ud}^{(8)} & 
(\bar u_p \gamma_\mu T^a u_r)(\bar d_s \gamma^\mu T^a d_t) 
\\\cline{1-2}
Q_{H\Box} & 
(H^\dag H)\Box(H^\dag H) &
Q_{dB} & 
(\bar q_p \sigma^{\mu\nu} d_r) H\, B_{\mu\nu} &
&
\\
Q_{H D} & 
\ \left(D^\mu H^\dag  H\right) \left(H^\dag D_\mu H\right) &
&&
&
\\\hline
\rowcolor{tablesColor}
 \multicolumn{2}{|c|}{$\Lag_6^{(4)}$ -- $X^2H^2$}& 
 \multicolumn{2}{c|}{$\Lag_6^{(7)}$ -- $\psi^2H^2 D$}& 
 \multicolumn{2}{c|}{$\Lag_6^{(8c)}$ -- $(\bar LL)(\bar RR)$}
\\\hline
Q_{H G}  & 
H^\dag H\, G^a_{\mu\nu} G^{a\mu\nu} &
Q_{H l}^{(1)} & 
(H^\dag i\overleftrightarrow{D}_\mu H)(\bar l_p \gamma^\mu l_r)&
Q_{le} & 
(\bar l_p \gamma_\mu l_r)(\bar e_s \gamma^\mu e_t)
\\
Q_{H\widetilde G} & 
H^\dag H\, \widetilde G^a_{\mu\nu} G^{a\mu\nu} &
Q_{H l}^{(3)} & 
(H^\dag i\overleftrightarrow{D}^i_\mu H)(\bar l_p \s^i \gamma^\mu l_r)&
Q_{lu} & 
(\bar l_p \gamma_\mu l_r)(\bar u_s \gamma^\mu u_t) 
\\
Q_{H W} & 
H^\dag H\, W^i_{\mu\nu} W^{I\mu\nu} &
Q_{H e} & 
(H^\dag i\overleftrightarrow{D}_\mu H)(\bar e_p \gamma^\mu e_r)&
Q_{ld} & 
(\bar l_p \gamma_\mu l_r)(\bar d_s \gamma^\mu d_t) 
\\
Q_{H\widetilde W} & 
H^\dag H\, \widetilde W^i_{\mu\nu} W^{i\mu\nu} &
Q_{H q}^{(1)} & 
(H^\dag i\overleftrightarrow{D}_\mu H)(\bar q_p \gamma^\mu q_r)&
Q_{qe} & 
(\bar q_p \gamma_\mu q_r)(\bar e_s \gamma^\mu e_t)
\\
Q_{H B} &  H^\dag H\, B_{\mu\nu} B^{\mu\nu} &
Q_{H q}^{(3)} & 
(H^\dag i\overleftrightarrow{D}^i_\mu H)(\bar q_p \s^i \gamma^\mu q_r)&
Q_{qu}^{(1)} & 
(\bar q_p \gamma_\mu q_r)(\bar u_s \gamma^\mu u_t)
\\
Q_{H\widetilde B} & 
H^\dag H\, \widetilde B_{\mu\nu} B^{\mu\nu} &
Q_{H u} & 
(H^\dag i\overleftrightarrow{D}_\mu H)(\bar u_p \gamma^\mu u_r)&
Q_{qu}^{(8)} & 
(\bar q_p \gamma_\mu T^a q_r)(\bar u_s \gamma^\mu T^a u_t) 
\\
Q_{H WB} & 
 H^\dag \s^i H\, W^i_{\mu\nu} B^{\mu\nu} &
Q_{H d} & 
(H^\dag i\overleftrightarrow{D}_\mu H)(\bar d_p \gamma^\mu d_r)&
Q_{qd}^{(1)} & 
(\bar q_p \gamma_\mu q_r)(\bar d_s \gamma^\mu d_t) 
\\
Q_{H\widetilde W B} & 
H^\dag \s^i H\, \widetilde W^i_{\mu\nu} B^{\mu\nu} &
Q_{H u d}+\hc & 
i(\widetilde H ^\dag D_\mu H)(\bar u_p \gamma^\mu d_r)&
Q_{qd}^{(8)} & 
(\bar q_p \gamma_\mu T^a q_r)(\bar d_s \gamma^\mu T^a d_t)
\\
\hline
\rowcolor{tablesColor}
 \multicolumn{2}{|c|}{$\Lag_6^{(5)}$ -- $\psi^2 H^3$} &
 \multicolumn{2}{c|}{$\Lag_6^{(8a)}$ -- $(\bar LL)(\bar LL)$}&
 \multicolumn{2}{c|}{$\Lag_6^{(8d)}$ -- $(\bar LR)(\bar RL)$, $(\bar LR)(\bar LR)$} 
 \\\hline
Q_{eH} & 
(H^\dag H)(\bar l_p e_r H) &
Q_{ll}  & 
(\bar l_p \gamma_\mu l_r)(\bar l_s \gamma^\mu l_t)&
Q_{ledq} & 
(\bar l_p^j e_r)(\bar d_s q_{tj})
\\
Q_{uH}  &
(H^\dag H)(\bar q_p u_r \widetilde H )&
Q_{qq}^{(1)} & 
(\bar q_p \gamma_\mu q_r)(\bar q_s \gamma^\mu q_t) &
Q_{quqd}^{(1)} & 
(\bar q_p^j u_r) \e_{jk} (\bar q_s^k d_t) 
\\
Q_{dH}  & 
(H^\dag H)(\bar q_p d_r H)&
Q_{qq}^{(3)} & 
(\bar q_p \gamma_\mu \s^i q_r)(\bar q_s \gamma^\mu \s^i q_t) &
Q_{quqd}^{(8)} & 
(\bar q_p^j T^a u_r) \e_{jk} (\bar q_s^k T^a d_t) 
\\
&&
Q_{lq}^{(1)} & 
(\bar l_p \gamma_\mu l_r)(\bar q_s \gamma^\mu q_t)&
Q_{lequ}^{(1)} & 
(\bar l_p^j e_r) \e_{jk} (\bar q_s^k u_t) 
\\
&&
Q_{lq}^{(3)} & 
(\bar l_p \gamma_\mu \s^i l_r)(\bar q_s \gamma^\mu \s^i q_t) &
Q_{lequ}^{(3)} & 
(\bar l_p^j \sigma_{\mu\nu} e_r) \e_{jk} (\bar q_s^k \sigma^{\mu\nu} u_t) 
\\\hline
\end{tabular}
\end{center}
\caption{\label{tab.Warsaw_basis}
$\Lag_6$ operators in the Warsaw basis~\cite{Grzadkowski:2010es}, categorized into eight classes $\Lag_6^{(n)}$ as in~\cite{Alonso:2013hga}. Only baryon number-conserving invariants are retained. The flavor indices $p,r,s,t$ are suppressed in the operators' names.}
\end{table}

\clearpage


\section{EWSB, field and parameter redefinitions}\label{sec.redefinitions}
This section reviews the Lagrangian manipulations that are required in order to compute physical processes in the SMEFT truncated at $\O(\Lambda^{-2})$.\footnote{ Within a Monte Carlo event generation, \smeftsim\ generally enables the computation of higher order corrections to a given observable, such as $\O(\Lambda^{-4})$ corrections stemming from the square of $\O(\Lambda^{-2})$ amplitudes (see Sec.~\ref{sec.MG_use}).  However, consistent results are only provided to $\O(\Lambda^{-2})$, as the SMEFT Lagrangian implemented is truncated at this order.
}The procedure described here largely overlaps with what reported eg. in~\cite{Grinstein:1991cd,Corbett:2012ja,Alonso:2013hga,Ghezzi:2015vva,Berthier:2015oma,Gauld:2015lmb,Passarino:2016pzb,Dedes:2017zog,Brivio:2017btx,Dawson:2018liq,Dawson:2018pyl,Brivio:2019myy,Cullen:2019nnr,Denner:2019vbn}.
\subsection{Higgs sector}
The operator $Q_H$ introduces a perturbation of the Higgs potential:
\begin{equation}
V(H) = -\frac{m^2}{2} H^\dag H +\lambda (H^\dag H)^2 - \frac{C_H}{\Lambda^2}(H^\dag H)^3   \,.
\end{equation} 
The true minimum of the potential, that triggers the electroweak symmetry breaking, is 
\begin{equation}\label{eq.vT}
\begin{aligned}
\langle H^\dag H \rangle  = \frac{v_T^2}{2} &\equiv \frac{v^2}{2}\left[1+\frac{3}{4\lambda}\frac{v^2}{\Lambda^2}C_H\right] + \O(\Lambda^{-4}) 
\\
&= \frac{v^2}{2}\left[1+\frac{3}{4\lambda}\C_H\right] + \O(\Lambda^{-4})\,,
\end{aligned}
\end{equation} 
with 
\begin{equation}
 v^2 = \frac{m^2}{2\lambda}\,.
\end{equation}
We have introduced the ``bar'' notation for Wilson coefficients:
\begin{equation}\label{eq.Cbar}
\bar C_\a \equiv \frac{v_T^2}{\Lambda^2}\, C_\a\,.
\end{equation}
Note that because $v= v_T + \O(\Lambda^{-2})$ and $\O(\Lambda^{-4})$ contributions are entirely neglected, the two quantities $v,v_T$ are interchangeable whenever they multiply a Wilson coefficent.\footnote{In fact, it would be more appropriate to define the $\C_\a$ notation with the parameter $\hv$ defined in Sec.~\ref{sec.inputs_ew}, rather than with $v_T$. However, as long as $\O(\Lambda^{-4})$ terms are neglected, all three $v$'s are formally identical when multiplying a Wilson coefficient.  }

The Higgs field $H$ is expanded around its vacuum expectation value (vev) as
\begin{equation}
H =\bmat -iG^+ \\ \dfrac{v_T+h+i G^0}{\sqrt2}\emat\,,
\end{equation}
with $G^+,G^0$ the charged and neutral Goldstone bosons and $h$ the physical Higgs boson.
In the broken phase, the kinetic terms of the scalar fields receive corrections from the operators $Q_{H\square},\,Q_{HD}$. As the scope of \smeftsim\ is limited to tree-level calculations, we choose to work in unitary gauge and neglect EFT effects in the Goldstone sector, both in the present discussion and in the code implementations. The Goldstone bosons case and the generalization of the gauge fixing procedure in the SMEFT were addressed in Refs.~\cite{Dedes:2017zog,Helset:2018fgq,Misiak:2018gvl,Passarino:2019yjx,Cullen:2019nnr,Denner:2019vbn,Helset:2020yio}.

Using integration by parts, the kinetic term of the physical Higgs boson takes the form
\begin{equation}\label{eq.dkH}
\Lag_{\rm Higgs} + \Lag_6 = \frac{1}{2}\de_\mu h \de^\mu h\left[
1-2\Delta\kappa_H\right]+\dots
\quad\text{ with }\quad
\Delta\kappa_H = \C_{H\square}-\frac{\C_{HD}}{4}\,,
\end{equation} 
and it is brought to the canonical normalization via the field redefinition
\begin{equation}\label{eq.h_redef}
h\to \left[1+\Delta\kappa_H\right]\, h\,.
\end{equation} 
This replacement is formally operated on the entire $\Lag_{\rm SMEFT}$. However, when applied to $\Lag_6$, its net effect is of $\O(\Lambda^{-4})$.  As we work at $\O(\Lambda^{-2})$, the replacement only needs to be performed on $\Lag_{\rm SM}$. This holds for all field and parameter redefinitions introduced in the following, unless otherwise specified. For the same reason, all quantities in a Wilson coefficient's prefactor are understood to be defined in the SM limit. 

The main consequence of~\eqref{eq.h_redef} is that the Wilson coefficients $C_{H\square},\,C_{HD}$ are recast into an overall rescaling of all SM Higgs couplings.
The resulting Higgs potential is
\begin{equation}\label{eq.VH_redef}
\begin{aligned}
V(H) &=
h^2\,\lambda v_T^2\left[1 +2\Delta\kappa_H- \frac{3}{2\lambda}\, \C_H\right]
+h^3\,\lambda v_T \left[1 +3\Delta\kappa_H - \frac{5}{2\lambda}\, \C_H\right]\\
&
+h^4\,\frac{\lambda}{4}\left[1 +4\Delta\kappa_H- \frac{15}{2\lambda}\, \C_H\right]
-\frac{3}{4}\frac{h^5}{v_T}\, \C_H 
-\frac{1}{8}\frac{h^6}{v_T^2}\,\C_H\,.
\end{aligned}
\end{equation} 
In the \feynrules\ implementation, the redefinitions of the physical Higgs field, Eq.~\eqref{eq.h_redef}, and of the vev, Eq.~\eqref{eq.vT}, are embedded in the definition of the Higgs doublet.

\subsection{Gauge sector}
Upon EWSB, the operators $Q_{HG},\, Q_{HW},\, Q_{HB},\, Q_{HWB}$ induce corrections to the kinetic terms of the gauge bosons. The first three lead to overall rescalings:
\begin{equation}
\Lag_{\rm gauge} + \Lag_6 = 
-\frac{1}{4}B_{\mu\nu} B^{\mu\nu} \left[1-2\C_{HB}\right] 
-\frac{1}{4}W^i_{\mu\nu} W^{i\mu\nu} \left[1-2\C_{HW}\right]
-\frac{1}{4}G^a_{\mu\nu} G^{a\mu\nu} \left[1-2\C_{HG}\right]
+ \dots
\end{equation} 
where the dots stand for all other interaction terms induced by $\Lag_6$.
The canonical normalization is easily restored at $\O(\Lambda^{-2})$, via the field redefinitions
\begin{align}\label{eq.BGW_redef}
G_\mu^a &\to G_\mu^a ( 1+ \C_{HG})\,,
&
W_\mu^i &\to W_\mu^i ( 1+ \C_{HW})\,,
&
B_\mu &\to B_\mu ( 1+ \C_{HB})\,.
\end{align}
In order to leave the covariant derivatives unchanged, the coupling constants need to be redefined at the same time. Neglecting $\O(\Lambda^{-4})$ corrections:
\begin{align}\label{eq.g_redef}
g_s &\to g_s  ( 1- \C_{HG})\,,
&
g_W &\to g_W  ( 1- \C_{HW})\,,
&
g_1 &\to g_1  ( 1- \C_{HB})\,.
\end{align}
The operator $Q_{HWB}$ introduces a kinetic mixing between the $B$ and $W^3$ fields of the form
\begin{equation}
-\frac{C_{HWB}}{2}\, \frac{v_T^2}{\Lambda^2}\, W_{\mu\nu}^3 B^{\mu\nu}\,.
\end{equation} 
The rotation~\cite{Alonso:2013hga}
\begin{equation}\label{eq.HWB_rotation}
\begin{pmatrix}W_\mu^3 \\ B_\mu\end{pmatrix} 
\to
\begin{pmatrix}1& -\bar C_{HWB} / 2 \\  - \bar C_{HWB} / 2 & 1\end{pmatrix}
\begin{pmatrix}W_\mu^3 \\ B_\mu\end{pmatrix}\,,
\end{equation}
removes this residual mixing and leads to fully canonical and diagonal kinetic terms.
Once Eqs.~\eqref{eq.BGW_redef},~\eqref{eq.g_redef},~\eqref{eq.HWB_rotation} have been applied,  the electric-charge eigenstates $W^\pm$ are obtained via the usual rotation
\begin{equation}
\begin{pmatrix}W_\mu^1 \\ W_\mu^2\end{pmatrix} 
=
\frac{1}{\sqrt2}
\begin{pmatrix}1& 1\\ i& -i\end{pmatrix}
\begin{pmatrix}W_\mu^+ \\ W_\mu^-\end{pmatrix}\,,
\end{equation} 
while the mass matrix of the neutral bosons is diagonalized by
\begin{align}\label{eq.BW_rotation}
\begin{pmatrix}W_\mu^3 \\ B_\mu\end{pmatrix} 
&= 
\frac{1}{\sqrt{g_1^2+g_W^2}}\begin{pmatrix}g_W& g_1\\ -g_1& g_W\end{pmatrix}
\begin{pmatrix}
    1& \dfrac{1}{2}\dfrac{g_W^2-g_1^2}{g_W^2+g_1^2}\bar C_{HWB}
    \\
    -\dfrac{1}{2}\dfrac{g_W^2-g_1^2}{g_W^2+g_1^2}\bar C_{HWB} & 1
    \end{pmatrix}
\begin{pmatrix}Z_\mu \\ A_\mu\end{pmatrix}\,,
\end{align}
The rightmost rotation is unitary up to $\O(\Lambda^{-4})$ corrections, and therefore does not reintroduce kinetic mixing at $d=6$. Equivalently,
\begin{equation}
\begin{pmatrix}W_\mu^3 \\ B_\mu\end{pmatrix} 
=
\begin{pmatrix}\cos\theta& \sin\theta\\ -\sin\theta& \cos\theta\end{pmatrix}
\begin{pmatrix}Z_\mu \\ A_\mu\end{pmatrix}\,,
\end{equation}
with a shifted weak mixing angle $\theta$ defined as
\begin{equation}\label{eq.tan_theta}
\tan\theta = \frac{g_1}{g_W} + \frac12\C_{HWB} \left(1-\frac{g_1^2}{g_W^2}\right)\,.
\end{equation} 
After all the coupling and field redefinitions have been applied, a generic covariant derivative has the form
\begin{align}
\label{eq.CD_redef}
D_\mu =&\, \partial_\mu + i Q \frac{g_1 g_W}{\sqrt{g_1^2+g_W^2}} A_\mu \left[1-\bar C_{HWB}\dfrac{g_1 g_W}{g_W^2+g_1^2}\right]
\\
&+ i\sqrt{g_1^2+g_W^2}\,Z_\mu\;\left[T_3-\frac{g_1^2}{g_1^2+g_W^2}\, Q
+\C_{HWB} \frac{g_1 g_W}{g_1^2+g_W^2} \left(T_3-\frac{g_W^2}{g_1^2+g_W^2}\, Q\right)
\right]
\nn\\
&+\dots
\nn\\
=&\, \partial_\mu + i Q \,g_W\, s_{\theta} \,A_\mu\left[1-\frac{1}{2}\frac{c_\theta}{s_\theta}\C_{HWB}\right]
+ i \frac{g_W}{c_\theta}\,Z_\mu \, \left(T_3- Qs^2_\theta\right)\left[1+\frac{1}{2}\frac{s_\theta}{c_\theta}\C_{HWB}\right]
\label{eq.CD_redef_theta}\\
&+\dots\nn
\end{align}
where $T_3$ denotes the eigenvalue of the 3$^{\rm rd}$ $SU(2)_L$ generator ($T_3=\pm 1/2$ for left-handed fields and $T_3=0$ for right-handed ones) and $Q=T_3+\mathbf{y}$ is the electric charge. We have also introduced the shorthand notation $s_\theta = \sin\theta,\,c_\theta=\cos\theta$, with $\theta$ defined as in Eq.~\eqref{eq.tan_theta}. 
The dots stand for potential gluon and $W^\pm$ terms, for which there are no residual $\Lag_6$ corrections.

Eq.~\eqref{eq.CD_redef} shows that the contributions from $Q_{HW},\,Q_{HB},\,Q_{HG}$ are fully reabsorbed in the definition of the fields and gauge couplings. As a consequence, these operators have no physical impact in the pure gauge sector, and they only contribute to Higgs-gauge interactions~\cite{Brivio:2017bnu}. 
On the other hand, the operator $Q_{HWB}$ introduces net modifications of all $\gamma$ and $Z$ couplings. In the former case the correction is a universal rescaling of the electromagnetic constant, while in the latter case the corrections depend on the field's charges. 
In particular, in the Higgs case ($T_3=-1/2,Q=0$) this implies a correction $\propto\C_{HWB}$ to the $Z$ mass term.
The physical interpretation of these contributions requires defining a set of input observables and is deferred to Section~\ref{sec.inputs}.

\vskip 1em

In \smeftsim, the redefinitions described in this subsection are applied simultaneously at the Lagrangian level in the \feynrules\ model. The coupling constants' rescaling in Eq.~\eqref{eq.g_redef} is implemented in the replacement list {\tt redefConst}. The field redefinitions are operated in the mass and charge eigenstate basis: the replacement list {\tt rotateGaugeB} implements the net mismatch between the series of rotations~\eqref{eq.BGW_redef},~\eqref{eq.HWB_rotation},~\eqref{eq.BW_rotation} and the usual SM rotations, i.e.
\begin{align}
G_\mu^a\ &\to (1+\C_{HG}) \;G_\mu^a\,,
\\
\binom{W^+_\mu}{W^-_\mu}&\to
(1+\C_{HW})
\binom{W^+_\mu}{W^-_\mu}\,,
\\
\binom{Z_\mu}{A_\mu}&\to 
R_{ZA}
\binom{Z_\mu}{A_\mu} \,,
\end{align}
with
\begin{equation}\label{eq.rot_ZA}
\begin{aligned}
R_{ZA}
&=\begin{pmatrix}
1+c_\theta^2 \,\C_{HW} + s_\theta^2\, \C_{HB} + \dfrac{s_{2\theta}}{2}\C_{HWB}
&
\dfrac{s_{2\theta}}{2}\left(\C_{HW}-\C_{HB}\right)-\dfrac{c_{2\theta}}{2}\C_{HWB}+\dfrac{\Delta s^2_\theta}{s_{2\theta}}
\\[4mm]
\dfrac{s_{2\theta}}{2}\left(\C_{HW}-\C_{HB}\right)-\dfrac{c_{2\theta}}{2}\C_{HWB}-\dfrac{\Delta s^2_\theta}{s_{2\theta}}
&
1+c_\theta^2\,\C_{HB} + s_\theta^2\, \C_{HW} - \dfrac{s_{2\theta}}{2}\C_{HWB}
\end{pmatrix}\,,
\end{aligned}
\end{equation} 
where  
\begin{align}\label{eq.ds2_redef}
\Delta s^2_\theta &=
\C_{HWB}\; g_1\, g_W\;\frac{g_W^2-g_1^2}{g_W^2+g_1^2}
=
\frac{s_{4\theta}}{4}\C_{HWB}  + \mathcal{O}(\Lambda^4)
\,,
\end{align} 
is the correction to the mixing angle stemming from Eq.~\eqref{eq.tan_theta}.


\section{Flavor assumptions}\label{sec.flavor}
The SMEFT Lagrangian defined in Sec.~\ref{sec.SMEFT} is not invariant under flavor rotations of the fermion fields, so the latter should always be defined in order to avoid ambiguities. 
In \smeftsim, the fields $q,l,u,d,e$ are defined in the mass basis of the charged leptons and of the up-type quarks, in which the Yukawa matrices in Eq.~\eqref{eq.LYukawa} take the form 
\begin{equation}\label{eq.Y_basis}
Y_d \equiv Y_d^{(d)} V^\dag\,,\qquad 
Y_u \equiv Y_u^{(d)}\,,\qquad
Y_l \equiv Y_l^{(d)}\,.
\end{equation}
The superscript $(d)$ denotes diagonal matrices and $V$ is the CKM matrix. This basis choice is consistently employed in the definition of both $\Lag_{\rm SM}$ and $\Lag_6$, and for all the flavor assumptions implemented in \smeftsim. The only special case are the \top\ and \topsl\ models, where quark mixing is entirely neglected by setting $V\equiv\mathbbm{1}$.

Upon EWSB, the Lagrangian can be written in terms of the fermionic mass eigenstates.
By definition the relation between the $SU(2)_L$ and mass bases is trivial for all fermion fields, except the left-handed quark doublet\footnote{For economy of notation, we use the same letters $u,d,e$ for the right-handed fields and for the mass eigenstates, both of them carrying flavor indices. To avoid ambiguities, the latter always carry $L,R$ subscripts, while the former don't. 
 }:
\begin{equation}\label{eq.Ffields_to_mass}
q_p = \binom{u_{L,p}}{V_{pr} d_{L,r}}\,,
\qquad 
l_p = \binom{\nu_{L,p}}{e_{L,p}}\,,
\qquad
u_p = u_{R,p}\,,
\qquad
d_p = d_{R,p}\,,
\qquad
e_p=e_{R,p}\,.
\end{equation}
In unitary gauge, the relevant terms in the SM Lagrangian are therefore
\begin{align}\label{eq.LW_ugauge}
\Lag_{\rm Fermions} &=  
- \frac{g_W}{\sqrt2} W^+_\mu \bar u_L \g^\mu \,V d_L   
- \frac{g_W}{\sqrt2} W^+_\mu \bar \nu_L \g^\mu \, e_L   
+\dots
\\
\Lag_{\rm Yukawa} &=  -\frac{v_T+h}{\sqrt{2}}\left[
 \bar d_{R} \, Y_{d}^{(d)} \, d_{L} 
+\bar u_{R} \, Y_{u}^{(d)} \,u_{L} 
+\bar e_{R} \, Y_{l}^{(d)} \,e_{L}\right]+\hc\,.
\end{align}
The CKM matrix is implemented in the Wolfenstein parameterization~\cite{Wolfenstein:1983yz}:
\begin{equation}
V = \bmat 
1 - \lambda_{CKM}^2/2 & \lambda_{CKM} & A \,\lambda_{CKM}^3(\rho- i \eta)
\\[2mm]
- \lambda_{CKM} & 1-\lambda^2_{CKM}/2 & A\, \lambda_{CKM}^2
\\[2mm]
A\, \lambda_{CKM}^3 ( 1-\rho-i\eta) & - A\, \lambda_{CKM}^2 & 1
\emat\,.
\end{equation}
The numerical values employed for the parameters are listed in Table~\ref{tab.defs_common}.

\vskip 1em
\smeftsim\ implements five alternative flavor scenarios: one with fully arbitrary indices, and four based on the implementation of different global symmetries. Three of these scenarios have been present since the first release, and two have been newly introduced in version~3.0.
The following sub-sections review in detail the properties of the $\Lag_6$ operators within each setup and provide the corresponding parameter counting. A dictionary between the different flavor assumptions is provided in Appendix~\ref{app.flavor_comparison}.

\subsection{\general: general flavor structure}\label{sec.flavor_general}
Without further assumptions on the flavor structure of the SMEFT, $\Lag_6$ contains the operators in Table~\ref{tab.Warsaw_basis}, summed over all possible flavor combinations:
\begin{align}
\label{eq.L2f_general}
\Lag_6^{(5,6)} &=  \frac{1}{\Lambda^2}\sum_\a\sum_{p,r=1}^{3} C_{\a,pr} Q_{\a,pr} +\hc\,,
&
\Lag_6^{(7)} &=  \frac{1}{\Lambda^2}\sum_\a\sum_{p,r=1}^{3} C_{\a,pr} Q_{\a,pr}\,,
\\
\label{eq.L4f_general}
\Lag_6^{(8a,8b,8c)} &=
\frac{1}{\Lambda^2}
\sum_\a
\sum_{p,r,s,t=1}^{3} C_{\a,prst} Q_{\a,prst}\,,
&
\Lag_6^{(8d)} &=
\frac{1}{\Lambda^2}
\sum_\a
\sum_{p,r,s,t=1}^{3} C_{\a,prst} Q_{\a,prst}+\hc\,.
\end{align}
Not all flavor combinations included in this way are independent, due to intrinsic symmetry properties of the effective operators. 
\smeftsim\ does not remove redundant terms from the sums in Eqs.~\eqref{eq.L2f_general},\eqref{eq.L4f_general}. Instead, the symmetry relations are enforced in the definition of the tensor Wilson coefficients $C_{\a,pr(st)}$: only a minimum number of independent parameters is defined for each operator, as reported in Appendix~\ref{app.parameters}, Tables~\ref{tab.defs_general},~\ref{tab.defs_general_indices}, and all the entries of $C_{\a,pr(st)}$ are functions of these parameters, consistent with the relations described below.

\paragraph{Classes 5 and 6.}
The operators in $\Lag_6^{(5,6)}$ are not Hermitian. Therefore each Wilson coefficient has 9 independent complex entries.
In total, this gives 198 real parameters (counting independently real and imaginary parts).

\paragraph{Class 7.}
All operators in $\Lag_6^{(7)}$, except $Q_{Hud}$, are Hermitian. In this case, the diagonal entries of the Wilson coefficients are real, and the off-diagonal ones are related by
\begin{equation}
C_{pr} = C_{rp}^*\,.
\end{equation}
In total, this class depends on 81 real parameters.

\paragraph{Class 8 a.}
All operators in $\Lag_6^{(8a)}$ are Hermitian. Moreover, each of the two currents that compose them is itself Hermitian. Therefore the following relation holds:
\begin{equation}\label{eq.4f_Hermitian}
C_{prst} = C_{rpts} = C_{rpst}^* = C_{prts}^*\,.
\end{equation}
In the operators $Q_{ll},Q_{qq}^{(1)},Q_{qq}^{(3)}$, the two currents contain the same fields, which leads to an additional exchange symmetry
\begin{equation}\label{eq.4f_samecurrents}
C_{prst} = C_{stpr}\,.
\end{equation}
Each of these three operators has then 15 real entries ($C_{pppp}, C_{pprr}, C_{prrp}$) and only 9 are independent, and 66 complex entries, 18 independent.
Operators $Q_{lq}^{(1)}$,$Q_{lq}^{(3)}$ have each 9 real entries, all independent, and 72 complex ones, only 36 independent.

In total, this class depends on 297 real parameters.

\paragraph{Class 8 b.}
All operators in $\Lag_6^{(8b)}$ are Hermitian and composed of two Hermitian currents, so relation~\eqref{eq.4f_Hermitian} holds for all Wilson coefficients in this class. Eq.~\eqref{eq.4f_samecurrents} is valid in addition for $C_{uu},C_{dd},C_{ee}$.

The operator $Q_{ee}$ is peculiar: because the $e$ field is a singlet under both $SU(2)_L$ and $SU(3)_c$, this term is invariant under Fierz rearranging.  This leads to the additional constraint
\begin{equation}
C_{prst} = C_{ptsr}\,.
\end{equation}
The coefficient $C_{ee}$ has then 15 real entries, 6 independent, and 66 complex entries, only 15 independent. 
The counting for the other operators is the same as for the invariants in class 8a, so $\Lag_6^{(8b)}$ has a total of 450 real parameters.

\paragraph{Class 8 c.}
All operators in $\Lag_6^{(8c)}$ are Hermitian and composed of two Hermitian currents, but no other symmetry is present.
Therefore only relation~\eqref{eq.4f_Hermitian} holds for all Wilson coefficients, leaving a total of 648 parameters.

\paragraph{Class 8 d.}
Finally, all operators in $\Lag_6^{8d}$ are non-Hermitian. No symmetry relation is present and this class has 810 independent real parameters.

\subsection{\Utf: maximal \texorpdfstring{$U(3)^5$}{U(3)\^{}5} symmetry}\label{sec.flavor_u35}
The number of independent parameters is considerably reduced if a flavor symmetry is assumed. The maximal symmetry available for the SM fermion fields is the symmetry of the kinetic terms~\cite{Chivukula:1987py}: $U(3)^5 = U(3)_q\times U(3)_u\times U(3)_d\times U(3)_l\times U(3)_e$.
Each field is assigned to a $\mathbf{3}$ representation of the associated group: denoting a generic $U(3)_\psi$ transformation by $\Omega_\psi$, the transformation rules are~\cite{Alonso:2013hga}
\begin{equation}\label{eq.u35_transform}
       q\mapsto \Omega_q\, q\,,
\qquad u\mapsto \Omega_u\, u\,, 
\qquad d\mapsto \Omega_d\, d\,,
\qquad l\mapsto \Omega_l\, l \,,
\qquad e\mapsto \Omega_e\, e\,.
\end{equation} 
Vector currents $\bar\psi_p\g^\mu\psi_r$ are trivially made invariant under $U(3)^5$ by imposing a $\d_{pr}$ contraction, that corresponds to the singlet composition of a $\mathbf{\bar{3}}$ and $\mathbf{3}$ representations. This is immediate to see applying the field transformations and using  $\Omega_\psi\Omega_\psi^\dag = \mathbbm{1} = \Omega_\psi^\dag \Omega_\psi$:
\begin{equation}
\bar\psi \g^\mu \psi \,\mapsto \,
\bar\psi\, \Omega_\psi^\dag \,\g^\mu\, \Omega_\psi\,\psi 
 \,=\, \bar\psi\,\g^\mu \psi, \qquad \psi=\{q,u,d,l,e\}\,.
\end{equation}
Chirality-flipping currents, with either scalar or tensor Lorentz structure, violate the flavor symmetry. To permit the introduction of fermion masses, it is customary to promote the Yukawa couplings to \emph{spurions} of the flavor symmetry, by assigning them transformation properties
\begin{equation}\label{eq.u35_Y_transform}
 Y_d\mapsto \Omega_d\, Y_d \,\Omega_q^\dag\,,\qquad 
 Y_u\mapsto \Omega_u\, Y_u \,\Omega_q^\dag\,,\qquad 
 Y_l\mapsto \Omega_e\, Y_l \,\Omega_l^\dag\,.
\end{equation} 
In this way the structures
\begin{equation}
\bar d \,Y_d\, q^i\,,\qquad \bar u \,Y_u\, q^i\,,\qquad \bar e \,Y_l\, l^i\,,
\end{equation}
are formally invariant. 
When the $U(3)^5$ symmetry is imposed, the flavor structure of each operator can be factored out of the Wilson coefficient, that becomes a scalar quantity:
\begin{align}
\Lag_6^{(5,6)} &= \frac{1}{\Lambda^2}\sum_\a\sum_{p,r=1}^3 C_\a\, X_{\a,pr}\, Q_{\a,pr} + \hc\,,
&
\Lag_6^{(7)} &= \frac{1}{\Lambda^2}\sum_\a\sum_{p,r=1}^3 C_\a\, X_{\a,pr}\, Q_{\a,pr}\,,
\\
\Lag_6^{(8a,8b,8c)} &= \frac{1}{\Lambda^2}\sum_\a\sum_{p,r,s,t=1}^3 C_\a\, X_{\a,prst}\, Q_{\a,prst}\,,
&
\Lag_6^{(8d)} &= \frac{1}{\Lambda^2}\sum_\a\sum_{p,r,s,t=1}^3  C_\a\, X_{\a,prst}\, Q_{\a,prst}+\hc\,.
\end{align}
In the construction of the $U(3)^5$ symmetric Lagrangian, we do not define  a power counting for insertions of the Yukawa couplings. Instead, we simply choose to retain the leading invariant structures for each operator, corresponding to no Yukawa insertions in $\Lag_6^{(7,8a,8b,8c)}$, one insertion in $\Lag_6^{(5,6)}$ and two insertions in $C_{Hud}$ and $\Lag_6^{(8d)}$.

\paragraph{Classes 5 and 6.}
All the operators in $\Lag_6^{(5,6)}$  require the insertion of a Yukawa coupling:
\begin{equation}
\begin{aligned}
X_{eH} = X_{eW} = X_{eB} &= Y_l^\dag = Y_l^{(d)}\,,
\\
X_{uH} = X_{uW} = X_{uB} = X_{uG} &= Y_u^\dag = Y_u^{(d)}\,,
\\
X_{dH} = X_{dW} = X_{dB} = X_{dG} &= Y_d^\dag = V Y_d^{(d)}\,,
\end{aligned}
\end{equation}
where the last equality in each line holds in the up-quarks mass basis, Eq.~\eqref{eq.Y_basis}.
Note that no net mixing among down-type quarks is induced in the mass basis, as the $V$ in the spurion cancels against $V^\dag$ in the $\bar q$ field, Eq.~\eqref{eq.Ffields_to_mass}. In fact, by construction all the operators in $\Lag_6^{(5,6)}$ have the same flavor structure as the SM Yukawas.

Because the operators are non-Hermitian, the associated $C_\a$ are complex. These classes therefore introduce 22 independent real parameters.

\paragraph{Class 7.}
All the currents appearing in the operators of class 7, except $Q_{Hud}$, are invariant under $U(3)$ with $X_{\a} = \mathbbm{1}$. This implies that the flavor structure of this class is exactly the same as in the SM kinetic terms. For instance, the charged quark current induced by the operator $Q_{Hq}^{(3)}$ is aligned with the SM one (Eq.~\eqref{eq.LW_ugauge}), that contains CKM mixing.\footnote{Note that this implies that, even though all Wilson coefficients are real, SM-sourced CP violation, due to the CKM phase in charged left-handed currents, is generally present in $\Lag_6^{(7)}$.}

In order to make $Q_{Hud}$ invariant, it is necessary to insert the spurion product 
\begin{equation}
X_{Hud} = Y_u Y_d^\dag = Y_u^{(d)} V \,Y_d^{(d)}\,.
\end{equation}
The number of independent real parameters is 9, as 7 out of 8 operators are Hermitian.

\paragraph{Class 8 a.}
Containing only vector currents, all the operators in $\Lag_6^{(8a)}$ are $U(3)^5$ invariant with the trivial flavor contraction $X_\a = \d_{pr}\d_{st}$.

The operators $Q_{ll}$, $Q_{qq}^{(1)}$, $Q_{qq}^{(3)}$ additionally admit the ``crossed'' contraction $X_\a^\prime = \d_{pt}\d_{sr}$. This is an independent structure that cannot be arbitrarily rearranged into $X_\a$: 
applying Fierz transformations in this case would introduce additional operators with $SU(2)_L$ triplet and $SU(3)_c$ octet contractions, see Sec.~\ref{sec.flavor_top}.
Therefore these operators are split into two invariants each, weighted by independent Wilson coefficients:
\begin{equation}\label{eq.cll_cll1}
\Lambda^2\Lag_6^{(8a)}=
\Big(C_{ll} \,\d_{pr}\d_{st} + C_{ll}^\prime \,\d_{pt}\d_{sr}\Big)Q_{ll,prst}
+\left(C_{qq}^{(1)} \,\d_{pr}\d_{st} + C_{qq}^{(1)\prime}\, \d_{pt}\d_{sr}\right)Q_{qq,prst}^{(1)}
+\dots
\end{equation}
where the dots stand for contributions from the other operators in $\Lag_6^{(8a)}$.\\
As all operators are Hermitian, $\Lag_6^{(8a)}$ contains 8 real parameters.

\paragraph{Class 8 b.}
All operators in $\Lag_6^{(8b)}$ are invariant with $X_\a = \d_{pr}\d_{st}$.

The operators $Q_{uu},Q_{dd}$ additionally admit independent crossed contractions $X_\a^\prime = \d_{pt}\d_{sr}$, and are treated analogously to $Q_{ll}, Q_{qq}^{(1),(3)}$. 
This is not the case for $Q_{ee}$ that, as mentioned above, is invariant under Fierz rearrangements: in this particular case the two flavor contractions are equivalent.
In total, there are 9 real parameters in $\Lag_6^{(8b)}$.

\paragraph{Class 8 c.}
All the operators in $\Lag_6^{(8c)}$ admit the invariant contraction is $X_\a = \d_{pr}\d_{st}$, leading to 8 independent real parameters.

\paragraph{Class 8 d.}
Finally, operators in $\Lag_6^{(8d)}$  require one Yukawa coupling insertion for each current. As they are not invariant under Fierz transformations, the operators $Q_{quqd}^{(1),(8)}$ admit two independent contractions, mapped to one another by interchanging the two $\bar q$ fields.\footnote{The two $X'_\a$ structures (4 real parameters) for $Q_{quqd}^{(1),(8)}$ were not included in previous versions of \smeftsim. I thank the authors of Ref.~\cite{Faroughy:2020ina} for pointing this out.}
\begin{equation}
\begin{aligned}
X_{ledq} &= (Y_l^\dag)_{pr} (Y_d)_{st} = (Y_l^{(d)})_{pr} (Y_d^{(d)} V^\dag)_{st}\,, 
\\
X_{quqd}^{(1)} = X_{quqd}^{(8)} &= (Y_u^\dag)_{pr} ( Y_d^\dag)_{st} = (Y_u^{(d)})_{pr}(VY_d^{(d)})_{st}\,, 
\\
X_{quqd}^{(1)\prime} = X_{quqd}^{(8)\prime} &= (Y_u^\dag)_{sr} ( Y_d^\dag)_{pt} = (Y_u^{(d)})_{sr}(VY_d^{(d)})_{pt}\,, 
\\
X_{lequ}^{(1)} = X_{lequ}^{(3)} &= (Y_l^\dag)_{pr}( Y_u^\dag)_{st} = (Y_l^{(d)})_{pr}( Y_u^{(d)})_{st}\,.
\end{aligned}
\end{equation}
Because the operators are non-Hermitian, there are 14 real parameters in $\Lag_6^{(8d)}$.

\subsection{\MFV: linear Minimal Flavor Violation}\label{sec.flavor_mfv}
The Minimal Flavor Violation ansatz~\cite{Chivukula:1987py,Hall:1990ac,D'Ambrosio:2002ex} assumes that the only sources of flavor and CP violation in $\Lag_{\rm SMEFT}$ are those already present in the SM, namely the Yukawa couplings and the CKM phase.

The requirement on CP violation implies that the Wilson coefficients of CP-odd bosonic operators scale with the Jarlskog invariant $J$~\cite{Jarlskog:1985ht,Jarlskog:1985cw}:
\begin{align}
&\{C_{\widetilde{G}},\, C_{\widetilde{W}},\, C_{H\widetilde{W}},\, C_{H\widetilde{B}},\, C_{H\widetilde{W}B},\, C_{H\widetilde{G}}\} \propto J\,,\\
&J = \im\left(V_{pr}V_{st}V^*_{pt}V^*_{sr}\right) 
\simeq \eta \,A^2\,\lambda_{CKM}^6 \left(1-\frac{\lambda_{CKM}^2}{2}\right) 
\simeq 3\times 10^{-5}\,.
\end{align}
As the $J$ suppression is stronger, for instance, than a loop factor, these coefficients can be safely neglected within the scope of \smeftsim. The corresponding operators are therefore not implemented in the \MFV\ version.
An analogous argument applies to sources of explicit CP violation in the fermion sector. In the Warsaw basis, these are the imaginary parts of the Wilson coefficients in $\Lag_6$, that are not defined either in the \smeftsim\ \MFV\ models.

The requirement on flavor violation is realized imposing a $U(3)^5$ symmetry on the fermion fields and allowing for arbitrary $U(3)^5$-invariant spurion insertions in the currents, that generate flavor violating effects. Such insertions are organized in an expansion in powers of the Yukawa couplings, that can be either resummed (obtaining a non-linear MFV formulation~\cite{Kagan:2009bn}) or treated as a truncated series.
\smeftsim\ adopts the latter option and retains contributions up to one power of $Y_l$ and up to 3 powers of $Y_u, Y_d$.
 
The relevant spurion structures at this order are 
\begin{align}
S^u  &= Y_u Y_u^\dag\,\sim(\mathbf{1},\mathbf{8},\mathbf{1}), 
&
S^u&\mapsto \Omega_u\, S^u\, \Omega_u^\dag\,,
\\
S^d &= Y_d Y_d^\dagger\,\sim(\mathbf{1},\mathbf{1},\mathbf{8}),
&
S^d &\mapsto \Omega_d\, S^d\, \Omega_d^\dag\,,
\\
S^{qu}  &= Y_u^\dag Y_u\,\sim(\mathbf{8},\mathbf{1},\mathbf{1}),
&
S^{qu}&\mapsto \Omega_q\, S^{qu}\, \Omega_q^\dag\,,
\\
S^{qd}  &= Y_d^\dag Y_d\,\sim(\mathbf{8},\mathbf{1},\mathbf{1}),
&
S^{qd}&\mapsto \Omega_q\, S^{qd}\, \Omega_q^\dag\,.
\end{align}
The first column indicates the spurions' representation under the $U(3)_q\times U(3)_u \times U(3)_d$ group, while the second provides the corresponding transformation rules. All of them are Hermitian and they satisfy
\begin{align}\label{eq.spurion_comm_1}
Y_u^\dag S^u &= S^{qu} Y_u^\dag\,,
&
Y_u S^{qu} &= S^{u} Y_u\,,
\\
\label{eq.spurion_comm_2}
Y_d^\dag S^d &= S^{qd} Y_d^\dag\,,
&
Y_d S^{qd} &= S^{d} Y_d\,.
\end{align}
In the mass basis of the up quarks (Eq.~\eqref{eq.Y_basis})
the spurions take the form
\begin{align}
S^u &\equiv (Y_u^{(d)})^2  \,,
&
S^{qu} &\equiv  (Y_u^{(d)})^2  = S^u \,,
\\
S^d &\equiv (Y_d^{(d)})^2 \,,
&
S^{qd} &\equiv  V( Y_d^{(d)})^2 V^\dag \,.
\end{align}
Additional relevant structures in this basis are
\begin{align}
S^{qu} Y_u^\dag &= (Y_u^{(d)})^3 \,,
&
S^{qu} Y_d^\dag &= (Y_u^{(d)})^2 V Y_d^{(d)}\,, 
\\
S^{qd} Y_u^\dag &= V(Y_d^{(d)})^2 V^\dag Y_u^{(d)} \,,
&
S^{qd} Y_d^\dag &= V (Y_d^{(d)})^3 \,.
\end{align}

\paragraph{Classes 5 and 6.}
With the power counting chosen, 
$\Lag_6^{(5,6)}$ take the form
\begin{equation}
\begin{aligned}
\Lambda^2\, \Lag_6^{(5)} = \sum_{p,r=1}^3 \;&\;\;
C_{eH} \,(Y_l^\dag )_{pr} Q_{eH,pr} \\[-2mm]
+&\left[C_{uH}^{(0)}Y_u^\dag +(\Delta^u C_{uH})\, S^{qu}Y_u^\dagger 
+ (\Delta^d C_{uH})\, S^{qd} Y_u^\dag\right]_{pr} Q_{uH,pr} \\
+&\left[C_{dH}^{(0)}Y_d^\dag +(\Delta^u C_{dH})\, S^{qu}Y_d^\dagger 
+ (\Delta^d C_{dH})\, S^{qd} Y_d^\dag\right]_{pr} Q_{dH,pr}\\
+&\hc\,,
\end{aligned}
\end{equation} 
\begin{equation}
\begin{aligned}
\Lambda^2\, \Lag_6^{(6)} =\sum_{p,r=1}^3\;\;&\;\; 
C_{eW} \,(Y_l^\dag )_{pr} Q_{eW,pr} + C_{eB} \,(Y_l^\dag )_{pr} Q_{eB,pr}\\[-3mm]
+&\left[C_{uG}^{(0)}Y_u^\dag +(\Delta^u C_{uG})\, S^{qu}Y_u^\dagger 
+ (\Delta^d C_{uG})\, S^{qd} Y_u^\dag\right]_{pr} Q_{uG,pr} \\
+&\left[C_{uW}^{(0)}Y_u^\dag +(\Delta^u C_{uW})\, S^{qu}Y_u^\dagger 
+ (\Delta^d C_{uW})\, S^{qd} Y_u^\dag\right]_{pr} Q_{uW,pr} \\
+&\left[C_{uB}^{(0)}Y_u^\dag +(\Delta^u C_{uB})\, S^{qu}Y_u^\dagger 
+ (\Delta^d C_{uB})\, S^{qd} Y_u^\dag\right]_{pr} Q_{uB,pr} \\
+&\left[C_{dG}^{(0)}Y_d^\dag +(\Delta^u C_{dG})\, S^{qu}Y_d^\dagger 
+ (\Delta^d C_{dG})\, S^{qd} Y_d^\dag\right]_{pr} Q_{dG,pr} \\
+&\left[C_{dW}^{(0)}Y_d^\dag +(\Delta^u C_{dW})\, S^{qu}Y_d^\dagger 
+ (\Delta^d C_{dW})\, S^{qd} Y_d^\dag\right]_{pr} Q_{dW,pr} \\
+&\left[C_{dB}^{(0)}Y_d^\dag +(\Delta^u C_{dB})\, S^{qu}Y_d^\dagger 
+ (\Delta^d C_{dB})\, S^{qd} Y_d^\dag\right]_{pr} Q_{dB,pr}\\
+&\hc\,,
\end{aligned}             
\end{equation} 
where the parameters $C_\a^{(0)},\, (\Delta^u C_\a),\, (\Delta^d C_\a)$ are real, scalar quantities.
The structures $Y_u^\dag S^u$, $Y_d^\dag S^d$ are also allowed for operators $Q_{uX},Q_{dX}$ respectively, but they are not independent due to Eqs.~\eqref{eq.spurion_comm_1},~\eqref{eq.spurion_comm_2}.
These two classes  contain a total of 27 real parameters.

\paragraph{Class 7.}
For the operators in class 7 we have
\begin{equation}
\begin{aligned}
\Lambda^2\, \Lag_6^{(7)} = \sum_{p,r=1}^3 \;&\;\; 
C_{Hl}^{(1)} \,\d_{pr}\, Q_{Hl,pr}^{(1)}
+
C_{Hl}^{(3)} \,\d_{pr}\, Q_{Hl,pr}^{(3)}
+
C_{He} \,\d_{pr} \,Q_{He,pr}
\\[-2mm]
+&
\left[C_{Hq}^{(1)(0)} \mathbbm{1} + (\Delta^u C^{(1)}_{Hq}) S^{qu}
+ (\Delta^d C^{(1)}_{Hq}) S^{qd}\right]_{pr} Q_{Hq,pr}^{(1)} \\
+&
\left[C_{Hq}^{(3)(0)} \mathbbm{1} + (\Delta^u C^{(3)}_{Hq}) S^{qu}
+ (\Delta^d C^{(3)}_{Hq}) S^{qd}\right]_{pr} Q_{Hq,pr}^{(3)} \\
+&
\left[C_{Hu}^{(0)} \mathbbm{1} + (\Delta C_{Hu}) S^{u}\right]_{pr} Q_{Hu,pr} \\
+&
\left[C_{Hd}^{(0)} \mathbbm{1} + (\Delta C_{Hd}) S^{d}\right]_{pr} Q_{Hd,pr}\\
+&
\bigg[
\left[C_{Hud}^{0} Y_u Y_d^\dag 
\right]_{pr} Q_{Hud,pr} +\hc \bigg]\,.
\end{aligned}
\end{equation} 
The total number of independent parameters in this class is 14.

\paragraph{Class 8 a.}
The operators of class 8a are composed of the same currents as those of class 7.
MFV corrections have therefore an analogous structure. With the power counting chosen, the independent contractions are
\begin{equation}
\begin{aligned}
\Lambda^2 \Lag_6^{(8a)} = \sum_{p,r,s,t=1}^3 &
\left[C_{ll}\, \d_{pr}\d_{st} + C_{ll}'\,\d_{pt}\d_{sr}\right] Q_{ll,prst}
\\[-2mm]
+&
\bigg[C_{qq}^{(1)(0)}\, \d_{pr}\d_{st}
+(\Delta^u C^{(1)}_{qq})\, S^{qu}_{pr}\d_{st}
+(\Delta^d C^{(1)}_{qq})\, S^{qd}_{pr}\d_{st}+
\nn\\
&
\;\; C_{qq}^{(1)\prime(0)}\, \d_{pt}\d_{sr}
+(\Delta^u C^{(1)\prime}_{qq})\, S^{qu}_{pt}\d_{sr}
+(\Delta^d C^{(1)\prime}_{qq})\, S^{qd}_{pt}\d_{sr}
\bigg] Q_{qq,prst}^{(1)}
\\
+&
\bigg[C_{qq}^{(3)(0)}\, \d_{pr}\d_{st}
+(\Delta^u C^{(3)}_{qq})\, S^{qu}_{pr}\d_{st}
+(\Delta^d C^{(3)}_{qq})\, S^{qd}_{pr}\d_{st}+
\nn\\
&
\;\; C_{qq}^{(3)\prime(0)}\, \d_{pt}\d_{sr}
+(\Delta^u C^{(3)\prime}_{qq})\, S^{qu}_{pt}\d_{sr}
+(\Delta^d C^{(3)\prime}_{qq})\, S^{qd}_{pt}\d_{sr}
\bigg] Q_{qq,prst}^{(3)}
\\
+&
\left[C_{lq}^{(1)(0)}\, \d_{pr}\d_{st}
+(\Delta^u C^{(1)}_{lq})\, \d_{pr} S^{qu}_{st}
+(\Delta^d C^{(1)}_{lq})\, \d_{pr} S^{qd}_{st}
\right]Q_{lq,prst}^{(1)}
\\
+&
\left[C_{lq}^{(3)(0)}\, \d_{pr}\d_{st}
+(\Delta^u C^{(3)}_{lq})\, \d_{pr} S^{qu}_{st}
+(\Delta^d C^{(3)}_{lq})\, \d_{pr} S^{qd}_{st}
\right]Q_{lq,prst}^{(3)}\,.
\end{aligned}
\end{equation} 
In the case of operators $Q_{ll},Q_{qq}^{(1),(3)}$ two possible flavor contractions are allowed, as discussed in Sec.~\ref{sec.flavor_u35}. In $Q_{qq}^{(1,3)}$ spurion insertions in the $(st)$, $(sr)$ currents are redundant by symmetry.
Class 8c contains 20 independent parameters.

\paragraph{Class 8 b.}
The MFV Lagrangian in class 8b has the form 
\begin{equation}
\begin{aligned}
\Lambda^2\Lag_6^{(8b)} =\sum_{p,r,s,t=1}^3\;&\;\;
C_{ee}\, \d_{pr} \d_{st} \, Q_{ee,prst}
\\[-3mm]
+& 
\left[C_{uu}^{(0)}\, \d_{pr}\d_{st}
+(\Delta C_{uu})\, S^{u}_{pr}\d_{st}
+C_{uu}^{\prime(0)}\, \d_{pt}\d_{sr}
+(\Delta C_{uu}^\prime)\, S^{u}_{pt}\d_{sr}
\right] Q_{uu,prst}
\\
+&
\left[C_{dd}^{(0)}\, \d_{pr}\d_{st}
+(\Delta C_{dd})\, S^{d}_{pr}\d_{st}
+C_{dd}^{\prime(0)}\, \d_{pt}\d_{sr}
+(\Delta C_{dd}^\prime)\, S^{d}_{pt}\d_{sr}
\right] Q_{dd,prst}
\\
+&
\left[C_{eu}^{(0)}\, \d_{st}
+(\Delta C_{eu})\,  S^{u}_{st} \right]\,\d_{pr}\, C_{eu,prst}
\\
+&
\left[C_{ed}^{(0)}\, \d_{st}
+(\Delta C_{ed})\,  S^{d}_{st} \right]\,\d_{pr}\, C_{ed,prst}
\\
+&
\left[C_{ud}^{(1)(0)}\, \d_{pr}\d_{st}
+(\Delta^u C^{(1)}_{ud})\, S^u_{pr} \d_{st}
+(\Delta^d C^{(1)}_{ud})\, \d_{pr} S^{d}_{st}
\right] C_{ud,prst}^{(1)}
\\
+&
\left[C_{ud}^{(8)(0)}\, \d_{pr}\d_{st}
+(\Delta^u C^{(8)}_{ud})\, S^u_{pr} \d_{st}
+(\Delta^d C^{(8)}_{ud})\, \d_{pr} S^{d}_{st}\right] C_{ud,prst}^{(8)}\,.
\end{aligned}
\end{equation} 
As in the $U(3)^5$ symmetric case, $Q_{uu},Q_{dd}$ admit two independent flavor contractions, and spurion insertions in only one of their currents is required, by symmetry.
Class 8b therefore contains a total of 19 independent parameters.

\paragraph{Class 8 c.}
For class 8c we have a total of 28 independent parameters:
\begin{equation}
\begin{aligned}
\Lambda^2\Lag_6^{(8c)} = \sum_{p,r,s,t=1}^{3} \;&\;\;
C_{le}\,  \d_{pr} \d_{st}\, Q_{le,prst}
\\[-3mm]
+&
\left[C_{lu}^{(0)}\, \d_{st}
+(\Delta C_{lu})\, S^{u}_{st}\right]\, \d_{pr}\, Q_{lu,prst}
\\
+&
\left[C_{ld}^{(0)}\, \d_{st}
+(\Delta C_{ld})\,  S^{d}_{st}\right]\, \d_{pr}\, Q_{ld,prst}
\\
+&
\left[C_{qe}^{(0)}\, \d_{pr}
+(\Delta^{u} C_{qe})\, S^{qu}_{pr} 
+(\Delta^{d} C_{qe})\, S^{qd}_{pr} \right]\,\d_{st}\, Q_{qe,prst}
\\
+&
\bigg[C_{qu}^{(1)(0)}\, \d_{pr}\d_{st}
+(\Delta_1^{u} C^{(1)}_{qu})\, S^{qu}_{pr} \d_{st}
+(\Delta_1^{d} C^{(1)}_{qu})\, S^{qd}_{pr} \d_{st}
+(\Delta_2 C^{(1)}_{qu})\, \d_{pr} S^{u}_{st}
+
\nn\\
&
\;\; C_{qu}^{(1)\prime(0)}\, (Y_u^\dag)_{pt}(Y_u)_{sr}
\bigg] Q_{qu,prst}^{(1)}
\\
+&
\bigg[C_{qu}^{(8)(0)}\, \d_{pr}\d_{st}
+(\Delta_1^{u} C^{(8)}_{qu})\, S^{qu}_{pr} \d_{st}
+(\Delta_1^{d} C^{(8)}_{qu})\, S^{qd}_{pr} \d_{st}
+(\Delta_2 C^{(8)}_{qu})\, \d_{pr} S^{u}_{st}
+
\nn\\
&
\;\; C_{qu}^{(8)\prime(0)}\, (Y_u^\dag)_{pt}(Y_u)_{sr}
\bigg] Q_{qu,prst}^{(8)}
\\
+&
\bigg[C_{qd}^{(1)(0)}\, \d_{pr}\d_{st}
+(\Delta_1^{u} C^{(1)}_{qd})\, S^{qu}_{pr} \d_{st}
+(\Delta_1^{d} C^{(1)}_{qd})\, S^{qd}_{pr} \d_{st}
+(\Delta_2 C^{(1)}_{qd})\, \d_{pr} S^{d}_{st}
+
\nn\\
&
\;\; C_{qd}^{(1)\prime(0)}\, (Y_d^\dag)_{pt}(Y_d)_{sr}
\bigg] Q_{qd,prst}^{(1)}
\\
+&
\bigg[C_{qd}^{(8)(0)}\, \d_{pr}\d_{st}
+(\Delta_1^{u} C^{(8)}_{qd})\, S^{qu}_{pr} \d_{st}
+(\Delta_1^{d} C^{(8)}_{qd})\, S^{qd}_{pr} \d_{st}
+(\Delta_2 C^{(8)}_{qd})\, \d_{pr} S^{d}_{st}
+
\nn\\
&
\;\; C_{qd}^{(8)\prime(0)}\, (Y_d^\dag)_{pt}(Y_d)_{sr}
\bigg] Q_{qd,prst}^{(8)}\,.
\end{aligned}
\end{equation} 
Note that the operators $O_{qu}^{(1),(8)}, Q_{qd}^{(1)(8)}$ admit a contraction $[pt][sr]$ with a Yukawa insertion in each current.

\paragraph{Class 8 d.}
Finally, the operators of class 8d have the structure
\begin{equation}
\begin{aligned}
\Lambda^2\Lag_6^{(8d)} = \sum_{p,r,s,t=1}^3\;&
\left[C_{ledq}^{(0)} \, Y_{d,st}
+(\Delta^u C_{ledq}) \, (Y_d S^{qu})_{st}
+(\Delta^d C_{ledq}) \, (Y_d S^{qd})_{st}\right]  \, (Y^\dag_l)_{pr}\, Q_{ledq,prst}
\\
+&
\bigg[C_{quqd}^{(1)} \,Y^\dag_{u,pr} Y^\dag_{d,st}
+ C_{quqd}^{(1)\prime} \,Y^\dag_{u,sr} Y^\dag_{d,pt}
\bigg] Q_{quqd,prst}^{(1)}
\\
+&
\bigg[C_{quqd}^{(8)} \,Y^\dag_{u,pr} Y^\dag_{d,st}
+ C_{quqd}^{(8)\prime} \,Y^\dag_{u,sr} Y^\dag_{d,pt}
\bigg] Q_{quqd,prst}^{(8)}
\\
+&
\left[C_{lequ}^{(1)(0)} \, Y^\dag_{u,st}
+(\Delta^{u} C^{(1)}_{lequ})\,(S^{qu}Y_u^\dag)_{st}
+(\Delta^{d} C^{(1)}_{lequ})\, (S^{qd}Y_u^\dag)_{st}
\right]\, (Y^\dag_l)_{pr}\, Q_{lequ,prst}^{(1)}
\\
+&
\left[C_{lequ}^{(3)(0)} \, Y^\dag_{u,st}
+(\Delta^{u} C^{(3)}_{lequ})\,(S^{qu}Y_u^\dag)_{st}
+(\Delta^{d} C^{(3)}_{lequ})\, (S^{qd}Y_u^\dag)_{st}
\right]\, (Y^\dag_l)_{pr}\, Q_{lequ,prst}^{(3)}\\
+&\hc\,,
\end{aligned}
\end{equation} 
The total number of parameters is 13.

\subsection{\top, \topsl: \texorpdfstring{$U(2)^3$}{U(2)\^{}3} symmetry in the quark sector}\label{sec.flavor_top} 
Two new sets of models have been introduced in version~3.0, that implement a flavor structure consistent with the recommendations of Ref.~\cite{AguilarSaavedra:2018nen} for the SMEFT interpretation of top quark measurements. 
The formalism builds upon~\cite{Barbieri:2011ci,Barbieri:2012uh,Blankenburg:2012nx} and  is defined by the following assumptions:
\begin{itemize}
\item quarks of the first two generations  and quarks of the 3rd are described by independent fields. We denote them respectively by 
($q_p,u_p,d_p)$ with $p=\{1,2\}$ and by ($Q,t,b$).

\item a symmetry $U(2)^3 = U(2)_q \times U(2)_u \times U(2)_d$ is imposed on the Lagrangian, under which only the light quarks transform:
\begin{equation}
q\mapsto \Omega_q q\,,\qquad 
u \mapsto \Omega_u u\,, \qquad 
d\mapsto \Omega_d d\,, \qquad
Q \mapsto Q\,, \qquad 
t\mapsto t\,,\qquad 
b \mapsto b\,.
\end{equation} 
\item mixing effects in the quark sector are neglected and $V_{CKM}\equiv \mathbbm{1}$ is assumed. \\                                                                          
This choice greatly simplifies the structure of the Lagrangian, as mixing between the light and heavy quarks can only be introduced through extra $U(2)$ spurions~\cite{Barbieri:2011ci,Faroughy:2020ina}.

\end{itemize}
With this notation, the SM Lagrangian is
\begin{equation}
\begin{aligned}
\Lag_{\rm fermions} &= i\bar q \slashed{D} q +  i\bar u\slashed{D} u + i\bar d \slashed{D} d
+  i\bar Q \slashed{D} Q+  i\bar t \slashed{D} t+  i\bar b \slashed{D} b &+ \text{ leptons}\,,
\\
\Lag_{\rm Yukawa}&=
-\bar d\, Y_d \,H^\dag q -\bar u \,Y_u\, \tilde H^\dag q 
-y_b\, \bar b\,H^\dag Q -y_t\, \bar t \, \tilde H^\dag Q &+ \text{ leptons}\,,
\end{aligned}
\end{equation} 
with the Yukawas of the light quarks 
$Y_u \equiv \diag(y_u,y_c)$, $Y_d \equiv \diag(y_d,y_s)$ 
promoted to spurions of $U(2)$
\begin{equation}
Y_u \mapsto \Omega_u\, Y_u\, \Omega_q^\dag\,,\qquad 
Y_d \mapsto \Omega_d\, Y_d\, \Omega_q^\dag\,,
\end{equation} 
while $y_t,\,y_b$ do not transform under any symmetry. As a consequence, only $(\bar LR), (\bar RL)$ currents with light quarks need to be weighted by Yukawa insertions.

It is convenient to construct a $U(2)^3$ invariant basis mapping the fermionic operators of Table~\ref{tab.Warsaw_basis} to the notation with 6 quark fields. We choose the set given in Table~\ref{tab.top_basis}, where, analogously to the $U(3)^5$ case, we retain the least Yukawa-suppressed $U(2)^3$-invariant contractions for each operator in the Warsaw basis.

In the lepton sector we consider two alternative ans\"atze:
\begin{itemize}
\item[(a)] a $U(1)_{l+e}^3 = U(1)_{e} \times U(1)_{\mu} \times U(1)_{\tau}$ symmetry under which the fields transform as
\begin{align}
l_1 &\mapsto e^{i\a_e} l_1\,,
&
l_2 &\mapsto e^{i\a_\mu} l_2\,,
&
l_3 &\mapsto e^{i\a_\tau} l_3\,,
\\
e_1 &\mapsto e^{i\a_e}e_1\,,
&
e_2 &\mapsto e^{i\a_\mu} e_2\,,
&
e_3 &\mapsto e^{i\a_\tau} e_3\,.
\end{align} 
This matches the ``baseline'' scenario in Ref.~\cite{AguilarSaavedra:2018nen} and corresponds to simple flavor-diagonality. It is implemented in the \top\ models.

\item[(b)] a $U(3)^2 = U(3)_l \times U(3)_e$ symmetry under which
\begin{equation}
l\mapsto \Omega_l l\,,\qquad e\mapsto \Omega_e e\,,\qquad Y_l \mapsto\Omega_e\, Y_l\, \Omega_l^\dag. 
\end{equation} 
In the lepton sector, this setup matches exactly the structure of the \Utf\ and \MFV\ models. It is more restrictive compared to $U(1)_{l+e}^3$ and contains fewer free parameters.\\ It is implemented in the \topsl\ models.
\end{itemize}

\noindent
In the $U(1)_{l+e}^3$ symmetric case, no transformation rule needs to be assigned to $Y_l$, as left- and right-handed leptons transform under the same symmetry. This implies that $(\bar LR),(\bar RL)$ lepton currents are weighted by $Y_l$ in the \topsl\ models but not in the \top\ ones.

\paragraph{Classes 5 and 6.}
The basis of quark operators for $\Lag_6^{(5)}$ and $\Lag_6^{(6)}$ in Table~\ref{tab.top_basis} is easily constructed splitting the quark currents for the first 2 and the 3rd generations. Insertions of $Y_u^\dag, Y_d^\dag$ in light quark currents, that are required for $U(2)^3$ invariance, are embedded in the operator definitions. $\Lag_6^{(5,6)}$ contain in total 32 real parameters (16 complex) coming from quark invariants.

When $U(1)_{l+e}^3$ is imposed (\top\ models) on the lepton fields, $Q_{eH,pr}, Q_{eW,pr},Q_{eB,pr}$ admit 3 independent contractions each, one per generation. 
When the more restrictive $U(3)^2$ is imposed (\topsl\ models), each operator is associated to only one complex Wilson coefficient.

The total number of real independent parameters in $\Lag_6^{(5,6)}$ is therefore 50 in the \top\ case and 38 in the \topsl\ case. The Lagrangian is
\begin{align}
\Lambda^2 \Lag_6^{(5)} &=
  C_{uH} Q_{uH} + C_{tH} Q_{tH}
+ C_{dH} Q_{dH}+ C_{bH} Q_{bH}
\nn\\
&+\begin{cases}
\sum_{p=1}^3\, (C_{eH})_{pp}\, Q_{eH,pp} & \quad U(1)_{l+e}^3\;[\top]
\\[2mm]
\sum_{p,r=1}^3 C_{eH} \, (Y_l^\dag)_{pr}\, Q_{eH,pr} & \quad U(3)^2\quad [\topsl]
\end{cases}
\\
&+\hc\,,
\nn\\
\Lambda^2 \Lag_6^{(6)} &=
 C_{uW} Q_{uW} + C_{uB} Q_{uB} + C_{uG} Q_{uG}
+ C_{tW} Q_{tW} + C_{tB} Q_{tB} + C_{tG} Q_{tG}
\nn\\
&+
C_{dW} Q_{dW} + C_{dB} Q_{dB} + C_{dG} Q_{dG}
+ C_{bW} Q_{bW} + C_{bB} Q_{bB} + C_{bG} Q_{bG}
\\
&+\begin{cases}
\sum_{p,r=1}^3\, (C_{eW})_{pp}\,Q_{eW,pp} + (C_{eB})_{pp}\, Q_{eB,pp} &
\quad U(1)_{l+e}^3\;[\top]
\\[2mm]
\sum_{p,r=1}^3C_{eW}\, (Y_l^\dag)_{pr}\, Q_{eW,pr} + C_{eB}\, (Y_l^\dag)_{pr}\,  Q_{eB,pr} &
\quad U(3)^2\quad[\topsl]
\end{cases}
\nn\\
&+\hc\,.
\nn
\end{align}

\clearpage
\thispagestyle{empty}
\begin{table}[h!]
\vspace*{-1cm}
\small
\hspace*{-2cm}
\scalebox{.76}{
\renewcommand{\arraystretch}{1.6}\begin{tabular}{|*4{>{$}c<{$}|>{$}l<{$}|}}
\hline
\rowcolor{tablesColor}
\multicolumn{8}{|c|}{$\Lag_6^{(5)} - \psi^2 H^3$}
\\\hline
Q_{uH}& (H^\dag H) (\bar q \, Y_u^\dag \, u \tilde H) 
& 
Q_{dH}& (H^\dag H) (\bar q\, Y_d^\dag\, d H)
& 
Q_{eH}& (H^\dag H) (\bar l_p e_r H)
&  & 
\\
Q_{tH}& (H^\dag H) (\bar Q  \tilde Ht)
&
Q_{bH}& (H^\dag H) (\bar Q H b)
& &
& &
\\\hline
\rowcolor{tablesColor}
\multicolumn{8}{|c|}{$\Lag_6^{(6)} - \psi^2 X H$}
\\\hline
Q_{eW}& (\bar l_p \s^{\mu\nu} e_r) \sigma^i H W_{\mu\nu}^i
&
Q_{uW}& (\bar q\, Y_u^\dag\, \s^{\mu\nu} u) \sigma^i \tilde H W_{\mu\nu}^i
&
Q_{uB}& (\bar q\, Y_u^\dag\, \s^{\mu\nu} u) \tilde H B_{\mu\nu}
&
Q_{uG}& (\bar q\, Y_u^\dag\, \s^{\mu\nu} T^a u) \tilde H G_{\mu\nu}^a
\\
Q_{eB}& (\bar l_p \s^{\mu\nu} e_r) H B_{\mu\nu}
&
Q_{tW}& (\bar Q \s^{\mu\nu} t) \sigma^i \tilde H W_{\mu\nu}^i
&
Q_{tB}& (\bar Q \s^{\mu\nu} t)  \tilde H B_{\mu\nu}
&
Q_{tG}& (\bar Q \s^{\mu\nu} T^a t) \tilde H G_{\mu\nu}^a
\\
Q_{dW}& (\bar q\, Y_d^\dag\, \s^{\mu\nu} d) \sigma^i H W_{\mu\nu}^i
&
Q_{dB}& (\bar q\, Y_d^\dag\, \s^{\mu\nu} d) H B_{\mu\nu}
&
Q_{dG}& (\bar q\, Y_d^\dag\, \s^{\mu\nu} T^a d) H G_{\mu\nu}^a
&&
\\
Q_{bW}& (\bar Q \s^{\mu\nu} b) \sigma^i H W_{\mu\nu}^i
&
Q_{bB}& (\bar Q \s^{\mu\nu} b) H B_{\mu\nu}
&
Q_{bG}& (\bar Q \s^{\mu\nu} T^a b)  H G_{\mu\nu}^a
&&
\\\hline
\rowcolor{tablesColor}
\multicolumn{8}{|c|}{$\Lag_6^{(7)} - \psi^2 H^2 D$}
\\\hline
Q_{Hl}^{(1)}& (H^\dag i\overleftrightarrow{D}_\mu H) (\bar l_p \g^\mu l_r)
&
Q_{Hl}^{(3)}& (H^\dag i\overleftrightarrow{D}_\mu^i H) (\bar l_p \s^i\g^\mu l_r)
&
Q_{He}&(H^\dag i\overleftrightarrow{D}_\mu H) (\bar e_p \g^\mu e_r)
&&
\\
Q_{Hq}^{(1)}& (H^\dag i\overleftrightarrow{D}_\mu H) (\bar q \g^\mu q)
&
Q_{Hq}^{(3)}& (H^\dag i\overleftrightarrow{D}^i_\mu H) (\bar q \s^i\g^\mu q)
&
Q_{Hu}& (H^\dag i\overleftrightarrow{D}_\mu H) (\bar u \g^\mu u)
&
Q_{Hd}& (H^\dag i\overleftrightarrow{D}_\mu H) (\bar d \g^\mu d)
\\
Q_{HQ}^{(1)}& (H^\dag i\overleftrightarrow{D}_\mu H) (\bar Q \g^\mu Q)
&
Q_{HQ}^{(3)}& (H^\dag i\overleftrightarrow{D}^i_\mu H) (\bar Q \s^i\g^\mu Q)
&
Q_{Ht}& (H^\dag i\overleftrightarrow{D}_\mu H) (\bar t \g^\mu t)
&
Q_{Hb}& (H^\dag i\overleftrightarrow{D}_\mu H) (\bar b \g^\mu b)
\\
Q_{Hud}& i(\tilde H^\dag D_\mu H) (\bar u \, Y_u Y_d^\dag\, \g^\mu d)
&
Q_{Htb}& i(\tilde H^\dag D_\mu H) (\bar t \g^\mu b)
&&
&&
\\\hline
\rowcolor{tablesColor}
\multicolumn{8}{|c|}{$\Lag_6^{(8a)} - (\bar LL)(\bar LL)$}
\\\hline
Q_{lq}^{(1)}& (\bar l_p \gamma_\mu l_r)(\bar q \gamma^\mu q)
&
Q_{lq}^{(3)}& (\bar l_p \s^i\gamma_\mu l_r)(\bar q \s^i\gamma^\mu q)
&
Q_{ll}&  (\bar l_p \gamma_\mu l_r)(\bar l_s\gamma^\mu l_t)
&&
\\
Q_{lQ}^{(1)}& (\bar l_p \gamma_\mu l_r)(\bar Q \gamma^\mu Q)
&
Q_{lQ}^{(3)}& (\bar l_p \s^i\gamma_\mu l_r)(\bar Q\s^i\gamma^\mu Q)
&
Q_{QQ}^{(1)}& (\bar Q \gamma_\mu Q)(\bar Q \gamma^\mu Q)
&
Q_{QQ}^{(8)}& (\bar Q T^a\gamma_\mu Q)(\bar Q T^a\gamma^\mu Q)
\\
Q_{qq}^{(1,1)}& (\bar q \gamma_\mu q)(\bar q \gamma^\mu q)
&
Q_{qq}^{(1,8)}& (\bar q T^a\gamma_\mu q)(\bar q T^a\gamma^\mu q)
&
Q_{qq}^{(3,1)}& (\bar q \s^i\gamma_\mu q)(\bar q \s^i\gamma^\mu q)
&
Q_{qq}^{(3,8)}& (\bar q \s^iT^a\gamma_\mu q)(\bar q \s^iT^a\gamma^\mu q)
\\
Q_{Qq}^{(1,1)}& (\bar Q \gamma_\mu Q)(\bar q \gamma^\mu q)
&
Q_{Qq}^{(1,8)}& (\bar Q T^a\gamma_\mu Q)(\bar q T^a\gamma^\mu q)
&
Q_{Qq}^{(3,1)}& (\bar Q \s^i\gamma_\mu Q)(\bar q \s^i\gamma^\mu q)
&
Q_{Qq}^{(3,8)}& (\bar Q \s^iT^a\gamma_\mu Q)(\bar q \s^iT^a\gamma^\mu q)
\\\hline
\rowcolor{tablesColor}
\multicolumn{8}{|c|}{$\Lag_6^{(8b)} - (\bar RR)(\bar RR)$}
\\\hline
Q_{eu}& (\bar e_p \gamma_\mu e_r)(\bar u \gamma^\mu u)
&
Q_{ed}& (\bar e_p \gamma_\mu e_r)(\bar d \gamma^\mu d)
&
Q_{ee}& (\bar e_p \gamma_\mu e_r)(\bar e_s \gamma^\mu e_t)
&&
\\
Q_{et}& (\bar e_p \gamma_\mu e_r)(\bar t \gamma^\mu t)
&
Q_{eb}& (\bar e_p \gamma_\mu e_r)(\bar b \gamma^\mu b)
&
Q_{tt}& (\bar t\gamma_\mu t)(\bar t\gamma^\mu t)
&
Q_{bb}& (\bar b\gamma_\mu b)(\bar b\gamma^\mu b)
\\
Q_{uu}^{(1)}& (\bar u \gamma_\mu u)(\bar u \gamma^\mu u)
&
Q_{uu}^{(8)}& (\bar u T^a\gamma_\mu u)(\bar u T^a\gamma^\mu u)
&
Q_{tu}^{(1)}& (\bar t \gamma_\mu t)(\bar u \gamma^\mu u)
&
Q_{tu}^{(8)}& (\bar t T^a\gamma_\mu t)(\bar u T^a\gamma^\mu u)
\\
Q_{dd}^{(1)}& (\bar d \gamma_\mu d)(\bar d \gamma^\mu d)
&
Q_{dd}^{(8)}& (\bar d T^a\gamma_\mu d)(\bar d T^a\gamma^\mu d)
&
Q_{bd}^{(1)}& (\bar b \gamma_\mu b)(\bar d \gamma^\mu d)
&
Q_{bd}^{(8)}& (\bar b T^a\gamma_\mu b)(\bar d T^a\gamma^\mu d)
\\
Q_{ud}^{(1)}& (\bar u \gamma_\mu u)(\bar d \gamma^\mu d)
&
Q_{ud}^{(8)}& (\bar u T^a\gamma_\mu u)(\bar d T^a\gamma^\mu d)
&
Q_{td}^{(1)}& (\bar t \gamma_\mu t)(\bar d \gamma^\mu d)
&
Q_{td}^{(8)}& (\bar t T^a\gamma_\mu t)(\bar d T^a\gamma^\mu d)
\\
Q_{ub}^{(1)}& (\bar u \gamma_\mu u)(\bar b \gamma^\mu b)
&
Q_{ub}^{(8)}& (\bar u T^a\gamma_\mu u)(\bar b T^a\gamma^\mu b)
&
Q_{tb}^{(1)}& (\bar t \gamma_\mu t)(\bar b \gamma^\mu b)
&
Q_{tb}^{(8)}& (\bar t T^a\gamma_\mu t)(\bar b T^a\gamma^\mu b)
\\
Q_{utbd}^{(1)}& (Y_uY_d^\dag)_{pr}(\bar u_p \gamma_\mu t)(\bar b\gamma^\mu d_r)
&
Q_{utbd}^{(8)}& (Y_uY_d^\dag)_{pr}(\bar u_p T^a\gamma_\mu t)(\bar b T^a\gamma^\mu d_r)
&&
&&
\\\hline
\rowcolor{tablesColor}
\multicolumn{8}{|c|}{$\Lag_6^{(8c)} - (\bar LL)(\bar RR)$}
\\\hline
Q_{lu}& (\bar l_p \gamma_\mu l_r)(\bar u \gamma^\mu u)
&
Q_{ld}& (\bar l_p \gamma_\mu l_r)(\bar d \gamma^\mu d)
&
Q_{qe}& (\bar q \gamma_\mu q)(\bar e_p \gamma^\mu e_r)
&
Q_{le}& (\bar l_p \gamma_\mu l_r)(\bar e_s \gamma^\mu e_t)
\\
Q_{lt}& (\bar l_p \gamma_\mu l_r)(\bar t \gamma^\mu t)
&
Q_{lb}& (\bar l_p \gamma_\mu l_r)(\bar b \gamma^\mu b)
&
Q_{Qe}& (\bar Q \gamma_\mu Q)(\bar e_p \gamma^\mu e_r)
&&
\\
Q_{qu}^{(1)}& (\bar q \gamma_\mu q)(\bar u \gamma^\mu u)
&
Q_{Qu}^{(1)}& (\bar Q \gamma_\mu Q)(\bar u \gamma^\mu u)
&
Q_{qt}^{(1)}& (\bar q \gamma_\mu q)(\bar t \gamma^\mu t)
&
Q_{Qt}^{(1)}& (\bar Q \gamma_\mu Q)(\bar t \gamma^\mu t)
\\
Q_{qu}^{(8)}& (\bar q T^a\gamma_\mu q)(\bar u T^a\gamma^\mu u)
&                    
Q_{Qu}^{(8)}& (\bar Q T^a\gamma_\mu Q)(\bar u T^a\gamma^\mu u)
&                    
Q_{qt}^{(8)}& (\bar q T^a\gamma_\mu q)(\bar t T^a\gamma^\mu t)
&                    
Q_{Qt}^{(8)}& (\bar Q T^a\gamma_\mu Q)(\bar t T^a\gamma^\mu t)
\\
Q_{qd}^{(1)}& (\bar q \gamma_\mu q)(\bar d \gamma^\mu d)
&
Q_{Qd}^{(1)}& (\bar Q \gamma_\mu Q)(\bar d \gamma^\mu d)
&
Q_{qb}^{(1)}& (\bar q \gamma_\mu q)(\bar b \gamma^\mu b)
&
Q_{Qb}^{(1)}& (\bar Q \gamma_\mu Q)(\bar b \gamma^\mu b)
\\
Q_{qd}^{(8)}& (\bar q T^a\gamma_\mu q)(\bar d T^a\gamma^\mu d)
&                   
Q_{Qd}^{(8)}& (\bar Q T^a\gamma_\mu Q)(\bar d T^a\gamma^\mu d)
&                   
Q_{qb}^{(8)}& (\bar q T^a\gamma_\mu q)(\bar b T^a\gamma^\mu b)
&                   
Q_{Qb}^{(8)}& (\bar Q T^a\gamma_\mu Q)(\bar b T^a\gamma^\mu b)
\\
Q_{qQtu}^{(1)}& (Y_u^\dag)_{pr}(\bar q_p \gamma_\mu Q)(\bar t\gamma^\mu u_r)
&
Q_{qQtu}^{(8)}& (Y_u^\dag)_{pr}(\bar q_p T^a\gamma_\mu Q)(\bar t T^a \gamma^\mu u_r)
&
Q_{qQbd}^{(1)}& (Y_d^\dag)_{pr}(\bar q_p \gamma_\mu Q)(\bar b\gamma^\mu d_r)
&
Q_{qQbd}^{(8)}& (Y_d^\dag)_{pr}(\bar q_p T^a\gamma_\mu Q)(\bar b T^a\gamma^\mu d_r)
\\\hline
\rowcolor{tablesColor}
\multicolumn{8}{|c|}{$\Lag_6^{(8d)} - (\bar LR)(\bar RL), (\bar LR)(\bar LR)$}
\\\hline
Q_{ledq}& (\bar l_p^j e_r)(\bar d \,Y_d\, q_{j})
&
Q_{lebQ}& (\bar l_p^j e_r)(\bar b Q_{j})
&
Q_{leQt}^{(1)}&  (\bar l_p^j e_r) \e_{jk} (\bar Q^k \, t)
&
Q_{leQt}^{(3)}&  (\bar l_p^j \sigma_{\mu\nu} e_r) \e_{jk} (\bar Q^k \sigma^{\mu\nu} t)
\\
Q_{lequ}^{(1)}&  (\bar l_p^j e_r) \e_{jk} (\bar q^k \,Y_u^\dag\, u)
&
Q_{lequ}^{(3)}&  (\bar l_p^j \sigma_{\mu\nu} e_r) \e_{jk} (\bar q^k\,Y_u^\dag\, \sigma^{\mu\nu} u)
&
Q_{QtQb}^{(1)}&  (\bar Q^j \, t) \e_{jk} (\bar Q^k\, b)
&
Q_{QtQb}^{(8)}&  (\bar Q^j \, T^a t) \e_{jk} (\bar Q^k\,T^a b)
\\
Q_{quqd}^{(1)}&  (\bar q^j \,Y_u^\dag\, u) \e_{jk} (\bar q^k\, Y_d^\dag\, d)
&
Q_{quqd}^{(8)}&  (\bar q^j \,Y_u^\dag\,T^a u) \e_{jk} (\bar q^k\, Y_d^\dag\, T^a d)
&
Q_{quqd}^{(1)\prime}&  (Y_u^\dag)_{sr} (Y_d^\dag)_{pt} (\bar q^j_p \, u_r) \e_{jk} (\bar q^k_s\, d_t)
&
Q_{quqd}^{(8)\prime}&  (Y_u^\dag)_{sr} (Y_d^\dag)_{pt}(\bar q^j_p \,T^a u_r) \e_{jk} (\bar q^k_s\, T^a d_t)
\\
Q_{Qtqd}^{(1)}&  (\bar Q^j \, t) \e_{jk} (\bar q^k\, Y_d^\dag\, d)
&
Q_{Qtqd}^{(8)}&  (\bar Q^j \, T^a t) \e_{jk} (\bar q^k\, Y_d^\dag\,T^a d)
&
Q_{quQb}^{(1)}&  (\bar q^j \,Y_u^\dag\, u) \e_{jk} (\bar Q^k\, b)
&
Q_{quQb}^{(8)}&  (\bar q^j \,Y_u^\dag\,T^a u) \e_{jk} (\bar Q^k\, T^a b)
\\
Q_{Quqb}^{(1)}&  (Y_u^\dag)_{pr}\,(\bar Q^j \, u_r) \e_{jk} (\bar q_p^k\, b)
&
Q_{Quqb}^{(8)}&  (Y_u^\dag)_{pr}\,(\bar Q^j \,T^a u_r) \e_{jk} (\bar q_p^k\,T^a b)
&
Q_{qtQd}^{(1)}&  (Y_d^\dag)_{pr}\,(\bar q_p^j \, t) \e_{jk} (\bar Q^k\, d_r)
&
Q_{qtQd}^{(8)}&  (Y_d^\dag)_{pr}\,(\bar q_p^j \,T^a t) \e_{jk} (\bar Q^k\,T^a d_r)
\\
\hline
\end{tabular}}
\caption{Basis of fermionic operators for the \top\ and \topsl\ flavor assumptions. Here $(q,u,d)$, $Y_u,Y_d$ denote quarks of the first 2 generations and their $2\times2$ Yukawa matrices. Quark fields of the 3rd generation are ($Q,t,b$). Flavor indices $p,r,s,t$ run over $\{1,2\}$ for light quarks and $\{1,2,3\}$ for leptons.  Whenever flavor indices are not specified, they are implicitly contracted within each current.}\label{tab.top_basis}
\end{table}
\clearpage

\paragraph{Class 7.}
Class 7 depends on 12 real parameters from quark operators, plus 9 (3) real parameters from lepton operators in the \top\ (\topsl) case.
$Q_{Hud}$ is defined with a $Y_u Y_d^\dag$ insertion to preserve $U(2)^3$, while $Q_{Htb}$ is independent of the Yukawas.
\begin{align}
\Lambda^2 \Lag_6^{(7)} &=
 C_{Hq}^{(1)} Q_{Hq}^{(1)} + C_{Hq}^{(3)} Q_{Hq}^{(3)} + C_{Hu} Q_{Hu}
+ C_{Hd} Q_{Hd} + \left[ C_{Hud} Q_{Hud} +\hc\right]
\nn\\
&+
C_{HQ}^{(1)} Q_{HQ}^{(1)} + C_{HQ}^{(3)} Q_{HQ}^{(3)} + C_{Ht} Q_{Ht}
+ C_{Hb} Q_{Hb} + \left[C_{Htb} Q_{Htb} +\hc\right]
\\
&+\begin{cases}
 \sum_{p=1}^3\,
(C_{Hl}^{(1)})_{pp}\, Q_{Hl,pp}^{(1)} + 
(C_{Hl}^{(3)})_{pp}\, Q_{Hl,pp}^{(3)}+ 
(C_{He})_{pp}\, Q_{He,pp}\,, &
\quad U(1)_{l+e}^3\;[\top]
\\[3mm]
\sum_{p,r=1}^3
C_{Hl}^{(1)}\,\d_{pr}\, Q_{Hl,pr}^{(1)} + 
C_{Hl}^{(3)}\, \d_{pr}\, Q_{Hl,pr}^{(3)}+ 
C_{He}\,\d_{pr}\, Q_{He,pr}\,,  &
\quad U(3)^2\quad[\topsl]
\end{cases}
\nn
\end{align}

\paragraph{Class 8 a.}
Class 8a contains 2 operators with 4 quarks. Mapping them to the formalism with 6 quark fields, each of them admits 5 independent $U(2)^3$ invariant contractions, that can be written 
\begin{align}\label{eq.qq_iijj}
Q_{qq}^{(1)}&=\sum_{p,r,s,t=1}^2\d_{pr}\d_{st} \, (\bar q_p \g_\mu q_r)  (\bar q_s \g^\mu q_t)\,,
&
Q_{Qq}^{(1)}&=\sum_{p,r=1}^2\d_{pr} \, (\bar q_p \g_\mu q_r)  (\bar Q \g^\mu Q)\,,
      \\
Q_{qq}^{(1)\prime}&=\sum_{p,r,s,t=1}^2\d_{pt}\d_{sr} \, (\bar q_p \g_\mu q_r)  (\bar q_s \g^\mu q_t)\,,
&
Q_{Qq}^{(1)\prime}&=\sum_{p,t=1}^2\d_{pt} \, (\bar q_p \g_\mu Q)  (\bar Q \g^\mu q_t)\,,
\\
Q_{QQ}^{(1)}&=(\bar Q \g_\mu Q)  (\bar Q \g^\mu Q)\,,
\label{eq.qq_3333}
\end{align}
and analogously for $Q_{qq}^{(3)}$.  In practice, for analyses involving top quark processes it is convenient to trade ``crossed'' flavor contractions, as well as $Q_{QQ}^{(3)}$, for operators with a color octet structure. This is motivated by top processes being largely dominated by QCD interactions in the SM.  The rotation is done using Fierz rearrangements and the completeness relations for $SU(2)$ and $SU(3)$
\begin{align}
\s^i_{jk} \s^i_{mn} &= 2 \d_{jn}\d_{mk} - \d_{jk}\d_{mn}\,,
\label{fierz.su2}\\
T^A_{ab}T^A_{cd} &= \frac{1}{2}\d_{ad}\d_{cb} -\frac{1}{6}\d_{ab}\d_{cd}\,.
\label{fierz.su3}
\end{align}
Consistent with the recommendations in Ref.~\cite{AguilarSaavedra:2018nen}, \smeftsim\ implements the invariants in Table~\ref{tab.top_basis}, that are related to those in Eqs.~\eqref{eq.qq_iijj}-\eqref{eq.qq_3333} and their $Q_{qq}^{(3)}$ counterparts as:
\begin{align}
\label{eq.top_relations_8a_1}
Q_{QQ}^{(1)} &= Q_{QQ}^{(1)} \,,
&
Q_{QQ}^{(3)} &= -\frac13 Q_{QQ}^{(1)} + 4Q_{QQ}^{(8)}\,,
\\[3mm]
Q_{qq}^{(1)} &= Q_{qq}^{(1,1)}\,,
&
Q_{qq}^{(1)\prime} &= \frac{1}{6} \left(Q_{qq}^{(1,1)} + Q_{qq}^{(3,1)}\right) + Q_{qq}^{(1,8)} +Q_{qq}^{(3,8)}\,,\\
Q_{qq}^{(3)} &= Q_{qq}^{(3,1)}\,,
&
Q_{qq}^{(3)\prime} &= \frac12 Q_{qq}^{(1,1)} - \frac16 Q_{qq}^{(3,1)} + 3 Q_{qq}^{(1,8)} - Q_{qq}^{(3,8)}\,,
\\[3mm]
Q_{Qq}^{(1)} &= Q_{Qq}^{(1,1)}\,,
&
Q_{Qq}^{(1)\prime} &= \frac{1}{6} \left(Q_{Qq}^{(1,1)} + Q_{Qq}^{(3,1)}\right) + Q_{Qq}^{(1,8)} +Q_{Qq}^{(3,8)}\,,\\
Q_{Qq}^{(3)} &= Q_{Qq}^{(3,1)}\,,
&
Q_{Qq}^{(3)\prime} &= \frac12 Q_{Qq}^{(1,1)} - \frac16 Q_{Qq}^{(3,1)} + 3 Q_{Qq}^{(1,8)} - Q_{Qq}^{(3,8)}\,.
\label{eq.top_relations_8a_2}
\end{align}
Eqs.~\eqref{eq.top_relations_8a_1}-\eqref{eq.top_relations_8a_2} can be written compactly as $\vec Q_{\rm Warsaw} = R \, \vec Q_{\rm top}$, with $\vec Q_{\rm Warsaw},\,\vec Q_{\rm top}$ the two ``operator vectors'' and $R$ a rotation matrix.
The relation among the Wilson coefficients is then derived equating the Lagrangian written in the two bases:
\begin{equation}
\begin{aligned}
\Lag &= \vec C_{\rm Warsaw}\cdot  \vec Q_{\rm Warsaw} = 
(\vec C_{\rm Warsaw})^T\,  R \, \vec Q_{\rm top}\\
&=  \vec C_{\rm top} \cdot\vec Q_{\rm top} \,,
\end{aligned}
\end{equation} 
with $\vec C_{\rm Warsaw},\,\vec C_{\rm top}$ the coefficients vectors. The solution is
\begin{equation}
\vec C_{\rm top} = R^T \,\vec C_{\rm Warsaw}\,,\qquad
\vec C_{\rm Warsaw} = (R^T)^{-1} \,\vec C_{\rm top} \,. 
\end{equation} 
Explicitly:
\begin{align}
C_{QQ}^{(1)}&= C_{QQ}^{(1)} - \frac{1}{3}C_{QQ}^{(3)}\,,
&
C_{QQ}^{(8)}&= 4 C_{QQ}^{(3)}\,,
\\[3mm]
C_{qq}^{(1,1)} &= C_{qq}^{(1)} +\frac16 C_{qq}^{(1)\prime} +\frac12 C_{qq}^{(3)\prime}\,,
&
C_{qq}^{(3,1)} &= C_{qq}^{(3)} +\frac16 C_{qq}^{(1)\prime} -\frac16 C_{qq}^{(3)\prime}\,,
\\
C_{qq}^{(1,8)} &= C_{qq}^{(1)\prime} +3 C_{qq}^{(3)\prime}\,,
&
C_{qq}^{(3,8)} &= C_{qq}^{(1)\prime} -C_{qq}^{(3)\prime}\,,
\\[3mm]
C_{Qq}^{(1,1)} &= C_{Qq}^{(1)} +\frac16 C_{Qq}^{(1)\prime} +\frac12 C_{Qq}^{(3)\prime}\,,
&
C_{Qq}^{(3,1)} &= C_{Qq}^{(3)} +\frac16 C_{Qq}^{(1)\prime} -\frac16 C_{Qq}^{(3)\prime}\,,
\\
C_{Qq}^{(1,8)} &= C_{Qq}^{(1)\prime} +3 C_{Qq}^{(3)\prime} \,,
&
C_{Qq}^{(3,8)} &= C_{Qq}^{(1)\prime} -C_{Qq}^{(3)\prime} \,,
\end{align}
and the inverse
\begin{align}
C_{QQ}^{(1)}&= C_{QQ}^{(1)} + \frac{1}{12}C_{QQ}^{(8)}\,,
&
C_{QQ}^{(3)}&= \frac14 C_{QQ}^{(8)}\,,
\\[3mm]
C_{qq}^{(1)} &= C_{qq}^{(1,1)}-\frac16 C_{qq}^{(1,8)}
\,,
&
C_{qq}^{(1)\prime} &= \frac14 C_{qq}^{(1,8)} +\frac34 C_{qq}^{(3,8)}\,,
\\
C_{qq}^{(3)} &=  C_{qq}^{(3,1)} -\frac16 C_{qq}^{(3,8)}\,,
&
C_{qq}^{(3)\prime} &= \frac14 C_{qq}^{(1,8)} -\frac14 C_{qq}^{(3,8)}\,,
\\[3mm]
C_{Qq}^{(1)} &= C_{Qq}^{(1,1)}-\frac16 C_{Qq}^{(1,8)}
\,,
&
C_{Qq}^{(1)\prime} &= \frac14 C_{Qq}^{(1,8)} +\frac34 C_{Qq}^{(3,8)}\,,
\\
C_{Qq}^{(3)} &=  C_{Qq}^{(3,1)} -\frac16 C_{Qq}^{(3,8)}\,,
&
C_{Qq}^{(3)\prime} &= \frac14 C_{Qq}^{(1,8)} -\frac14 C_{Qq}^{(3,8)}\,.
\end{align}
The operator $Q_{ll}$ admits 2 independent contractions in the $U(3)^2$ symmetric case (Eq.~\ref{eq.cll_cll1}) and 9 in the $U(1)_{l+e}^3$ case. We choose them as the $(C_{ll})_{prst}$ entries with indices $prst$ in the set
\begin{equation}\label{eq.Pll}
P_{ll}=  \{1111, 2222, 3333, 1122, 1133, 2233, 1221, 1331, 2332\}\,.                                                                    \end{equation} 
Operators $Q_{lq}^{(1),(3)}, Q_{lQ}^{(1),(3)}$ admit 3 (1) contractions each in the \top\ (\topsl) case.\\
The Lagrangian is therefore
\begin{align}
\Lambda^2 \Lag_6^{(8a)} &=
C_{qq}^{(1,1)} Q_{qq}^{(1,1)} + C_{qq}^{(1,8)} Q_{qq}^{(1,8)} + C_{qq}^{(3,1)} Q_{qq}^{(3,1)} + C_{qq}^{(3,8)} Q_{qq}^{(3,8)} + C_{QQ}^{(1)} Q_{QQ}^{(1)}
\nn\\
&+ C_{Qq}^{(1,1)} Q_{Qq}^{(1,1)} + C_{Qq}^{(1,8)} Q_{Qq}^{(1,8)} + C_{Qq}^{(3,1)} Q_{Qq}^{(3,1)} + C_{Qq}^{(3,8)} Q_{Qq}^{(3,8)} + C_{QQ}^{(8)} Q_{QQ}^{(8)}
\\[2mm]
&+\begin{cases}
\sum_{p=1}^3 (C_{lq}^{(1)})_{pp}  Q_{lq,pp}^{(1)}
+(C_{lq}^{(3)})_{pp}  Q_{lq,pp}^{(3)}
+(C_{lQ}^{(1)})_{pp}  Q_{lQ,pp}^{(1)}
+(C_{lQ}^{(3)})_{pp}  Q_{lQ,pp}^{(3)}\\
+\sum_{prst\in P_{ll}}\, (C_{ll})_{prst}\, Q_{ll,prst} \,,
\hspace*{7cm} U(1)_{l+e}^3\;[\top]
\\[3mm]
\sum_{p,r=1}^3 C_{lq}^{(1)} \,\d_{pr}\, Q_{lq,pr}^{(1)} 
+C_{lq}^{(3)} \,\d_{pr}\, Q_{lq,pr}^{(3)} 
+C_{lQ}^{(1)} \,\d_{pr}\, Q_{lQ,pr}^{(1)} 
+C_{lQ}^{(3)} \,\d_{pr}\, Q_{lQ,pr}^{(3)} \\
+\sum_{p,r,s,t=1}^3 \left(C_{ll}\, \d_{pr} \d_{st} + C_{ll}' \,\d_{pt}\d_{sr}\right) Q_{ll,prst} \,,
\hspace*{4.2cm}\quad U(3)^2\quad[\topsl]
\end{cases}
\nn
\end{align}
Note that the allowed flavor contractions in the $U(1)_{l+r}^3$ and $U(3)^2$ cases are the same, but the different symmetry properties generally lead to different relative normalizations. For instance, considering the (1111) and (1122) entries, one has
\begin{equation}
\Lambda^2 \Lag_6^{(8a)} \supset 
\begin{cases}
(C_{ll})_{1111} Q_{1111} + (C_{ll})_{1122} Q_{1122} +\dots
&\quad U(1)_{l+e}^3\;[\top]
\\
\left(C_{ll}+C_{ll}'\right)\, Q_{1111} + 2C_{ll}\, Q_{1122} +\dots
&
\quad U(3)^2\quad[\topsl]
\end{cases}
\end{equation}
where the relative 2 between the $C_{ll}$ contributions to $Q_{1122}$ and $Q_{1111}$ is due to $U(3)^2$ requiring to sum over both the 1122 and 2211 contractions, that are equivalent for this particular operator. 
In total, $\Lag_6^{(8a)}$ contains 31 independent real parameters in the \top\ case and 16 in the \topsl\ case.

\paragraph{Class 8 b.}
A basis rotation analogous to the one performed in $\Lag_6^{(8a)}$ is applied to $Q_{uu},Q_{ud}$ in $\Lag^{(8b)}_6$. No modification is needed for $Q_{ud}^{(1),(8)}$ as in this case the color octet contraction is already manifest.
The set of 5 independent $U(2)^3$-invariant contractions in the Warsaw basis is in this case
\begin{align}
Q_{uu} &= \sum_{p,r,s,t=1}^2 \d_{pr}\d_{st} \, (\bar u_p \g_\mu u_r)(\bar u_s\g^\mu u_t) \,,
&
Q_{tu} &= \sum_{p,r=1}^2 \d_{pr} \, (\bar u_p \g_\mu u_r)(\bar t\g^\mu t) \,,
\\
Q_{uu}^{\prime} &= \sum_{p,r,s,t=1}^2 \d_{pt}\d_{sr} \, (\bar u_p \g_\mu u_r)(\bar u_s\g^\mu u_t) \,,
&
Q_{tu}^{\prime} &= \sum_{p,t=1}^2 \d_{pt} \, (\bar u_p \g_\mu t)(\bar t\g^\mu u_t) \,,
\\
Q_{tt} &= (\bar t \g_\mu t)(\bar t\g^\mu t) \,,
\end{align}
and analogously for the $Q_{dd}$ counterparts. Using Fierz transformations and Eqs.~\eqref{fierz.su2},\eqref{fierz.su3}: 
\begin{align}
Q_{tt} &= Q_{tt}\,,
&
Q_{bb} &= Q_{bb}\,,
\\
Q_{uu} &= Q_{uu}^{(1)}\,,
&
Q_{uu}^{\prime} &= \frac13 Q_{uu}^{(1)} + 2 Q_{uu}^{(8)}\,,
\\
Q_{dd} &= Q_{dd}^{(1)}\,,
&
Q_{dd}^{\prime} &= \frac13 Q_{dd}^{(1)} + 2 Q_{dd}^{(8)}\,,
\\
Q_{tu} &= Q_{tu}^{(1)}\,,
&
Q_{tu}^{\prime} &= \frac13 Q_{tu}^{(1)} + 2 Q_{tu}^{(8)}\,,
\\
Q_{bd} &= Q_{bd}^{(1)}\,,
&
Q_{bd}^{\prime} &= \frac13 Q_{bd}^{(1)} + 2 Q_{bd}^{(8)}\,,
\end{align}
where the operators on the right-hand side of the equations are defined in Table~\ref{tab.top_basis}.
The relations among Wilson coefficients are
\begin{align}
C_{tt} &=   C_{tt}\,,
&
C_{bb} &=   C_{bb}\,,
\\
C_{uu}^{(1)} &=  C_{uu}+\frac13 C_{uu}'\,,
&
C_{uu}^{(8)} &= 2 C_{uu}' \,,
\\
C_{dd}^{(1)} &= C_{dd}+\frac13 C_{dd}'  \,,
&
C_{dd}^{(8)} &= 2 C_{dd}'\,,
\\
C_{tu}^{(1)} &= C_{tu}+\frac13 C_{tu}' \,,
&
C_{tu}^{(8)} &= 2 C_{tu}' \,,
\\
C_{bd}^{(1)} &= C_{bd}+\frac13 C_{bd}' \,,
&
C_{bd}^{(8)} &= 2 C_{bd}' \,,
\end{align}
and the inverse
\begin{align}
C_{tt} &=   C_{tt}\,,
&
C_{bb} &=   C_{bb}\,,
\\
C_{uu} &=  C_{uu}^{(1)}-\frac16 C_{uu}^{(8)}\,,
&
C_{uu}^{\prime} &= \frac12 C_{uu}^{(8)} \,,
\\
C_{dd} &= C_{dd}^{(1)}-\frac16 C_{dd}^{(8)}  \,,
&
C_{dd}^{\prime} &=  \frac12 C_{dd}^{(8)}\,,
\\
C_{tu} &= C_{tu}^{(1)}-\frac16 C_{tu}^{(8)} \,,
&
C_{tu}^{\prime} &= \frac12 C_{tu}^{(8)} \,,
\\
C_{bd} &= C_{bd}^{(1)}-\frac16 C_{bd}^{(1)} \,,
&
C_{bd}^{\prime} &= \frac12 C_{bd}^{(8)} \,.
\end{align}
The operator $Q_{ee}$ admits 6 independent contractions in the \top\ case, with indices that we choose in the set
\begin{equation}\label{eq.Pee}
P_{ee}=  \{1111, 2222, 3333, 1122, 1133, 2233\}\,.                                                     \end{equation}
In the $U(3)^2$ case, there is instead only 1 available contraction.
Each of the operators $Q_{lu}^{(1),(3)},Q_{ld}^{(1),(3)}$ is mapped into 6 (2) independent invariants in the \top\ (\topsl) case. The Lagrangian for class 8a has the form
\begin{align}
\Lambda^2 \Lag_6^{(8b)} &=
C_{uu}^{(1)} Q_{uu}^{(1)} + C_{uu}^{(8)} Q_{uu}^{(8)}  + C_{tt} Q_{tt}
+C_{tu}^{(1)} Q_{tu}^{(1)} + C_{tu}^{(8)} Q_{tu}^{(8)}
\nn\\
&+ C_{dd}^{(1)} Q_{dd}^{(1)} + C_{dd}^{(8)} Q_{dd}^{(8)} + C_{bb} Q_{bb}
+C_{bd}^{(1)} Q_{bd}^{(1)} + C_{bd}^{(8)} Q_{bd}^{(8)}
\nn\\
&+C_{ud}^{(1)} Q_{ud}^{(1)} + C_{ud}^{(8)} Q_{ud}^{(8)} 
+C_{td}^{(1)} Q_{td}^{(1)} + C_{td}^{(8)} Q_{td}^{(8)}
+C_{ub}^{(1)} Q_{ub}^{(1)} + C_{ub}^{(8)} Q_{ub}^{(8)}
+C_{tb}^{(1)} Q_{tb}^{(1)} + C_{tb}^{(8)} Q_{tb}^{(8)}
\nn\\
&+\left[
 C_{utbd}^{(1)} Q_{utbd}^{(1)}
+C_{utbd}^{(8)} Q_{utbd}^{(8)}
+\hc\right]
\\[2mm]
&+\begin{cases}
\sum_{p=1}^3 
 (C_{eu})_{pp}  Q_{eu,pp}
+(C_{ed})_{pp}  Q_{ed,pp}
+(C_{et})_{pp}  Q_{et,pp}
+(C_{eb})_{pp}  Q_{eb,pp}\\
+\sum_{prst\in P_{ee}}\, (C_{ee})_{prst}\, Q_{ee,prst}\,, 
\hspace{6.5cm} U(1)_{l+e}^3\;[\top]
\\[3mm]
\sum_{p,r=1}^3 
 C_{eu}  \,\d_{pr}\,   Q_{eu,pr}
+C_{ed}  \,\d_{pr}\,   Q_{ed,pr}
+C_{et}  \,\d_{pr}\,   Q_{et,pr}
+C_{eb}  \,\d_{pr}\,   Q_{eb,pr}\\
+\sum_{p,r,s,t=1}^3 C_{ee}\, \d_{pr} \d_{st}  Q_{ee,prst}\,, 
\hspace*{6.5cm}U(3)^2\quad[\topsl]
\end{cases}
\nn
\end{align}
and it depends on 40 (27) real independent parameters in the \top\ (\topsl) case.

\paragraph{Class 8 c.}
No basis rotation is required in $\Lag_6^{(8c)}$, and the quark currents are mapped directly. 
In the lepton sector, $Q_{le}$ admits 1 independent contraction in the $U(3)^2$ case (neglecting the subleading contribution $\propto Y_l^2$) and 12 in the $U(1)_{l+e}^3$ case. We choose those with indices $prst$ in the set
\begin{align}\label{eq.Ple}
P_{le} &= P_{le}^h \cup P_{le}^{nh}\,,\\
P_{le}^h &= \{1111,2222,3333,1122,1133,2233,2211,3311,3322\}\,,\quad
P_{le}^{nh} = \{1221,1331,2332\}\,,                                                                        
\nn
\end{align}
where the contractions in $P_{le}^h$ are Hermitian and those in $P_{le}^{nh}$ are not.
The operators $Q_{qQtu}^{(1),(8)},\,Q_{qQbd}^{(1),(8)}$ are not Hermitian and therefore the associated Wilson coefficients are complex.

The Lagrangian reads
\begin{align}
\Lambda^2 \Lag_6^{(8c)} &=
 C_{qu}^{(1)} Q_{qu}^{(1)} 
+C_{qt}^{(1)} Q_{qt}^{(1)} 
+C_{Qu}^{(1)} Q_{Qu}^{(1)} 
+C_{Qt}^{(1)} Q_{Qt}^{(1)}
\nn\\
&
+C_{qu}^{(8)} Q_{qu}^{(8)} 
+C_{qt}^{(8)} Q_{qt}^{(8)} 
+C_{Qu}^{(8)} Q_{Qu}^{(8)} 
+C_{Qt}^{(8)} Q_{Qt}^{(8)} 
\nn\\
&
+C_{qd}^{(1)} Q_{qd}^{(1)} 
+C_{qb}^{(1)} Q_{qb}^{(1)} 
+C_{Qd}^{(1)} Q_{Qd}^{(1)} 
+C_{Qb}^{(1)} Q_{Qb}^{(1)} 
\\
&
+C_{qd}^{(8)} Q_{qd}^{(8)} 
+C_{qb}^{(8)} Q_{qb}^{(8)} 
+C_{Qd}^{(8)} Q_{Qd}^{(8)} 
+C_{Qb}^{(8)} Q_{Qb}^{(8)} 
\nn\\
&
+\left[
 C_{qQtu}^{(1)} Q_{qQtu}^{(1)}+C_{qQtu}^{(8)} Q_{qQtu}^{(8)}
+C_{qQbd}^{(1)} Q_{qQbd}^{(1)}+C_{qQbd}^{(8)} Q_{qQbd}^{(8)}
+\hc\right]
\nn\\[2mm]
&+\begin{cases}
\sum_{p=1}^3 
 (C_{lu})_{pp}  Q_{lu,pp}
+(C_{ld})_{pp}  Q_{ld,pp}
+(C_{lt})_{pp}  Q_{lt,pp}
+(C_{lb})_{pp}  Q_{lb,pp}\\
\qquad 
+(C_{qe})_{pp}  Q_{qe,pp}
+(C_{Qe})_{pp}  Q_{Qe,pp}
+\sum_{prst\in P_{le}^h}\, (C_{le})_{prst}\, Q_{le,prst} 
& U(1)_{l+e}^3\;[\top]
\\
\qquad
+\left[\sum_{prst\in P_{le}^{nh}}\, (C_{le})_{prst}\, Q_{le,prst} + \hc\right]\,,
\\[3mm]
\sum_{p,r=1}^3 
 C_{lu}  \,\d_{pr}\,   Q_{lu,pr}
+C_{ld}  \,\d_{pr}\,   Q_{ld,pr}
+C_{lt}  \,\d_{pr}\,   Q_{lt,pr}
+C_{lb}  \,\d_{pr}\,   Q_{lb,pr}\\
\qquad
+C_{qe}  \,\d_{pr}\,   Q_{qe,pr}
+C_{Qe}  \,\d_{pr}\,   Q_{Qe,pr}
+\sum_{p,r,s,t=1}^3 C_{le}\, \d_{pr} \d_{st}  Q_{le,prst}\,, 
&U(3)^2\quad[\topsl]
\end{cases}
\nn
\end{align}
and it depends on 54 (31) independent real parameters in the \top\ (\topsl) case.

\paragraph{Class 8 d.}
Finally, the operators in $\Lag_6^{(8d)}$ are also mapped directly to the notation with 6 quark fields. $U(2)^3$ invariance requires an insertion of a light Yukawa couplings for each $(\bar q u)$ or $(\bar q d)$ current and an insertion of $Y_l$ for each $(\bar le)$ current, as indicated in Table~\ref{tab.top_basis}.

This class includes a total of 64 (40) real parameters in the \top\ (\topsl) case:
\begin{align}
\Lambda^2 \Lag_6^{(8d)} &=
 C_{quqd}^{(1)} Q_{quqd}^{(1)} 
+C_{quqd}^{(8)} Q_{quqd}^{(8)} 
+C_{quqd}^{(1)\prime} Q_{quqd}^{(1)\prime} 
+C_{quqd}^{(8)\prime} Q_{quqd}^{(8)\prime} 
+C_{QtQb}^{(1)} Q_{Qtqb}^{(1)} 
+C_{QtQb}^{(8)} Q_{Qtqb}^{(8)} 
\nn\\
&
+C_{Qtqd}^{(1)} Q_{Qtqb}^{(1)} 
+C_{quQb}^{(1)} Q_{quQb}^{(1)} 
+C_{Quqb}^{(1)} Q_{Quqb}^{(1)} 
+C_{qtQd}^{(1)} Q_{qtQd}^{(1)} 
\\
&
+C_{Qtqd}^{(8)} Q_{Qtqb}^{(8)} 
+C_{quQb}^{(8)} Q_{quQb}^{(8)} 
+C_{Quqb}^{(8)} Q_{Quqb}^{(8)} 
+C_{qtQd}^{(8)} Q_{qtQd}^{(8)} 
\nn
\\[2mm]
&+\begin{cases}
\sum_{p=1}^3 
 (C_{lequ}^{(1)})_{pp}  Q_{lequ,pp}^{(1)}
+(C_{leQt}^{(1)})_{pp}  Q_{leQt,pp}^{(1)}
+(C_{lequ}^{(3)})_{pp}  Q_{lequ,pp}^{(3)}
\\
\quad
+(C_{leQt}^{(3)})_{pp}  Q_{leQt,pp}^{(3)}
+ (C_{ledq})_{pp}  Q_{ledq,pp}
+(C_{lebQ})_{pp}  Q_{lebQ,pp}\,,
\hspace*{3cm} U(1)_{l+e}^3\;[\top]
\\[3mm]
\sum_{p,r=1}^3 
 C_{lequ}^{(1)} \, (Y_l^\dag)_{pr}\, Q_{lequ,pr}^{(1)}
+C_{leQt}^{(1)} \, (Y_l^\dag)_{pr}\, Q_{leQt,pr}^{(1)}
+C_{lequ}^{(3)} \, (Y_l^\dag)_{pr}\, Q_{lequ,pr}^{(3)}
\\
\quad
+C_{leQt}^{(3)} \, (Y_l^\dag)_{pr}\, Q_{leQt,pr}^{(3)}
+C_{ledq}\, (Y_l^\dag)_{pr}\,  Q_{ledq,pr}
+C_{lebQ}\, (Y_l^\dag)_{pr}\,  Q_{lebQ,pr}\,,
\hspace*{1.2cm} U(3)^2\quad[\topsl]
\end{cases}
\nn\\
&+\hc
\nn
\end{align}

\subsection{Comparison with the literature}\label{sec.flavor_comparison}
We conclude this section with a comparison of the parameterizations presented in this section with other recent results in the literature. 
As a quantitative reference, Table~\ref{tab.counting} summarizes the number of independent real parameters for each class of $\Lag_6$ operators and flavor setup.

\begin{table}[t]\centering  
 \renewcommand{\arraystretch}{1.2}
\begin{tabular}{|c*5{|cc}|}
\toprule
&\multicolumn{2}{p{2.2cm}|}{\centering\general} & 
 \multicolumn{2}{p{2.2cm}|}{\centering\Utf} &
 \multicolumn{2}{p{2.2cm}|}{\centering\MFV} & 
 \multicolumn{2}{p{2.2cm}|}{\centering\top}& 
 \multicolumn{2}{p{2.2cm}|}{\centering\topsl}\\\cline{2-11}
 
 & all & \cancel{CP}&  all & \cancel{CP}& all & \cancel{CP}&all & \cancel{CP}&all & \cancel{CP}\\\midrule 
 
$\Lag_6^{(1)}$& 
4& 2&
4& 2&
2& -&
4& 2&
4& 2
\\
$\Lag_6^{(2,3)}$& 
3& -&
3& -&
3& -&
3& -&
3& -
\\
$\Lag_6^{(4)}$& 
8& 4&
8& 4&
4& -&
8& 4&
8& 4
\\\midrule
$\Lag_6^{(5)}$& 
54& 27&
6 & 3&
7 & -&
14& 7&
10& 5
\\
$\Lag_6^{(6)}$& 
144& 72&
16 & 8&
20 & -&
36 & 18&
28 & 14
\\
$\Lag_6^{(7)}$& 
81& 30&
9 & 1&
14& -&
21& 2&
15& 2
\\
$\Lag_6^{(8a)}$& 
297& 126&
8 & -&
10& -&
31& -&
16& -
\\
$\Lag_6^{(8b)}$& 
450& 195&
9&  -&
19& -&
40& 2&
27& 2
\\
$\Lag_6^{(8c)}$& 
648& 288&
8&  -&
28& -&
54& 4&
31& 4
\\
$\Lag_6^{(8d)}$& 
810& 405&
14& 7&
13& -&
64& 32&
40& 20
\\\midrule
tot& 
2499& 1149&
85&  25&
120& -&
275& 71&
182& 53
\\
\bottomrule
\end{tabular}
\caption{Number of independent real parameters in each class of dimension 6 operators, for the 5 flavor structures implemented in \smeftsim.}\label{tab.counting}
\end{table}

\vskip 2em
Compared to previous versions of \smeftsim~\cite{Brivio:2017btx}, the following changes were made:
\begin{itemize}
\item the dependence on the CKM matrix in currents involving left-handed down quarks was neglected in the effective operators defined in the \general\ and \MFV\ versions. It has been restored in  version~3.0.

\item four parameters corresponding to the real and imaginary parts of $C_{quqd}^{(1)\prime}\,$, $C_{quqd}^{(8)\prime}$ were missing in the \Utf\ and \MFV\ models, and have now been included. 

\item the \MFV\ models have been modified: all Yukawas are now retained in the spurions, instead of only $y_t,y_b$. Moreover, the Lagrangian is now organized according to a power counting in the quark Yukawas, that led to some flavor-violating terms (eg. $\Delta^u C_{Hud},\Delta_1^u C_{quqd}^{(1)}\dots$) being dropped, and others (eg. $\Delta^d C_{uH}, C_{qu}^{(1)\prime}\dots$) being added.

\item versions \top\ and \topsl\ are new in version~3.0.
\end{itemize}

The \Utf\ and \MFV\ models can be compared, for instance, to the $U(3)^5$ spurion analyses presented in Refs.~\cite{Bordone:2019uzc,Faroughy:2020ina}.
Ref.~\cite{Bordone:2019uzc} contains an exhaustive classification of all the flavor spurions associated with SM fermion currents in the presence of a $U(3)^5$ symmetry. In their notation, $S^u, S^d$ correspond to $\Delta_U,\Delta_D$ respectively, while both $S^{qu},S^{qd}$ are mapped to $\Delta_Q$. The structure $Y_uY_d^\dag$ corresponds to $\Delta_{UD}$ and, since we only retain linear insertions of $Y_l$, $\Delta_L=\Delta_E=0$ in \smeftsim.
Any other spurion leads to baryon  and/or lepton number non-conservation, and therefore does not have an equivalent in the Lagrangian considered here.

Ref.~\cite{Faroughy:2020ina} presented a detailed classification of all the $U(3)^5$ and $U(2)^5$ invariant structures in the Warsaw basis. In the $U(3)^5$ case, their results can be directly compared with the parameterizations of the \Utf\ and \MFV\ models in \smeftsim, while the $U(2)^3$ case can be compared (in the quark sector) to the \top\ and \topsl\ models. We find complete agreement in the characterization of the structures, and the operator countings are consistent once a few differences in the organization of the invariants are taken into account:
\begin{itemize}
\item the $U(3)^5$ and $U(2)^5$ Lagrangians in Ref.~\cite{Faroughy:2020ina} are organized according to a power counting in the Yukawas, while for the \Utf, \top\ and \topsl\ models in \smeftsim\ we simply choose to retain the leading invariant for each operator in the Warsaw basis.

\item in the \MFV\ models we retain terms up to order $(Y_l^1, (Y_d+Y_u)^3)$. This choice is different from the power counting in Ref.~\cite{Faroughy:2020ina}, that truncates at  ($Y_l^1,Y_d^1 Y_u^2$).

\item the Lagrangian of the \MFV\ models includes spurions $\propto Y_d^2$, that were neglected in Ref.~\cite{Faroughy:2020ina}.

\item in the $U(2)$ case, different symmetries were chosen for the lepton sector: $U(2)^3$ in Ref.~\cite{Faroughy:2020ina} vs $U(1)_{l+e}^3$ and $U(3)^2$ in the \top\ and \topsl\ models.
\end{itemize}

The structure of the \top\ and \topsl\ versions builds upon those of Refs.~\cite{AguilarSaavedra:2018nen,Hartland:2019bjb,Brivio:2019ius}.
The main difference compared to these works is that in \smeftsim\ the parameterization has been systematically extended to all operators of the Warsaw basis, including at the same time CP violating terms, interactions that do not involve the top quark, and spurion insertions of the light quark Yukawas.

\section{Input parameters}\label{sec.inputs}
Once the kinetic terms have been canonically normalized and the flavor structure has been fixed, the Lagrangian parameters can be assigned numerical values, with a procedure that is sometimes referred to as ``fixing an input parameter scheme'' or ``finite renormalization''.
This section revisits this procedure in the SM and in the SMEFT, using a general formalism that accounts for terms up to arbitrary EFT order. They can be applied to both tree level and loop calculations but, in the latter case, this procedure needs to be combined with the usual renormalization to reabsorb UV divergences. 
In Sections~\ref{sec.inputs_ew},~\ref{sec.inputs_fermions} these formulas are applied to the Warsaw basis case, to recover the known tree-level results, see eg.~\cite{Alonso:2013hga,Berthier:2015oma,Berthier:2015gja,Brivio:2017bnu,Brivio:2017vri,Brivio:2017btx,Dawson:2018dxp}. Aspects specific to the NLO case have been discussed in~\cite{Ghezzi:2015vva,Gauld:2015lmb,Passarino:2016pzb,Hartmann:2016pil,Dawson:2018pyl,Cullen:2020zof}.

\vskip 1em
The Lagrangian parameters are fixed imposing a set of defining conditions that relate them to  (pseudo-)observables: for a Lagrangian with $N$ independent parameters $ g = \{g_i\dots g_N\}$, $M\geq N$~independent input observables $\O = \{\O_1\dots \O_M\}$  need to be selected. Computing each $\O_n$ in the theory at a chosen perturbative order, one obtains relations 
\begin{equation}\label{eq.renorm_conditions}
\O_n = F^{(0)}_n \left( g\right)\,,\quad n=1\dots M\,,
\end{equation} 
where $F^{(0)}_n$ denotes a function of the parameters $g$.
If $M=N$ and~\eqref{eq.renorm_conditions} is an invertible system of equations, the solution
\begin{equation}\label{eq.renorm_conditions_sol}
g_i = K_i^{(0)}(\O)\,,\quad i=1\dots N\,,
\end{equation} 
fixes $g_i$ as a function $K_i^{(0)}$ of $\O$. The numerical values of the parameters $g$ are then univocally determined by measurements of $\O$.

In the SM case, one has 19 independent parameters, that we can classify as
\begin{equation}\label{eq.SM_pars}
\begin{aligned}
&\a_s,\bar\theta,
&&
\text{QCD}\\
&g_1, g_W, v,\lambda,
&&
\text{EW+Higgs}
\\
&y_e, y_\mu, y_\tau, y_u, y_c, y_t, y_d, y_s, y_b,
&&
\text{Yukawas}
\\
&\theta_{12},\theta_{13},\theta_{23},\d\,,
&& 
\text{CKM}
\end{aligned} 
\end{equation} 
where we have introduced the CP-violating $\theta$ angle of QCD ($\bar\theta$) and the CKM angles and CP phase ($\theta_{12},\theta_{13},\theta_{23}$ and $\delta$ respectively).
The procedure outlined above is most often employed to determine the value of  $\bar\theta$ and of the EW+Higgs and Yukawa parameters. On the other hand, the determination of the CKM parameters and of $\a_s$ usually relies on a large number of observables: in these cases, the system~\eqref{eq.renorm_conditions} is not invertible and the parameters' values are extracted via a global fit.

When transitioning from the SM to the SMEFT, a large number of additional parameters enters the Lagrangian, namely the cutoff $\Lambda$ and the Wilson coefficients $C_\a$. Fixing their numerical values in terms of measured observables is obviously still an open challenge (and indeed the ultimate goal of the present work), so these quantities are necessarily left free in the Lagrangian. 
Nevertheless, they play a role in  the finite renormalization procedure, because the observables $\O$ employed to fix the SM quantities generically receive contributions from higher dimensional operators. Working order by order in the EFT expansion, the relations in~\eqref{eq.renorm_conditions} are modified into\footnote{Terms suppressed by odd powers of $\Lambda$ are omitted here, as they typically contribute to $B$ and/or $L$ violating observables, that are not relevant for the extraction of SM parameters. Nevertheless, the results derived in this section directly generalize to the case where these contributions are retained.}
\begin{equation}\label{eq.renorm_conditions_SMEFT}
\O_n = F_n^{(0)}(g) + \frac{1}{\Lambda^2}F_n^{(2)}\left( g, C\right) + \frac{1}{\Lambda^4}F_n^{(4)}(g, C) + \dots
\end{equation} 
where  $C$ here generically represents the set of relevant Wilson coefficients, that can be associated to operators of any dimension.
In cases where the system of Eq.~\eqref{eq.renorm_conditions} can be inverted, \eqref{eq.renorm_conditions_SMEFT} can also be solved expanding around the SM solution. The result has the general form:
\begin{equation}\label{eq.renorm_conditions_sol_SMEFT}
g_i = K_i^{(0)}(\O) + \frac{1}{\Lambda^2}K^{(2)}_i (\O,C) + \frac{1}{\Lambda^4}K^{(4)}_i (\O,C)+\dots
\end{equation} 
where $K_i^{(0)}(\O)$ is the SM solution and the following $K$ terms are SMEFT corrections that depend on the Wilson coefficients.
The leading term in the solution~\eqref{eq.renorm_conditions_sol_SMEFT} is defined imposing that the SM relation holds:
\begin{equation}
\O_n  \equiv F_n^{(0)}(K^{(0)}(\O))\,.
\end{equation} 
The explicit form of the remaining $K$ terms is found inserting Eq.~\eqref{eq.renorm_conditions_sol_SMEFT} into~\eqref{eq.renorm_conditions_SMEFT}, expanding in $\Lambda$ and requiring that SMEFT corrections cancel order by order  in the resulting expression. Iteratively, one finds
\begin{align}\label{eq.theta6}
K^{(2)}_i &= - (J^{-1})_{in} \;F_n^{(2)}\,,
\\
K^{(4)}_i &= - (J^{-1})_{in}\,\left[F_n^{(4)} + 
\frac{\de F^{(2)}_n}{\de g_k} K_k^{(2)}
+ \frac{1}{2}\frac{\de^2F_n^{(0)}}{\de g_k \de g_j}
K_k^{(2)} \;K_j^{(2)}\right]\,,
\label{eq.theta8}
\\ & \vdots\nn
\\
K^{(d)}_i &= - (J^{-1})_{in} \left[
F_n^{(d)} +\sum_{\substack{m<d\\m+d_1+\dots d_D=d}} \frac{1}{D!}\frac{\de^D F_n^{(m)}}{\de g_{i_1}\cdots \de g_{i_D}} K_{i_1}^{(d_1)}\cdots K_{i_D}^{(d_D)}
\right]\,,
\label{eq.thetaD}
\end{align}
 where $ (J^{-1})_{in} = \de g_i/\de F_n^{(0)}$ is the inverse of the Jacobian matrix
 \begin{equation}\label{eq.Jacobian}
 J_{ni}= \frac{\de F_n^{(0)}}{\de g_i}\,,
 \end{equation} 
and in Eq.~\eqref{eq.thetaD} the sum runs over all possible terms with $m<d$ and such that $m+d_1+\dots+d_D=d$. 
All functions and derivatives appearing explicitly in Eqs~\eqref{eq.theta6}-\eqref{eq.thetaD} are evaluated at the SM solution for the parameters ${g\equiv K^{(0)}(\O)}$ and the indices $n,k,j,i_{1}\dots i_{D}$ are implicitly contracted internally and summed over.

A generic \emph{predicted} observable $\P$ inherits a dependence on the corrections $F_n^{(d\geq 2)}$ to the input quantities. Analogous to $\O$, $\P$ will have the generic form
\begin{equation}\label{eq.predicted_obs_def}
\P  = P^{(0)}(g) + \frac{1}{\Lambda^2}P^{(2)}(g,C) + \frac{1}{\Lambda^4}P^{(4)}(g,C)+\dots
\end{equation} 
where $P^{(0)}$ is the SM expression and $P^{(d\geq 2)}$ encode direct EFT contributions to $\P$, induced by effective operators entering the relevant Feynman diagrams. Calculating $\P$ in the SMEFT starting from input quantities $\O$ means inserting the expressions of $g$ in Eq.~\eqref{eq.renorm_conditions_sol_SMEFT} into Eq.~\eqref{eq.predicted_obs_def}. This operation introduces ``indirect'' EFT contributions, that are a direct consequence of the $F_n^{(d\geq 2)}$ terms in Eq.~\eqref{eq.renorm_conditions_SMEFT}. The dependence on the latter quantities can be made explicit:
\begin{equation}\label{eq.predicted_obs}
\P = P^{(0)}+\frac{1}{\Lambda^2}\left[P^{(2)} - A_n F_n^{(2)}\right]
+\frac{1}{\Lambda^4}\left[P^{(4)}  - A_n F_n^{(4)}
- A^{(2)}_n F_n^{(2)} + B_{mn} F_m^{(2)} F_n^{(2)}\right]+\dots
\end{equation} 
where the $m,n$ indices are summed over and, as above, all functions are implicitly evaluated at $g\equiv K^{(0)}(\O)$. The coefficients $A,B$ are found via chain differentiation:
\begin{align}\label{eq.inputs_An}
A_n &= \frac{\de P^{(0)}}{\de F_n^{(0)}} = 
\frac{\de P^{(0)}}{\de g_i} (J^{-1})_{in}
\,,
\\
\label{eq.inputs_A6n}
A^{(2)}_n & = \frac{\de P^{(2)}}{\de F_n^{(0)}}-
\frac{\de P^{(0)}}{\de F_m^{(0)}}\frac{\de F_m^{(2)}}{\de F_n^{(0)}} 
=
\frac{\de P^{(2)}}{\de g_i} (J^{-1})_{in}
-A_m \frac{\de F_m^{(2)}}{\de g_k}(J^{-1})_{kn}
\,,
\\
\label{eq.inputs_Bmn}
B_{mn} &=
\frac12 \frac{\de^2P^{(0)}}{\de F_m^{(0)} \de F_n^{(0)}} -\frac{1}{2}\frac{\de P^{(0)}}{\de F_p^{(0)}}\frac{\de^2F_p^{(0)}}{\de F_m^{(0)}\de F_n^{(0)}}
=
\nn\\
&= \frac12\frac{\de^2 P^{(0)}}{\de g_i \de g_j} (J^{-1})_{im}(J^{-1})_{jn}
-
\frac12 A_p \frac{\de ^2 F_p^{(0)}}{\de g_i \de g_j} (J^{-1})_{im}(J^{-1})_{jn}
\,.
\end{align}
Here $A_n$ and the first term in $A_n^{(2)}$ account for linear $K^{(2)}$ corrections to $g$ in the $P^{(0)}$ and $P^{(2)}$ function respectively. The first term in $B_{mn}$ contains double $K^{(2)}$ insertions\footnote{Here ``double insertions'' refers to any contribution quadratic in the $\Lag_6$ coefficients. This includes contributions from the square of a diagram with one EFT insertion, as well as from the interference between SM and EFT diagrams with two EFT vertices, or EFT diagrams with a single interaction $\propto C^2$. The latter generally stem from field or parameter redefinitions in the Lagrangian. } in $P^{(0)}$, while the second terms of $A_n^{(2)}$ and $B_{mn}$ both stem from $K^{(4)}$ contributions in $P^{(0)}$.

The net effect of the finite renormalization procedure is that all the EFT corrections to input measurements are recast into corrections to predicted quantities:
if $\P$ is an input observable $\P\equiv\O_q$, all EFT corrections in Eq.~\eqref{eq.predicted_obs} cancel order by order in the EFT. This happens by construction and follows trivially from the defining conditions imposed.  It can be checked explicitly: in this case $P^{(d)}=F_q^{(d)}$ and assuming that $\O$ is a set of independent quantities,  also $\de F_q^{(0)}/\de F_n^{(0)} = \d_{qn}$ and $ \de^2F_q^{(0)}/\de F_m^{(0)}\de F_n^{(0)} =0$. This immediately leads to
\begin{align}
P^{(2)}-A_n F_n^{(2)} &=0\,,
&
B_{mn} &=0\,,
\\
A^{(2)}_n &=0\,,
&
P^{(4)}-A_n F_n^{(4)} &=0
\,.
\end{align}

Eq.~\eqref{eq.predicted_obs} provides a dictionary between different input parameter schemes: comparing sets $\O$ and $\O'$, the difference in the predicted $\P$ is
\begin{align}
\P(\O)-\P(\O') &=
\frac{1}{\Lambda^2}\left[- A_n F_n^{(2)} +  A'_n F_n^{(2)\prime}\right]
+
\frac{1}{\Lambda^4}\bigg[- A_n F_n^{(4)} +  A'_n F_n^{(4)\prime}+
\nn\\
&
- A_n^{(2)} F_n^{(2)} + A_n^{(2)\prime} F_n^{(2)\prime}
+ B_{mn} F_m^{(2)} F_n^{(2)} - B'_{mn} F_m^{(2)\prime} F_n^{(2)\prime} 
\bigg]+\dots
\end{align}
which is easily evaluated via Eqs.~\eqref{eq.inputs_An}-\eqref{eq.inputs_Bmn}. This result is consistent with those in the Appendix of Ref.~\cite{Brivio:2017bnu} and in Ref.~\cite{Baglio:2018bkm}.

\subsection{Implementation in \smeftsim}\label{sec.inputs_smeftsim}

\smeftsim\ implements the finite renormalization procedure via replacements of the form\footnote{Ref.~\cite{Brivio:2017bnu} used the notation $\bar g_i \to \hat g_i + \d g_i$ from Ref.~\cite{Berthier:2015oma}. This is completely equivalent to the one used here, dropping the bars.}
\begin{equation}\label{eq.shift_generic}
g_i \to \hat g_i +\d g_i\,,
\end{equation} 
where $\hat g_i$ satisfies the SM relation $\hat g_i = K_i^{(0)}(\O)$ and $\d g_i$ encodes all the dependence on the Wilson coefficients.
In the \feynrules\ models, these replacements are operated at the Lagrangian level via the lists {\tt redefConst} (applied simultaneously to the redefinitions in Eq.~\eqref{eq.g_redef}) and {\tt redefVev}, and the hats are subsequently dropped in the notation. In this way, all the SM parameters appearing explicitly in the final $\Lag_{\rm SMEFT}$ are hatted quantities, i.e. they are conveniently defined in the exact same way as in the SM and their numerical value is directly defined by the input observables chosen.

The shifts $\d g$ appear explicitly in the interaction terms, and they are responsible for propagating input shifts corrections to the computed processes. 
By construction, the dependence on $\d g$ themselves is universal, while their expressions in terms of Wilson coefficients are fixed by the input scheme choice:
\begin{align}\label{eq.def_dg}
\d g_i &= \frac{1}{\Lambda^2}K_i^{(2)}(\O) = -\frac{1}{\Lambda^2}(J^{-1})_{in} F_n^{(2)}\,.
\end{align}
As noted in Sec.~\ref{sec.redefinitions}, because we work at order $\Lambda^{-2}$,  the replacements of Eq.~\eqref{eq.shift_generic} only need to be performed on $\Lag_{\rm SM}$ and only linear terms in $\d g$ need to be retained.
Moreover, one can replace $v_T\to \hv$  in the $\bar C_\a$ notation, Eq.~\eqref{eq.Cbar}.

\vskip 1em

This procedure is implemented for parameters listed in the Higgs, EW and Yukawa sectors in~\eqref{eq.SM_pars}, as described below. 
Eq.~\eqref{eq.renorm_conditions_SMEFT} makes manifest that the extraction of SM parameters from global fits can become problematic when generalized to the SMEFT. Whenever this set of equations is not invertible, it is not possible to find a simple form for $g_i$ that expands around the SM solution. A consistent treatment of EFT corrections to such input observables would require to extract simultaneously $g_i$ and $C_\a$, which can be very unpractical or even unfeasible, in the presence of blind directions.  

In the case of the CKM parameters, this issue has been overcome in Ref.~\cite{Descotes-Genon:2018foz}, where an optimal set of 4 input measurements was proposed, that allows for a treatment of the CKM angles and phase analogous to that of EW parameters. Its implementation is left for future versions of \smeftsim.

The case of $\a_s$ poses a bigger challenge. The strong coupling constant can be determined from a particularly vast range of processes~\cite{PDG2020}, and its extraction is often correlated to that of other physical quantities, such as parton distribution functions (PDFs). A proof-of-concept analysis of SMEFT effects on the PDFs determination was presented in Ref.~\cite{Carrazza:2019sec}, that explored the consequences of including four-fermion operators in a fit to deep-inelastic scattering data. Further studies are needed in order to define an optimal strategy for the treatment of SMEFT contributions in this context. For the time being, input shift corrections associated to the determination of $\a_s$ are omitted in \smeftsim.

\begin{table}[t]
\parbox{.4\tw}{
\renewcommand{\arraystretch}{2.3}
\begin{tabular}{|>{$}c<{$}@{ = }>{$}p{4cm}<{$}|}
\hline
\Delta \kappa_H&\C_{H\square}-\dfrac{\C_{HD}}{4}
\\
\Delta m_Z^2& \dfrac{\C_{HD}}{2} + \dfrac{2g_1 g_W}{g_1^2+g_W^2}\C_{HWB}
\\
\Delta \aem& - \dfrac{2g_1 g_W}{g_1^2+g_W^2}\C_{HWB}
\\
\Delta m_h^2&2\Delta\kappa_H - \dfrac{3}{2\lambda}\C_H
\\[2mm]
\hline
\end{tabular}
}~\parbox{.4\tw}{
\renewcommand{\arraystretch}{3.1}
\begin{tabular}{|l|>{$}c<{$}@{ = }>{$}p{5.cm}<{$}|}
\hline
\general&
\Delta G_F &
(\C_{Hl}^{(3)})_{11}+(\C_{Hl}^{(3)})_{22}-(\C_{ll})_{1221}
\\
\parbox{1.7cm}{\Utf, \MFV,\\ \topsl}&
\Delta G_F  & 
2\C_{Hl}^{(3)} - \C_{ll}^\prime
\\
\top&
\Delta G_F  &
\C_{Hl,11}^{(3)} + \C_{Hl,22}^{(3)} - \dfrac{\C_{ll,1221}}{2}
\\[2mm]
\hline
\end{tabular}}
\caption{Expressions of input parameter shifts and the kinetic correction $\Delta\kappa_H$ (defined in~\eqref{eq.dkH}) in terms of Wilson coefficients. The left column is common to all flavor versions, while $\Delta G_F$ varies as indicated in the right column. We use the notation $\C_\a = C_\a (\hv^2/\Lambda^2)$.}\label{tab.input_shifts_expr}
\end{table}

\subsection{Higgs and EW sectors}\label{sec.inputs_ew}
The electroweak sector of the SM contains 4 independent quantities, that can be chosen as $g=\{g_1,g_W,v,\lambda\}$. The 4 (pseudo-)observables needed to fix their values are usually taken in the set
$$\{\aew,\, G_F,\, m_Z,\, m_W,\, m_h\}\,.$$
While $m_h$ always needs to be retained in order to fix $\lambda\,$, the choice of the 3 remaining inputs is free, and several combinations have been adopted in the literature. \smeftsim\ implements the two alternative schemes \ascheme\ and \mwscheme, providing independent \ufo\ models for both.

The fine structure constant $\aew(0)$ is taken to be measured in Thomson scattering\footnote{
As we work at tree level, only direct SMEFT corrections to Thomson scattering (i.e. to the determination of $\aem(0)$) are included here. The determination of $\aem(m_Z)$ at one loop in the SMEFT is another major open problem, as potential EFT contributions in the running have not been estimated to date. The main challenge in this task is posed by non-perturbative effects, particularly those arising as $\aem$ runs through the hadronic resonances region.
}, 
the Fermi constant $G_F$ measured in muon decays $\mu^-\to e^- \nu_\mu\bar\nu_e$, and $m_W,m_Z,m_h$ are defined as the bosons' pole masses, see Ref.~\cite{Brivio:2017btx} and references therein for further details. 
With these definitions, at tree level:\footnote{The normalization of $\Delta G_F$ has been modified compared to previous \smeftsim\ versions in order to homogenize the notation with the remaining shifts.}
\begin{align}
\aew &=
\frac{1}{4\pi}\frac{ g_W^2 g_1^2}{ g_W^2 + g_1^2}\left[1 + \Delta\aem\right]\,,
&
G_F &=
\frac{1}{\sqrt2 v_T^2}\left[1 + \Delta G_F\right]\,,
\label{eq.GF}\\
m_W^2 &= \frac{v_T^2}{4} g_W^2\,,
&
m_Z^2 &= 
\frac{v_T^2}{4}( g_W^2 + g_1^2)\left[1 + \Delta m_Z^2\right]\,,
\label{eq.mz}
\\
m_h^2&=
2v_T^2\lambda \left[1 + \Delta m_h^2\right]\,.
\label{eq.mh}
\end{align}
The  $\Delta$ quantities are dimensionless and  defined in Table~\ref{tab.input_shifts_expr}: $\Delta G_F$ is inferred computing the muon decay width at tree level in the SMEFT, while the remaining shifts can be read from the relevant Lagrangian terms. In particular, the contributions in $\C_{HWB}$ to $\Delta\aem,\,\Delta m_Z^2$ follow directly from Eq.~\eqref{eq.CD_redef} and $\Delta m_h^2$ follows from Eq.~\eqref{eq.VH_redef}.

The relations~\eqref{eq.GF}-\eqref{eq.mh} can be directly mapped to the notation of Eq.~\eqref{eq.renorm_conditions_SMEFT}: for instance
\begin{equation}
F_{G_F}^{(0)} = \frac{1}{\sqrt2 v_T^2}
\,,\qquad
\frac{1}{\Lambda^2}F_{G_F}^{(2)} = F_{G_F}^{(0)} \Delta G_F\,,
\end{equation} 
and analogously for the other observables.

\subsubsection{\texorpdfstring{\ascheme}{(a\_ew, mZ, GF)} scheme} \label{sec.inputs_ew_a}

Solving 3 of the 4 Eqs. in~\eqref{eq.GF},~\eqref{eq.mz}, plus Eq.~\eqref{eq.mh}, gives expressions for the SM parameters of the form of~\eqref{eq.renorm_conditions_sol_SMEFT}. 
Let us choose the input quantities $\O^{(\a)}=\{\aem,m_Z^2,G_F,m_h^2\}$. 

It is convenient to define the vector of SM parameters as $g = \{g_1^2, g_W^2, v_T^2, \lambda\}$.
The SM solutions $\hat g_i \equiv K_i^{(0)}(\O)$ are then
\begin{equation}
\hat g_1^2 =  \frac{4\pi\aew}{\hcw^2}\,,\qquad
\hat g_W^2 =  \frac{4\pi\aew}{\hsw^2}\,,\qquad
\hat v^2 =  \frac{1}{\sqrt2 G_F}\,,\qquad
\hat\lambda = \frac{m_h^2 G_F}{\sqrt2}
\,,
\end{equation} 
having defined the weak angle $\hat\theta$ as
\begin{equation}
\hsw^2=\sin^2\hat\theta \equiv \frac{\hat g_1^2}{\hat g_1^2+\hat g_W^2}
=\frac{1}{2}\left[1-\sqrt{1-\frac{2\sqrt2 \pi\aew}{G_F m_Z^2}}\right].
\end{equation}
The Jacobian $J=\de\O^{(\a)}/\de g$ defined in Eq.~\eqref{eq.Jacobian} takes the form
\begin{equation}
J = \frac14\bmat
\hcw^4/\pi\hspace*{3mm}&
\hsw^4/\pi\hspace*{3mm}&
& 
\\
\hat v^2\hspace*{3mm}& \hat v^2\hspace*{3mm}& \hat g_W^2/\hcw^2\hspace*{3mm}&
\\
& & -2\sqrt2/ \hat v^4& 
\\
& & 8\hat\lambda & 8 \hat v^2
\emat\,.
\end{equation} 
Taking the inverse and plugging it in Eq.~\eqref{eq.theta6}, one obtains explicit expressions for the parameter shifts defined as in~\eqref{eq.def_dg}:
\begin{align}\label{eq.g1_ascheme}
g_1^2 &=\hat g_1^2\left[1+2\frac{\d g_1}{\hat g_1}\right]\,,
&
\frac{\d g_1}{\hat g_1} 
&=
\frac{\hsw^2}{2c_{2\hat\theta}}\left(\Delta m_Z^2+\Delta G_F\right) - \frac{\hcw^2}{2c_{2\hat\theta}}\Delta\aem\,,
\\
g_W^2 &= \hat g_W^2\left[1+2\frac{\d g_W}{\hat g_W}\right]\,,
&
\frac{\d g_W}{\hat g_W} 
&=
 -\frac{\hcw^2}{2c_{2\hat\theta}}\left(\Delta m_Z^2+\Delta G_F\right) + \frac{\hsw^2}{2c_{2\hat\theta}}\Delta\aem\,,
\\
v_T^2 &=\hat v^2\left[1+2\frac{\d v}{\hat v}\right]\,,
&
\frac{\d v}{\hat v}
&=
\frac{\Delta G_F}{2}\,,
\\
\lambda &= \hat\lambda\left[1+\frac{\d \lambda}{\hat \lambda}\right]\,,
&
\frac{\d \lambda}{\hat \lambda}
&=
-\Delta G_F-\Delta m_h^2\,.
\label{eq.lam_ascheme}
\end{align}
It can be convenient, as a shorthand notation, to define a shift for $\sin^2\theta$. In the input schemes considered here, this is always a predicted quantity, that can be expressed as
\begin{equation}\label{eq.dsth2}
\d s^2_\theta= 2\hcw^2\hsw^2\left(\frac{\d g_1}{\hat g_1}-\frac{\d g_W}{\hat g_W}\right)+ \Delta s_\theta^2\,,
\end{equation} 
with  $\Delta s_\theta^2$ defined in Eq.~\eqref{eq.ds2_redef}.

With this input scheme choice, $m_W$ is also a predicted quantity and its expression can be derived from Eq.~\eqref{eq.predicted_obs}. From Eq.~\eqref{eq.mz}, we have that $P_{m_W^2}^{(2)}=0$, so
\begin{equation}
\begin{aligned}\label{eq.dmw_alphascheme}
m_W^2 &= P_{m_W^2}^{(0)} -\frac{1}{\Lambda^2}\frac{\de P_{m_W^2}^{(0)}}{\de g_i}(J^{-1})_{in} F_n^{(2)}
\\
&=
\frac{\hat v^2\hat g_W^2}{4}\left[ 1
-
\frac{\hcw^2}{c_{2\hat\theta}}\,\Delta m_Z^2 + 
\frac{\hsw^2}{c_{2\hat\theta}}( \Delta \aem- \Delta G_F)\right]
\\
&=\frac{\hat v^2\hat g_W^2}{4}\left[ 1 + 2\frac{\d v}{\hat v} + 2\frac{\d g_W}{\hat g_W}\right]
\\
&=\hat m_W^2\left[1+2\frac{\d m_W}{\hat m_W}\right]\,,
\end{aligned}
\end{equation} 
where we defined the shift 
\begin{align}
\label{eq.dmw_alphascheme_expr}
\frac{\d m_W}{\hat m_W}&=
\frac{\d v}{\hat v} + \frac{\d g_W}{\hat g_W} = 
\\
&=
-\frac{s_{2\hat\theta}}{4c_{2\hat\theta}}\left[\frac{1}{2}\frac{\hcw}{\hsw}\C_{HD} + 2\C_{HWB} + \frac{\hsw}{\hcw}\left((\C_{Hl}^{(3)})_{11}+(\C_{Hl}^{(3)})_{22}-(\C_{ll})_{1221}\right)\right]
\,.
\nn
\end{align}
The second line was evaluated with generic flavor indices for $\Delta G_F$, and it can be easily mapped to other flavor structures with the dictionary in App.~\ref{app.flavor_comparison}.
Finally, it is worth noting that electromagnetic interactions do not receive any corrections in this scheme:
\begin{equation}
\frac{\d e}{\hat e} = \hcw^2\,\frac{\d g_1}{\hat g_1} +\hsw^2\,\frac{\d g_W}{\hat g_W} +\frac{1}{2}\Delta\aem= 0\,,
\end{equation} 
consistent with $\aem$ being an input quantity. 

\subsubsection{\texorpdfstring{\mwscheme}{(mW, mZ, GF)} scheme} \label{sec.inputs_ew_mw}
Choosing the input observables $\O^{(m_W)} = \{m_W^2, m_Z^2, G_F, m_h^2\}$, the SM expressions $\hat g_i\equiv K_i^{(0)}(\O)$ for the relevant parameters are
\begin{equation}
\hat g_1^2 = 4\sqrt2 G_F m_Z^2\hsw^2\,,\qquad
\hat g_W^2 = 4\sqrt2 G_F m_W^2\,,\qquad
\hat v^2 =\frac{1}{\sqrt2 G_F} \,,\qquad
\hat \lambda = \frac{m_h^2G_F}{\sqrt2}\,,\qquad
\end{equation}
with the weak angle  defined by
\begin{equation}
\hsw^2 \equiv\frac{\hat g_1^2}{\hat g_1^2 + \hat g_W^2} =  1-\frac{m_W^2}{m_Z^2}\,.
\end{equation}
The Jacobian $J=\de\O^{(m_W)}/\de g$ takes the form
\begin{equation}
J =\frac14 \bmat
& \hat v^2& \hat g_W^2\hspace*{3mm}&
\\
\hat v^2\hspace*{3mm}& \hat v^2 \hspace*{3mm}& \hat g_W^2/\hcw^2 \hspace*{3mm}& 
\\
& & -2\sqrt2/ \hat v^4 & 
\\
& & 8\hat\lambda & 8\hat v^2
\emat\,,
\end{equation} 
and from Eq.~\eqref{eq.theta6} one has
\begin{align}
g_1^2 &= \hat g_1^2 \left[1+2\frac{\d g_1}{\hat g_1}\right]\,,
&
\frac{\d g_1}{\hat g_1}
&=
 -\frac 12\left[\Delta G_F + \frac{\Delta m_Z^2}{\hsw^2}\right]\,,
\\
g_W^2 &= \hat g_W^2\left[1+2\frac{\d g_W}{\hat g_W}\right]\,,
&
\frac{\d g_W}{\hat g_W}
&=
- \frac{\Delta G_F}{2}\,,
\\
v_T^2 &= \hat v^2\left[1+2\frac{\d v}{\hat v}\right]\,,
&
\frac{\d v}{\hat v}
&=
\frac{\Delta G_F}{2}\,,
\label{eq.v_shift}
\\
\lambda &= \hat\lambda\left[1-\frac{\d \lambda}{\hat \lambda}\right]\,,
&
\frac{\d \lambda}{\hat \lambda}
&=
-\Delta G_F-\Delta m_h^2\,.
\end{align}
With this input scheme choice, $\aem$ is now a predicted quantity. From Eq.~\eqref{eq.predicted_obs}:
\begin{equation}
\begin{aligned}\label{eq.da_mwscheme}
\aem &= P_{\aem}^{(0)} +\frac{1}{\Lambda^2}\left[\frac{\de P_{\aem}^{(0)}}{\de g_i}(J^{-1})_{in} F_n^{(2)} + P_{\aem}^{(2)}\right]
\\
&=
\frac{1}{4\pi}\frac{\hat g_1^2\hat g_W^2}{\hat g_1^2 + \hat g_W^2}\left[ 1
-
\Delta G_F -\frac{\hcw^2}{\hsw^2} \,\Delta m_Z^2+\Delta \aem\right]
\\
&=\frac{1}{4\pi}\frac{\hat g_1^2\hat g_W^2}{\hat g_1^2 + \hat g_W^2}\left[ 1 +2\hcw^2\, \frac{\d g_1}{\hat g_1} 
+2\hsw^2\, \frac{\d g_W}{\hat g_W}+\Delta\aem\right]
\\
&=\frac{\hat e^2}{4\pi}\left[1+2\frac{\d e}{\hat e}\right]\,.
\end{aligned}
\end{equation} 
It can be instructive to write the final form of the Higgs potential, once the input shifts are applied onto Eq.~\eqref{eq.VH_redef}. For both input schemes considered here, the result is
\begin{equation}\label{eq.VH_redef_inputs}
\begin{aligned}
V(H)+\Lag_6 &=
h^2\,\hat\lambda\, \hv^2
+h^3\,\hat\lambda\, \hv \left[1-\frac{\Delta G_F}{2}  +\Delta\kappa_H - \frac{1}{\hat\lambda}\, \C_H\right]
\\
&
+h^4\,\frac{\hat\lambda}{4}\left[1 -\Delta G_F +2\Delta\kappa_H- \frac{6}{\hat\lambda}\, \C_H\right]
-\frac{3}{4}\frac{h^5}{\hv}\, \C_H 
-\frac{1}{8}\frac{h^6}{\hv^2}\,\C_H\,.
\end{aligned}
\end{equation}

\subsection{Yukawa sector}\label{sec.inputs_fermions}
To fix the SM Yukawa couplings, we take fermion masses as input quantities. From the propagators' poles, at tree level, we have
\begin{equation}\label{eq.Mf_shift}
M_\psi = \frac{v_T}{\sqrt 2}\left[Y_\psi^{(d)} - \Delta M_\psi\right]\,,
\end{equation} 
with $Y_\psi^{(d)}$ diagonal.  In the \top, \topsl\ cases the index $\psi$ runs over $\psi = \{l,u,d,t,b\}$ so that $M_l,Y_l,\Delta M_l$ are $3\times 3$ tensors, $M_{u,d},Y_{u,d},\Delta M_{u,d}$ are $2\times 2$ and $M_{t,b},Y_{t,b},\Delta M_{t,b}$ are scalar quantities. In the other flavor setups $\psi = \{l,u,d\}$ and all quantities are $3\times 3$ matrices. The SMEFT corrections $\Delta M_\psi$ are given in Table~\ref{tab.dmf} for each flavor assumption.
\begin{table}[t] 
\small
\renewcommand{\arraystretch}{2.3}
\hspace*{-15mm}\begin{tabular}{|l|*3{>{$}l<{$}}|}
\hline
&
\Delta M_l& \Delta M_u& \Delta M_d
\\\hline
\general&
\dfrac{1}{2} \C_{e H}^\dag&
\dfrac{1}{2} \C_{u H}^\dag&
\dfrac{1}{2} \C_{d H}^\dag V
\\
\Utf&
\dfrac{1}{2}\C_{e H}^* Y_l^{(d)}&
\dfrac{1}{2}\C_{u H}^* Y_u^{(d)}&
\dfrac{1}{2}\C_{d H}^* Y_d^{(d)}
\\
\MFV&
\dfrac{1}{2}\C_{e H} Y_l^{(d)}
&
\dfrac{1}{2}Y_u^{(d)}\left[\C_{uH}^{(0)} + (\Delta^u\C_{uH}) S^{qu} +(\Delta^d\C_{uH})S^{qd}\right]
&
\dfrac{1}{2}Y_d^{(d)}V^\dag\left[\C_{dH}^{(0)} + (\Delta^u\C_{dH}) S^{qu} +(\Delta^d\C_{dH}) S^{qd}\right]V
\\
\top&
\dfrac12(\C_{e H})_{pp}^* 
&
\dfrac12\C_{u H}^* Y_u^{(d)}  \hspace*{1cm} (u,c)
&
\dfrac12\C_{d H}^* Y_d^{(d)}  \hspace*{1cm} (d,s)
\\
& &
\dfrac12\C_{t H}^*  \hspace*{1.75cm} (t)
&
\dfrac12\C_{b H}^*  \hspace*{1.7cm} (b)
\\
\topsl&
\dfrac12\C_{e H}^* Y_l^{(d)}
&
\dfrac12\C_{u H}^* Y_u^{(d)} \hspace*{1cm} (u,c)
&
\dfrac12\C_{d H}^* Y_d^{(d)} \hspace*{1cm} (d,s)
\\
& &
\dfrac12\C_{t H}^* \hspace*{1.75cm} (t) 
&
\dfrac12\C_{b H}^* \hspace*{1.7cm} (b)
\\
\hline
\end{tabular}
\caption{SMEFT corrections to the fermion mass matrices for each flavor assumption, see Sec.~\ref{sec.flavor} for all definitions. All Wilson coefficients are scalar quantities, except in the \general\ case, where they are $3\times3$ matrices. In the \MFV\ case all parameters are real.  In the \top\ and \topsl\ cases $\Delta M_{u,d}, Y_{u,d}$ are $2\times2$ matrices for the first two generations, and the mass term for the $t,b$ quarks are independent.
We use the notation $\C_\a = C_\a(\hat v^2/\Lambda^2)$ and the results are given in the mass basis of the up-quarks and charged leptons.}\label{tab.dmf}
\end{table} 
The SM solutions are
\begin{equation}\label{eq.Yuk_SMsol}
\hat Y_\psi^{(d)} =  \frac{\sqrt2}{\hat v}\,M_\psi\,,
\end{equation} 
and the shifts $\d Y_\psi$ have the form
\begin{align}\label{eq.dY}
Y_\psi^{(d)}&\to \hat Y_\psi^{(d)} + \d Y_\psi^{(d)}\,,
&
\d Y_\psi^{(d)} &= - \frac{\Delta G_F}{2}\,\hat Y_\psi^{(d)} + \Delta M_\psi\,,
\end{align}
where $\Delta G_F$ enters via Eq.~\eqref{eq.v_shift} and $\Delta M_\psi$ is non-diagonal and non-Hermitian in general. 
The expressions~\eqref{eq.Yuk_SMsol},~\eqref{eq.dY} can be easily generalized to setups where $M_\psi$ is not diagonal, by applying the appropriate flavor rotations to both sides of the equations.

The net effect of the finite renormalization procedure is that $\Delta M_\psi$ corrections to the fermion mass terms are recast into corrections to the $h\bar\psi\psi$ couplings. In unitary gauge, the Lagrangian resulting from the replacements~\eqref{eq.h_redef},~\eqref{eq.dY} is
\begin{align}
\Lag_{\rm Yukawa} + \Lag_6^{(5)} =& 
-\dfrac{\bar{d}_{R,p}\,d_{L,r}}{\sqrt 2} \left[
\hat v \,\hat Y_d^{(d)}
+ h\, \hat Y_d^{(d)}\left(1-\frac{\Delta G_F }{2}+\Delta\kappa_H\right)
-  \left(2h+\frac{3h^2}{\hat v}+ \frac{h^3}{\hat v^2}\right)\,  \Delta M_d
\right]_{pr}
\nn\\
&
-\dfrac{\bar{u}_{R,p}\,u_{L,r}}{\sqrt 2} \left[
\hat v \,\hat Y_u^{(d)}
+ h\, \hat Y_u^{(d)}\left(1-\frac{\Delta G_F }{2}+\Delta\kappa_H\right)
-  \left(2h+\frac{3h^2}{\hat v}+ \frac{h^3}{\hat v^2}\right)\, \Delta M_u
\right]_{pr}
\nn\\
&
-\dfrac{\bar{e}_{R,p}\,e_{L,r}}{\sqrt 2} \left[
\hat v \, \hat  Y_l^{(d)}
+ h\, \hat Y_l^{(d)}\left(1-\frac{\Delta G_F }{2}+\Delta\kappa_H\right)
-  \left(2h+\frac{3h^2}{\hat v}+ \frac{h^3}{\hat v^2}\right) \,\Delta M_l  
\right]_{pr}
\nn\\
&+\hc
\label{eq.Yukawas_redefined}
\end{align}
In the \top, \topsl\ models analogous terms with $t,b$ quarks are also present.

\vskip 1em
In the \feynrules\ implementation, the common shifts $\d v,\, \d\lambda,\, \d Y_\psi$ are automatically replaced with the corresponding expressions in terms of Wilson coefficients. On the other hand, the dependence on the EW shifts $\d g_1,\d g_W$ is left explicit in the Lagrangian, as it is identical for all EW input schemes. Once an inputs set is selected, these shifts can be traded for  Wilson coefficients expressions: in \mathematica\ this is done via the replacement lists {\tt alphaShifts} or {\tt MwShifts}.
In the \ufo\ models all shifts are replaced with the Wilson coefficient expressions.


\section{SM loop-generated Higgs interactions}\label{sec.Hloops}
Because \smeftsim\ is designed as a tree-level model, it cannot reproduce processes that only occur at 1-loop.  In fact, estimating SMEFT corrections to observables that are genuinely loop-generated both in the SM and at $d=6$ level is beyond the scope of \smeftsim. 
Nevertheless, there are cases where a 1-loop SM processes receives \emph{tree} $\Lag_6$ corrections. This notably happens in a few relevant Higgs production and decay channels.

In order to enable an estimate of interference terms between $\Lag_6$ and SM diagrams for the processes $gg\to h$, $h\to\g\g$, $h\to Z\g$, \smeftsim\ implements effective SM interactions obtained in the large $m_t$ limit. This formally corresponds to matching the SM onto an EFT (we will refer to this as ``top-EFT'') where the top quark has been integrated out. The advantage of this approach is that the top loops are effectively reduced to point vertices that can be inserted in tree diagrams. The obvious caveat is that the top-EFT is only valid in a limited kinematic region, as discussed below.

\smeftsim~3.0 contains $h\g\g$, $hZ\g$, $hgg$, $hggg$ and $hgggg$ interactions with contributions up to $\O(m_t^{-2})$, i.e. $d=7$ in the top-EFT. The implemented Lagrangian is:
\begin{equation}\label{eq.L_SMHloop}
\Lag_{\rm SM h loop} = g_{H\g\g} \O_{\g\g} + g_{HZ\g} O_{Z\g}+
 g_{Hgg}^{(1)} \O_{gg}^{(1)}+ \dfrac{1}{m_t^2} \sum_{i=2}^5 g_{Hgg}^{(i)} \O_{gg}^{(i)}\,,
\end{equation}
where
\begin{align}
\O_{\g\g} &= A_{\mu\nu} A^{\mu\nu} \dfrac{h}{v}\,,
&
\O_{z\g} &= Z_{\mu\nu} A^{\mu\nu} \dfrac{h}{v}\,,
& 
\O_{gg}^{(1)} &= G_{\mu\nu}^a G^{a\mu\nu} \dfrac{h}{v}\,,
\\
\O_{gg}^{(2)} &= D_\s G^a_{\mu\nu} D^\s G^{a\mu\nu} \dfrac{h}{v}\,,
&
\O_{gg}^{(3)} &= f_{abc} G_{\mu}^{a\nu} G_{\nu}^{b\s}G_\s^{c\mu}\dfrac{h}{v}\,,\\
\O_{gg}^{(4)} &= D^\mu G^a_{\mu\nu} D_{\s} G^{a\s\nu} \dfrac{h}{v}\,,
&
\O_{gg}^{(5)} &= G^{a\mu\nu} D_\nu D^\s G^{a}_{\s\mu}\dfrac{h}{v}\,.
\end{align}
The corresponding $g$ coefficients are fixed via a 1-loop matching procedure of the SM onto the top-EFT. 
For the $h\g\g$ and $hZ\g$ interactions we use the results from Refs.~\cite{Ellis:1975ap,Shifman:1979eb,Bergstrom:1985hp,Manohar:2006gz}, that include loops of both top quarks and $W$ bosons: 
\begin{align}
\label{eq.gHaa}
g_{H\g\g} &= \dfrac{e^2}{8\pi^2}\left[
I_w\left(\frac{m_h^2}{4 m_W^2}\right)+\frac43 I_f\left(\frac{m_h^2}{4m_t^2},0\right)\right]\,,\\
\label{eq.gHza}
g_{HZ\g} &= \frac{e^2}{4\pi^2} \left[t_\theta I_w^Z\left(\frac{m_h^2}{4 m_W^2},\frac{m_Z^2}{4m_W^2}\right) + \left(\frac12-\frac{4}{3}\sw^2\right) I_f\left(\frac{m_h^2}{4 m_t^2},\frac{m_Z^2}{4m_t^2}\right)\right]\,.
\end{align}
The loop functions $I_f,I_w,I_w^Z$ are evaluated in the limit where the Higgs boson is on-shell and higher order corrections are simply obtained via Taylor-expansion, retaining terms up to $\mathcal{O}(m_t^{-2} m_W^{-6})$:
\begin{align}
I_f(a,b) &= \int_0^1\int_0^{1-x} \frac{1-4xy}{1-4(a-b)xy-4by(1-y)} dy dx \\
&=\frac13 +\frac{7a}{90} + \frac{11 b}{90} + \O(a^2,b^2)\,,
\label{eq.If_expansion}\\[4mm]
I_w(a) &= \int_0^1\int_0^{1-x}\frac{-4+6xy+4axy}{1-4axy}dydx\\
&=-\frac{7}{4} - \frac{11 a}{30} - \frac{19 a^2}{105} - \frac{58 a^3}{525} +\O(a^4)\,,\\[4mm]
I_w^Z(a,b) &= \frac{1}{t_\theta^2}\int_0^1\int_0^{1-x} 
\frac{(5-t_\theta^2+2a(1-t_\theta^2))xy +t_\theta^2-3}{1-4(a-b)xy-4by(1-y)}dydx\\
&=
\frac{1}{24}\left(11-\frac{31}{t_\theta^2}\right)
+ \frac{11}{36}\left(\frac15-\frac{1}{t_\theta^2}\right)a  
+\frac{1}{9}\left(\frac75-\frac{4}{t_\theta^2}\right)b 
+ \frac{2}{315}\left(9-\frac{31}{t_\theta^2}\right) a b
+ \frac{19}{126}\left(\frac15-\frac{1}{t_\theta^2}\right) a^2 +
\nn\\
&\quad
+ \frac{1}{70}\left(\frac{53}{9}-\frac{17}{t_\theta^2}\right)b^2
+ \frac{29}{315}\left(\frac15-\frac{1}{t_\theta^2}\right) a^3
+ \frac{10}{63}\left(\frac{43}{125}-\frac{1}{t_\theta^2}\right) b^3 
+
\nn\\
&\quad
+\frac{4}{105}\left(\frac{4}{5}-\frac{3}{t_\theta^2}\right) a^2 b  
+\frac{1}{315}\left(\frac{67}{5}-\frac{43}{t_\theta^2}\right) ab^2
+\O(a^4,b^4)\,.
\end{align}
For the Higgs-gluon interactions, the matching has been performed in Refs~\cite{Neill:2009tn,Harlander:2013oja,Dawson:2014ora} up to dimension 7 in the top expansion:\footnote{There is a sign difference in the definition of $\O_3$ compared to Refs.~\cite{Neill:2009tn,Harlander:2013oja}. The sign of $C_3$ is also affected by the sign in the covariant derivative definition, that was taken with the opposite convention in Ref.~\cite{Dawson:2014ora}.}
\begin{align}
\label{eq.gHgg1}
g_{Hgg}^{(1)} &= 
\frac{g_s^2}{48\pi^2} +\O(g_s^4)\,,\\
\label{eq.gHgg2}
g_{Hgg}^{(2)} &= -\frac{7g_s^2}{2880\pi^2}+\O(g_s^4)\,,\\
\label{eq.gHgg3}
g_{Hgg}^{(3)} &= \frac{g_s^3}{240\pi^2}+\O(g_s^5)\,,\\
\label{eq.gHgg4}
g_{Hgg}^{(4)} &= \frac{g_s^2}{1440\pi^2}+\O(g_s^4)\,,\\
\label{eq.gHgg5}
g_{Hgg}^{(5)} &= \frac{g_s^2}{80\pi^2}+\O(g_s^4)\,.
\end{align}

Note that the $d=7$ operators produce interactions with one Higgs and up to 6 gluon legs. While the full gauge-invariant Lagrangian is implemented in the \feynrules\ models, only vertices with up to 4 gluon legs ($hgg, hggg, hgggg$) were exported to the \ufo s. The Feynman rules of the $hggggg$ and $hgggggg$ vertices are extremely complex both in the color and Lorentz structures, to the point that their inclusion  makes the Monte Carlo event generation computationally challenging. They are available upon request. 

\subsection{Validity of the approximations used}
The Higgs interactions described in this section are implemented to the specific purpose of enabling the simulation of Higgs production and decay processes. In general, these vertices \emph{should not} be inserted into other arbitrary processes.
In \madgraph, the insertions can be controlled at the diagram generation level via the interaction order {\tt SMHLOOP}~=~1 that is assigned to all the $g$ couplings in the Lagrangian~\eqref{eq.L_SMHloop}, see also Sec.~\ref{sec.int_orders}.

The following limitations should also be kept in mind:

\begin{itemize}
\begin{figure}[t]
\hspace*{-2.5cm}
\includegraphics[height=11cm,trim=0pt 0pt 9.5cm 0cm, clip]{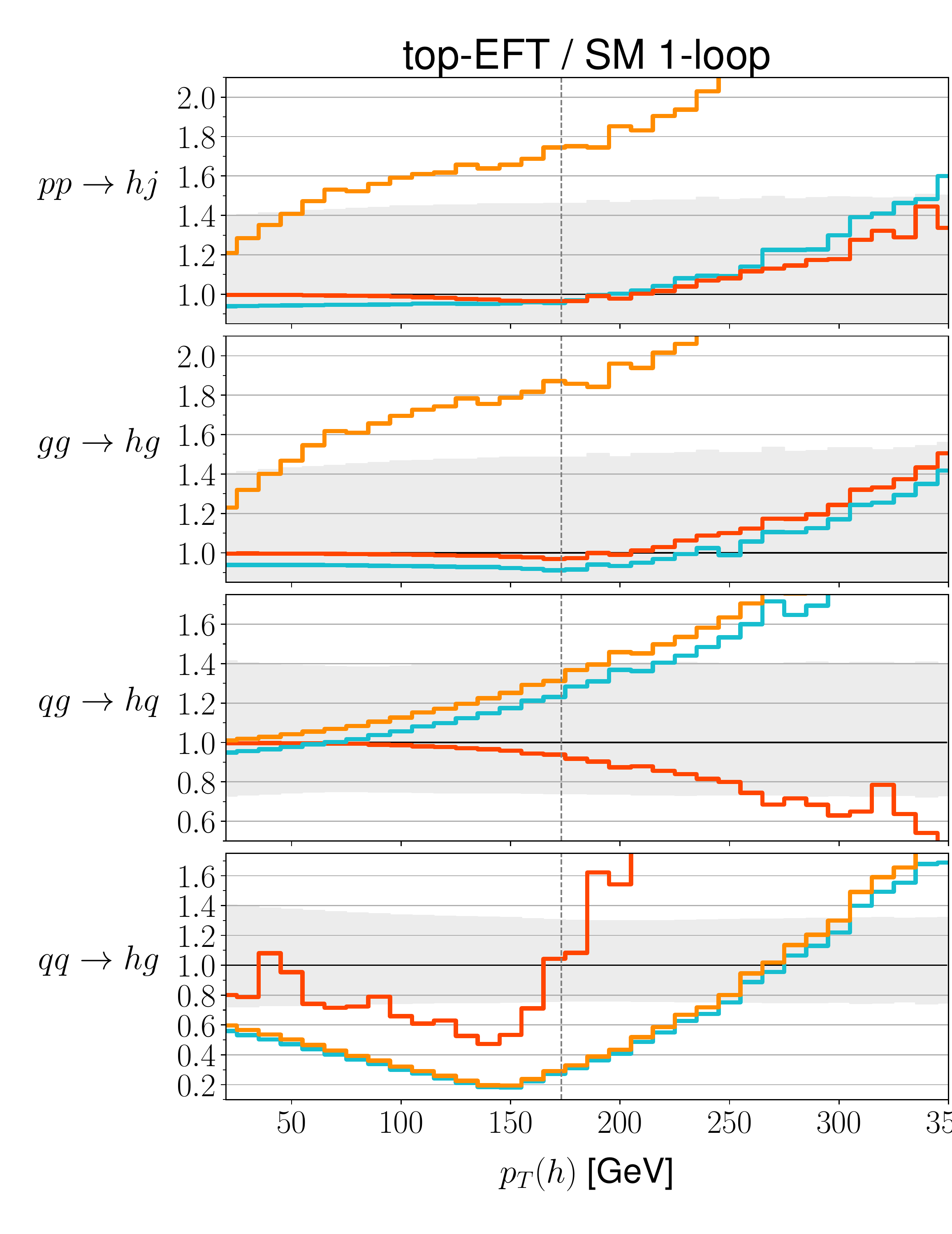}~
\includegraphics[height=11cm]{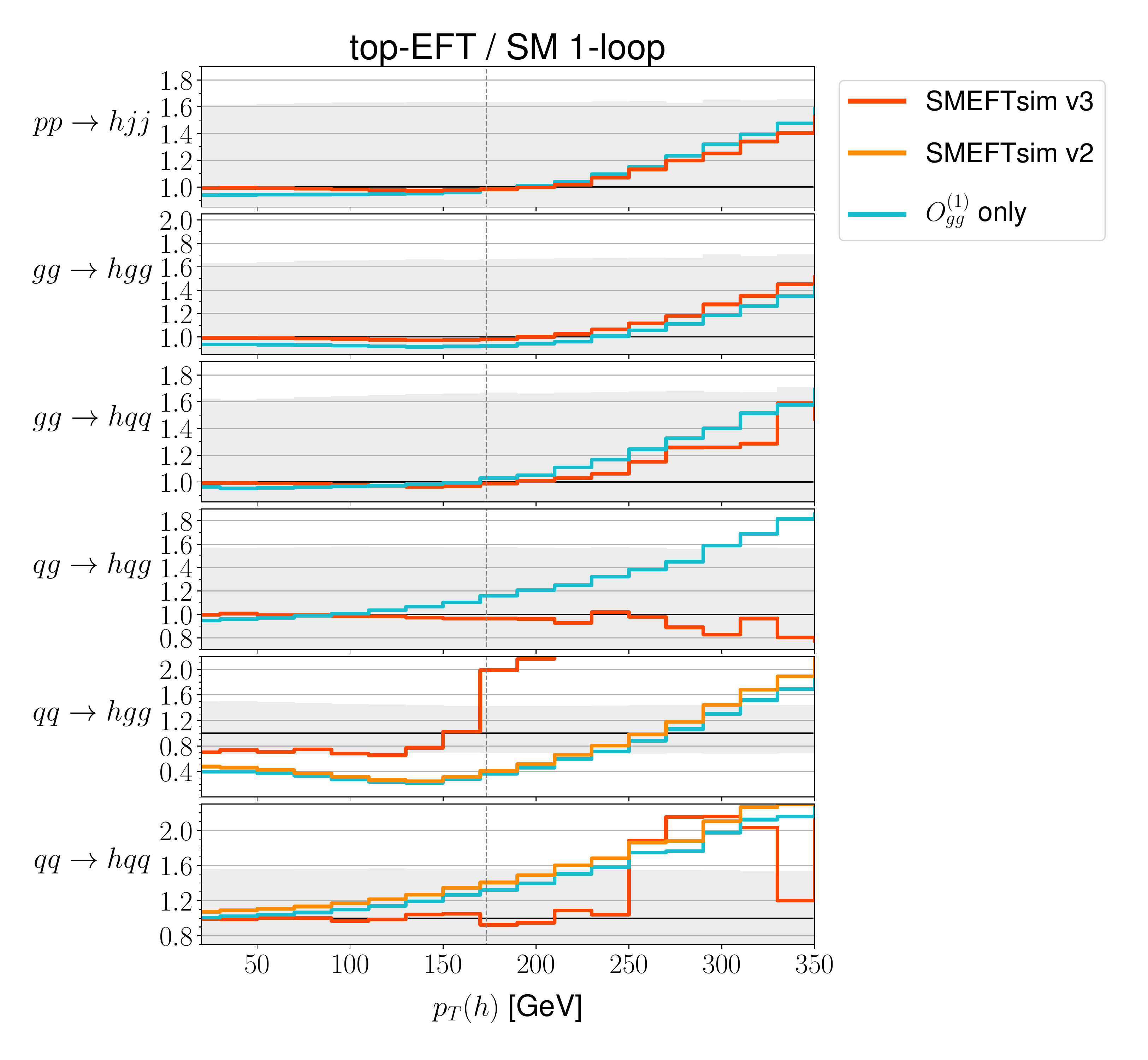}
\caption{Top-EFT predictions for $d\sigma/d p_T(h)$ in $pp\to hj$  (left) and $pp\to hjj$ (right), normalized to the SM 1-loop results, see text for the calculation details. In both figures, the first panel shows the combined result, while the lower ones give the breakdown into the contributing channels. Note that here $q$ generically denotes a light quark or antiquark.  
The statistical uncertainties are plotted in color, and they are not visible in most cases.
For reference, the gray bands show the systematic uncertainty on the SM event generation and the vertical dashed line marks $p_T=m_t$, where the top-EFT is expected to break down. \smeftsim~v2 curves for $pp\to hjj$, $gg\to hgg$, $gg\to hqq$, $qg\to hqg$ lie above the plotted range.   }\label{fig.hj_hjj_comparison}
\end{figure}

\item The implementation relies on the top-EFT formalism, that is only valid when the momentum  $q$ flowing through the effective vertex is $q<m_t$. This condition is always fulfilled for $gg\to h$ with no extra jets, for which the top-EFT reproduces the 1-loop SM cross-section within an accuracy of a few permille. With more complex final states, a validity threshold is present and it can translate differently in terms of measured observables, depending on the process.

Fig.~\ref{fig.hj_hjj_comparison} shows the relative deviation of the top-EFT predictions from the 1-loop SM results for $d\sigma/d p_T(h)$ in $pp \to hj$ and $pp\to hjj$, as obtained at parton level with \madgraph, in a 4-flavor scheme and neglecting all electroweak contributions. The SM prediction was obtained generating 100000 events for each $pp\to hj$ channel and for $qq\to hqq$, and 50000 events for the remaining $pp\to hjj$ channels, with the {\tt loop\_sm} \ufo. The associated PDF and scale uncertainties were estimated with the \madgraph\ functionalities~\cite{Frederix:2011ss}, and their combination in quadrature is shown for reference as a grey band.  
The top-EFT predictions were obtained reweighting the events with \smeftsim. The lines in color compare three different implementations of the top-EFT: including all operators up to $d=7$ (red), including only the $d=5$ operator $\O_{gg}^{(1)}$ (blue) and including only the $ggh$ vertex as in the previous \smeftsim\ versions (orange), see the next subsection.  The statistical uncertainty associated to each line is shown as a colored band surrounding the solid curves. Because the statistical errors associated to the reweighted histogram and to the original one are fully correlated, in most cases, the uncertainty on their ratio cancels and it is not visible on the plot. Uncertainties due to the reweighting procedure itself have been neglected, in the absence of a prescription for their estimation.

For both $pp\to hj$ and $pp\to hjj$, the total cross section is dominated by $gg$- and $qg$-initiated  channel, for which the $m_t\to\infty$ approximation breaks down roughly at $p_T(h)\simeq 250$~GeV~\cite{Baur:1989cm,DelDuca:2003ba,Keung:2009bs,Harlander:2013oja,Dawson:2014ora,Buschmann:2014sia}. 
Within the top-EFT validity regime, the $d=7$ implementation reproduces the SM 1-loop result within an accuracy of few~\%, see also Ref.~\cite{Dawson:2014ora}. The large $m_t$ approximation fails most significantly in $qq$-initiated processes that, nevertheless, give  a negligible contribution to the total cross section. This behavior is due to the quarks' PDFs preferring significantly larger $x$ compared to the gluon one, which leads to large $\hat s$ contributions being suppressed for $gg$ and $qg$ initial states, but not for $qq$~\cite{Keung:2009bs}.

\item The operators $O_{Hgg}^{(i)}$ form a complete basis up to $\O(m_t^{-2})$ for Higgs interactions with up to 4 gluons~\cite{Gracey:2002he,Neill:2009tn}. 
This means that, within the regime of validity of the top-EFT, \smeftsim\ can reproduce 1-loop SM Higgs production in gluon fusion with up to 2 jets. 

Processes $gg\to h+nj$ with $n\geq3$ cannot be fully reproduced with \smeftsim, even with the inclusion of $hggggg$, $ hgggggg$ vertices, because a complete matching to $\O(m_t^{-2})$ onto these vertices would require $d=9$ top-EFT operators.

\item In addition to the validity of the large $m_t$ approximation, the implementation of $h\g\g,\,hZ\g$ assumes an on-shell Higgs in the parameterization of the loop function. 

\end{itemize}

\subsection{Comparison to previous versions of \smeftsim}\label{sec.Hloops_comparison}
Previous \smeftsim\ versions only implemented the $hgg$, $h\g\g$ and $hZ\g$ vertices, while interactions with higher numbers of gluons were omitted. 
In version~3.0, all the vertices induced by the operator $\O_{Hgg}^{(1)}$ and vertices with up to 5 legs (4 gluons) from $\O_{Hgg}^{(2,3,4,5)}$ are included.

Moreover,  the $hgg$ interaction was previously parameterized in the on-shell Higgs limit, analogously to $h\g\g,\,hZ\g$, via a coupling~\cite{Bergstrom:1985hp,Dawson:1990zj,Manohar:2006gz}
$$g_{Hgg} = \frac{g_s^2}{16\pi^2}I_f\left(\frac{m_h^2}{4m_t^2},0\right)\,,$$
and the loop functions $I_f,\,I_w,\,I_w^Z$ were expanded up to $\O(m_t^{-6}m_W^{-6})$, which is formally equivalent to a matching up to $d=11$ in the top-EFT for an on-shell Higgs boson.
In version~3.0 this is replaced by a consistent matching up to $\O(m_t^{-2}m_W^{-6})$ that does not rely on the on-shell assumption. The two parameterizations are completely equivalent up to $\O(m_t^{-4})$ corrections for on-shell Higgs production with no extra jets:
the Feynman rule of the $ggh$ interaction is
\begin{align}
\parbox{3.5cm}{\includegraphics[width=3.5cm]{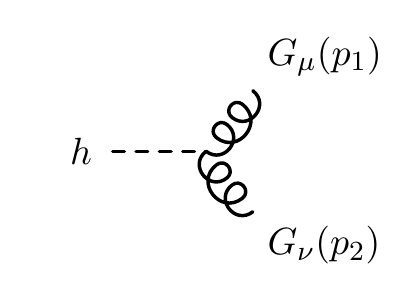}}=&\frac{4i}{v}\left[p_{1}^\nu p_{2}^\mu - \eta^{\mu\nu} p_{1}\cdot p_{2}\right]\left[
g_{Hgg}^{(1)}
-\frac{p_{1}\cdot p_{2}}{m_t^2}g_{Hgg}^{(2)}
+\frac{p_{1}^2+p_{2}^2}{4m_t^2}g_{Hgg}^{(5)}
\right]+
\nn
\\[-3mm]&
-\frac{2i}{v}\frac{
p_{1}^\mu p_{1}^\nu\, p_{2}^2 + p_{2}^\mu p_{2}^\nu\, p_{1}^2 
- \eta^{\mu\nu} p_{1}^2\, p_{2}^2
- p_{1}^\mu p_{2}^\nu\, p_{1}\cdot p_{2} 
}{m_t^2}g_{Hgg}^{(4)}\,,
\end{align}
with the momenta $p_1,\,p_2$ taken to be incoming.
In the limit $p_{1,2}^2=0,\,p_h^2=(p_1+p_2)^2=m_h^2$
\begin{equation}
\parbox{3.5cm}{\includegraphics[width=3.5cm]{diagrams.pdf}}\quad\stackrel{\text{on-shell}}{\longrightarrow}\quad 
\frac{4i}{v}\left[p_{1}^\nu p_{2}^\mu - \eta^{\mu\nu} \frac{m_h^2}{2}\right]\left[g_{Hgg}^{(1)} - \frac{m_h^2}{2m_t^2}g_{Hgg}^{(2)}\right]
+\frac{i}{v}\frac{m_h^2}{m_t^2}\, p_1^\mu p_2^\nu \, g_{Hgg}^{(4)}\,.
\end{equation}
The last term vanishes for external gluons and, using Eqs.~\eqref{eq.gHgg1}, \eqref{eq.gHgg2},
\begin{equation}
g_{Hgg}^{(1)} - \frac{m_h^2}{2m_t^2}g_{Hgg}^{(2)} = 
\frac{g_s^2}{16\pi^2}\left[\frac{1}{3} + \frac{7}{90}\frac{m_h^2}{4m_t^2}\right]\,.
\end{equation}
The terms in brackets reproduce the expansion of the top loop $I_f$ (Eq.~\eqref{eq.If_expansion}) up to $\O(m_t^{-2})$.


\section{Propagator corrections}\label{sec.propagators}
Mass terms and decay widths of the SM particles generally receive corrections from $\Lag_6$ operators.
In order to compute amplitudes consistently at $\O(\Lambda^{-2})$, these corrections need to be included in the propagators.

In unitary gauge the propagator of a generic unstable vector $V$, scalar $S$ or fermion $\psi$ has the form
\begin{align}
\label{def.PV}
P_{V}^{\mu\nu} &= \frac{i}{q^2-m_V^2+i m_V \Gamma_V}\left(-\eta^{\mu\nu} + \frac{q^\mu q^\nu}{m_V^2}\right)\,,\\
P_S &= \frac{i}{q^2-m_S^2+i m_S \Gamma_S}\label{def.PS}\,,\\
P_\psi &= \frac{i(\slashed{q}+m_\psi)}{q^2-m_\psi^2+i m_\psi \Gamma_\psi}\,,
\label{def.PF}
\end{align}
In the SMEFT we can write, for each particle,
\begin{equation}
m= m^{\rm SM} + \d m,\qquad 
\Gamma = \Gamma^{\rm SM} + \d \Gamma\,,
\end{equation}
where the shifts $\d m,\, \d \Gamma$ collect all the contributions from $d\geq6$ operators.
The corresponding propagator expressions be expanded to linear order in the shifts~\cite{Berthier:2016tkq}
\begin{equation}\label{eq.P_linear}
P_V^{\mu\nu} = P_V^{\mu\nu,SM} +\Delta P_V^{\mu\nu},\qquad
P_S = P_S^{\rm SM} +\Delta P_S\,,\qquad
P_\psi = P_\psi^{\rm SM} +\Delta P_\psi\,,
\end{equation}
where the expressions for $P^{\rm SM}$ are given by Eqs.~\eqref{def.PV}-\eqref{def.PF} replacing $ m\to m^{\rm SM}$, $\Gamma\to\Gamma^{\rm SM}$. The corrections read
\begin{align}
\Delta P_{V}^{\mu\nu} &=- P_V^{\mu\nu,SM} \frac{i m_V^{\rm SM}}{D_V(q^2)} \bigg[\d\Gamma_V
+2i\left(1- \frac{i\Gamma_V^{\rm SM}}{2m_V^{\rm SM}}\right)\d m_V
\bigg] - 2i\frac{q^\mu q^\nu}{D_V(q^2)} \frac{\d m_V}{(m_V^{\rm SM})^3}\,,
\label{eq.DPV}\\
\Delta P_S &=- P_S^{\rm SM} \frac{i m_S^{\rm SM}}{D_S(q^2)}
\bigg[\d\Gamma_S
+2i\left(1- \frac{i\Gamma_S^{\rm SM}}{2m_S^{\rm SM}}\right)\d m_S
\bigg] \,,
\label{eq.DPS}\\
\Delta P_\psi &=- P_\psi^{\rm SM} \frac{i m_\psi^{\rm SM}}{D_\psi(q^2)}
\bigg[\d\Gamma_\psi
+2i\left(1- \frac{i\Gamma_\psi^{\rm SM}}{2m_\psi^{\rm SM}}\right)\d m_\psi
\bigg] + \frac{i \d m_\psi}{D_{\psi}(q^2)} \,,
\label{eq.DPF}
\end{align}
with the shorthand notation 
\begin{equation}
D(q^2) = q^2-(m^{\rm SM})^2+i m^{\rm SM} \,\Gamma^{\rm SM}\,.
\end{equation}
Note that, since $\Delta P\propto D(q^2)^{-1}$, propagator corrections are expected to be relevant in the on-shell kinematic region and suppressed when the particle is largely off-shell.
The dominant contributions are therefore approximated by the on-shell expressions:\footnote{Longitudinal contributions for vector bosons were neglected here.}
\begin{align}
\left.\Delta P_V^{\mu\nu}\right|_{q^2=m^2} &=
\left. P_V^{\mu\nu}\right|_{q^2=m^2}\left[-\frac{\d \Gamma_V}{\Gamma_V^{\rm SM}} - \left(1+\frac{2i m_V^{\rm SM}}{\Gamma_V^{\rm SM}}\right)\frac{\d m_V}{m_V^{\rm SM}}\right]\,,
\label{eq.Delta_PV_onshell}
\\
\left.\Delta P_S\right|_{q^2=m^2} &=
\left. P_S\right|_{q^2=m^2}\left[-\frac{\d \Gamma_S}{\Gamma_S^{\rm SM}} - \left(1+\frac{2i m_S^{\rm SM}}{\Gamma_S^{\rm SM}}\right)\frac{\d m_S}{m_S^{\rm SM}}\right]\,,
\\
\left.\Delta P_\psi\right|_{q^2=m^2} &=
\left. P_\psi\right|_{q^2=m^2}\left[-\frac{\d \Gamma_\psi}{\Gamma_\psi^{\rm SM}} - \left(1+\frac{2i m_\psi^{\rm SM}}{\Gamma_S^{\rm SM}}\right)\frac{\d m_\psi}{m_\psi^{\rm SM}}\right]+\frac{\d m_\psi}{m_\psi^{\rm SM} \Gamma_\psi^{\rm SM}}\,,
\end{align}
with
\begin{equation}
\left. P_V^{\mu\nu}\right|_{q^2=m^2} = 
\frac{-\eta^{\mu\nu}}{m_V \Gamma_V}\,,
\qquad
\left. P_S\right|_{q^2=m^2} = 
\frac{1}{m_S \Gamma_S}\,,
\qquad
\left. P_\psi \right|_{q^2=m^2} = 
\frac{\slashed{q}+m_\psi}{m_\psi \Gamma_\psi}\,.
\end{equation}

\subsection{Implementation in \smeftsim}
\label{sec.propagators_smeftsim}
\begin{figure}[t]
\hspace*{-1.5cm}
\includegraphics[width=.55\tw]{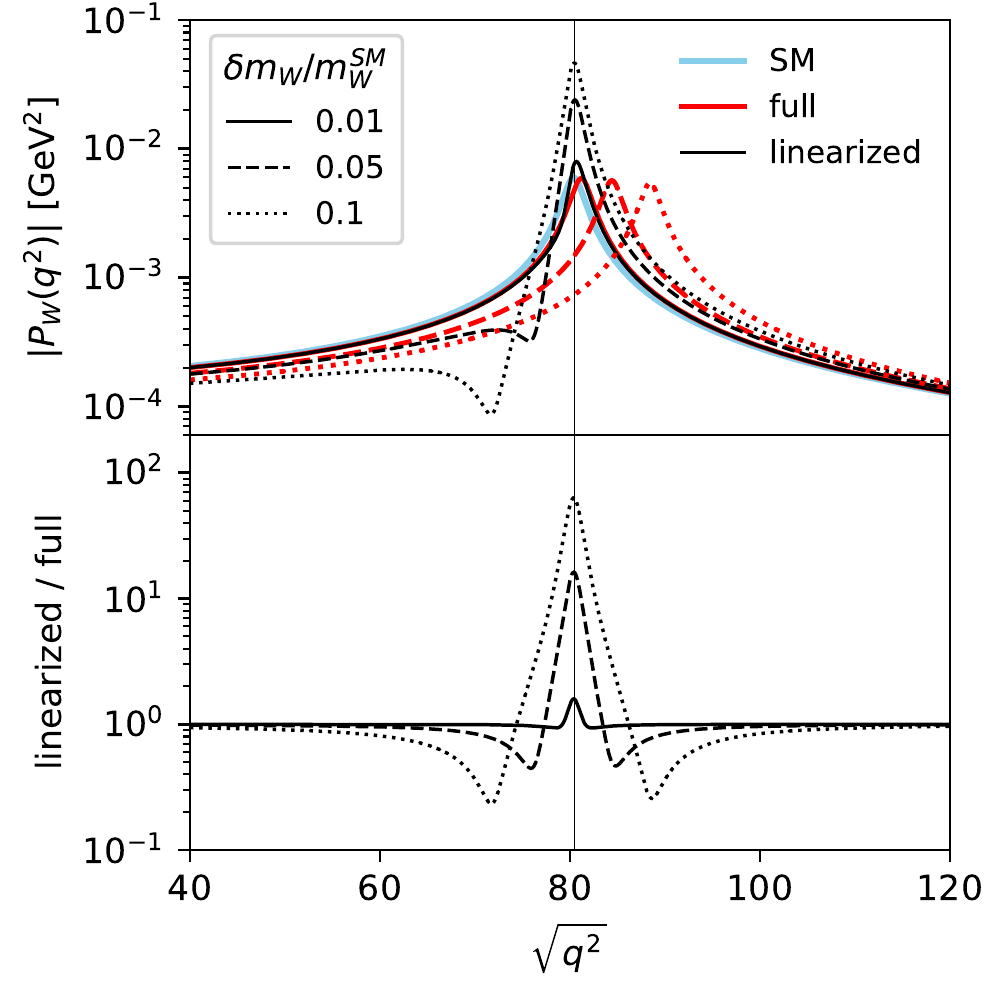}
\includegraphics[width=.55\tw]{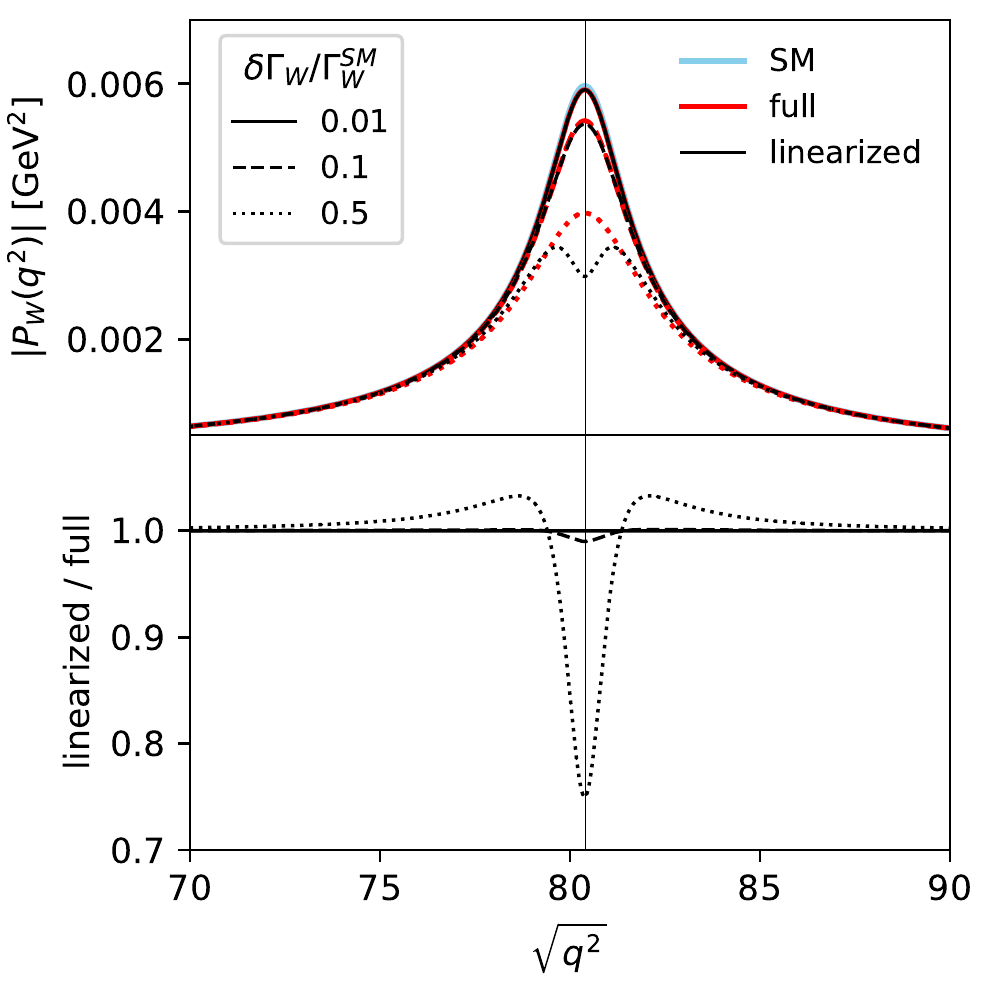}
\caption{Illustrative comparison of full and linearized $W$ propagators varying $\d m_W$ (left) and $\d \Gamma_W$ (right). The upper panels show  the absolute value of the form factor of the transverse propagator $|P_W(q^2)|$ as a function of $\sqrt{q^2}$. Red lines are obtained modifying $m_W,\Gamma_W$ in the full propagator, Eq.~\eqref{def.PV}, while black lines are obtained linearizing out $\d m_W,\d\Gamma_W$ corrections as in Eqs.~\eqref{eq.P_linear},\eqref{eq.DPV}. The lower panels show the ratio between the two curves.
The thin vertical line marks $\sqrt{q^2}= m_W^{\rm SM}=\unit[80.387]{GeV}$. The light blue curve in the upper plots marks for reference the SM behavior ($\d \Gamma_W=0=\d m_W$). Solid, dashed and dotted lines correspond to three different sizes of $\d m_W$ and $\d\Gamma_W$, with the same coding in upper and lower panels. In the right panel, the black, red and light blue solid lines are indistinguishable.
}\label{plot.linearProp}
\end{figure}

\smeftsim~3.0 implements propagator corrections for the $Z,W,h$ bosons and for the top quark.
The user has two alternative options for including them in SMEFT predictions:
\begin{itemize}
\item[(a)] using the linearized propagator expressions of Eqs.~\eqref{eq.P_linear}-\eqref{eq.DPF}. In this case the pole of the propagator remains located at $m^{\rm SM}$, and the dependence on the Wilson coefficients, stemming both from $\d m$ and $\d \Gamma$, is linear at the amplitude level. This option is selected fixing {\tt linearPropCorrections} = 1 (or any value $\neq 0$) in the {\tt param\_card}.

\item[(b)] using the propagator expressions in Eqs.~\eqref{def.PV}-\eqref{def.PF}, with shifted masses. In this case the pole of the propagator is located at $m = m^{\rm SM} + \d m$ while width corrections are entirely dropped.\footnote{The implementation of EFT-corrected decay widths has been avoided to prevent potential conflicts with the treatment of widths in Monte Carlo generators.}
The dependence on the Wilson coefficients is generally non-linear, as contributions $\propto 1/C_\a$  are induced in the amplitude. 
This is the default option and it's selected with {\tt linearPropCorrections} = 0.
\end{itemize}
While option (a) is recommended for consistency of the EFT expansion, we caution the user that the linearization can be problematic, particularly in the presence of \emph{mass} corrections. Formally, expanding around the complex pole of the propagator is not a gauge-invariant operation~\cite{Veltman:1963th,Stuart:1991xk,Grunewald:2000ju}. 
Numerically, significantly large discrepancies between methods (a) and (b) can occur, as illustrated in Figure~\ref{plot.linearProp} (left) for the case of the $W$ boson. Using linearized propagators leads to sizeable numerical distortions already for $\d m_W/m_W^{\rm SM}\gtrsim$~few~\%, as the dashed curves show. 
For comparison, the linear approximation works very well for width corrections up to $\O(10\%)$, see Fig.~\ref{plot.linearProp} (right). This is partially due to the $W$ boson being narrow in the SM.
Adopting a \mwscheme\ input scheme is convenient in this respect, because mass correction effects are entirely avoided. For a more general discussion of the theoretical advantages of this scheme choice, see eg. Refs.~\cite{Ghezzi:2015vva,Gauld:2015lmb,Gauld:2016kuu,Berthier:2016tkq,Hartmann:2016pil,Brivio:2017vri,Brivio:2017bnu,Brivio:2017btx}.

\bigskip
Option (a) is implemented in \smeftsim\ following the method outlined in Ref.~\cite{Brooijmans:2018xbu}, contribution 12:\footnote{I thank O. Mattelaer for pointing me to this reference.} four dummy fields $Z',W',h',t'$ ({\tt Z1,W1,H1,t1}) are introduced, with propagators $\Delta P_Z^{\mu\nu}$, $\Delta P_W^{\mu\nu}$, $\Delta P_h$, $\Delta P_t$.
The numerical values assigned to $m^{\rm SM},\,\Gamma^{\rm SM}$ appearing in these expressions are those of the nominal mass and widths of the dummy fields. The latter, in turn, are internal parameters and defined as equal to the SM masses and widths of the corresponding dynamical particles. The $\delta m,\,\delta\Gamma$ parameters are also defined as internal parameters, function of the relevant SM couplings and Wilson coefficients.

The dummy states have the same SM interactions as $Z,W,h,t$ and do not enter $d=6$ operators:  
\smeftsim\ contains copies of all SM vertices, with one or more of the standard fields replaced by its dummy counterpart. 
Vertices with $n$ dummy legs are proportional to $n$ powers of a flag parameter {\tt propCorr} and have interaction order {\tt NPprop} = $n$. The numerical value of {\tt propCorr} is set to $ 0 (1)$ if  {\tt linearPropCorrections} $ = 0$ (a non-zero value).

In this way, for instance, linearized $Z$-propagator corrections to $pp\to \mu^+ \mu^-$ can be estimated computing the $pp\to Z\to \mu^+ \mu^-$ and  $pp\to Z'\to\mu^+ \mu^-$ amplitudes, and using the interaction order {\tt NPprop} to isolate the pure SM/interference/quadratic contributions as detailed in Sec.~\ref{sec.propagators_madgraph}.

Note that linearized  propagator corrections are available only in the \ufo\ models, as the propagators are modified directly in the {\tt propagators.py} file and not in \feynrules.\footnote{The expressions in {\tt propagators.py} differ by an overall $i$ factor from those in~\eqref{eq.DPV},\eqref{eq.DPS}, because an $i$ is conventionally added by {\tt ALOHA} upon parsing the \ufo\ model~\cite{Christensen:2013aua}.}

\paragraph{Mass and width corrections implemented.}\label{sec.propagator_corr_implemented}
All the mass and width shifts implemented in \smeftsim\ are computed to $\O(\Lambda^{-2})$, i.e. linearly in the Wilson coefficients.
Because $m_Z$, $m_h$ and $m_t$ are taken as input parameters,
\begin{equation}
\d m_Z\equiv 0,\qquad \d m_h\equiv 0, \qquad \d m_t\equiv 0\,.
\end{equation}
The $\d m_W$ correction is non-vanishing only in the \ascheme\ scheme, and the expression was given in Eq.~\eqref{eq.dmw_alphascheme_expr}.
Decay width corrections for the $Z,W$ bosons and for the top quark are defined as
\begin{align}
\d \Gamma_Z &= \Gamma_Z^{SM,\text{best}} \left[\frac{\d \Gamma_Z}{\Gamma_Z^{\rm SM}}\right]_\text{tree}\!\!\!\!\!,
&
\d \Gamma_W &= \Gamma_W^{SM,\text{best}} \left[\frac{\d \Gamma_W}{\Gamma_W^{\rm SM}}\right]_\text{tree}\!\!\!\!\!,
&
\d \Gamma_t &= \Gamma_t^{SM,\text{best}} \left[\frac{\d \Gamma_t}{\Gamma_t^{\rm SM}}\right]_\text{tree}\!\!\!\!\!,
\end{align}
with
\begin{align}
\Gamma_Z^{SM,\text{best}}&=\unit[2.4952]{GeV}, 
&\Gamma_W^{SM,\text{best}}&=\unit[2.085]{GeV},
&\Gamma_t^{SM,\text{best}}=\unit[1.33]{GeV},
\end{align}
the loop-improved SM predictions~\cite{PDG2020,Gao:2012ja}. These are free parameters in the models, that can be modified by the user.
The quantities $\left[\d\Gamma/\Gamma^{\rm SM}\right]_\text{tree}$ are calculated at tree level (both numerator and denominator) using the width computation tools in \feynrules~\cite{Alwall:2014bza}. They include all 2-body decays and are extracted in the limit $V_{CKM}=\mathbbm{1}$, with all fermion masses set to zero,  except those of the $b$ and $t$ quarks.  Analytic expressions are given in Appendix~\ref{app.dwidthExpressions}. 

The correction to the total Higgs width is computed using individual $K$-factors for each decay channel, as in Ref.~\cite{Brivio:2019myy}:\footnote{This  normalization choice is due to radiative corrections affecting the various channels in significantly different ways. For comparison, in the case of $Z,W$ decays, using individual $K$-factors leads to variations $\lesssim 2\%$ in the Wilson coefficient dependence compared to an overall rescaling. The top case is trivial, as there is only one relevant decay channel $t\to b W^+$. In the Higgs case, due to the heterogeneity  of the relevant decay processes, the discrepancy between the two normalizations is of order 20-50\%. }
\begin{equation}\label{eq.GammaH_def}
\d \Gamma_h = \Gamma_h^{SM,\text{best}}\sum_f 
\Br_{h\to f}^{SM,\text{best}}
\left[\frac{\d\Gamma_{h\to f}}{\Gamma_{h\to f}^{\rm SM}}\right]_{\rm tree},
\hspace{3mm}
\Gamma_h^{SM,\text{best}} = \unit[4.07]{MeV}\,,
\end{equation}
with $f$ running over the set $\{\g\g, Z\g, gg, b\bar b, c\bar c, \tau^+\tau^-\}$ plus the allowed 4-fermion channels. In the \smeftsim\ implementation, only 4-fermion decays proceeding via charged currents ($h\to WW^*\to 4f$) are retained, in order to simplify the analytic expressions. Channels mediated by neutral bosons ($h\to ZZ^*, Z\g^*, \g^*\g^*,g^*g^*\to 4f$) give subdominant corrections, that are estimated in a $3-5\%$ change to the dependence on $\C_{HW},\C_{HB},\C_{HD}$ and  a  change $\lesssim1\%$ for the other Wilson coefficients~\cite{Brivio:2019myy}.

\begin{table}[t]\centering
\renewcommand{\arraystretch}{1.5}
\begin{tabular}{|*3{>{$}c<{$}>{$}l<{$}|}}
\toprule
f& \Br_{h\to f}^{SM,\text{best}}&
f& \Br_{h\to f}^{SM,\text{best}}&
f& \Br_{h\to f}^{SM,\text{best}}
\\\midrule
\g\g & 2.27 \times 10^{-3}
&
Z\g & 1.541 \times 10^{-3}
&
gg & 0.0818
\\
\bar bb & 0.5809
&
\bar cc & 0.02884
&
 \tau^+ \tau^-&0.06256
\\
ll\bar\nu\nu& 0.0256
&
\bar uu\bar d d& 0.1097 
&
l^-\bar\nu \bar du + \hc& 0.1062
\\\bottomrule
\end{tabular}
\caption{Numerical values of the Higgs boson branching ratios employed in the definition of $\d \Gamma_h$, Eq.~\eqref{eq.GammaH_def}. The values for 2-body decays are taken from Ref.~\cite{LHCHXSWG-tables}, with $m_h=\unit[125.09]{GeV}$. The values in the last line are estimated computing the partial widths with {\tt Prophecy4f 2.0}~\cite{Bredenstein:2006rh} and normalizing their sum to 
$\Br_{h\to 4f}^{SM,\text{best}} = 0.24161$
~\cite{LHCHXSWG-tables}. They include only charged current contributions and are summed over all allowed flavor combinations.}
\label{tab.Hwidths_SM}
\end{table}

$\Gamma_h^{SM,\text{best}}$ is a free parameter in the models and can be modified by the user. The best-fit branching ratios, instead, are embedded numerically in the $\d\Gamma_h$ expressions and cannot be changed.
The values employed are reported in Table~\ref{tab.Hwidths_SM}. 
The relative deviations $\left[\d\Gamma_{h\to f}/\Gamma_{h\to f}^{\rm SM}\right]_\text{tree}$ for 2-body decays are computed with the \feynrules\ tools, retaining the full dependence on all the relevant fermion masses and Yukawa couplings. If a given Yukawa coupling $y_f$ is set to zero in the {\tt param\_card}, all contributions to $\d\Gamma_h$ originating from the $h\to f\bar f$ decay channel are dropped.
For the $h\to 4f$  channels we take the analytic results of  Ref.~\cite{Brivio:2019myy}, that neglect \emph{all} fermion masses and quark mixings.
Note that the results in Ref.~\cite{Brivio:2019myy} were given for the $U(3)^5$ flavor symmetric case, and they have been generalized to the other flavor assumptions in \smeftsim. 
Full analytic results for SMEFT corrections are reported in Appendix~\ref{app.dwidthExpressions}.


\section{Usage in \mathematica}\label{sec.Mathematica_use}
The \feynrules\ files in \smeftsim\ can be imported in \mathematica~\cite{Mathematica} and used to print out analytic expressions for the Feynman rules and Lagrangian terms. A template \mathematica\ notebook is available at the GitHub repository, with examples of usage of the code in different setups.
The functionalities are the standard \feynrules\ ones. However, since  \smeftsim\ is an unusually complex model, some recommendations are in order. 

Before importing the model, the user must specify a flavor setup and EW input scheme choice. For instance, after loading \feynrules:
\begin{lstlisting}
SetDirectory["PATH/TO/SMEFTSIM/DIR/"];

Flavor = U35;
Scheme = MwScheme;
LoadModel["SMEFTsim_main.fr"];
\end{lstlisting}
Allowed options for {\tt Flavor} are \general, \Utf, \MFV, \top, \topsl. Allowed options for {\tt Scheme} are {\tt alphaScheme, MwScheme}.
The loading time varies between flavor assumptions and can take up to a few seconds.\footnote{All the timings indicated in this section refer to a four-core laptop, with \feynrules\ calculations parallelized.}

The \feynrules\ code is split over different files that contain the required operators, parameters and Lagrangian definitions. The implementation is such that only the objects matching the selected flavor structure and EW input scheme are defined upon loading. In all models, the following Lagrangians are defined (definitions were given in Section~\ref{sec.SMEFT}):
\begin{itemize}
\item {\tt LGauge} $=\Lag_{\rm gauge}$. Contains the SM terms plus the linearized SMEFT corrections due to field redefinitions and input parameter shifts.
\item {\tt LGaugeP}. Same as {\tt LGauge}, but with at least one $W$ or $Z$  boson replaced with the corresponding dummy field {\tt W1, Z1}.
\item {\tt LHiggs} $=\Lag_{\rm Higgs}$.  Contains the SM terms plus the linearized SMEFT corrections due to field redefinitions and input parameter shifts.
\item {\tt LHiggsP}. Same as {\tt LHiggs}, but with at least one $W$, $Z$ or Higgs boson replaced with the corresponding dummy field {\tt W1, Z1, H1}.
\item {\tt LFermions} $=\Lag_{\rm fermions}$. Contains the SM terms plus the linearized SMEFT corrections due to field redefinitions and input parameter shifts.
\item {\tt LFermionsP}. Same as {\tt LFermions}, but with at least one top quark or $W,Z$ boson replaced with the corresponding dummy field {\tt t1, W1, Z1}.
\item {\tt LYukawa}  $=\Lag_{\rm Yukawa}$. Contains the SM terms plus the linearized SMEFT corrections due to field redefinitions and input parameter shifts.
\item {\tt LYukawaP}. Same as {\tt LYukawa}, but with at least one top quark or Higgs boson replaced with the corresponding dummy field {\tt t1, H1}.
\item {\tt LSM} $=\Lag_{\rm SM}$. The SM Lagrangian without any SMEFT correction.
\item {\tt LSMlinear}. The SM lagrangian plus the linearized SMEFT corrections due to field redefinitions and input parameters shifts.
\item {\tt LSMloop} $=\Lag_{\rm SMhloop}$ as defined in Eq.~\eqref{eq.L_SMHloop}.
\item {\tt LSMloopP}. Same as {\tt LSMloop}, but with at least one Higgs or $Z$  boson replaced with the corresponding dummy field {\tt H1, Z1}.
\item {\tt LSMincl} = {\tt LSMlinear + LSMloop} 
\item {\tt L6cl[n]} $=\Lag_6^{(n)}$ with {\tt n} = 1\dots 8. 
\\ For class 4, the definition is split into {\tt L6cl4, L6cl4cpv} containing only the CP-even and -odd terms respectively.
\\ For class 8, sub-Lagrangians {\tt L6cl8a} \dots {\tt L6cl8d} are defined in addition.
\item {\tt L6no4f} $=\sum_{n=1}^7 \Lag_6^{(n)}$
\item {\tt L6} $=\sum_{n=1}^8 \Lag_6^{(n)}$
\item {\tt LSMEFT} $=$ {\tt LSMincl + L6}
\end{itemize}
The last 3 Lagrangians contain extremely long expressions. It is strongly recommended to use them with care and avoid calling these variables unless strictly necessary.

The parameters notation in the code is provided in Appendix~\ref{app.parameters}.
In addition, the following parameters lists are defined in all models:
\begin{itemize}
\item {\tt WC6}. The list of all Wilson coefficients.\\
In the \general\ model the list {\tt WC6indices} is defined in addition. In this case {\tt WC6} contains eg. {\tt cHuIm11, cHuIm33}, while {\tt WC6indices} contains {\tt cHu[ff1\_,ff2\_]} with blank flavor indices.
\item {\tt shifts}. The list of all shift parameters, such as {\tt dGf, dMZ2, dgw, dg1} etc. The complete list is given in Table~\ref{tab.defs_bosonic}.
\item {\tt d6pars}. List of all SMEFT quantities, including Wilson coefficients with and without free indices, and shifts. 
\end{itemize}
Two handy functions are also defined:
\begin{itemize}
 \item {\tt LinearWC[x\_]}. Expands the expression {\tt x} to linear order in the SMEFT parameters (Wilson coefficients and shifts).
 \item {\tt SMlimit[x\_]}. Returns the SM limit of the expression {\tt x}, setting to zero all the {\tt d6pars}.
\item {\tt relativeVariation[x\_]}. Returns {\tt x / SMlimit[x]}.
\item {\tt SimplifyWC[x\_]}. Returns the expression {\tt x} in a form that collects the contributions from each SMEFT parameter (both Wilson coefficients and shifts).
\end{itemize}
By default, all the Feynman rules are printed out in an input scheme-independent form. The expressions in terms of Wilson coefficients are recovered via a replacement rule that should be applied with {\tt ReplaceRepeated}, as in:\\[2mm]
\includegraphics[height=1.8cm]{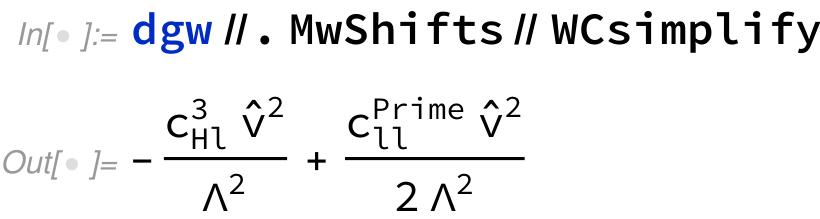}\\
if the model was imported with {\tt Scheme = MwScheme},
or 
\\[2mm]
\includegraphics[height=1.8cm]{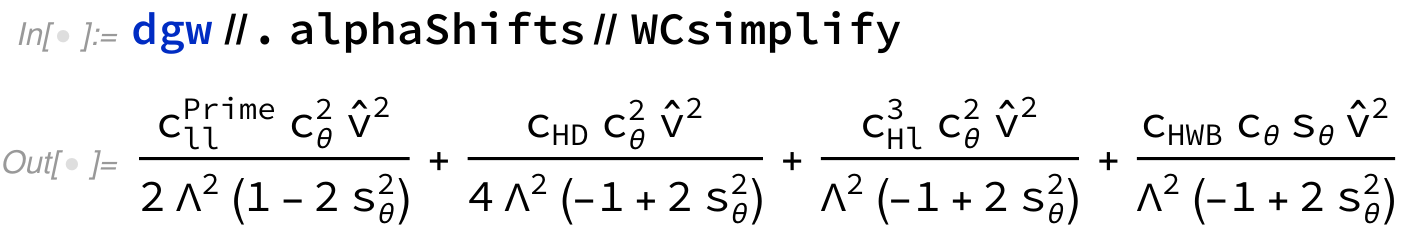}\\
if the model was imported with {\tt Scheme = alphaScheme}. For consistency, only the appropriate replacement list is defined in each case.  All Feynman rules are printed in unitary gauge. It is possible to switch to Feynman gauge by changing the flag {\tt FeynmanGauge} to {\tt True} at any time. However, we caution the user that the SMEFT contributions to the Goldstone and ghost Lagrangians are not fully implemented. In particular, gauge fixing terms have been omitted and the Goldstone kinetic terms are \emph{not} canonically normalized.

The operators' names start with {\tt O} and their definitions carry free flavor indices (eg. {\tt OHu[ff1\_,ff2\_]}). In this way they can be shared by multiple setups (the \general, \Utf\ and \MFV\ models all use a set of definitions, and the \top, \topsl\ models share a separate one). The distinction between flavor assumption is coded through the flavor contractions in the Lagrangian definition, contained in {\tt SMEFTsim\_d6\_lagrangian.fr}. Therefore it is recommended to isolate each operator through its Wilson coefficient. For instance, the Feynman rules of the operator $Q_{Hu}$ with a $U(3)^5$ flavor symmetry can be printed out via:\\[2mm]
\includegraphics[width=35cm]{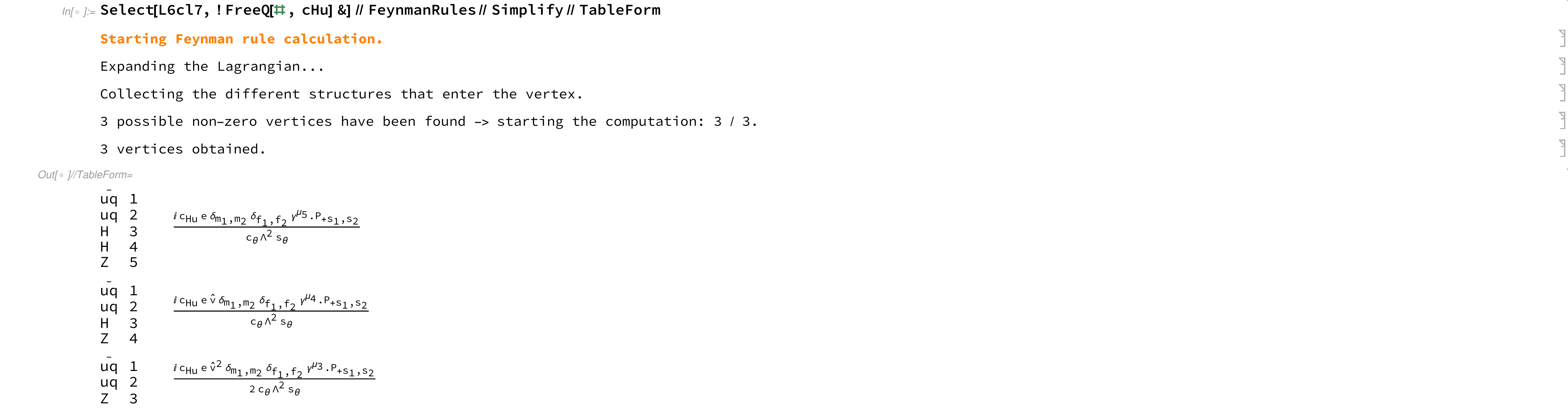}\\[2mm]
Alternatively, one can reproduce the flavor contraction structure explicitly. In this case:\\[2mm]
\includegraphics[width=35cm]{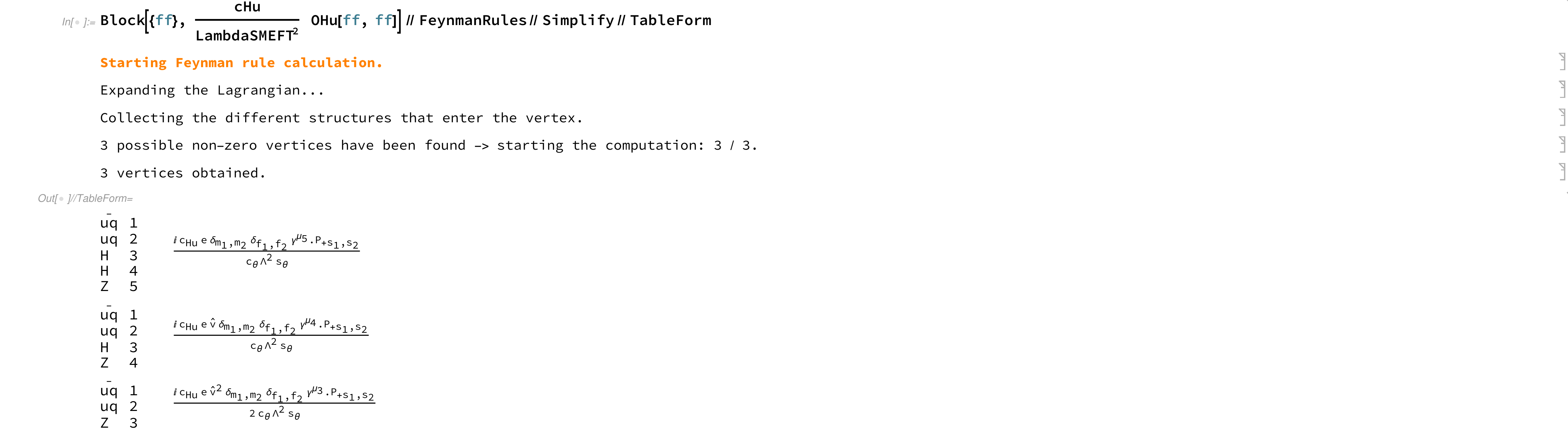}\\[2mm]
or print the result for one flavor entry only:
\\[2mm]
\includegraphics[width=35cm]{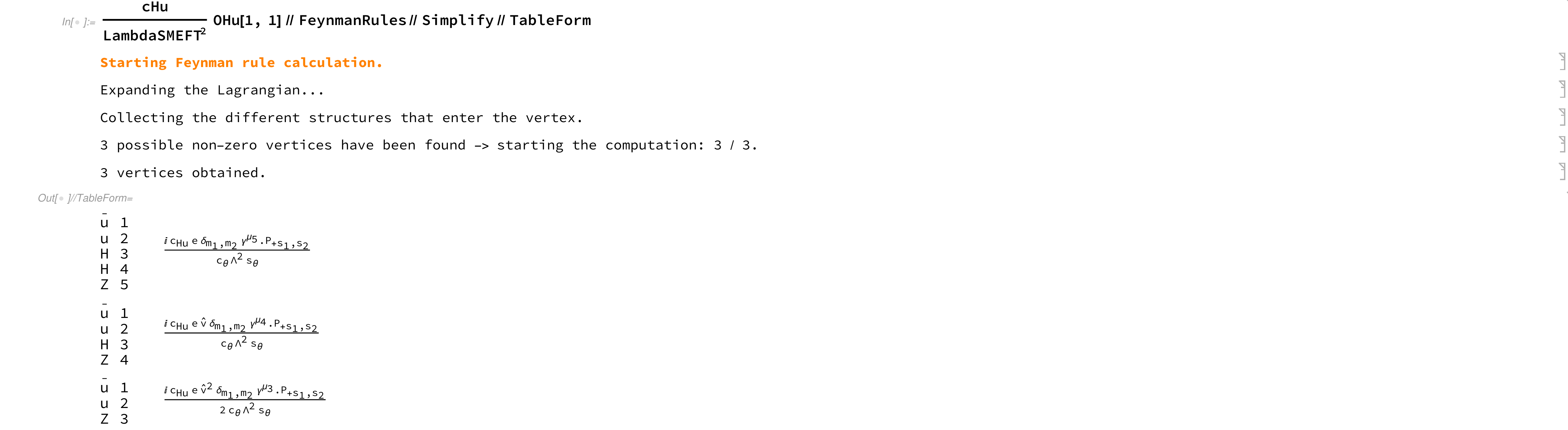}

The Feynman rules of the SM Higgs loop Lagrangian are quite complex, especially in the Higgs-gluon operators' case. Their evaluation with the {\tt FeynmanRules} command can be extremely slow and take up to a few hours for the most complex vertices. In order to facilitate their access, the file {\tt SMEFTsim\_SMHloop\_FRs.nb} is provided, that contains pre-exported expressions. They can be accessed from another notebook via
\begin{lstlisting}
 SetDirectory["PATH/TO/SMEFTSIM/DIR"];
 NotebookEvaluate["SMEFTsim_SMHloop_FRs.nb"];
\end{lstlisting}
Without producing any output, this will define the objects {\tt lhloop5S, lhloop5PS}.
The former is a list of all Feynman rules from $\Lag_{\rm SMhloop}$, for vertices with up to 5 legs. The latter is the same, but with at least one Higgs boson replaced by the dummy field {\tt H1}. For instance, the first entry is the $h\g\g$ vertex:\\[2mm]
\includegraphics[height=1.7cm]{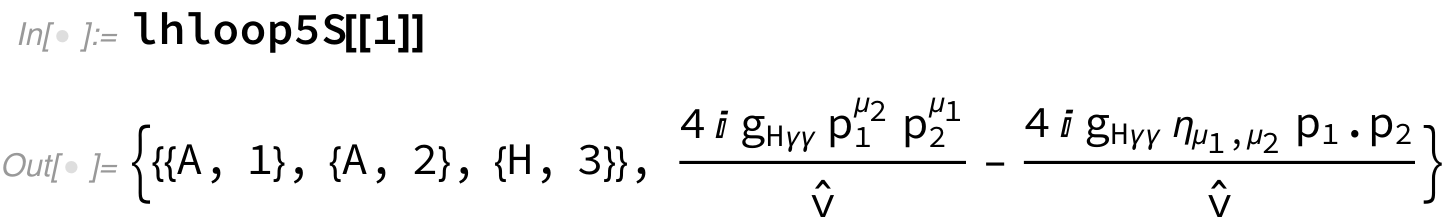}\\
and
\\[2mm]
\includegraphics[height=1.7cm]{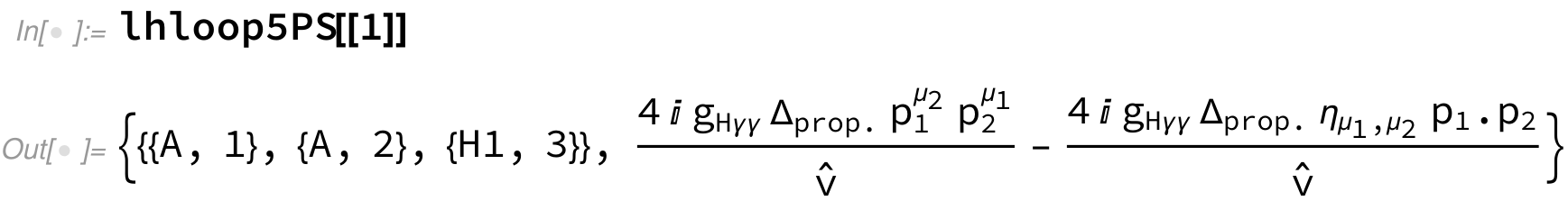}\\
is explicitly proportional to the {\tt propCorr} parameter (shown here as $\Delta_{\rm prop.}$), see Sec.~\ref{sec.propagators_smeftsim},~\ref{sec.propagators_madgraph}.

Finally, it is \emph{not} recommended to export the \ufo\ models independently, unless only a small subset of operators is included. The \ufo s provided in the GitHub repository have been exported in a specific, optimized way and the python files have been manipulated a posteriori in order to introduce the modified form of the propagators and to define the Higgs decay width such that the fermionic decay contributions to $\d\Gamma_h$ is removed whenever the corresponding Yukawa coupling is set to zero.
The notebook with the original export procedure is available upon request.


\section{Usage in \madgraph}\label{sec.MG_use}
This section provides recommendations for the use of \smeftsim\ in \madgraph. It is in no way meant as a manual for the functionalities of \madgraph\ itself, for which we defer the reader to the appropriate references, see eg.~\cite{Alwall:2014hca,MG_launchpad,MG_refs}.

\vskip 1em

The \smeftsim\ package provides 10 pre-exported \ufo\ models, one for each flavor setup and input parameter scheme. Each of them contains the full $\Lag_{\rm SMEFT}$ defined in Sec.~\ref{sec.SMEFT} and~\ref{sec.flavor} and $\Lag_{\rm SMhloop}$ defined in Sec.~\ref{sec.Hloops}. 
The manipulations and redefinitions described in Sec.~\ref{sec.redefinitions} and~\ref{sec.inputs} have been consistently applied. 
The vertices contained in the models are derived in unitary gauge and the ghost fields have been removed: \smeftsim\ is designed for LO event generation and does not support the NLO syntax.
A list of the SMEFT parameters defined in the codes is provided in Appendix~\ref{app.parameters}, with a mapping to the notation used in this notes. All the \ufo\ models have been validated following the recommendations in Ref.~\cite{Durieux:2019lnv}, as detailed in Appendix~\ref{app.validation}.

Although the selection of an appropriate model is of course up to the taste of the user, each flavor setup is meant to optimize the parameterization of a certain class of effects in the SMEFT.
For instance, the \top\ and \topsl\ models are designed to single out the couplings of the top and bottom quarks~\cite{AguilarSaavedra:2018nen}, and they only differ in that the \top\ case provides more freedom to distinguish the lepton flavors. The \Utf\ models allow one to work with a minimal number of parameters and are recommended for flavor-blind processes or whenever the flavor structure can be assumed to be strictly  SM-like. At the other side of the spectrum, the \general\ models provide maximal freedom and can be used to study flavor-violating processes or to realize arbitrary flavor structures beyond those implemented.
An operator by operator comparison of the different flavor structures is provided in Appendix~\ref{app.flavor_comparison}.

As discussed in Sec.~\ref{sec.propagators}, the use of the \mwscheme\ input scheme is particularly recommended for processes involving $W$ bosons, as it avoids the problematic introduction of SMEFT corrections to the $W$ pole mass.
The \ascheme\ and \mwscheme\ input scheme implementations are expected to give results that differ most significantly in the dependence on the Wilson coefficients $C_{HWB},\, C_{HD}, \, C_{Hl}^{(3)},\, C_{ll}'$ (or $(C_{Hl}^{(3)})_{11},\, (C_{Hl}^{(3)})_{22},\,(C_{ll})_{1221}$ in models with explicit flavor indices) and in the presence/absence of corrections to either $m_W$ or $\aem$, as described in Sec.~\ref{sec.inputs_ew}. Further numerical differences affecting both EFT and SM predictions can be present, due to the different definition of the SM parameters in the two cases. These are generally subleading.

\subsection{Parameter cards and restrictions}\label{sec.param_rst_cards}
The model parameters are grouped in blocks, that are explicitly shown in the parameter cards.
Besides the usual ones ({\tt SMINPUTS, MASS, DECAY, YUKAWA, CKMBLOCK}) , the parameter card of each \smeftsim\ \ufo\ model contains the blocks:
\begin{description}
\item[\tt SMEFTCUTOFF] -- the parameter $\Lambda$. By default this is $\unit[1]{TeV}$.
\item[\tt SMEFT] -- the CP-conserving Wilson coefficients, with default value 0.
\item[\tt SMEFTCPV]-- the CP-violating Wilson coefficients, with default value 0.\\ This block is absent in the \MFV\ models.
\item[\tt SMEFTFV] -- the $(\Delta C_\a)$  parameters of the \MFV\ setup, with default value 0.\\ This block is only present in \MFV\ models.
\item[\tt SWITCHES] -- the parameter {\tt linearPropCorrections}, that can be used to switch ON/OFF the linearization of SMEFT corrections in the propagators. The default value is 0 (OFF).
\end{description} 
The {\tt SMINPUTS} block contains $G_F$, $\a_s$ and either $\aem$ or $m_W$ depending on the input scheme.

\vskip 1em
\noindent
The use of restriction cards allows one to reduce the number of diagrams generated for a given process. 
Two restriction cards are provided by default with each \ufo:
\begin{itemize}
\item {\tt restrict\_massless.dat.}
The masses and Yukawa couplings of all fermions, except the top and bottom quarks, are set to 0. The CKM matrix is set to the identity.
The Wilson coefficients are set to arbitrary numerical values.

\item {\tt restrict\_SMlimit\_massless.dat.} As in {\tt restrict\_massless.dat}, but with all Wilson coefficients set to 0.
\end{itemize}
The restrictions should be applied at the stage where the model is imported, eg.:
\begin{lstlisting}
import model SMEFTsim_U35_MwScheme_UFO-massless
\end{lstlisting}
In this way, all the parameters that are set to either 0 or 1 in the restriction are fixed to their value and cannot be edited further. Sets of parameters that are assigned an identical value in the restriction are fixed to be identical: while their numerical value can still be edited, they cannot be disentangled from one another.  Diagrams that are proportional to a vanishing parameter will not be generated. 

The use of one of the {\tt massless} restrictions is recommended for LHC studies, because it simplifies significantly the calculations. 
There are of course several possible strategies for the use of these restrictions in \madgraph: for instance, one can create a modified version of {\tt restrict\_SMlimit\_massless.dat} turning on one Wilson coefficient with some arbitrary value $\neq0$. Importing the model with this modified restriction allows one to generate events with the chosen coefficient only, while all the other operators are forbidden. 
Alternatively, if the model is imported with {\tt restrict\_massless.dat}, all the Wilson coefficients are retained: all the allowed SMEFT diagrams will be generated and all the parameters can be freely edited at the event generation stage. Note that, to achieve this, all the Wilson coefficients in {\tt restrict\_massless.dat} are assigned different non-vanishing and non-unitary values\footnote{In previous versions of \smeftsim, the Wilson coefficients in the restriction cards were all set to the special value {\tt 9.999999e-01}, that in principle allows one to set the parameters to {\tt 1} without \emph{fixing} their value. However, this syntax is not fully supported by \madgraph, and is occasionally source of unexpected numerical behavior in \ufo\ models with a very large number of parameters, such as \smeftsim\ \general\ or \MFV.   }, that will need to be changed prior to the event generation. To simplify this operation, a  ``restricted'' parameter card {\tt param\_card\_massless.dat} is provided in the \ufo, where all the Wilson coefficients are set to 0. This card can be directly copied in the {\tt PROC/Cards/} directory of the exported process and modified at will.

\subsection{Interaction orders}\label{sec.int_orders}
A standard feature of \ufo\ models is that every coupling parameter is assigned an interaction order, i.e. a ``flag'' that provides control on the number of coupling insertions in generated Feynman diagrams. Each parameter carries an arbitrary number of interaction orders.

\subsubsection{Definitions}
In the \smeftsim\ \ufo\ models the interaction orders are assigned as reported in Table~\ref{tab.interaction_orders}.
\begin{table}[t]\centering
\renewcommand{\arraystretch}{1.4}
\begin{tabular}{|>{\tt}lc|l|}
\toprule
\multicolumn{2}{|l|}{\rm\bf Order}& \bf Parameters assigned \\
\midrule
QED = 1& &$e, \,g_W, \,g_1, \,Y_\psi$  \\
QED = 2&  &$G_F,  \,\hat \lambda, \,\aem$   \\
QED = -1& &$\hat v,\, \Lambda$  \\
\midrule
QCD = 1&  &$g_s$  \\
QCD = 2&  &$\a_s$   \\
\midrule
SMHLOOP = 1& &$g_{H\g\g},\, g_{HZ\g},\,g_{Hgg}^{(1)}\dots g_{Hgg}^{(5)}$\\
\midrule
NP = 1&  &all Wilson coefficients and shifts, except {\tt propCorr}\\
NPcpv = 1 &$(\star)$& all CP-violating Wilson coefficients\\

NPfv = 1&$(\star)$& all ($\Delta  C)$ parameters in \MFV\ models\\

NPc[a] = 1&$(\star)$& all the CP components and flavor indices of the Wilson coefficient $C_\a$\\

NPprop = 1&$(\star)$& {\tt propCorr} \\

\bottomrule
\end{tabular}
\caption{Interaction orders defined in the \smeftsim\ \ufo\ models. Those marked with a $(\star)$ are new in version~3.0. The string {\tt [a]} stands for a generic root name of a Wilson coefficient, as listed in Appendix~\ref{app.parameters}.}\label{tab.interaction_orders}
\end{table}

The orders {\tt QED} and {\tt QCD} are assigned as customary in the standard SM \ufo\ implementations, with the exceptions of the SMEFT cutoff $\Lambda$, that has been assigned {\tt QED=-1} such that the combination $(\hat v/\Lambda)$ is order-less, and of the Wilson coefficient $C_H$, that has been assigned {\tt QED=1}. This prevents the $C_H$ correction to the $h^3$ interaction, that is proportional to $\hat v^3 C_H/\Lambda^2$ (see Eq.~\eqref{eq.VH_redef_inputs}), from having overall order {\tt QED = -1}. 

The interaction order {\tt SMHLOOP} labels the SM loop-generated Higgs interactions introduced in Sec.~\ref{sec.Hloops}. Since by definition they are proportional to the SM gauge couplings, the $g_{Hgg}^{(k)}$ parameters additionally carry {\tt QCD=2} and the $g_{H\g\g},\,g_{HZ\g}$ parameters carry {\tt QED=2}.

The interaction order {\tt NP} (New Physics) is assigned to all the Wilson coefficients and shifts indistinctly.
In addition, starting from version~3.0, individual interaction orders have been introduced for each effective operators. The same order  {\tt NPc[a]} is assigned to all the associated CP-conserving and violating parameters, irrespective of the flavor indices carried. For instance, in the \top\ models, the parameters $\re (C_{eH})_{pp},\,\im(C_{eH})_{pp}$ for $p=\{1,2,3\}$ all have order {\tt NPceH=1}. In the \Utf, \MFV, \top\ and \topsl\ models, distinct interaction orders are assigned to independent flavor contractions. For instance $C_{ll}$ and $C_{ll}'$ have orders {\tt NPcll} and {\tt NPcll1} respectively. In the \top\ models, the parameters $(C_{ll})_{pprr}$ have order {\tt NPcll}, while the $(C_{ll})_{prrp}$ contractions have order {\tt NPcll1}, etc. In most cases the label {\tt [a]} coincides with the name root of the associated Wilson coefficient, that can be read off from the tables in Appendix~\ref{app.parameters}.
If in doubt, the user can resort to the {\tt.fr} source files or check explicitly the {\tt couplings.py} file to identify the exact orders assigned to a given parameter or coupling. 

All the CP-violating parameters, that belong to the {\tt SMEFTCPV} block, have an order {\tt NPcpv=1}. Analogously, all the $(\Delta C_\a)$ quantities in the \MFV\ models, that belong to the {\tt SMEFTFV} block, have an order {\tt NPfv=1}.

Finally, the order {\tt NPprop} labels the interactions of the dummy fields $W',Z',h',t'$ carrying linearized propagator corrections, see Sec.~\ref{sec.propagators_smeftsim}. It is carried by a dummy internal parameter {\tt propCorr} that only takes values 0/1, when the {\tt linearPropCorrections} switch is set to 0/a nonzero value. Its application is discussed in the next subsection.
By default, the interaction order {\tt NPprop} is ``switched off'', as it is assigned an upper limit of 0 interactions, that can be lifted as shown below. No upper limit is set for the other orders. 

The interaction orders {\tt SMHLOOP, NP, NPprop, NPcpv, NPfv} have been assigned hierarchy 99.  \madgraph\ will therefore generally avoid insertions of the associated vertices, unless these orders are specified. 

\subsubsection{Recommended use}
Interaction orders are specified at the stage of process generation in \madgraph, eg:
\begin{lstlisting}
generate p p > mu+ mu- SMHLOOP=0 NPprop=0 NP=1 NP^2==1 
\end{lstlisting}
where {\tt =} is equivalent to {\tt <=}, while {\tt ==} selects uniquely the order specified. The syntax {\tt XX=n} acts at the amplitude level, i.e. it specifies the total number of couplings with order {\tt XX} to be inserted in each Feynman diagram. The syntax {\tt XX\^{}2} acts instead at the \emph{squared} amplitude level. 
This functionality works very nicely for EFT studies, as it allows one to disentangle contributions at different orders in the expansion.

Although \emph{a priori} \smeftsim\ can be used for computations to any allowed order in $\Lambda$, it implements the SMEFT Lagrangian consistently expanded only up to $\O(\Lambda^{-2})$.
This means that any \smeftsim\ prediction beyond this order is necessarily incomplete in the Effective Theory. It is worth noting that this statement does not concern only higher dimensional operators in $\Lag_{\rm SMEFT}$, but also affects the dependence on some of the Wilson coefficients in $\Lag_6$. 
For instance, it was stressed at multiple stages in Sections~\ref{sec.redefinitions} and~\ref{sec.inputs} that terms of order $\Lambda^{-4}$ or higher were neglected in the field and parameter redefinitions performed, as well as in the treatment of input parameters. The impact of these $\Lag_6$ contributions has been discussed in Ref.~\cite{Hays:2020scx} for the case of $\O(\Lambda^{-4})$ corrections to $1\to2$ decays, using the geoSMEFT formalism~\cite{Helset:2020yio}.

Complete results truncated at $\O(\Lambda^{-2})$ can be obtained with the syntax {\tt NP<=1 NP\^{}2<=1}, that retains only SM plus SM-$\Lag_6$ interference contributions. Contributions of order $\Lambda^{-4}$ stemming from the square of an $\O(\Lambda^{-2})$ amplitude, although incomplete, are also commonly included in the SMEFT calculations: they are selected with {\tt NP=1 NP\^{}2==2} or just {\tt NP==1}. For the reasons above, it is generally recommended to use the specification {\tt NP<=1} (or {\tt NP=1}) for any process, to limit the number of EFT insertions to one per Feynman diagram.
A generic observable computed in this way will have the form
\begin{equation}\label{eq.generic_obs}
\s = \s_{\rm SM} + \sum_\a \s_\a \C_a + \sum_{\a,\b} \s_{\a\b} \C_\a \C_\b\,,
\end{equation} 
where  $\s_{\rm SM},\,\s_\a,\,\s_{\a\b}$ intuitively denote  the SM, interference and quadratic contributions respectively. 
$\s_{\rm SM}$ and $\s_{\a\a}$ are always positive quantities, while $\s_\a,\,\s_{\a\beta}$ with $\a\neq\beta$ can take negative values.
Table~\ref{tab.int_orders_syntax} shows examples of how the interaction order syntax can be used to disentangle these contributions, for a simple case with two Wilson coefficients. The expressions directly generalize if three or more parameters are present.

 \begin{table}[t]\centering
 \begin{tabular}{|>{\tt}l|*6{c}|}
 \toprule
          & $\s_{\rm SM}$& $\s_\a$& $\s_\b$& $\s_{\a\a}$& $\s_{\b\b}$& $\s_{\a\b}$\\\midrule
  NP=0&  \checkmark& & & & & 
  \\
  NP<=1&  \checkmark& \checkmark& \checkmark& \checkmark& \checkmark& \checkmark
  \\
  NP==1& & & & \checkmark& \checkmark& \checkmark
  \\
  NP<=1 NP\^{}2<=1& \checkmark & \checkmark& \checkmark& & &
  \\
  NP<=1 NP\^{}2==1& & \checkmark& \checkmark& & &
  \\
  NP<=1 NPc[a]\^{}2<=1& \checkmark& \checkmark& & & &\checkmark
  \\
  NP<=1 NPc[a]\^{}2<=1 NPc[b]\^{}2<=1& \checkmark& \checkmark& \checkmark& & &\checkmark  
  \\
  NP<=1 NPc[a]==1& &  \checkmark& &  \checkmark& &
  \\  
  NP<=1 NPc[a]\^{}2==1& &  \checkmark& &  & &\checkmark
  \\
  NP<=1 NPc[a]\^{}2==2& &  & & \checkmark & &
  \\
  NP<=1 NP\^{}2==1 NPc[a]\^{}2==1& &  \checkmark& &  & &  
  \\
  NP<=1 NP\^{}2==2 NPc[a]\^{}2==1& & & &  & & \checkmark
  \\\bottomrule
 \end{tabular}
 \caption{Examples of interaction-order syntax that select different EFT contributions to a generic observable with the dependence given in Eq.~\eqref{eq.generic_obs}, for the case of 2 Wilson coefficients $C_\a,\,C_\b$. }\label{tab.int_orders_syntax}
 \end{table}

Finally, as discussed in Sec.~\ref{sec.Hloops}, the loop-generated SM Higgs couplings implemented in \smeftsim\ are defined in the $m_t\to\infty$ limit, and their use should be limited to (on-shell) Higgs production and decay processes. Outside of this regime, it is strongly recommended to use {\tt SMHLOOP=0}.

\subsection{Propagator corrections and decay widths} \label{sec.propagators_madgraph}     
As discussed in Sec.~\ref{sec.propagators}, SMEFT corrections are generally present in the propagators of unstable particles, due to $d=6$ operators modifying their masses and/or decay widths. Sec.~\ref{sec.propagators_smeftsim} outlined two alternative methods for estimating these contributions in a given process. In the following we illustrate how they can be implemented in \madgraph.

\subsubsection{Method (a): linearized corrections}

\smeftsim~3.0 offers the possibility to linearize propagator corrections as in Sec.~\ref{sec.propagators} for the $Z,\,W,\,h$ bosons and for the top quark.
The implementation relies on the introduction of dummy fields $Z',\,W',\,h',\,t'$ whose couplings carry interaction order {\tt NPprop}, which allows one to single them out.
For instance, the syntax
\begin{lstlisting}
generate p p > e+ e- SMHLOOP=0 NP==1 NPprop=0
\end{lstlisting}
selects all diagrams with one effective operator in a \emph{vertex} (including 4-point $\bar qq e^+e^-$ interactions) but none in propagators, while
\begin{lstlisting}
generate p p > e+ e- SMHLOOP=0 NP=0 NPprop<=2 NPprop^2==2
generate p p > e+ e- SMHLOOP=0 NP=0 NPprop==2
\end{lstlisting}
extract the pure SM-$\Lag_6$ interference ($\propto \d\Gamma_Z$) and the quadratic ($\propto (\d\Gamma_Z)^2$) contributions respectively from corrections to the $Z$ propagator\footnote{Remember that {\tt NPprop} counts the number of dummy \emph{vertices}, so, in this case, the order specified is twice the number of dummy propagators.}, and exclude EFT insertions in vertices.
Due to the absence of additional interaction orders, there is unfortunately no equivalent to Table~\ref{tab.int_orders_syntax} in this case: propagator corrections from different operators and EFT orders cannot be disentangled at this level.

\underline{Important}: in order to avoid unwanted insertions of the dummy fields in standard process generations, the functionality described here has to be activated in 2 steps: (i)~in the file {\tt coupling\_order.py}, the {\tt expansion\_order} option for the order {\tt NPprop} has to be set to a number $\geq 2$ (recommended: 99). The default is 0, which forbids dummy interactions completely.
(ii)~The parameter {\tt linearPropCorrections} in the {\tt param\_card} has to be set to a non-zero value. If this is not the case, dummy vertices will be included in the diagrams, but they will be idle, as they are proportional to {\tt propCorr} = $0/1$ for {\tt linearPropCorrections}~=~0/non-zero.

\subsubsection{Method (b): full corrections}
As an alternative to linearization, propagator corrections can be estimated following more canonical procedures. This generally means computing processes with the propagator forms in Eqs.~\eqref{def.PV}-\eqref{def.PF}, with mass and decay parameters that depend explicitly on the Wilson coefficients, to either linear or quadratic order.
The resulting process will thus exhibit a non-polynomial dependence on the SMEFT parameters.

The most relevant caveat here is that the implementation of the Wilson coefficient dependence is necessarily different for masses and widths. In the former case, it is possible to define mass parameters as internal and assign them an analytic expression, eg {\tt MW = MWsm + dMW}, with {\tt dMW} defined as in~\eqref{eq.dmw_alphascheme}. In \smeftsim\ the {\tt dMW} term is only included when {\tt linearPropCorrections} $=0$, and switched off otherwise.
Note also that the expression of a generic $\delta m$ is extracted at the Lagrangian level and is purely of $\O(\Lambda^{-2}$).

On the other hand,
due to how Monte Carlo generators and their interface to parton shower or decay modules are structured, decay widths cannot be defined as internal parameters in \ufo\ models. 
Therefore the only way their SMEFT expressions can be inserted in the calculation is by letting \madgraph\ compute them, by setting the relevant widths to {\tt Auto} in the {\tt param\_card}~\cite{Alwall:2014bza}. The on-the-fly calculation  will include all allowed 2-body decays as well as higher multiplicity decays estimated to be numerically relevant, and it will rely on the pre-computed decay results collected in the file {\tt decays.py}, which include $\O(\Lambda^{-4})$ terms.\footnote{Since only $1\to 2$ decays are included, these results consistently stem from the square of $\O(\Lambda^{-2})$ amplitudes. Dummy fields are not included in the pre-computed decay widths. } Note that, with this procedure, the decay widths will need to be re-evaluated every time the value of a relevant Wilson coefficient is modified.

With both mass and width corrections evaluated as above, the denominator of a generic propagator has the form
\begin{equation}\label{eq.propagator_denom_full}
q^2 - (m^{\rm SM} + \d m)^2 + i (m^{\rm SM}+\d m)(\Gamma^{\rm SM} + \d^{(1)} \Gamma + \d^{(2)} \Gamma)\,,
\end{equation} 
where $\d m$ and $\d^{(1)}\Gamma$ are of $\O(\Lambda^{-2})$ and $\d^{(2)}\Gamma$ is of $\O(\Lambda^{-4})$. Eq.~\eqref{eq.propagator_denom_full} contains therefore terms up to $\O(\Lambda^{-6})$. An observable computed for a process with $k$ internal lines corrected in this way, will contain terms up to $\O(\Lambda^{-12\, k})$ at the denominator.

In principle this functional dependence can be reconstructed fitting the appropriate rational function to a sufficient number of benchmark points. 
However, it is recommended to reduce the proliferation of higher-order terms in the propagators, by evaluating mass and width corrections separately and by treating propagator corrections to different internal states individually, whenever possible. This is achieved avoiding to switch on at the same time the {\tt linearPropCorrections} flag (that turns on $\d m_W$) and the {\tt Auto} computation of a decay width, or of two decay widths simultaneously.

\subsection{Example: Higgs production and decay including \texorpdfstring{$W,Z$}{W,Z} propagator corrections}
As a practical example for the use of \smeftsim\ in \madgraph, we compute SMEFT corrections to Higgs production and decay processes that are mediated by $W,Z$ exchange, as an illustration of the propagator corrections feature. 

\subsubsection{STXS for \texorpdfstring{$\bar qq\to h \bar qq$}{q qbar -> h q qbar}}
We consider two bins of the stage 1.1 Simplified Template Cross Section (STXS) parameterization~\cite{deFlorian:2016spz,Badger:2016bpw,Berger:2019wnu,Amoroso:2020lgh}, for the EW $\bar qq\to h\bar qq$ production channel at low Higgs $p_T$.\\
They are defined by the cuts~\cite{Berger:2019wnu}:
\begin{center}
 \begin{tabular}{l|ll}
 VBF-like  & $ \unit[350]{GeV} <  m_{jj}\quad$& 
 $p_T(h) < \unit[200]{GeV}$,\,$|y_h|<2.5$
  \\
 VH-like & $ \unit[60]{GeV} <  m_{jj} <\unit[120]{GeV}$& $p_T(h) < \unit[200]{GeV}$,\,$|y_h|<2.5$.
\end{tabular}
\end{center}
with $y_h$ the rapidity of the Higgs boson.
In each bin, the Higgs production cross section in the SMEFT can be parameterized as:
\begin{equation}
 \s_{SMEFT} = {\s_{\rm SM}} + \sum_\a \s_\a\C_\a + \O(\Lambda^{-4})\,.
\end{equation}
Table~\ref{tab.MG_example_VBF_VH} reports the values of $\s_\a/\s_{\rm SM}$ for the relevant fermionic operators, computed at parton level using \smeftsim\ in the \Utf\ flavor-symmetric, \mwscheme\ input scheme version. 
The following procedure was followed: 
\begin{enumerate}
\item 50000 events are generated for each bin in \madgraph, for $\sqrt{s}=\unit[13]{TeV}$. The syntax used is
\begin{lstlisting}
 import model SMEFTsim_U35_MwScheme_UFO-vbf
 generate q q > h q q QCD=0 NP=0 SMHLOOP=0 NPprop=0
\end{lstlisting}
where the {\tt -vbf} flag indicates that the model is imported with a custom restriction card {\tt restrict\_vbf.dat}, that in this case sets to zero the masses and Yukawa couplings of all fermions except the bottom and top quarks, as well as all the Wilson coefficients that are known not to contribute to the process. The remaining ones are set to a random non-zero value in this card. The STXS defining cuts in $p_T(h)$ and $y_h$ cuts are implemented at the level of the {\tt run\_card.dat} in \madgraph, while the invariant mass cuts are applied when analyzing the events a posteriori.
This gives the tree-level SM cross sections
\begin{equation}
 \s_{\rm SM}^{VBF} = \unit[1.56]{pb}\,,\quad \s_{\rm SM}^{VH} = \unit[0.67]{pb}\,.
\end{equation}

\item The events are reweighted using the reweight module in \madgraph~\cite{Mattelaer:2016gcx}. Individual weights are computed for each Wilson coefficient, splitting contributions from operator insertions in the vertices (labeled as ``direct'') and from insertions in the $W,Z$ propagators (labeled as ``propagator'').  This is done setting each coefficient to 1 and the SMEFT cutoff scale {\tt LambdaSMEFT} to 1~TeV.
For instance, for the $C_{ll}'$ parameter, the {\tt reweight\_card.dat} for the direct contributions is
\begin{lstlisting}[style=MG_wide]
change process q q > h q q  QCD=0 NP=1 NP^2==1 NPprop=0 SMHLOOP=0
launch --rwgt_name=SMEFTsim-cll1-direct
set cll1 1
set cHl1 0
set cHl3 0
set cHe 0
set cHq1 0
set cHq3 0
set cHu 0
set cHd 0
done
\end{lstlisting}
For estimating the pure propagator contributions the first two lines are replaced with
\begin{lstlisting}[style=MG_wide]
change process q q > h q q  QCD=0 NP=0 NPprop=2 NPprop^2==2 SMHLOOP=0
launch --rwgt_name=SMEFTsim-cll1-propagator
\end{lstlisting}
Analyzing the reweighted events gives $\left(\s_\a\frac{\hv^2}{\Lambda^2}\right)$ for each $C_\a$. The numbers in Table~\ref{tab.MG_example_VBF_VH} are finally obtained dividing by $\s_{\rm SM}$ and normalizing to $\C_\a$. 

\end{enumerate}

\begin{table}[t]\centering
\renewcommand{\arraystretch}{1.3}
\begin{tabular}{|>{$}c<{$}|*2{>{$}r<{$}}|*2{>{$}r<{$}}|*2{>{$}r<{$}}|}
\toprule
&\multicolumn{2}{c}{$\bar qq\to h\bar qq$ VBF-like}&
\multicolumn{2}{|c}{$\bar qq\to h\bar qq$ VH-like}
&\multicolumn{2}{|c|}{$h\to e^+ e^-\mu^+\mu^-$}
\\
\hline
&\text{direct} & \text{propagators} & \text{direct}& \text{propagators}&\text{direct}& \text{propagators}\\
\midrule
\C_{He}&
&
5.32	\cdot 10^{-5}
&
&
0.0526
&
-1.724
&
0.153
\\
\C_{Hl}^{(1)}&
&
5.32\cdot 10^{-5}
&
&
0.0526
&
2.144
&0.153
\\
\C_{Hl}^{(3)}&
-6
&
1.351	\cdot 10^{-3}
&
-6
&
\cellcolor{tablesColor!80}
1.258
&
-3.856
&
\cellcolor{tablesColor!80}
1.147
\\
\C_{Hq}^{(1)}&
0.109	
&
-1.363\cdot 10^{-4}
&
-0.197
&
\cellcolor{tablesColor!80}
-0.135
&
&
\cellcolor{tablesColor!80}
-0.39
\\
\C_{Hq}^{(3)}&
-5.345	
&
-1.423 \cdot 10^{-3}
&
25.66
&
-1.329
&
&
\cellcolor{tablesColor!80}
-1.353
\\
\C_{Hu}&
-0.323	
&
-7.092\cdot 10^{-5}
&
1.926
&
-0.070
&
&
-0.203
\\
\C_{Hd}&
0.103	
&
5.24\cdot 10^{-5}
&
-0.608
&
0.0518
&
&
0.150
\\

\C_{ll}^\prime&
3	
&
-1. \cdot 10^{-3}
&
3
&
\cellcolor{tablesColor!80}
-0.936
&
3
&
\cellcolor{tablesColor!80}
-0.839
\\\bottomrule
\end{tabular}
\caption{Values of $\s_a/\s_{\rm SM}$ for the relevant fermionic Wilson coefficients $\C_\a$ contributing to $\bar qq\to h \bar qq$ and $h\to e^+e^-\mu^+\mu^-$. For the first two columns $\s$ is the cross section in the VBF-like and VH-like STXS bins defined in the text, while in the third the numbers refer to the partial decay width. The results are given for \mwscheme\ inputs with a $U(3)^5$ flavor symmetry, and neglecting all fermion masses. The ``direct'' contributions stem from operator insertions in vertices, including parameter shifts, while the ``propagator'' ones stem from the corrections to the $W,Z$ decay width in internal propagators. The lines highlighted in color are those for which the latter are most relevant.}\label{tab.MG_example_VBF_VH}
\end{table}
The results show that propagator corrections are negligible in the VBF regime, where the relative SMEFT corrections to the cross section is
\begin{equation}
\frac{\s_{SMEFT}}{\s_{\rm SM}} = 1 - 3 \cdot 10^{-4} \,\frac{\d\Gamma_Z}{\Gamma_Z^{\rm SM}}
 - 7 \cdot 10^{-4}\, \frac{\d\Gamma_W}{\Gamma_W^{\rm SM}} + \text{direct}\,.
\end{equation}
This is expected, as the intermediate bosons in $t$-channel are mostly off-shell in this process.
In the VH bin, on the other hand, one vector boson can be on-shell, which enhances the propagator effects. In this case, the relative SMEFT correction is
\begin{equation}
\frac{\s_{SMEFT}}{\s_{\rm SM}} =1 - 0.29 \,\frac{\d\Gamma_Z}{\Gamma_Z^{\rm SM}}
 - 0.65 \, \frac{\d\Gamma_W}{\Gamma_W^{\rm SM}} + \text{direct}\,,
\end{equation}
where the numerical prefactors reflect the proportions of $W$ and $Z$ bosons produced. In fact, the largest numerical effects in Tab.~\ref{tab.MG_example_VBF_VH} are observed in the operators entering $\d\Gamma_W$.

\subsubsection{\texorpdfstring{$h\to e^+e^-\mu^+\mu^-$}{h -> e+ e- mu+ mu-}}
An analysis of the $Z$-mediated Higgs decay $h\to e^+ e^-\mu^+\mu^-$ was performed following a procedure analogous to the one described for $\bar qq\to h\bar qq$. In the decay case, one $Z$ boson is always on-shell, leading to significant contributions from the intermediate $Z$ propagator. The relative SMEFT correction to the decay width is found to be
\begin{equation}
\frac{\Gamma_{SMEFT}}{\Gamma_{\rm SM}} =1 - 0.84 \,\frac{\d\Gamma_Z}{\Gamma_Z^{\rm SM}}
 + \text{direct}\,.
\end{equation}
The breakdown into fermionic Wilson coefficients is given in Table~\ref{tab.MG_example_VBF_VH} and it agrees with the analytic results of Ref.~\cite{Brivio:2019myy}.

\section{Summary}\label{sec.conclusion}
The \smeftsim\ package contains models in \feynrules\ and in the \ufo\ format, that implement the complete Warsaw basis of dimension six operators, under different flavor assumptions and with different choices of the input quantities for the EW sector. Its main scope is the Monte Carlo simulation of LHC processes in the SMEFT, but it can also be employed for simple analytic calculations, exploiting the \feynrules\ interface in \mathematica.

This work reviewed the theoretical elements that are implemented in \smeftsim\ and presented the improvements in version~3.0. The most significant changes compared to previous releases are the addition of two new flavor structures for top quark physics, the implementation of a brand new tool for the inclusion of SMEFT corrections in the propagator of unstable particles and the general improvement of the code, particularly of the parameterization of Higgs-gluon interactions in the SM.

As in previous versions, \smeftsim~3.0 supports the {\tt WCxf} exchange format~\cite{Aebischer:2017ugx}. The corresponding interface will be updated shortly after the code release. Finally, support for the translation of the \ufo\ models to {\tt python3} will be provided in the near future.

\acknowledgments
I would like to thank Sebastian Bruggisser, Ana Cueto, Sally Dawson, Gauthier Durieux, Saskia Falke, Gino Isidori, Olivier Mattelaer, Luca Merlo, Ken Mimasu and Mike Trott for several conversations that inspired and improved this work,  Tyler Corbett and Pietro Govoni for their valuable comments on the manuscript. 
The author acknowledges support by the state of Baden-W\"urttemberg through  bwHPC  and  the  German  Research  Foundation  (DFG)  through  grant  no  INST39/963-1  FUGG  (bwForCluster  NEMO).

\appendix


\section{Analytic expressions of decay width corrections}\label{app.dwidthExpressions}
This appendix reports analytic expressions for the decay widths of the $W, Z, h, t$ particles in the SMEFT, that are implemented in the tool documented in Sec.~\ref{sec.propagators_smeftsim}.  

Only linear terms in the Wilson coefficients are retained.
CKM mixing and all fermion masses except $m_b$ and $m_t$ are neglected, unless otherwise specified.
Results are reported in the flavor-general setup, and they can be mapped to the symmetric scenarios using the tables in Appendix~\ref{app.flavor_comparison}.
We use an input-scheme independent notation. Scheme-specific results can be obtained replacing the generic shifts $\d g_W, \d g_1, \d m_W^2$ with the expressions reported in Sec.~\ref{sec.inputs_ew_a} or~\ref{sec.inputs_ew_mw}.
The quantities $\Delta G_F, \Delta m_Z^2, \Delta\kappa_H$ are defined in Table~\ref{tab.input_shifts_expr}.
Finally, we use the $\C$ notation defined in Eq.~\eqref{eq.Cbar} and the hat notation defined in Sec.~\ref{sec.inputs}.

\subsection{\texorpdfstring{$Z$}{Z} boson}
The $Z$ boson couplings to a fermion pair $\psi\bar \psi$ are
\begin{align}
 g_{\psi L} &= T_3^\psi-Q_\psi \hsw^2\,,
 &
 g_{\psi R} &= -Q_\psi \hsw^2\,,
\end{align}
where $T_3^\psi=\pm 1/2$ is the isospin eigenvalue and $Q_\psi$ is the electric charge of the fermion $\psi$.
We also define:
\begin{align}
 \Delta_{\nu,pp}^L &=(\C_{Hl}^{(1)})_{pp}- (\C_{Hl}^{(3)})_{pp}\,,
 \\
 \Delta_{e,pp}^L &=(\C_{Hl}^{(1)})_{pp}+ (\C_{Hl}^{(3)})_{pp}\,,
 &
 \Delta_{e,pp}^R &=(\C_{He})_{pp}\,,
 \\
 \Delta_{u,pp}^L &=(\C_{Hq}^{(1)})_{pp}- (\C_{Hq}^{(3)})_{pp}\,,
 &
 \Delta_{u,pp}^R &=(\C_{Hu})_{pp}\,,
 \\ 
 \Delta_{d,pp}^L &=(\C_{Hq}^{(1)})_{pp}+ (\C_{Hq}^{(3)})_{pp} \,,
 &
 \Delta_{d,pp}^R &=(\C_{Hd})_{pp}\,.
\end{align}
At tree level in the SM, the partial decay width of the $Z$ boson into a $\bar \psi_p \psi_p$ pair, with  
 $\psi=\{\nu,l^-,u,d\}$ and flavor $p$, is 
\begin{equation}
 \Gamma_{Z\to \bar \psi_p \psi_p}^{\rm SM} = \frac{G_F m_Z^3}{3\sqrt2 \pi}N_C^\psi\left[g_{\psi L}^2+g_{\psi R}^2\right]\,,
\end{equation}
where  $N_c^\psi$ is the number of colors of the fermion species $\psi$. 

The relative SMEFT correction to a partial width can be inferred differentiating in the $g_{\psi L}, g_{\psi R}$ couplings and inserting the expressions of their SMEFT shifts:
\begin{align}
 \d g_{\psi L} &= g_{\psi L}\, \d g_Z - Q_\psi\, \d s_\theta^2 - \frac{\Delta_{\psi,pp}^L}{2}\,,
 \\
 \d g_{\psi R} &= g_{\psi R}\, \d g_Z - Q_\psi\, \d s_\theta^2 - \frac{\Delta_{\psi,pp}^R}{2}\,,
\end{align}
with 
\begin{equation}
\d g_Z  = \frac{\d g_W}{\hat g_W} + \frac{\d s_\theta^2}{2\hcw^2} + \frac{\hsw}{2\hcw}\C_{HWB}\,.
\end{equation}
Using the expression of $\d s_\theta^2$ provided in Eq.~\eqref{eq.dsth2}, one obtains
\begin{align}
 \frac{\d\Gamma_{Z\to \bar\psi_p \psi_p}}{\Gamma_{Z\to \bar\psi_p \psi_p}^{\rm SM}}
 &=
 2\hcw^2\left[1+ 2\hsw^2 Q_\psi \frac{g_{\psi L}+g_{\psi R}}{g_{\psi L}^2+g_{\psi R}^2}\right]\frac{\d g_W}{\hat g_W}
 +2\hsw^2\left[1- 2\hcw^2 Q_\psi \frac{g_{\psi L}+g_{\psi R}}{g_{\psi L}^2+g_{\psi R}^2}\right]\frac{\d g_1}{\hat g_1}
 \nn\\
 &
 + \hsdw \left[1 - \hcdw Q_\psi \frac{g_{\psi L}+g_{\psi R}}{g_{\psi L}^2+g_{\psi R}^2}\right]\C_{HWB}
-\frac{g_{\psi L}\Delta_{\psi,pp}^L+g_{\psi R}\Delta_{\psi,pp}^R}{g_{\psi L}^2+g_{\psi R}^2}\,.
\label{eq.dWZff}
\end{align}
Flavor violating decays are absent at $\O(\Lambda^{-2})$.
As $m_b\neq0$ is retained, the $Z\to\bar bb$ result contains additional terms. The partial width expression in the SM is
\begin{equation}
 \Gamma_{Z\to \bar bb}^{\rm SM} = \frac{G_F m_Z^3}{\sqrt2 \pi}\left[g_{dL}^2+g_{dR}^2\right]\sqrt{1-4x_b^2}\left[
 1-
 x_b^2\left(1-\frac{6\,g_{dL}\, g_{dR}}{g_{dL}^2+g_{dR}^2}\right)
 \right]\,,
\end{equation}
with $x_b = m_b/m_Z$.
The relative SMEFT correction is
\begin{align}
 \frac{\d\Gamma_{Z\to  \bar bb}}{\Gamma_{Z\to \bar bb}^{\rm SM}}
 &= \frac{\d\Gamma_{Z\to \bar bb}}{\Gamma_{Z\to \bar bb}^{\rm SM}}\bigg|_{m_b= 0}+
 \frac{3 x_b^2  }{(g_{dL}^2+g_{dR}^2)(1-x_b^2)+6x_b^2\, g_{dL}\, g_{dR}}\,
 \frac{g_{dL}^2-g_{dR}^2}{g_{dL}^2+g_{dR}^2}\;\times
 \nn\\
 &
 \times\bigg[
 -\frac13 \hsdw^2 (g_{dL}-g_{dR})
 \left(\frac{\d g_W}{\hat g_W}- \frac{\d g_1}{\hat g_1} -\frac{\hcdw}{\hsdw}\C_{HWB}\right)
+g_{dL} \Delta_{d,33}^R-g_{dR}\Delta_{d,33}^L
\bigg]
\nn\\
&- \frac{6\sqrt2 \,x_b(g_{dL}^2-g_{dR}^2)}{(g_{dL}^2+g_{dR}^2)(1-x_b^2)+6x_b^2\, g_{dL}\, g_{dR}}\left[\hcw(\C_{dW})_{33} + \hsw(\C_{dB})_{33}\right]\,,
 \end{align}
where the first term stands for the contributions in Eq.~\eqref{eq.dWZff}.
The relative SMEFT correction to the total decay width is finally obtained as
\begin{align}
 \frac{\d\Gamma_{Z}}{\Gamma_Z^{\rm SM}}
  &=\sum_{f} \left[\frac{\d\Gamma_{Z\to f}}{\Gamma_{Z\to f}^{\rm SM}}\right] \Br_{Z\to f}^{\rm SM}\,,
 \end{align}
 with $f$ running over all the allowed fermion pairs and $\Br_{Z\to f}^{\rm SM}, \Gamma_Z^{\rm SM}$ computed  directly from the tree level expressions.

\subsection{\texorpdfstring{$W$}{W} boson}
At tree level in the SM, the partial decay width of the $W^+$ boson into a fermion pair $f^+ = \{ e^+_p\nu_p, \bar d_p u_p\}$ with flavor $p$ is
\begin{equation}
 \Gamma_{W^+\to f^+}^{\rm SM} = \frac{G_F m_W^3}{6\sqrt2\pi} N_C^{f}\,.
 \end{equation}
with $N_C^{f}=\{1,3\}$ the number of colors. Only decays into same-generation fermions are considered here, as CKM mixing is neglected.
The relative SMEFT correction for each channel is
\begin{align}
  \frac{\d\Gamma_{W^+\to l^+_p \nu_p}}{\Gamma_{W^+\to l^+_p \nu_p}^{\rm SM}} &=
  2\frac{\d g_W}{\hat g_W}+\frac{\d m_W}{\hat m_W} + 2 (\C_{Hl}^{(3)})_{pp}\,,
  \\
  \frac{\d\Gamma_{W^+\to u_p\bar d_p}}{\Gamma_{W^+\to u_p\bar d_p}^{\rm SM}} &=
   2\frac{\d g_W}{\hat g_W}+\frac{\d m_W}{\hat m_W} + 2 (\C_{Hq}^{(3)})_{pp}\,.
\end{align}
The total $W^+$ decay width in the SM is
\begin{equation}
 \Gamma_W = \sum_{p=1}^3\Gamma_{W^+\to l^+_p \nu_p} + \sum_{p=1}^2\Gamma_{W^+\to u_p\bar d_p}\,.
\end{equation}
Since in this case the branching ratios are simple rational numbers, the relative SMEFT correction simplifies into
\begin{align}
 \frac{\d\Gamma_W}{\Gamma_W^{\rm SM}}
  &= \frac{1}{9}\sum_{p=1}^3\frac{\d\Gamma_{W^+\to l^+_p\nu_p}}{\Gamma_{W^+\to l^+_p\nu_p}^{\rm SM}}
  +
  \frac{1}{3}\sum_{p=1}^2\frac{\d\Gamma_{W^+\to \bar d_pu_p}}{\Gamma_{W^+\to \bar d_pu_p}^{\rm SM}} 
\\
&=
  2\frac{\d g_W}{\hat g_W}+\frac{\d m_W}{\hat m_W} + \frac29\sum_{p=1}^3 (\C_{Hl}^{(3)})_{pp} + \frac23\sum_{p=1}^2 (\C_{Hq}^{(3)})_{pp}\,.
\end{align}

\subsection{Higgs boson}
The SM partial widths for two-body Higgs decays are:
\begin{align}
 \Gamma_{h\to \bar \psi_p \psi_p}^{\rm SM} &=\frac{m_h\, y_{\psi_p}^2}{16\pi}N_C^\psi\left[1-4x_{\psi_p}\right]^{3/2}\,,
 \\
 \Gamma_{h\to \g\g}^{\rm SM} &= \frac{m_h^3}{4\pi \hv^2}g_{H\g\g}^2\,,
 \\
 \Gamma_{h\to Z\g}^{\rm SM} &= \frac{m_h^3}{8\pi \hv^2}g_{HZ\g}^2\left[1-x_Z^2\right]^3\,,
 \\
 \Gamma_{h\to gg}^{\rm SM} &=\frac{m_h^3}{2\pi \hv^2}\left[2g_{Hgg}^{(1)}-\frac{1}{x_t^2}g_{Hgg}^{(2)}\right]^2\,,
\end{align}
where $x_{\psi_p} = m_{\psi_p}/m_h$ for a generic fermion $\psi_p$, $x_Z=m_Z/m_h$, $x_t=m_t/m_h$ and $y_{\psi_p}\equiv(Y_\psi)_{pp}$ is the relevant Yukawa coupling.
The relative SMEFT corrections are
\begin{align}
\label{eq.dGammah_ff}
 \frac{\d\Gamma_{h\to \bar\psi_p \psi_p}}{\Gamma_{h\to \bar\psi_p \psi_p}^{\rm SM}} &=
 1-\Delta G_F + 2\Delta \kappa_H - 2\frac{(\C_{\psi H})_{pp}}{y_{f_p}}\,,
 \\
 \frac{\d\Gamma_{h\to \g\g}}{\Gamma_{h\to \g\g}^{\rm SM}} &=\frac{2}{g_{H\g\g}^2}\left[\hcw^2\C_{HB} + \hsw^2\C_{HW} - \hsw\hcw \C_{HWB}\right]\,,
 \\
 \frac{\d\Gamma_{h\to Z\g}}{\Gamma_{h\to Z\g}^{\rm SM}} &=\frac{2}{g_{HZ\g}^2}\left[\hsdw(\C_{HW} - \C_{HB}) - \hcdw \C_{HWB}\right]\,,
 \\
 \frac{\d\Gamma_{h\to gg}}{\Gamma_{h\to gg}^{\rm SM}} &=
 \frac{4\C_{HG}}{2 g_{Hgg}^{(1)}-g_{Hgg}^{(2)}/x_t^2}\,,
\end{align}
with the $h\g\g,hZ\g,hgg$ couplings defined in Sec.~\ref{sec.Hloops}. The \smeftsim\ implementation retains the masses of the tau lepton, charm and bottom quarks in Eq.~\eqref{eq.dGammah_ff}.
Four-body decays were included neglecting neutral current contributions, CKM mixing and all fermion masses.
The analytic expressions were taken from Ref.~\cite{Brivio:2019myy} and generalized to all flavor setups. 

For each individual decay channel, the partial decay width in the SM is 
\begin{equation}
 \Gamma_{h\to f_+f_-}^{\rm SM} 
  =N_C^{f_+} N_C^{f_-}\frac{m_W^8 G_F^3}{m_h } \left(4.65\times 10^{-4}\right)\,,
\end{equation}
with $f_+=\{l^+_p\nu_p, \bar d_p u_p\}$, $f_-=\{\bar\nu_pl^-_p, \bar u_p d_p\}$ and $N_C^{f_+},\, N_C^{f-_-}$ the appropriate color multiplicities.
The numerical factor comes from the phase space integration, that is performed taking $\hat m_W =\unit[80.387]{GeV}$. 

Summing over all allowed flavor combinations, the relative SMEFT corrections are:
\begin{align}
\frac{\d\Gamma_{h\to l^+\nu \bar\nu l^-} }{
\Gamma_{h\to l^+\nu \bar\nu l^-}^{\rm SM}}
&= F_1 + \left(\frac43-0.588\right)\sum_{p=1}^3 (\C_{Hl}^{(3)})_{pp}\,,
\\
\frac{\d\Gamma_{h\to \bar u d \bar du} }{
\Gamma_{h\to \bar u d\bar du}^{\rm SM}}
&= F_1 + \left(2-0.882\right)\sum_{p=1}^2 (\C_{Hq}^{(3)})_{pp}\,,
\\
\frac{\d\Gamma_{h\to l^+\nu \bar ud+\hc} }{
 \Gamma_{h\to l^+\nu \bar ud+\hc}^{\rm SM}}
&= F_1 + \left(\frac23-0.294\right)\sum_{p=1}^3 (\C_{Hl}^{(3)})_{pp} + \left(1-0.441\right)\sum_{p=1}^2 (\C_{Hq}^{(3)})_{pp}\,,
\end{align}
with 
\begin{equation}
 F_1 = -\Delta G_F + 2\Delta\kappa_H +4\,\frac{\d g_W}{\hat g_W}
  - 0.463\, \frac{\d\Gamma_W}{\Gamma_W^{\rm SM}}
  -9.643 
  \,\frac{\d m_W}{m_W}- 1.487\,\C_{HW}\,.
\end{equation}
The SMEFT correction to the total Higgs width is finally estimated as 
\begin{equation}
\frac{\d \Gamma_h}{\Gamma_f^{\rm SM}} = \sum_f 
\Br_{h\to f}^{SM,\text{best}}
\left[\frac{\d\Gamma_{h\to f}}{\Gamma_{h\to f}^{\rm SM}}\right]\,,
\end{equation}
with $f=\{\bar bb,\bar cc,\tau^+\tau^-,\g\g,Z\g,gg,l^+\nu\bar\nu l^-,\bar u d\bar d u,l^+\nu \bar ud+\hc\}$ and the branching ratio values in Table~\ref{tab.Hwidths_SM}.

\subsection{Top quark}
To a very good approximation, the top quark decays exclusively to $Wb$. The SM width is
\begin{equation}
 \Gamma_t^{\rm SM} = \Gamma_{t\to W^+ b}^{\rm SM} =
 \frac{G_F  m_t^3}{8\sqrt2\pi}\left[
 (1-x_b^2)^2+(1+x_b^2) x_W^2-2 x_W^4
 \right]\sqrt{\lambda(1,x_b^2,x_W^2)}\,,
\end{equation}
with $x_b= m_b/m_t,\,x_W=\hat m_W/m_t$ and $\lambda(a,b,c)=a^2+b^2+c^2-2ab-2ac-2bc$.
The relative SMEFT correction is:
\begin{align}
 \frac{\d\Gamma_t}{\Gamma_t^{\rm SM}} = \frac{\d\Gamma_{t\to W^+ b}}{\Gamma_{t\to W^+ b}^{\rm SM}} &=
 2\frac{\d g_W}{\hat g_W}+2(\C_{Hq}^{(3)})_{33}
 +\frac{6x_W }{(1-x_b^2)^2+(1+x_b^2) x_W^2-2 x_W^4}\bigg[
 x_W x_b(\re\C_{Hud})_{33}
 +\nn\\&
 -\sqrt{2}
 (\re\C_{uW})_{33} \left(1-x_W^2-x_b^2\right)
 +\sqrt2 (\re\C_{dW})_{33}\, x_b\left(1+x_W^2-x_b^2\right)
 \bigg]+
 \nn\\
 &+2\left[
 \frac{(1-x_b^2)^2-3(1+x_b^2)x_W^2+2x_W^4}{\lambda(1,x_b^2,x_W^2)}
 -
 \frac{2(1-x_b^2)^2+(1+x_b)^2x_W^2}{(1-x_b^2)^2+(1+x_b^2) x_W^2-2 x_W^4}
 \right]\frac{\d m_W}{\hat m_W}\,.
\end{align}
Decaying the $W$ in final state does not lead to any additional contribution to $\d\Gamma_t/\Gamma_t^{\rm SM}$. This happens because the $W$ is always on-shell, so its decay essentially factorizes out: using the narrow width approximation, one trivially has
\begin{equation}
 \Gamma_t = \sum_{f}  \Gamma_{t\to W^+ b} \cdot \Br_{W^+\to f} = \Gamma_{t\to W^+ b}\,.
\end{equation}
Note that this conclusion only holds at the SM--$\Lag_6$ interference level, while at $\O(\Lambda^{-4})$ additional SMEFT corrections arise through  contact vertices $(\bar t b)(\bar\nu l), (\bar t b)(\bar du)$.

\section{What's new in version~3.0}\label{app.new}
Here we briefly summarize the most significant updates and features introduced in \smeftsim~3.0, compared to previous versions:
\begin{itemize}
\item The flavor assumptions \top\ and \topsl\ described in Sec.~\ref{sec.flavor_top} have been added.

\item The flavor structure of all models has been generally improved. See Sec.~\ref{sec.flavor_comparison} for details.

\item The treatment of propagator corrections described in Sec.~\ref{sec.propagators} has been implemented, enabling the estimate of linearized EFT corrections to the $W,Z,H,t$ widths and to the $W$ mass.

\item The treatment of SM loop-generated Higgs interactions has been improved, particularly in the Higgs-gluon case. See Sec.~\ref{sec.Hloops} for the general treatment and Sec.~\ref{sec.Hloops_comparison} for a detailed comparison with previous implementations.

\item All interaction vertices with up to 6 legs are now included in all \ufo\ models. In the previous version, only 4-point functions were retained.

\item The numerical values of the SM parameters have been updated, see Table~\ref{tab.defs_common}. The default value has been set to 0 for all Wilson coefficients.

\item All complex Wilson coefficients are expressed in terms of their real and imaginary parts, rather than absolute values and phases.

\item Individual interaction orders have been defined for each operator. Additionally, interaction orders {\tt NPcpv, NPprop, NPfv} have been added, to provide more control on each class of EFT contributions. See Sec.~\ref{sec.int_orders} for further details.

\item In the \ufo\ models, the SMEFT parameters have been organized in parameters blocks: {\tt SMEFTcutoff} contains only $\Lambda$ and {\tt SMEFT} ({\tt SMEFTcpv}) contain CP conserving (violating) Wilson coefficients. 
In the \MFV\ models the flavor-violating $\Delta C_\a$ parameters are contained in the additional {\tt SMEFTFV} block. See Sec.~\ref{sec.param_rst_cards}.

\item The normalization of $\Delta G_F$ ({\tt dGf}) has been modified. This is explicit in the \feynrules\ Lagrangian but does not have any consequence for the \ufo\ models.
\end{itemize}


\clearpage
\section{Conversion tables between flavor assumptions}\label{app.flavor_comparison}
This Appendix collects the results of Sec.~\ref{sec.flavor} and compares the flavor structure of the fermionic operators across the five setups considered.
Tables~\ref{tab.dictionary_56}~--~\ref{tab.dictionary_8d} provide a dictionary between the different models: in order to translate between two flavor assumptions it is sufficient to exchange the corresponding expressions within each table block.   All structures are given explicitly in terms of diagonal Yukawa matrices and of the CKM matrix $V$. In the \top\ and \topsl\ cases, $Y_u$ and $Y_d$ are $2\times2$ matrices and $V=\mathbbm{1}$ is assumed.

\newcommand{\prL}{\text{\small $p,r=\{1,2\}$}}
\newcommand{\prH}{\text{\small $p=r=3$}}
\newcommand{\stL}{\text{\small $s,t=\{1,2\}$}}
\newcommand{\stH}{\text{\small $s=t=3$}}
\newcommand{\ptL}{\text{\small $p,t=\{1,2\}$}}
\newcommand{\ptH}{\text{\small $p=t=3$}}
\newcommand{\srL}{\text{\small $s,r=\{1,2\}$}}
\newcommand{\srH}{\text{\small $s=r=3$}}
\newcommand{\prstL}{\text{\small $p,r,s,t=\{1,2\}$}}
\newcommand{\prstH}{\text{\small $p=r=s=t=3$}}

\begin{table}[h!]\centering 
\renewcommand{\arraystretch}{1.35}
\begin{tabular}{|r>{$}p{6.5cm}<{$}>{$}p{5cm}<{$}|}
\toprule
\general&
(C_{eH})_{pr} 
&\\
\Utf, \MFV, \topsl
&
C_{eH}\, (Y_{l}^{(d)})_{pr}
&\\
\top
&(C_{eH})_{pp}\, \d_{pr} 
&\\
\midrule
%
\general&
(C_{uH})_{pr}
&\\
\Utf&
C_{uH}\,(Y_{u}^{(d)})_{pr} 
&\\
\MFV&
\left[
C_{uH}^{(0)}\, Y_u^{(d)} 
+ (\Delta^u C_{uH})\, (Y_u^{(d)})^3 
+(\Delta^d C_{uH})\, V (Y_d^{(d)})^2 V^\dag Y_u^{(d)}\right]_{pr}
&\\
\top, \topsl&
C_{uH}\,(Y_{u}^{(d)})_{pr} \qquad \prL\,,
& C_{tH} \qquad \prH
\\
\midrule
%
\general&
(V^\dag C_{dH})_{pr}
&\\
\Utf&
C_{dH}\,(Y_{d}^{(d)})_{pr} 
&\\
\MFV&
\left[
C_{dH}^{(0)}\, Y_d^{(d)} 
+ (\Delta^u C_{dH})\, V^\dag (Y_u^{(d)})^2V Y_d^{(d)}
+(\Delta^d C_{dH})\,  (Y_d^{(d)})^3\right]_{pr}
&\\
\top, \topsl&
C_{dH}\,(Y_{d}^{(d)})_{pr}\qquad \prL\,,
& C_{bH}\qquad \prH
\\
\midrule
\general & 
(C_{eW})_{pr}
&\text{Same as } (C_{eH})_{pr}\,.
\\
\midrule
\general & 
(C_{eB})_{pr}
&\text{Same as } (C_{eH})_{pr}\,.
\\
\midrule
\general & 
(C_{uW})_{pr}
&\text{Same as } (C_{uH})_{pr}\,.
\\
\midrule
\general & 
(C_{uB})_{pr}
&\text{Same as } (C_{uH})_{pr}\,.
\\
\midrule
\general & 
(C_{uG})_{pr}
&\text{Same as } (C_{uH})_{pr}\,.
\\
\midrule
\general & 
(V^\dag C_{dW})_{pr}
&\text{Same as } (V^\dag C_{dH})_{pr}\,.
\\
\midrule
\general & 
(V^\dag C_{dB})_{pr}
&\text{Same as } (V^\dag C_{dH})_{pr}\,.
\\
\midrule
\general & 
(V^\dag C_{dG})_{pr}
&\text{Same as } (V^\dag C_{dH})_{pr}\,.
\\\bottomrule
\end{tabular}
\caption{Conversion table among the 5 flavor assumptions considered for the operators in $\Lag_6^{(5)}$, $\Lag_6^{(6)}$.}\label{tab.dictionary_56}
\end{table}

\newpage
\begin{table}[h!]\centering 
\renewcommand{\arraystretch}{1.4}
\begin{tabular}{|r>{$}p{6.5cm}<{$}>{$}p{6.cm}<{$}|}
\toprule
\general&
(C_{Hl}^{(1)})_{pr} 
&
\\
\Utf, \MFV, \topsl
&
C_{Hl}^{(1)}\, \d_{pr}
&\\
\top
&(C_{Hl}^{(1)})_{pp}\, \d_{pr} 
&\\
\midrule
\general &
(C_{Hl}^{(3)})_{pr}&
\text{Same as } (C_{Hl}^{(1})_{pr}\,.
\\
\midrule
\general &
(C_{He})_{pr}&
\text{Same as } (C_{Hl}^{(1})_{pr}\,.
\\
\midrule
\general&
(C_{Hq}^{(1)})_{pr} 
&
\\
\Utf
&
C_{Hq}^{(1)}\, \d_{pr}
&\\
\MFV&
\left[C_{Hq}^{(1)(0)}\, \mathbbm{1}
+(\Delta^u C_{Hq}^{(1)})\, (Y_u^{(d)})^2 + (\Delta^d C_{Hq}^{(1)})\, V (Y_d^{(d)})^2 V^\dag
\right]_{pr}
&\\
\top, \topsl
&C_{Hq}^{(1)}\, \d_{pr}\qquad \prL\,,
& C_{HQ}^{(1)}\qquad \prH
\\
\midrule
\general &
(C_{Hq}^{(3)})_{pr}&
\text{Same as } (C_{Hq}^{(1})_{pr}\,.
\\
\midrule
\general&
(C_{Hu})_{pr} 
&
\\
\Utf
&
C_{Hu}\, \d_{pr}
&\\
\MFV&
\left[C_{Hu}^{(0)}\, \mathbbm{1}
+(\Delta C_{Hu})\, (Y_u^{(d)})^2 \right]_{pr}
&\\
\top, \topsl
&C_{Hu}\, \d_{pr}\qquad \prL\,,
& C_{Ht}\qquad \prH
\\
\midrule
\general &
(C_{Hd})_{pr}&
\text{Same as  }(C_{Hu})_{pr}
\\
&&
\text{with } Y_u^{(d)}\to Y_d^{(d)}, C_{Ht}\to C_{Hb}\,.
\\
\midrule
\general&
(C_{Hud})_{pr} 
&\\
\Utf
&
C_{Hud}\, \left[Y_u^{(d)} V Y_d^{(d)} \right]_{pr}
&\\
\MFV&
C_{Hud}^{(0)}\, \left[Y_u^{(d)} V Y_d^{(d)} \right]_{pr}
&\\
\top, \topsl
&C_{Hud}\, \left[Y_u^{(d)} Y_d^{(d)}\right]_{pr} \qquad \prL\,,
& C_{Htb}\qquad \prH
\\\bottomrule
\end{tabular}
\caption{Conversion table among the 5 flavor assumptions considered, for the operators in $\Lag_6^{(7)}$ .}\label{tab.dictionary_7}
\end{table}

\newpage
\begin{table}[h!]\centering 
\renewcommand{\arraystretch}{1.5}
\begin{tabular}{|r>{$}p{6.5cm}<{$}>{$}p{6.cm}<{$}|}
\toprule
\general&
(C_{ll})_{prst} 
&\\
\Utf, \MFV, \topsl
&
C_{ll}\, \d_{pr}\d_{st} + C_{ll}^\prime\, \d_{pt}\d_{sr}
&\\
\top&
\frac12 (C_{ll})_{prst} \,(\d_{pr}\d_{st} + \d_{pt}\d_{sr}) , \quad  prst\in P_{ll}
&
\\
\midrule
\general&
(C_{qq}^{(1)})_{prst} 
&\\
\Utf
&
C_{qq}^{(1)}\, \d_{pr}\d_{st} + C_{qq}^{(1)\prime}\, \d_{pt}\d_{sr} 
&\\
\MFV&
\multicolumn{2}{l|}{$
C_{qq}^{(1)(0)}\, \d_{pr}\d_{st} 
+(\Delta^u C_{qq}^{(1)}) \left[(Y_u^{(d)})^2\right]_{pr} \d_{st}
+(\Delta^d C_{qq}^{(1)}) \left[V(Y_d^{(d)})^2V^\dag\right]_{pr}\d_{st}
$}\\
&
\multicolumn{2}{l|}{$
+ C_{qq}^{(1)\prime(0)}\, \d_{pt}\d_{sr} 
+(\Delta^u C_{qq}^{(1)\prime}) \left[(Y_u^{(d)})^2\right]_{pt} \d_{sr}
+(\Delta^d C_{qq}^{(1)\prime}) \left[V(Y_d^{(d)})^2V^\dag\right]_{pt} \d_{sr}
$}\\
\top, \topsl&
\multicolumn{2}{l|}{$
\left(C_{qq}^{(1,1)}-\frac16 C_{qq}^{(1,8)}\right) \, \d_{pr}\d_{st}
+
\frac14\left( C_{qq}^{(1,8)}+3 C_{qq}^{(3,8)}\right) \, \d_{pt}\d_{sr}
\qquad\;\, \prstL\,,
$}
\\
&
\left(\frac12 C_{Qq}^{(1,1)}-\frac{1}{12} C_{Qq}^{(1,8)}\right) \, \d_{pr}
& \prL\,,\,\stH\,,
\\
&
\frac18\left(C_{Qq}^{(1,8)}+ 3 C_{Qq}^{(3,8)}\right) \, \d_{pt}
& \ptL\,,\, \srH\,,
\\
&
C_{QQ}^{(1)}+\frac{1}{12} C_{QQ}^{(8)}
&\prstH
\\
\midrule
\general&
(C_{qq}^{(3)})_{prst} 
&\\
\Utf
&
C_{qq}^{(3)}\, \d_{pr}\d_{st} + C_{qq}^{(3)\prime}\, \d_{pt}\d_{sr} 
&\\
\MFV&
\multicolumn{2}{l|}{$
C_{qq}^{(3)(0)}\, \d_{pr}\d_{st} 
+(\Delta^u C_{qq}^{(3)}) \left[(Y_u^{(d)})^2\right]_{pr} \d_{st}
+(\Delta^d C_{qq}^{(3)}) \left[V(Y_d^{(d)})^2V^\dag\right]_{pr}\d_{st}
$}\\
&
\multicolumn{2}{l|}{$
+ C_{qq}^{(3)\prime(0)}\, \d_{pt}\d_{sr} 
+(\Delta^u C_{qq}^{(3)\prime}) \left[(Y_u^{(d)})^2\right]_{pt} \d_{sr}
+(\Delta^d C_{qq}^{(3)\prime}) \left[V(Y_d^{(d)})^2V^\dag\right]_{pt} \d_{sr}
$}\\
\top, \topsl&
\multicolumn{2}{l|}{$
\left(C_{qq}^{(3,1)}-\frac16 C_{qq}^{(3,8)}\right) \, \d_{pr}\d_{st}
+
\frac14\left(C_{qq}^{(1,8)} - C_{qq}^{(3,8)}\right) \, \d_{pt}\d_{sr}
\qquad\;\;\; \prstL\,,
$}\\
&
\left(\frac12 C_{Qq}^{(3,1)}-\frac{1}{12} C_{Qq}^{(3,8)}\right) \, \d_{pr}
& \prL\,,\, \stH\,,
\\
&
\frac18\left(C_{Qq}^{(1,8)}- C_{Qq}^{(3,8)}\right) \, \d_{pt}
& \ptL\,,\, \srH\,,
\\
&
\frac14 C_{QQ}^{(8)}
& \prstH
\\
\midrule
\general&
(C_{lq}^{(1)})_{prst} 
&
\\
\Utf
&
C_{lq}^{(1)}\, \d_{pr}\d_{st} 
&\\
\MFV&
\multicolumn{2}{l|}{$
C_{lq}^{(1)(0)}\, \d_{pr}\d_{st} +(\Delta^u C_{lq}^{(1)})\d_{pr} \left[(Y_u^{(d)})^2\right]_{st} +(\Delta^d C_{lq}^{(1)})\d_{pr} \left[V(Y_d^{(d)})^2V^\dag\right]_{st}
$}
\\
\top&
(C_{lq}^{(1)})_{pp}\, \d_{pr}\d_{st}\qquad \stL\,,
&
(C_{lQ}^{(1)})_{pp}\, \d_{pr}\qquad \stH
\\
\topsl&
C_{lq}^{(1)}\, \d_{pr}\d_{st}\qquad\quad\;\; \stL\,,
&
C_{lQ}^{(1)}\, \d_{pr}\qquad\quad\;\;\stH
\\
\midrule
\general&
(C_{lq}^{(3)})_{prst}&
\text{Same as } (C_{lq}^{(1)})_{prst}\,.
\\\bottomrule
\end{tabular}
\caption{Conversion table among the 5 flavor assumptions considered, for the operators in $\Lag_6^{(8a)}$. The set $P_{ll}$ is defined in Eq.~\eqref{eq.Pll}.}\label{tab.dictionary_8a}
\end{table}

\newpage
\begin{table}[h!]\centering 
\renewcommand{\arraystretch}{1.4}
\begin{tabular}{|r>{$}p{6.5cm}<{$}>{$}p{6.3cm}<{$}|}
\toprule
\general&
(C_{ee})_{prst} 
&\\
\Utf, \MFV, \topsl
&
\frac 12 C_{ee}\, (\d_{pr}\d_{st}+ \d_{pt}\d_{sr})
&\\
\top&
\multicolumn{2}{l|}{$
\frac14 (C_{ee})_{prst} \,(\d_{pr}\d_{st} +3 \d_{pt}\d_{sr}),\qquad  prst\in P_{ee}
$
}
\\
\midrule
\general&
(C_{uu})_{prst} 
&
\\
\Utf&
C_{uu} \, \d_{pr}\d_{st} + C_{uu}^\prime \, \d_{pt}\d_{sr}
&\\
\MFV&
\multicolumn{2}{l|}{$
C_{uu}^{(0)} \, \d_{pr}\d_{st} + (\Delta C_{uu})\, \left[(Y_u^{(d)})^2\right]_{pr}\d_{st}
+
C_{uu}^{(0)\prime} \, \d_{pt}\d_{sr} + (\Delta C_{uu}^\prime)\, \left[(Y_u^{(d)})^2\right]_{pt}\d_{sr}
$}
\\
\top&
\left(C_{uu}^{(1)}-\frac16 C_{uu}^{(8)}\right)\, \d_{pr}\d_{st}
+\frac12 C_{uu}^{(8)}\, \d_{pt}\d_{sr}
& \prstL\,,
\\
&
\left(\frac12 C_{tu}^{(1)}-\frac{1}{12} C_{tu}^{(8)}\right)\, \d_{pr}
& \prL\,,\, \stH\,,
\\
&
\frac14 C_{tu}^{(8)}\, \d_{pt}
& \ptL\,,\, \srH\,,
\\
&
C_{tt}
& \prstH
\\
\midrule
\general&
(C_{dd})_{prst}
& 
\text{Same as } (C_{uu})_{prst} \text{ with } Y_u^{(d)}\to Y_d^{(d)}\,,
\\
&
& C_{tu}^{(1),(8)}\to C_{bd}^{(1),(8)}\,,\, C_{tt}\to C_{bb}\,.
\\
\midrule
\general&
(C_{eu})_{prst} 
&\\
\Utf&
C_{eu} \, \d_{pr}\d_{st}
&\\
\MFV&
C_{eu}^{(0)} \, \d_{pr}\d_{st} + (\Delta C_{eu})\, \d_{pr} \left[(Y_u^{(d)})^2\right]_{st}
&\\
\top&
(C_{eu})_{pp}\, \d_{pr}\d_{st} 
\qquad \stL\,,
&
(C_{et})_{pp}\, \d_{pr}
\qquad \stH\,,
\\
\topsl&
C_{eu}\, \d_{pr}\d_{st} 
\qquad\quad\;\;\, \stL\,,
&
C_{et}\, \d_{pr}
\qquad\quad\;\;\, \stH
\\
\midrule
\general&
(C_{ed})_{prst}
& 
\text{Same as } (C_{eu})_{prst} 
\\[-2mm]
&
&\text{with } Y_u^{(d)}\to Y_d^{(d)}\,, C_{et}\to C_{eb}\,.
\\\midrule
\general&
(C_{ud}^{(1)})_{prst} 
&
\\
\Utf&
C_{ud}^{(1)} \, \d_{pr}\d_{st}
&\\
\MFV&
\multicolumn{2}{l|}{$
C_{ud}^{(1)(0)} \, \d_{pr}\d_{st} + (\Delta^u C_{ud}^{(1)}) \, \left[(Y_u^{(d)})^2\right]_{pr}\d_{st} + (\Delta^d C_{ud}^{(1)}) \,
\d_{pr} \left[(Y_d^{(d)})^2\right]_{st}
$}\\
\top, \topsl&
C_{ud}^{(1)}\, \d_{pr}\d_{st}\qquad\quad \prstL\,,
&
C_{tb}^{(1)}\, \qquad\quad \prstH\,,
\\
&
C_{ub}^{(1)}\, \d_{pr} \qquad\qquad \prL\,,\, \stH\,,
&
C_{td}^{(1)}\, \d_{st} \qquad \stL\,,\, \prH
\\
&
C_{utbd}^{(1)}\, \left[Y_u^{(d)}Y_d^{(d)}\right]_{pt}
\;\; \ptL\,,\,\srH\,,
&
\\
&
C_{utbd}^{(1)\star}\, \left[Y_d^{(d)}Y_u^{(d)}\right]_{sr}
\;\; \srL\,,\,\ptH\,,
&
\\
\midrule
\general&
(C_{ud}^{(8)})_{prst}&
\text{Same as } (C_{ud}^{(1)})_{prst}\,.
\\
\bottomrule
\end{tabular}
\caption{Conversion table among the 5 flavor assumptions considered, for the operators in $\Lag_6^{(8b)}$. The set $P_{ee}$ is defined in Eq.~\eqref{eq.Pee}.}\label{tab.dictionary_8b}
\end{table}

\newpage
\begin{table}[h!]\centering 
\renewcommand{\arraystretch}{1.4}
\vspace*{-8mm}
\scalebox{.97}{\begin{tabular}{|r>{$}p{6.5cm}<{$}>{$}p{6.3cm}<{$}|}
\toprule
\general&
(C_{le})_{prst} 
&\\
\Utf, \MFV, \topsl
&
C_{le}\, \d_{pr}\d_{st} 
&\\
\top&
(C_{le})_{prst}\, \qquad \text{\small $prst \in P_{le}$}
&\\
\midrule
\general&
(C_{lu})_{prst} 
&
\\
\Utf
&
C_{lu}\, \d_{pr}\d_{st} 
&\\
\MFV&
C_{lu}^{(0)}\, \d_{pr}\d_{st} + (\Delta C_{lu})\d_{pr}\left[(Y_u^{(d)})^2\right]_{st}
&\\
\top&
(C_{lu})_{pp} \, \d_{pr}\d_{st}\qquad \stL\,,
&
(C_{lt})_{pp} \, \d_{pr}\qquad \stH\,,
\\
\topsl&
C_{lu} \, \d_{pr}\d_{st}\qquad\quad\;\; \stL\,,
&
C_{lt} \, \d_{pr}\qquad\quad\;\; \stH
\\
\midrule
\general&
(C_{ld})_{prst}&
\text{Same as } (C_{lu})_{prst}
\\[-2mm]
&&\text{ with } Y_u^{(d)}\to Y_d^{(d)}\,,\, C_{lt}\to C_{lb}\,.
\\
\midrule
\general&
(C_{qe})_{prst}
&\\
\Utf&
C_{qe} \, \d_{pr}\d_{st}
&\\
\MFV&
\multicolumn{2}{l|}{$
C_{qe}^{(0)}\, \d_{pr}\d_{st} + (\Delta^u C_{qe}) \left[(Y_u^{(d)})^2\right]_{pr} \d_{st} + (\Delta^d C_{qe}) \left[V(Y_d^{(d)})^2V^\dag\right]_{pr} \d_{st}
$}\\
\top&
(C_{qe})_{pp} \, \d_{pr}\d_{st}\qquad \prL\,,
&
(C_{Qe})_{pp} \, \d_{st}\qquad \prH\,,
\\
\topsl&
C_{qe} \, \d_{pr}\d_{st}\qquad\quad\;\; \prL\,,
&
C_{Qe} \, \d_{st}\qquad\quad\;\; \prH
\\
\midrule
\general&
(C_{qu}^{(1)})_{prst}
&
\\
\Utf&
C_{qu}^{(1)} \, \d_{pr}\d_{st}
&\\
\MFV&
\multicolumn{2}{l|}{$
C_{qu}^{(1)(0)} \, \d_{pr}\d_{st} + (\Delta_1^u C_{qu}^{(1)})\, \left[(Y_u^{(d)})^2\right]_{pr} \d_{st}
+ (\Delta_1^d C_{qu}^{(1)})\, \left[V(Y_d^{(d)})^2V^\dag\right]_{pr} \d_{st}
$}\\
&
\multicolumn{2}{l|}{$
+ (\Delta_2 C_{qu}^{(1)})\,\d_{pr} \left[(Y_u^{(d)})^2\right]_{st}
+ C_{qu}^{(1)\prime(0)} \, (Y_u^{(d)})_{pt} (Y_u^{(d)})_{sr}
$}
\\
\top, \topsl&
C_{qu}^{(1)}\, \d_{pr}\d_{st}\qquad\quad \prstL\,,
&
C_{Qt}^{(1)}\qquad\qquad\quad\; \prstH\,,
\\
&
C_{qt}^{(1)}\, \d_{pr}\qquad\qquad \prL\,\,, \stH\,,
&
C_{Qu}^{(1)}\, \d_{st}\qquad\quad\;\;\; \stL\,\,, \prH
\\
&
C_{qQtu}^{(1)}\, \left[Y_u^{(d)}\right]_{pt}
\;\;\,
\ptL\,\,,\srH
&
C_{qQtu}^{(1)\star}\, \left[Y_u^{(d)}\right]_{sr}
\;
\srL\,\,,\ptH
\\
\midrule
\general&
(C_{qu}^{(8)})_{prst}&
\text{Same as } (C_{qu}^{(1)})_{prst}\,.
\\
\midrule
\general&
(C_{qd}^{(1)})_{prst}
&
\\
\Utf&
C_{qd}^{(1)} \, \d_{pr}\d_{st}
&\\
\MFV&
\multicolumn{2}{l|}{$
C_{qd}^{(1)(0)} \, \d_{pr}\d_{st} + (\Delta_1^u C_{qd}^{(1)})\, \left[(Y_u^{(d)})^2\right]_{pr} \d_{st}
+ (\Delta_1^d C_{qd}^{(1)})\, \left[V(Y_d^{(d)})^2V^\dag\right]_{pr} \d_{st}
$}\\
&
\multicolumn{2}{l|}{$
+ (\Delta_2 C_{qd}^{(1)})\,\d_{pr} \left[(Y_d^{(d)})^2\right]_{st}
+ C_{qd}^{(1)\prime(0)} \, \left[V Y_d^{(d)}\right]_{pt} \left[Y_d^{(d)}V^\dag\right]_{sr}
$}
\\
\top, \topsl&
C_{qd}^{(1)}\, \d_{pr}\d_{st}\qquad \prstL\,,
&
C_{Qb}^{(1)}\qquad\quad\;\; \prstH\,,
\\
&
C_{qb}^{(1)}\, \d_{pr}\qquad\quad \prL\,,\,\stH\,,
&
C_{Qd}^{(1)}\, \d_{st}\qquad \stL\,,\,\prH
\\
\midrule
\general&
(C_{qd}^{(8)})_{prst}&
\text{Same as } (C_{qd}^{(1)})_{prst}\,.
\\\bottomrule
\end{tabular}}
\caption{Conversion table among the 5 flavor assumptions considered, for the operators in $\Lag_6^{(8c)}$. The set $P_{le}$ is defined in Eq.~\eqref{eq.Ple}.}\label{tab.dictionary_8c}
\end{table}

\newpage
\begin{table}[h!]\centering 
\renewcommand{\arraystretch}{1.5}
\begin{tabular}{|r>{$}p{6.5cm}<{$}>{$}p{5cm}<{$}|}
\toprule
\general&
(C_{ledq})_{prst}
&\\
\Utf&
C_{ledq}\, \left[Y_l^{(d)}\right]_{pr}\left[Y_d^{(d)} V^\dag\right]_{st} 
&\\
\MFV&
\multicolumn{2}{l|}{$
C_{ledq}^{(0)}\, \left[Y_l^{(d)}\right]_{pr}\left[Y_d^{(d)} V^\dag\right]_{st} 
+(\Delta^u C_{ledq})\, \left[Y_l^{(d)}\right]_{pr}\left[Y_d^{(d)} V^\dag(Y_u^{(d)})^2\right]_{st}
$}\\
&
\multicolumn{2}{l|}{$
+(\Delta^d C_{ledq})\, \left[Y_l^{(d)}\right]_{pr}\left[(Y_d^{(d)})^3 V^\dag\right]_{st}
$}\\
\top&
(C_{ledq})_{pp}\, \d_{pr}\left[Y_d^{(d)} \right]_{st}
\qquad\, \stL\,,
&
(C_{lebQ})_{pp}\, \d_{pr} \qquad\quad \stH
\\
\topsl&
C_{ledq}\, \left[Y_l^{(d)}\right]_{pr}\left[Y_d^{(d)} \right]_{st}\quad\; \stL\,,
&
C_{lebQ}\, \left[Y_l^{(d)}\right]_{pr} \qquad \stH
\\
\midrule
\general&
(C_{lequ}^{(1)})_{prst} 
&
\\
\Utf
&
C_{lequ}^{(1)}\, \left[Y_l^{(d)}\right]_{pr}\left[Y_u^{(d)}\right]_{st} 
&\\
\MFV&
\multicolumn{2}{l|}{$
C_{lequ}^{(1)(0)}\, \left[Y_l^{(d)}\right]_{pr}\left[Y_u^{(d)}\right]_{st} 
+(\Delta^u C_{lequ}^{(1)})\, \left[Y_l^{(d)}\right]_{pr}\left[(Y_u^{(d)})^3\right]_{st}
$}\\
&
\multicolumn{2}{l|}{$
+(\Delta^d C_{lequ}^{(1)})\, \left[Y_l^{(d)}\right]_{pr}\left[V(Y_d^{(d)})^2 V^\dag Y_u^{(d)}\right]_{st}
$}\\
\top&
(C_{lequ}^{(1)})_{pp}\, \d_{pr}\left[Y_u^{(d)} \right]_{st}
\qquad\, \stL\,,
&
(C_{leQt}^{(1)})_{pp}\, \d_{pr} \qquad\quad \stH
\\
\topsl&
C_{lequ}^{(1)}\, \left[Y_l^{(d)}\right]_{pr}\left[Y_u^{(d)} \right]_{st}
\quad\; \stL\,,
&
C_{leQt}^{(1)}\, \left[Y_l^{(d)}\right]_{pr} \qquad \stH
\\
\midrule
\general&
(C_{lequ}^{(3)})_{prst}&
\text{Same as } (C_{lequ}^{(1)})_{prst}\,.
\\
\midrule
\general&
(C_{quqd}^{(1)})_{prst} 
&
\\
\Utf, \MFV
&
\multicolumn{2}{l|}{$
C_{quqd}^{(1)}\, \left[Y_u^{(d)}\right]_{pr}\left[VY_d^{(d)} \right]_{st} 
+
C_{quqd}^{(1)\prime}\, \left[Y_u^{(d)}\right]_{sr}\left[VY_d^{(d)} \right]_{pt} 
$}\\
\top, \topsl&
\multicolumn{2}{l|}{$
C_{quqd}^{(1)}\, \left[Y_u^{(d)}\right]_{pr}\left[Y_d^{(d)} \right]_{st} +
C_{quqd}^{(1)\prime}\, \left[Y_u^{(d)}\right]_{sr}\left[Y_d^{(d)} \right]_{pt} 
\qquad \prstL\,,
$}
\\&
C_{QtQb}^{(1)}& \prstH\,,
\\
&
C_{quQb}^{(1)}\,  \left[Y_u^{(d)}\right]_{pr}
&
\prL\,, \,\stH\,,
\\&
C_{Qtqd}^{(1)}\,  \left[Y_d^{(d)}\right]_{st}
&
\stL\,,\,\prH\,,
\\
&
C_{Quqb}^{(1)}\,  \left[Y_u^{(d)}\right]_{sr}
&
\srL\,,\, \ptH\,,
\\&
C_{qtQd}^{(1)}\,  \left[Y_d^{(d)}\right]_{pt}
&
\ptL\,,\,\srH
\\
\midrule
\general&
(C_{quqd}^{(8)})_{prst}&
\text{Same as } (C_{quqd}^{(1)})_{prst}\,.
\\\bottomrule
\end{tabular}
\caption{Conversion table among the 5 flavor assumptions considered, for the operators in $\Lag_6^{(8d)}$.}\label{tab.dictionary_8d}
\end{table}

\clearpage

\section{Parameter definitions in the code implementation}
\label{app.parameters}
This Appendix provides tables to facilitate the interpretation of the \feynrules\ and \ufo\ implementations in terms of the theory discussion in the main text.

Table~\ref{tab.defs_common} lists the external parameters that are are defined in all \smeftsim\ models, specifying the corresponding code name and default numerical value.
Table~\ref{tab.defs_bosonic} shows the nomenclature used for the Wilson coefficients of the bosonic operators and for the shift quantities defined in Sec.~\ref{sec.inputs_smeftsim}. Tables~\ref{tab.defs_general}~--~\ref{tab.defs_topsl} do the same for the Wilson coefficients of fermionic operators, for each flavor assumption. As a common rationale, primes are replaced by {\tt 1} in the code name and  real and imaginary parts are specified by  with {\tt Re, Im} suffixes. If needed, flavor indices are fully specified and appended at the very end of the code names.
In the \MFV\ models, the coefficients $\Delta_n^q(C_\a)$ are denoted {\tt Delta[n][q]c[a]}.

Although the correspondence between parameters names is most often direct, some notational changes were necessary, particularly in the \top\ and \topsl\ implementations. Most notably the lowercase $q$ has been replaced with {\tt j} in all the parameters' and operators' names, as the $q/Q$ distinction between light and heavy quark fields is problematic for non-case-sensitive interfaces.
Analogously, the coefficient $C_{Hb}$ is denoted as {\tt cHbq} to avoid conflict with {\tt cHB}, while  $C_{bB}$ is denoted as {\tt cbBB}, distinct from {\tt cbb}. The internal parameter $C_{tH}$ is denoted as {\tt ctHH} to avoid conflict with the cosine of the weak angle {\tt cth}.

\begin{table}[h!] 
\renewcommand{\arraystretch}{1.3}
\hspace*{-2cm}
\begin{tabular}{|>{$}c<{$}>{\tt}c@{\!\!}r@{ }p{8mm}c|>{$}c<{$}>{\tt}cr@{ }p{5mm}c|}
\toprule\rowcolor{tablesColor}
\multicolumn{10}{|c|}{Common parameters defined in \smeftsim}\\
\toprule
\text{parameter} &  \ufo & \multicolumn{2}{c}{default value}  & &  \text{parameter} & \ufo & \multicolumn{2}{c}{default value} &\\
\midrule
G_F& Gf& $1.1663787 \times 10^{-5}$&   $\unit{GeV^{-2}}$& \cite{Mohr:2012tt,PDG2020}&   
\a_s(m_Z)& aS& 0.1179&&\cite{PDG2020}\\
\aem(m_Z)\, (^\star)& aEW& 1/127.95& & \cite{Mohr:2012tt,PDG2020}& 
m_W\,(^{\star\star})& MW& 80.387& GeV&\cite{Aaltonen:2013iut}
\\\midrule

m_d& MD& $4.67 \times 10^{-3}$& GeV& \cite{PDG2020}&     y_d \,\hv/\sqrt2& ymdo& $4.67 \times 10^{-3}$& GeV&\cite{PDG2020}\\          
m_s& MS& 0.093& GeV& \cite{PDG2020}&                     y_s \,\hv/\sqrt2& yms& 0.093& GeV&\cite{PDG2020}\\    
m_b& MB& 4.18& GeV&  \cite{PDG2020}&                     y_b \,\hv/\sqrt2& ymb& 4.18& GeV&\cite{PDG2020}\\                     
m_u & MU& $2.16 \times 10^{-3}$& GeV& \cite{PDG2020}&    y_u \,\hv/\sqrt2& ymup& $2.16 \times 10^{-3}$& GeV&\cite{PDG2020}\\
m_c & MC& 1.27& GeV&  \cite{PDG2020}&                    y_c \,\hv/\sqrt2& ymc& 1.27& GeV&\cite{PDG2020}\\                  
m_t & MT& 172.76& GeV&  \cite{PDG2020}&                  y_t \,\hv/\sqrt2& ymt& 172.76& GeV&\cite{PDG2020}\\
m_e&     Me&  $5.11 \times 10^{-4}$&  GeV&  \cite{PDG2020}&    y_e \,\hv/\sqrt2& yme& $5.11 \times 10^{-4}$& GeV&\cite{PDG2020}\\
m_\mu&   MMU& 0.10566& GeV&  \cite{PDG2020}&             y_\mu \,\hv/\sqrt2& ymm& 0.10566& GeV&\cite{PDG2020}\\            
m_\tau&  MTA&  1.777& GeV& \cite{PDG2020}&               y_\tau \,\hv/\sqrt2&ymtau& 1.777& GeV&\cite{PDG2020}\\             
\midrule

m_Z &  MZ& 91.1876& GeV&   \cite{Z-pole,Mohr:2012tt,PDG2020}&           
m_h & MH& 125.09& GeV&\cite{PDG2020}\\
\Gamma_Z^{SM,\text{best}}&  WZ& 2.4952& GeV& \cite{PDG2020}& 
\Gamma_h^{SM,\text{best}}& WH&   $4.07 \times 10^{-3}$& GeV&\cite{deFlorian:2016spz}\\
\Gamma_W^{SM,\text{best}}&  WW& 2.085& GeV& \cite{PDG2020}& 
\Gamma_t^{SM,\text{best}}&  WT&   1.33& GeV&\cite{Gao:2012ja}\\

\midrule
\lambda_{CKM}&  CKMlambda& 0.22650& & \cite{PDG2020}&               
A&  CKMA& 0.790& &\cite{PDG2020}\\
\rho& CKMrho& 0.141& &  \cite{PDG2020}&                             
\eta&  CKMeta& 0.357&&\cite{PDG2020}\\

\midrule
\Lambda & LambdaSMEFT& $1$& TeV& &&&& & 
\\
\bottomrule
\end{tabular}
\flushleft
{\small

$(^*)$ only in models with \ascheme\ inputs. 
$(^{\star\star})$ only in models with \mwscheme\ inputs.
}
\caption{Common external parameters defined in all models, with the corresponding name in the code and default numerical value. The latter have been updated compared to previous \smeftsim\ versions, according to the references indicated.}\label{tab.defs_common}
\end{table}

\begin{table}[h!]\centering 
\renewcommand{\arraystretch}{1.3}
\begin{tabular}{|p{1cm}|*4{p{1.cm}>{\tt}p{1.5cm}|}}
\toprule\rowcolor{tablesColor}
\multicolumn{9}{|c|}{Bosonic SMEFT parameters in \smeftsim}\\
\toprule
$\Lag_6^{(1)}$&
$C_G$& cG& 
$C_{\widetilde{G}}$& cGtil&
$C_W$& cW& 
$C_{\widetilde{W}}$& cWtil\\

\midrule
$\Lag_6^{(2,3)}$&
$C_H$& cH& 
$C_{H\square}$& cHbox&
$C_{HD}$& cHDD& 
 &\\
 
\midrule
\multirow{2}{*}{$\Lag_6^{(4)}$}&
$C_{HG}$& cHG& 
$C_{HW}$& cHW&
$C_{HB}$& cHB& 
$C_{HWB}$& cHWB\\
&
$C_{H\widetilde G}$& cHGtil& 
$C_{H\widetilde W}$& cHWtil&
$C_{H\widetilde B}$& cHBtil& 
$C_{H\widetilde WB}$& cHWBtil\\
\bottomrule
\end{tabular}

\vspace*{2mm}

\begin{tabular}{|*4{p{1.35cm}>{\tt}p{1.5cm}|}}
\rowcolor{tablesColor}\toprule
\multicolumn{8}{|c|}{Shift parameters in \smeftsim}\\
\toprule
$\Delta m_Z^2$ & dMZ2 &
$\Delta m_h^2$ & dMH2 &
$\Delta G_F$ & dGf &
$\Delta \kappa_H$ & dkH 
\\\midrule
$\d \Gamma_Z$ & dWZ &
$\d \Gamma_W$ & dWW &
$\d \Gamma_h$ & dWH &
$\d \Gamma_t$ & dWT 
\\\midrule
$\d g_1/\hat g_1$ & dg1&
$\d g_W/\hat g_W$ & dgw&
$\d m_W$ & dMW&
&
\\
\bottomrule
\end{tabular}
\caption{Upper block: Wilson coefficients for the 15 bosonic operators. They are common to all flavor versions except \MFV, where the CP violating $C_{\widetilde G}, C_{\widetilde W}, C_{H\widetilde G},C_{H\widetilde W}, C_{H\widetilde B}, C_{H\widetilde WB}$ are not defined. Lower block: shift parameters defined in \smeftsim, see Sec.~\ref{sec.inputs_smeftsim} for details. $\d m_W $ is defined only in models with the \ascheme\ input scheme.}\label{tab.defs_bosonic}
\end{table}

\clearpage \thispagestyle{empty}
\begin{table}[h!]\centering 
\vspace*{-1.5cm}\scalebox{.88}{
\begin{tabular}{|c|*2{>{$}c<{$}>{\tt}c>{\tt}l|}}
\toprule\rowcolor{tablesColor}
\multicolumn{7}{|c|}{Fermionic SMEFT parameters in \smeftsim\ \general}\\
\toprule
class&
\rm parameter & \ufo & [pr(st)]& \rm parameter & \ufo &[pr(st)]\\
\midrule
\multirow{3}{*}{$\Lag_6^{(5)}$}&
\re (C_{eH})_{pr} & ceHRe[pr] &  [2f-NH-R]&      
\im (C_{eH})_{pr} & ceHIm[pr] &  [2f-NH-I]       \\
&
\re (C_{uH})_{pr} & cdHRe[pr] &  [2f-NH-R]&      
\im (C_{uH})_{pr} & cdHIm[pr] &  [2f-NH-I]       \\
&                                    
\re (C_{dH})_{pr} & cdHRe[pr] &  [2f-NH-R]&      
\im (C_{dH})_{pr} & cdHIm[pr] &  [2f-NH-I]       \\
\midrule                            

\multirow{8}{*}{$\Lag_6^{(6)}$}&
\re (C_{eW})_{pr} & ceWRe[pr] &  [2f-NH-R]&      
\im (C_{eW})_{pr} & ceWIm[pr] &  [2f-NH-I]       \\
&                                    
\re (C_{eB})_{pr} & ceBRe[pr] &  [2f-NH-R]&      
\im (C_{eB})_{pr} & ceBIm[pr] &  [2f-NH-I]       \\
&                                    
\re (C_{uG})_{pr} & cuGRe[pr] &  [2f-NH-R]&      
\im (C_{uG})_{pr} & cuGIm[pr] &  [2f-NH-I]       \\
&                                    
\re (C_{uW})_{pr} & cuWRe[pr] &  [2f-NH-R]&      
\im (C_{uW})_{pr} & cuWIm[pr] &  [2f-NH-I]       \\
&                                    
\re (C_{uB})_{pr} & cuBRe[pr] &  [2f-NH-R]&      
\im (C_{uB})_{pr} & cuBIm[pr] &  [2f-NH-I]       \\
&                                    
\re (C_{dG})_{pr} & cdGRe[pr] &  [2f-NH-R]&      
\im (C_{dG})_{pr} & cdGIm[pr] &  [2f-NH-I]       \\
&                                    
\re (C_{dW})_{pr} & cdWRe[pr] &  [2f-NH-R]&      
\im (C_{dW})_{pr} & cdWIm[pr] &  [2f-NH-I]       \\
&                                    
\re (C_{dB})_{pr} & cdBRe[pr] &  [2f-NH-R]&      
\im (C_{dB})_{pr} & cdBIm[pr] &  [2f-NH-I]       \\

\midrule

\multirow{8}{*}{$\Lag_6^{(7)}$}&
\re (C_{Hl}^{(1)})_{pr} & cHl1Re[pr] &  [2f-H-R]&      
\im (C_{Hl}^{(1)})_{pr} & cHl1Im[pr] &  [2f-H-I] \\
&
\re (C_{Hl}^{(3)})_{pr} & cHl3Re[pr] &  [2f-H-R] &
\im (C_{Hl}^{(3)})_{pr} & cHl3Im[pr] &  [2f-H-I]   \\
&
\re (C_{He})_{pr} & cHeRe[pr] &  [2f-H-R]        &
\im (C_{He})_{pr} & cHeIm[pr] &  [2f-H-I]\\
&
\re (C_{Hq}^{(1)})_{pr} & cHq1Re[pr] &  [2f-H-R] &
\im (C_{Hq}^{(1)})_{pr} & cHq1Im[pr] &  [2f-H-I]\\
&
\re (C_{Hq}^{(3)})_{pr} & cHq3Re[pr] &  [2f-H-R] &
\im (C_{Hq}^{(3)})_{pr} & cHq3Im[pr] &  [2f-H-I]\\
&
\re (C_{Hu})_{pr} & cHuRe[pr] &  [2f-H-R]       &
\im (C_{Hu})_{pr} & cHuIm[pr] &  [2f-H-I]\\
&
\re (C_{Hd})_{pr} & cHdRe[pr] &  [2f-H-R]        &
\im (C_{Hd})_{pr} & cHdIm[pr] &  [2f-H-I] \\
&
\re (C_{Hud})_{pr} & cHudRe[pr] &  [2f-NH-R]       &
\im (C_{Hud})_{pr} & cHudIm[pr] &  [2f-NH-I]       \\

\midrule
\multirow{5}{*}{$\Lag_6^{(8a)}$}&
\re (C_{ll})_{prst} & cllRe[prst] &  [4f-H-S-R]   &
\im (C_{ll})_{prst} & cllIm[prst] &  [4f-H-S-I]   \\
&
\re (C_{qq}^{(1)})_{prst} & cqq1Re[prst] &  [4f-H-S-R]   &
\im (C_{qq}^{(1)})_{prst} & cqq1Im[prst] &  [4f-H-S-I]   \\
&
\re (C_{qq}^{(3)})_{prst} & cqq3Re[prst] &  [4f-H-S-R]   &
\im (C_{qq}^{(3)})_{prst} & cqq3Im[prst] &  [4f-H-S-I]   \\
&
\re (C_{lq}^{(1)})_{prst} & clq1Re[prst] &  [4f-H-R]   &
\im (C_{lq}^{(1)})_{prst} & clq1Im[prst] &  [4f-H-I]   \\
&
\re (C_{lq}^{(3)})_{prst} & clq3Re[prst] &  [4f-H-R]   &
\im (C_{lq}^{(3)})_{prst} & clq3Im[prst] &  [4f-H-I]   \\

\midrule
\multirow{7}{*}{$\Lag_6^{(8b)}$}&
\re (C_{ee})_{prst} & ceeRe[prst] &  [4f-ee-R]   &
\im (C_{ee})_{prst} & ceeIm[prst] &  [4f-ee-I]   \\
&
\re (C_{uu})_{prst} & cuuRe[prst] &  [4f-H-S-R]   &
\im (C_{uu})_{prst} & cuuIm[prst] &  [4f-H-S-I]   \\
&
\re (C_{dd})_{prst} & cddRe[prst] &  [4f-H-S-R]   &
\im (C_{dd})_{prst} & cddIm[prst] &  [4f-H-S-I]   \\
&
\re (C_{eu})_{prst} & ceuRe[prst] &  [4f-H-R]   &
\im (C_{eu})_{prst} & ceuIm[prst] &  [4f-H-I]   \\
&
\re (C_{ed})_{prst} & cedRe[prst] &  [4f-H-R]   &
\im (C_{ed})_{prst} & cedIm[prst] &  [4f-H-I]   \\
&
\re (C_{ud}^{(1)})_{prst} & cud1Re[prst] &  [4f-H-R]   &
\im (C_{ud}^{(1)})_{prst} & cud1Im[prst] &  [4f-H-I]   \\
&
\re (C_{ud}^{(8)})_{prst} & cud8Re[prst] &  [4f-H-R]   &
\im (C_{ud}^{(8)})_{prst} & cud8Im[prst] &  [4f-H-I]   \\

\midrule
\multirow{8}{*}{$\Lag_6^{(8c)}$}&
\re (C_{le})_{prst} & cleRe[prst] &  [4f-H-R]   &
\im (C_{le})_{prst} & cleIm[prst] &  [4f-H-I]   \\
&
\re (C_{lu})_{prst} & cluRe[prst] &  [4f-H-R]   &
\im (C_{lu})_{prst} & cluIm[prst] &  [4f-H-I]   \\
&
\re (C_{ld})_{prst} & cldRe[prst] &  [4f-H-R]   &
\im (C_{ld})_{prst} & cldIm[prst] &  [4f-H-I]   \\
&
\re (C_{qe})_{prst} & cqeRe[prst] &  [4f-H-R]   &
\im (C_{qe})_{prst} & cqeIm[prst] &  [4f-H-I]   \\
&
\re (C_{qu}^{(1)})_{prst} & cqu1Re[prst] &  [4f-H-R]   &
\im (C_{qu}^{(1)})_{prst} & cqu1Im[prst] &  [4f-H-I]   \\
&
\re (C_{qu}^{(8)})_{prst} & cqu8Re[prst] &  [4f-H-R]   &
\im (C_{qu}^{(8)})_{prst} & cqu8Im[prst] &  [4f-H-I]   \\
&
\re (C_{qd}^{(1)})_{prst} & cqd1Re[prst] &  [4f-H-R]   &
\im (C_{qd}^{(1)})_{prst} & cqd1Im[prst] &  [4f-H-I]   \\
&
\re (C_{qd}^{(8)})_{prst} & cqd8Re[prst] &  [4f-H-R]   &
\im (C_{qd}^{(8)})_{prst} & cqd8Im[prst] &  [4f-H-I]   \\

\midrule
\multirow{5}{*}{$\Lag_6^{(8d)}$}&
\re (C_{ledq})_{prst} & cledqRe[prst] &  [4f-NH-R]   &
\im (C_{ledq})_{prst} & cledqIm[prst] &  [4f-NH-I]   \\
&
\re (C_{quqd}^{(1)})_{prst} & cquqd1Re[prst] &  [4f-NH-R]   &
\im (C_{quqd}^{(1)})_{prst} & cquqd1Im[prst] &  [4f-NH-I]   \\
&
\re (C_{quqd}^{(8)})_{prst} & cquqd8Re[prst] &  [4f-NH-R]   &
\im (C_{quqd}^{(8)})_{prst} & cquqd8Im[prst] &  [4f-NH-I]   \\
&
\re (C_{lequ}^{(1)})_{prst} & clequ1Re[prst] &  [4f-NH-R]   &
\im (C_{lequ}^{(1)})_{prst} & clequ1Im[prst] &  [4f-NH-I]   \\
&
\re (C_{lequ}^{(3)})_{prst} & clequ3Re[prst] &  [4f-NH-R]   &
\im (C_{lequ}^{(3)})_{prst} & clequ3Im[prst] &  [4f-NH-I]   \\

\bottomrule                                                    
\end{tabular}}
\caption{The 2484 independent parameters in $\Lag_6^{(5,6,7,8)}$  defined in the \general\ flavor model, see Sec.~\ref{sec.flavor_general}. The indices strings {\tt[pr]}, {\tt[prst]} take values in the sets indicated, that are defined in Table~\ref{tab.defs_general_indices}.}\label{tab.defs_general}
\end{table}

\begin{table}[h!]\centering 
\begin{tabular}{|ll|}
\toprule
\tt [2f-NH-R] &\tt 11, 22, 33, 12, 13, 23, 21, 31, 32\\[2mm]
\tt [2f-NH-I] &\tt 11, 22, 33, 12, 13, 23, 21, 31, 32\\[2mm]
\tt [2f-H-R]  &\tt 11, 22, 33, 12, 13, 23\\[2mm]
\tt [2f-H-I]  &\tt 12, 13, 23\\[2mm]
\tt [4f-ee-R] & 
\tt 1111, 1122, 1133, 2222, 2233, 3333, \\&
\tt 1112, 1113, 1123, 1212, 1213, 1222, 1232, 1233, 1313, 1322, 1323, 1333,  \\&
\tt
 2223, 2323, 3323
\\[2mm]

\tt [4f-ee-I] & 
\tt 1112, 1113, 1123, 1212, 1213, 1222, 1232, 1233, 1313, 1322, 1323, 1333, \\&
\tt2223, 2323, 3323
\\[2mm]

\tt [4f-H-S-R] & 
\tt 1111, 1122, 1133, 2222, 2233, 3333, 1221, 1331, 2332, \\&
\tt 1112, 1113, 1123, 1212, 1213, 1222, 1232, 1233, 1313, 1322,  1323, 1333,\\&
\tt 2223, 2323, 3323, 1231, 1223, 1332
\\[2mm]

\tt [4f-H-S-I] & 
\tt 1112, 1113, 1123, 1212, 1213, 1222, 1232, 1233, 1313, 1322,  1323, 1333,\\&
\tt 2223, 2323, 3323, 1231, 1223, 1332
\\[2mm]

\tt [4f-H-R] & 
\tt 1111, 1122, 1133, 2222, 2233, 3333, 1221, 1331, 2332, 2211, 3311, 3322,\\&
\tt
1112, 1113, 1123, 1212, 1213, 1222, 1232, 1233, 1313, 1322, 1323, 1333,\\&
\tt
2223, 2323, 3323, 1231, 1223, 1332,
1211, 1311, 1312, 1321, 2212, 2213,\\&
\tt
2311, 2312, 2313, 2321, 2322, 2331, 2333, 3312, 3313\\[2mm]

\tt [4f-H-I] & 
\tt

1112, 1113, 1123, 1212, 1213, 1222, 1232, 1233, 1313, 1322, 1323, 1333,\\&
\tt
2223, 2323, 3323, 1231, 1223, 1332,
1211, 1311, 1312, 1321, 2212, 2213,\\&
\tt
2311, 2312, 2313, 2321, 2322, 2331, 2333, 3312, 3313, 2211, 3311, 3322
\\
\bottomrule                                                    
\end{tabular}
\caption{Sets of indices implemented for each category of fermionic operator in the \general\ model.}\label{tab.defs_general_indices}
\end{table}

\begin{table}[h!]\centering 
\renewcommand{\arraystretch}{1.3}
\hspace*{-1.3cm}\begin{tabular}{|c|*4{>{$}c<{$}@{\hspace*{6mm}}>{\tt}c|}}
\toprule\rowcolor{tablesColor}
\multicolumn{9}{|c|}{Fermionic SMEFT parameters in \smeftsim\ \Utf}\\
\toprule
\multirow{2}{*}{$\Lag_6^{(5)}$}
&\re C_{eH} & ceHRe &          \re C_{uH} & cuHRe &      \re C_{dH} & cdHRe  & &\\
&\im C_{eH} & ceHIm &          \im C_{uH} & cuHIm &      \im C_{dH} & cdHIm  & &\\
\midrule
\multirow{4}{*}{$\Lag_6^{(6)}$}
&\re C_{eW} & ceWRe &     \re C_{eB} & ceBRe &       \re C_{uG} & cuGRe &   \re C_{uW} & cuWRe \\
&\im C_{eW} & ceWIm &     \im C_{eB} & ceBIm &       \im C_{uG} & cuGIm &   \im C_{uW} & cuWIm \\

&\re C_{uB} & cuBRe &     \re C_{dG} & cdGRe &       \re C_{dW} & cdWRe &   \re C_{dB} & cdBRe \\
&\im C_{uB} & cuBIm &     \im C_{dG} & cdGIm &       \im C_{dW} & cdWIm &   \im C_{dB} & cdBIm \\

\midrule
\multirow{3}{*}{$\Lag_6^{(7)}$}
&C_{Hl}^{(1)} & cHl1 &    C_{Hq}^{(1)} & cHq1 &  C_{Hu} & cHu &  C_{He}  & cHe \\
&C_{Hl}^{(3)} & cHl3 &    C_{Hq}^{(3)} & cHq3 &  C_{Hd} & cHd &  &\\
& \re C_{Hud} & cHudRe & \im C_{Hud}& cHudIm& & & &\\

\midrule
\multirow{2}{*}{$\Lag_6^{(8a)}$}

&C_{ll}  & cll  &   C_{qq}^{(1)}       & cqq1  &   C_{qq}^{(3)}       & cqq3  &  C_{lq}^{(1)} & clq1 \\
&C_{ll}' & cll1 &   C_{qq}^{(1)\prime} & cqq11 &   C_{qq}^{(3)\prime} & cqq31 &  C_{lq}^{(3)} & clq3 \\

\midrule

\multirow{3}{*}{$\Lag_6^{(8b)}$}

&C_{uu}  & cuu  &   C_{dd}  & cdd  &   C_{eu} & ceu &  C_{ud}^{(1)} & cud1 \\
&C_{uu}' & cuu1 &   C_{dd}' & cdd1 &   C_{ed} & ced &  C_{ud}^{(8)} & cud8 \\

&C_{ee} & cee & & & & & & \\

\midrule

\multirow{2}{*}{$\Lag_6^{(8c)}$}
&C_{le}  & cle &    C_{lu}  & clu &  C_{qu}^{(1)}  & cqu1 &  C_{qd}^{(1)}  & cqd1 \\
&C_{qe}  & cqe &    C_{ld}  & cld &  C_{qu}^{(8)}  & cqu8 &  C_{qd}^{(8)}  & cqd8 \\

\midrule

\multirow{2}{*}{$\Lag_6^{(8d)}$}
&\re C_{quqd}^{(1)}  & cquqd1Re & \re C_{quqd}^{(8)}  & cquqd8Re 
&\im C_{quqd}^{(1)}  & cquqd1Im  &\im C_{quqd}^{(8)}  & cquqd8Im  \\
&\re C_{quqd}^{(1)'}  & cquqd11Re & \re C_{quqd}^{(8)'}  & cquqd81Re 
&\im C_{quqd}^{(1)'}  & cquqd11Im  &\im C_{quqd}^{(8)'}  & cquqd81Im  \\ 

&  \re C_{lequ}^{(1)} &clequ1Re & \re C_{lequ}^{(3)}  & clequ3Re
&  \im C_{lequ}^{(1)} &clequ1Im&  \im C_{lequ}^{(3)}  & clequ3Im 
\\
& \re C_{ledq}  & cledqRe  & \im C_{ledq}  & cledqIm & & & & \\

\bottomrule                                                    
\end{tabular}
\caption{The 70 independent parameters in $\Lag_6^{(5,6,7,8)}$   defined in the \Utf\ model, see Sec.~\ref{sec.flavor_u35}.}\label{tab.defs_U35}
\end{table}

\clearpage \thispagestyle{empty}
\begin{table}[h!]
\renewcommand{\arraystretch}{1.3}
\hspace*{-1.5cm}\scalebox{.88}{
\begin{tabular}{|c|*4{>{$}c<{$}@{\hspace*{4mm}}>{\tt}c|}}
\toprule\rowcolor{tablesColor}
\multicolumn{9}{|c|}{Fermionic SMEFT parameters in \smeftsim\ \MFV}\\
\toprule
\multirow{2}{*}{$\Lag_6^{(5)}$}
&C_{eH} & ceH &     C_{uH}^{(0)} & cuH0 &  C_{dH}^{(0)} & cdH0  & &\\
&(\Delta^u C_{uH}) & DeltaucuH &   (\Delta^d C_{uH}) & DeltadcuH     
&(\Delta^u C_{dH}) & DeltaucdH &   (\Delta^d C_{dH}) & DeltadcdH \\
\midrule
\multirow{4}{*}{$\Lag_6^{(6)}$}
& C_{eW} & ceW & C_{eB} & ceB &  C_{uG}^{(0)} & cuG0 &   C_{uW}^{(0)} & cuW0 \\

&C_{uB}^{(0)} & cuB0 &  C_{dG}^{(0)} & cdG0 &   C_{dW}^{(0)} & cdW0 &   C_{dB}^{(0)} & cdB0 \\

&(\Delta^u C_{uG}) & DeltaucuG &   (\Delta^d C_{uG}) & DeltadcuG     
&(\Delta^u C_{dG}) & DeltaucdG &   (\Delta^d C_{dG}) & DeltadcdG
\\
&(\Delta^u C_{uW}) & DeltaucuW &   (\Delta^d C_{uW}) & DeltadcuW     
&(\Delta^u C_{dW}) & DeltaucdW &   (\Delta^d C_{dW}) & DeltadcdW
\\
&(\Delta^u C_{uB}) & DeltaucuB &   (\Delta^d C_{uB}) & DeltadcuB     
&(\Delta^u C_{dB}) & DeltaucdB &   (\Delta^d C_{dB}) & DeltadcdB
\\
\midrule
\multirow{3}{*}{$\Lag_6^{(7)}$}
&C_{Hl}^{(1)} & cHl1 &    C_{Hq}^{(1)(0)} & cHq10 &  C_{Hu}^{(0)} & cHu0 &  C_{He}  & cHe \\
&C_{Hl}^{(3)} & cHl3 &    C_{Hq}^{(3)(0)} & cHq30 &  C_{Hd}^{(0)} & cHd0 & C_{Hud}^{(0)} & cHud0
\\
&(\Delta^u C_{Hq}^{(1)}) & DeltaucHq1 &   (\Delta^d C_{Hq}^{(1)}) & DeltadcHq1
&(\Delta^u C_{Hq}^{(3)}) & DeltaucHq3 &   (\Delta^d C_{Hq}^{(3)}) & DeltadcHq3
\\
&(\Delta C_{Hu}) & DeltacHu &   (\Delta C_{Hd}) & DeltacHd
& & & &
\\
\midrule
\multirow{2}{*}{$\Lag_6^{(8a)}$}

&C_{ll}  & cll  &   C_{qq}^{(1)(0)}       & cqq10  &   C_{qq}^{(3)(0)}       & cqq30  &  C_{lq}^{(1)(0)} & clq10 \\
&C_{ll}' & cll1 &   C_{qq}^{(1)\prime(0)} & cqq110 &   C_{qq}^{(3)\prime(0)} & cqq310 &  C_{lq}^{(3)(0)} & clq30 \\

&(\Delta^u C_{qq}^{(1)}) & Deltaucqq1 &   (\Delta^d C_{qq}^{(1)}) & Deltadcqq1
&(\Delta^u C_{qq}^{(3)}) & Deltaucqq3 &   (\Delta^d C_{qq}^{(3)}) & Deltadcqq3
\\
&(\Delta^u C_{qq}^{(1)\prime}) & Deltaucqq11 &   (\Delta^d C_{qq}^{(1)\prime}) & Deltadcqq11
&(\Delta^u C_{qq}^{(3)\prime}) & Deltaucqq31 &   (\Delta^d C_{qq}^{(3)\prime}) & Deltadcqq31
\\
&(\Delta^u C_{lq}^{(1)}) & Deltauclq1 &   (\Delta^d C_{lq}^{(1)}) & Deltadclq1
&(\Delta^u C_{lq}^{(3)}) & Deltauclq3 &   (\Delta^d C_{lq}^{(3)}) & Deltadclq3
\\
\midrule

\multirow{3}{*}{$\Lag_6^{(8b)}$}

&C_{uu}^{(0)}  & cuu0  &   C_{dd}^{(0)}  & cdd0  &   C_{eu}^{(0)} & ceu0 &  C_{ud}^{(1)(0)} & cud10 \\
&C_{uu}^{\prime(0)} & cuu10 &   C_{dd}^{\prime(0)} & cdd10 &   C_{ed}^{(0)} & ced0 &  C_{ud}^{(8)(0)} & cud80 \\

&C_{ee} & cee & (\Delta C_{eu})& Deltaceu& (\Delta C_{ed})& Deltaced & & \\

&(\Delta^u C_{uu}) & Deltaucuu &   (\Delta^d C_{dd}) & Deltadcdd
&(\Delta^u C_{uu}^{\prime}) & Deltaucuu1 &   (\Delta^d C_{dd}^{\prime}) & Deltadcdd1
\\
&(\Delta^u C_{ud}^{(1)}) & Deltaucud1 &   (\Delta^d C_{ud}^{(1)}) & Deltadcud1
&(\Delta^u C_{ud}^{(8)}) & Deltaucud8 &   (\Delta^d C_{ud}^{(8)}) & Deltadcud8
\\
\midrule

\multirow{2}{*}{$\Lag_6^{(8c)}$}
&C_{le}  & cle & 
C_{lu}^{(0)}  & clu0 &  
C_{ld}^{(0)} & cld0 &
C_{qe}^{(0)}  & cqe0 \\
& 
C_{qu}^{(1)(0)}  & cqu10 &
C_{qd}^{(1)(0)} & cqd10&
C_{qu}^{(8)(0)}  & cqu80 &  
C_{qd}^{(8)(0)}  & cqd80 \\
&
C_{qu}^{(1)\prime(0)} & cqu110& 
C_{qd}^{(1)\prime(0)} & cqd110& 
C_{qu}^{(8)\prime(0)} & cqu810& 
C_{qd}^{(8)\prime(0)} & cqd810 
\\
&(\Delta_1^u C_{qu}^{(1)}) & Delta1ucqu1 &   (\Delta_1^d C_{qu}^{(1)}) & Delta1dcqu1
&(\Delta_2 C_{qu}^{(1)}) & Delta2cqu1 &   (\Delta C_{lu}) & Deltaclu 
\\
&(\Delta_1^u C_{qu}^{(8)}) & Delta1ucqu8 &   (\Delta_1^d C_{qu}^{(8)}) & Delta1dcqu8
&(\Delta_2 C_{qu}^{(8)}) & Delta2cqu8 &   (\Delta C_{ld}) & Deltacld 
\\
&(\Delta_1^u C_{qd}^{(1)}) & Delta1ucqd1 &   (\Delta_1^d C_{qd}^{(1)}) & Delta1dcqd1
&(\Delta_2 C_{qd}^{(1)}) & Delta2cqd1 &  (\Delta^u C_{qe}) & Deltaucqe
\\
&(\Delta_1^u C_{qd}^{(8)}) & Delta1ucqd8 &   (\Delta_1^d C_{qd}^{(8)}) & Delta1dcqd8
&(\Delta_2 C_{qd}^{(8)}) & Delta2cqd8 &  (\Delta^d C_{qe}) & Deltadcqe
\\
\midrule

\multirow{2}{*}{$\Lag_6^{(8d)}$}
&C_{lequ}^{(1)(0)} & clequ10 & C_{lequ}^{(3)(0)}  & clequ30 &  C_{ledq}^{(0)}  & cledq0  & &\\
&C_{quqd}^{(1)} & cquqd1 & C_{quqd}^{(8)}  & cquqd8
&C_{quqd}^{(1)'} & cquqd11 & C_{quqd}^{(8)'}  & cquqd81
\\
& (\Delta^uC_{ledq}) & Deltaucledq & (\Delta^dC_{ledq}) & Deltadcledq& & & &
\\
&(\Delta^u C_{lequ}^{(1)}) & Deltauclequ1 &   (\Delta^d C_{lequ}^{(1)}) & Deltadclequ1
&(\Delta^u C_{lequ}^{(3)}) & Deltauclequ3 &   (\Delta^d C_{lequ}^{(3)}) & Deltadclequ3
\\
\bottomrule                                                    
\end{tabular}}
\caption{The 121 independent parameters in $\Lag_6^{(5,6,7,8)}$  defined in the \MFV\ model, see Sec.~\ref{sec.flavor_mfv}.}\label{tab.defs_MFV}
\end{table}

\clearpage \thispagestyle{empty}
\begin{table}[h!] 
\renewcommand{\arraystretch}{1.2}
\vspace*{-1.5cm}
\hspace*{-1cm}\scalebox{.78}{
\begin{tabular}{|c|*4{>{$}c<{$}@{\hspace*{6mm}}>{\tt}c|}}
\toprule\rowcolor{tablesColor}
\multicolumn{9}{|c|}{Fermionic SMEFT parameters in \smeftsim\ \top}\\
\toprule
\multirow{3}{*}{$\Lag_6^{(5)}$}
&\re (C_{eH})_{pp} & ceHRe[pp] &  \im (C_{eH})_{pp} & ceHIm[pp] & & & &\\
&\re C_{uH} & cuHRe &     \im C_{uH} & cuHIm &      \re C_{tH} & ctHRe& \im C_{tH} & ctHIm\\
&\re C_{dH} & cdHRe &     \im C_{dH} & cdHIm &      \re C_{bH} & cbHRe& \im C_{bH} & cbHIm\\

\midrule
\multirow{7}{*}{$\Lag_6^{(6)}$}
&\re (C_{eW})_{pp} & ceWRe[pp] &  \im (C_{eW})_{pp} & ceWIm[pp]   
&\re (C_{eB})_{pp} & ceBRe[pp] &  \im (C_{eB})_{pp} & ceBIm[pp]\\

&\re C_{uG} & cuGRe &     \im C_{uG} & cuGIm &      \re C_{tG} & ctGRe& \im C_{tG} & ctGIm\\
&\re C_{dG} & cdGRe &     \im C_{dG} & cdGIm &      \re C_{bG} & cbGRe& \im C_{bG} & cbGIm\\

&\re C_{uW} & cuWRe &     \im C_{uW} & cuWIm &      \re C_{tW} & ctWRe& \im C_{tW} & ctWIm\\
&\re C_{dW} & cdWRe &     \im C_{dW} & cdWIm &      \re C_{bW} & cbWRe& \im C_{bW} & cbWIm\\

&\re C_{uB} & cuBRe &     \im C_{uB} & cuBIm &      \re C_{tB} & ctBRe& \im C_{tB} & ctBIm\\
&\re C_{dB} & cdBRe &     \im C_{dB} & cdBIm &      \re C_{bB} & cbBRe& \im C_{bB} & cbBIm\\

\midrule

\multirow{4}{*}{$\Lag_6^{(7)}$}
&(C_{Hl}^{(1)})_{pp} & cHl1[pp] &   (C_{Hl}^{(3)})_{pp} & cHl3[pp] &  
(C_{He})_{pp} & cHe[pp] & &\\

&C_{Hq}^{(1)} & cHj1 &   C_{HQ}^{(1)} & cHQ1 &    C_{Hu} & cHu & C_{Ht} & cHt \\
&C_{Hq}^{(3)} & cHj3 &   C_{HQ}^{(3)} & cHQ3 &    C_{Hd} & cHd & C_{Hb} & cHbq \\

& \re C_{Hud} & cHudRe & \im C_{Hud}& cHudIm & \re C_{Htb} & cHtbRe & \im C_{Htb}& cHtbIm\\

\midrule
\multirow{5}{*}{$\Lag_6^{(8a)}$}

&(C_{ll})_{prst}  & cll[prst]  & \multicolumn{6}{l|}{\tt prst = $\{$1111, 2222, 3333, 1122, 1133, 2233, 1221, 1331, 2332$\}$}\\
& (C_{lq}^{(1)})_{pp} & clj1[pp] & (C_{lQ}^{(1)})_{pp} & cQl1[pp] & 
 (C_{lq}^{(3)})_{pp} & clj3[pp] & (C_{lQ}^{(3)})_{pp} & cQl3[pp] \\

&C_{qq}^{(1,1)} & cjj11  & C_{qq}^{(1,8)} & cjj18  & C_{Qq}^{(1,1)} & cQj11  & C_{Qq}^{(1,8)} & cQj18  \\ 
&C_{qq}^{(3,1)} & cjj31  & C_{qq}^{(3,8)} & cjj38  & C_{Qq}^{(3,1)} & cQj31  & C_{Qq}^{(3,8)} & cQj38  \\ 
&C_{QQ}^{(1)}    & cQQ1  & C_{QQ}^{(8)}& cQQ8& & & & \\

\midrule

\multirow{8}{*}{$\Lag_6^{(8b)}$}
& (C_{ee})_{prst}& cee[prst]& \multicolumn{6}{l|}{\tt prst = $\{$1111, 2222, 3333, 1122, 1133, 2233$\}$}\\

& (C_{eu})_{pp}& ceu[pp]& (C_{et})_{pp}& cte[pp]
& (C_{ed})_{pp}& ced[pp]& (C_{eb})_{pp}& cbe[pp]\\

&C_{uu}^{(1)}  & cuu1  &   C_{uu}^{(8)}  & cuu8  &C_{dd}^{(1)}  & cdd1  &   C_{dd}^{(8)}  & cdd8        \\   

&C_{tu}^{(1)}  & ctu1  &   C_{tu}^{(8)}  & ctu8 &  C_{bd}^{(1)}  & cbd1  &   C_{bd}^{(8)}  & cbd8  \\   
&C_{tt}& ctt & & & C_{bb}& cbb & &  \\

& C_{ud}^{(1)} & cud1 & C_{td}^{(1)} & ctd1 & C_{ub}^{(1)} & cbu1 & C_{tb}^{(1)} & ctb1\\
& C_{ud}^{(8)} & cud8 & C_{td}^{(8)} & ctd8 & C_{ub}^{(8)} & cbu8 & C_{tb}^{(8)} & ctb8\\

&\re ( C_{utbd}^{(1)})& cutbd1Re&
\im ( C_{utbd}^{(1)})& cutbd1Im&
\re ( C_{utbd}^{(8)})& cutbd8Re&
\im ( C_{utbd}^{(8)})& cutbd8Im
\\

\midrule

\multirow{8}{*}{$\Lag_6^{(8c)}$}
& (C_{le})_{prst}& cle[prst]& \multicolumn{6}{l|}{\tt prst = $\{$1111,2222,3333,1122,1133,2233,2211,3311,3322,1221,1331,2332$\}$}\\

& (C_{lu})_{pp}& clu[pp]& (C_{lt})_{pp}& ctl[pp]
& (C_{ld})_{pp}& cld[pp]& (C_{lb})_{pp}& cbl[pp]\\

& (C_{qe})_{pp}& cje[pp]& (C_{Qe})_{pp}& cQe[pp]& & & &\\

&C_{qu}^{(1)}  & cju1 &  C_{Qu}^{(1)}  & cQu1 &  C_{qt}^{(1)}  & ctj1 &  C_{Qt}^{(1)}  & cQt1 \\
&C_{qu}^{(8)}  & cju8 &  C_{Qu}^{(8)}  & cQu8 &  C_{qt}^{(8)}  & ctj8 &  C_{Qt}^{(8)}  & cQt8 \\
&\re (C_{qQtu}^{(1)})& cjQtu1Re& 
\im (C_{qQtu}^{(1)})& cjQtu1Im& 
\re (C_{qQtu}^{(8)})& cjQtu8Re& 
\im (C_{qQtu}^{(8)})& cjQtu8Im
\\

&C_{qd}^{(1)}  & cjd1 &  C_{Qd}^{(1)}  & cQd1 &  C_{qb}^{(1)}  & cbj1 &  C_{Qb}^{(1)}  & cQb1 \\
&C_{qd}^{(8)}  & cjd8 &  C_{Qd}^{(8)}  & cQd8 &  C_{qb}^{(8)}  & cbj8 &  C_{Qb}^{(8)}  & cQb8 \\
&\re (C_{qQbd}^{(1)})& cjQbd1Re& 
\im (C_{qQbd}^{(1)})& cjQbd1Im& 
\re (C_{qQbd}^{(8)})& cjQbd8Re& 
\im (C_{qQbd}^{(8)})& cjQbd8Im
 
\\

\midrule

\multirow{4}{*}{$\Lag_6^{(8d)}$}
& \re (C_{ledq})_{pp}& cledjRe[pp]& \im (C_{ledq})_{pp}& cledqIm[pp] 
& \re (C_{lebQ})_{pp}& clebQRe[pp] & \im (C_{lebQ})_{pp}& clebQIm[pp]\\

& \re (C_{lequ}^{(1)})_{pp}& cleju1Re[pp] & \im (C_{lequ}^{(1)})_{pp}& cleju1Im[pp] 
& \re (C_{leQt}^{(1)})_{pp}& cleQt1Re[pp] & \im (C_{leQt}^{(1)})_{pp}& cleQt1Im[pp] \\
 
& \re (C_{lequ}^{(3)})_{pp}& cleju3Re[pp] & \im (C_{lequ}^{(3)})_{pp}& cleju3Im[pp]
& \re (C_{leQt}^{(3)})_{pp}& cleQt3Re[pp] & \im (C_{leQt}^{(3)})_{pp}& cleQt3Im[pp]\\

&\re C_{quqd}^{(1)}  & cjujd1Re & \im C_{quqd}^{(1)}  & cjujd1Im& \re C_{QtQb}^{(1)}  & cQtQb1Re& \im C_{QtQb}^{(1)}  & cQtQb1Im \\

&\re C_{quqd}^{(8)}  & cjujd8Re & \im C_{quqd}^{(8)}  & cjujd8Im& \re C_{QtQb}^{(8)}  & cQtQb8Re& \im C_{QtQb}^{(8)}  & cQtQb8Im \\

&\re C_{quqd}^{(1)\prime}  & cjujd11Re & \im C_{quqd}^{(1)\prime}  & cjujd11Im
&\re C_{quqd}^{(8)\prime}  & cjujd81Re & \im C_{quqd}^{(8)\prime}  & cjujd81Im \\

&\re C_{Qtqd}^{(1)}  & cQtjd1Re & \im C_{Qtqd}^{(1)}  & cQtjd1Im&
\re C_{quQb}^{(1)}  & cjuQb1Re& \im C_{quQb}^{(1)}  & cjuQb1Im \\
&\re C_{Qtqd}^{(8)}  & cQtjd8Re & \im C_{Qtqd}^{(8)}  & cQtjd8Im&
\re C_{quQb}^{(8)}  & cjuQb8Re& \im C_{quQb}^{(8)}  & cjuQb8Im \\

&\re C_{qtQd}^{(1)}  & cjtQd1Re & \im C_{qtQd}^{(1)}  & cjtQd1Im&
\re C_{Quqb}^{(1)}  & cQujb1Re& \im C_{Quqb}^{(1)}  & cQujb1Im \\
&\re C_{qtQd}^{(8)}  & cjtQd8Re & \im C_{qtQd}^{(8)}  & cjtQd8Im&
\re C_{Quqb}^{(8)}  & cQujb8Re& \im C_{Quqb}^{(8)}  & cQujb8Im \\

\bottomrule                                                    

\end{tabular}}
\caption{The 260 independent parameters in $\Lag_6^{(5,6,7,8)}$ defined in the \top\ model, see Sec.~\ref{sec.flavor_top}. Lepton flavor indices {\tt[pp]} always run over {\tt $\{$11, 22, 33$\}$}. Indices {\tt[prst]} take the values indicated in line.}\label{tab.defs_top}
\end{table}

\clearpage \thispagestyle{empty}
\begin{table}[h!]\centering 
\renewcommand{\arraystretch}{1.2}
\vspace*{-1cm}
\scalebox{.85}{
\begin{tabular}{|c|*4{>{$}c<{$}@{\hspace*{6mm}}>{\tt}c|}}
\toprule\rowcolor{tablesColor}
\multicolumn{9}{|c|}{Fermionic SMEFT parameters in \smeftsim\ \topsl}\\
\toprule
\multirow{3}{*}{$\Lag_6^{(5)}$}
&\re C_{eH} & ceHRe &  \im C_{eH} & ceHIm & & & &\\
&\re C_{uH} & cuHRe &  \im C_{uH} & cuHIm &      \re C_{tH} & ctHRe& \im C_{tH} & ctHIm\\
&\re C_{dH} & cdHRe &  \im C_{dH} & cdHIm &      \re C_{bH} & cbHRe& \im C_{bH} & cbHIm\\

\midrule
\multirow{7}{*}{$\Lag_6^{(6)}$}
&\re C_{eW} & ceWRe &  \im C_{eW} & ceWIm   
&\re C_{eB} & ceBRe &  \im C_{eB} & ceBIm\\

&\re C_{uG} & cuGRe &     \im C_{uG} & cuGIm &      \re C_{tG} & ctGRe& \im C_{tG} & ctGIm\\
&\re C_{dG} & cdGRe &     \im C_{dG} & cdGIm &      \re C_{bG} & cbGRe& \im C_{bG} & cbGIm\\

&\re C_{uW} & cuWRe &     \im C_{uW} & cuWIm &      \re C_{tW} & ctWRe& \im C_{tW} & ctWIm\\
&\re C_{dW} & cdWRe &     \im C_{dW} & cdWIm &      \re C_{bW} & cbWRe& \im C_{bW} & cbWIm\\

&\re C_{uB} & cuBRe &     \im C_{uB} & cuBIm &      \re C_{tB} & ctBRe& \im C_{tB} & ctBIm\\
&\re C_{dB} & cdBRe &     \im C_{dB} & cdBIm &      \re C_{bB} & cbBRe& \im C_{bB} & cbBIm\\

\midrule

\multirow{4}{*}{$\Lag_6^{(7)}$}
&C_{Hl}^{(1)} & cHl1 &   C_{Hl}^{(3)} & cHl3 &  C_{He} & cHe & &\\

&C_{Hq}^{(1)} & cHj1 &   C_{HQ}^{(1)} & cHQ1 &    C_{Hu} & cHu & C_{Ht} & cHt \\
&C_{Hq}^{(3)} & cHj3 &   C_{HQ}^{(3)} & cHQ3 &    C_{Hd} & cHd & C_{Hb} & cHbq \\

& \re C_{Hud} & cHudRe & \im C_{Hud}& cHudIm & \re C_{Htb} & cHtbRe & \im C_{Htb}& cHtbIm\\

\midrule
\multirow{5}{*}{$\Lag_6^{(8a)}$}

&C_{ll}  & cll  & C_{ll}' & cll1& C_{QQ}^{(1)}    & cQQ1  & C_{QQ}^{(8)}& cQQ8 \\
& C_{lq}^{(1)} & clj1 & C_{lQ}^{(1)} & cQl1 & 
  C_{lq}^{(3)} & clj3 & C_{lQ}^{(3)} & cQl3 \\

&C_{qq}^{(1,1)} & cjj11  & C_{qq}^{(1,8)} & cjj18  & C_{Qq}^{(1,1)} & cQj11  & C_{Qq}^{(1,8)} & cQj18  \\ 
&C_{qq}^{(3,1)} & cjj31  & C_{qq}^{(3,8)} & cjj38  & C_{Qq}^{(3,1)} & cQj31  & C_{Qq}^{(3,8)} & cQj38  \\

\midrule

\multirow{8}{*}{$\Lag_6^{(8b)}$}
& C_{eu}& ceu& C_{et}& cte
& C_{ed}& ced& C_{eb}& cbe\\

&C_{uu}^{(1)}  & cuu1  &   C_{uu}^{(8)}  & cuu8  &C_{dd}^{(1)}  & cdd1  &   C_{dd}^{(8)}  & cdd8        \\   

&C_{tu}^{(1)}  & ctu1  &   C_{tu}^{(8)}  & ctu8 &  C_{bd}^{(1)}  & cbd1  &   C_{bd}^{(8)}  & cbd8  \\   
&C_{tt}& ctt &  C_{bb}& cbb & C_{ee}& cee&&  \\

& C_{ud}^{(1)} & cud1 & C_{td}^{(1)} & ctd1 & C_{ub}^{(1)} & cbu1 & C_{tb}^{(1)} & ctb1\\
& C_{ud}^{(8)} & cud8 & C_{td}^{(8)} & ctd8 & C_{ub}^{(8)} & cbu8 & C_{tb}^{(8)} & ctb8\\
&\re (C_{qQbd}^{(1)})& cjQbd1Re& 
\im (C_{qQbd}^{(1)})& cjQbd1Im& 
\re (C_{qQbd}^{(8)})& cjQbd8Re& 
\im (C_{qQbd}^{(8)})& cjQbd8Im
 
\\

\midrule

\multirow{8}{*}{$\Lag_6^{(8c)}$}
& C_{le}& cle& C_{qe}& cje& C_{Qe}& cQe& &\\

& C_{lu}& clu& C_{lt}& ctl
& C_{ld}& cld& C_{lb}& cbl\\

&C_{qu}^{(1)}  & cju1 &  C_{Qu}^{(1)}  & cQu1 &  C_{qt}^{(1)}  & ctj1 &  C_{Qt}^{(1)}  & cQt1 \\
&C_{qu}^{(8)}  & cju8 &  C_{Qu}^{(8)}  & cQu8 &  C_{qt}^{(8)}  & ctj8 &  C_{Qt}^{(8)}  & cQt8 \\
&\re (C_{qQtu}^{(1)})& cjQtu1Re& 
\im (C_{qQtu}^{(1)})& cjQtu1Im& 
\re (C_{qQtu}^{(8)})& cjQtu8Re& 
\im (C_{qQtu}^{(8)})& cjQtu8Im
\\
&C_{qd}^{(1)}  & cjd1 &  C_{Qd}^{(1)}  & cQd1 &  C_{qb}^{(1)}  & cbj1 &  C_{Qb}^{(1)}  & cQb1 \\
&C_{qd}^{(8)}  & cjd8 &  C_{Qd}^{(8)}  & cQd8 &  C_{qb}^{(8)}  & cbj8 &  C_{Qb}^{(8)}  & cQb8 \\
&\re (C_{qQbd}^{(1)})& cjQbd1Re& 
\im (C_{qQbd}^{(1)})& cjQbd1Im& 
\re (C_{qQbd}^{(8)})& cjQbd8Re& 
\im (C_{qQbd}^{(8)})& cjQbd8Im
\\
\midrule

\multirow{4}{*}{$\Lag_6^{(8d)}$}
& \re C_{ledq}& cledjRe & \im C_{ledq}& cledqIm 
& \re C_{lebQ}& clebQRe & \im C_{lebQ}& clebQIm\\

& \re C_{lequ}^{(1)}& cleju1Re & \im C_{lequ}^{(1)}& cleju1Im 
& \re C_{leQt}^{(1)}& cleQt1Re & \im C_{leQt}^{(1)}& cleQt1Im \\
 
& \re C_{lequ}^{(3)}& cleju3Re & \im C_{lequ}^{(3)}& cleju3Im 
& \re C_{leQt}^{(3)}& cleQt3Re & \im C_{leQt}^{(3)}& cleQt3Im\\

&\re C_{quqd}^{(1)}  & cjujd1Re & \im C_{quqd}^{(1)}  & cjujd1Im& \re C_{QtQb}^{(1)}  & cQtQb1Re& \im C_{QtQb}^{(1)}  & cQtQb1Im \\

&\re C_{quqd}^{(8)}  & cjujd8Re & \im C_{quqd}^{(8)}  & cjujd8Im& \re C_{QtQb}^{(8)}  & cQtQb8Re& \im C_{QtQb}^{(8)}  & cQtQb8Im \\

&\re C_{quqd}^{(1)\prime}  & cjujd11Re & \im C_{quqd}^{(1)\prime}  & cjujd11Im
&\re C_{quqd}^{(8)\prime}  & cjujd81Re & \im C_{quqd}^{(8)\prime}  & cjujd81Im \\

&\re C_{Qtqd}^{(1)}  & cQtjd1Re & \im C_{Qtqd}^{(1)}  & cQtjd1Im&
\re C_{quQb}^{(1)}  & cjuQb1Re& \im C_{quQb}^{(1)}  & cjuQb1Im \\
&\re C_{Qtqd}^{(8)}  & cQtjd8Re & \im C_{Qtqd}^{(8)}  & cQtjd8Im&
\re C_{quQb}^{(8)}  & cjuQb8Re& \im C_{quQb}^{(8)}  & cjuQb8Im \\

&\re C_{qtQd}^{(1)}  & cjtQd1Re & \im C_{qtQd}^{(1)}  & cjtQd1Im&
\re C_{Quqb}^{(1)}  & cQujb1Re& \im C_{Quqb}^{(1)}  & cQujb1Im \\
&\re C_{qtQd}^{(8)}  & cjtQd8Re & \im C_{qtQd}^{(8)}  & cjtQd8Im&
\re C_{Quqb}^{(8)}  & cQujb8Re& \im C_{Quqb}^{(8)}  & cQujb8Im \\

\bottomrule                                                    

\end{tabular}}
\caption{The 167 independent parameters in $\Lag_6^{(5,6,7,8)}$ defined in the \topsl\ model, see Sec.~\ref{sec.flavor_top}.}\label{tab.defs_topsl}
\end{table}


\clearpage
\section{Comparison to other SMEFT UFO models}\label{app.UFO_comparison}
In this section we compare \smeftsim\ with other \ufo\ models dedicated to SMEFT studies, and provide a mapping of the common parameters. For the time being, the comparison is restricted to \dimsixtop~\cite{dim6top,AguilarSaavedra:2018nen} and \smeftatnlo~\cite{smeftatnlo,Degrande:2020evl}, that are both based on the Warsaw basis.

For each model we summarize the main features and provide conversion tables with the parameters defined in \smeftsim.
To our knowledge, \smeftsim\ is currently the only publicly available \ufo\ model that implements linearized SMEFT corrections to propagators.

\subsection{\dimsixtop}
\dimsixtop~\cite{dim6top,AguilarSaavedra:2018nen} contains LO \ufo\ models dedicated to EFT studies in the top sector.
Here we refer specifically to {\tt dim6top\_LO\_UFO} and {\tt dim6top\_LO\_UFO\_each\_coupling\_order} in the version published in May 2020. 

\begin{description}

\item[Flavor structure.]
\dimsixtop\ is based on the recommendations provided in Ref.~\cite{AguilarSaavedra:2018nen}, and it assumes a $U(2)^3$ flavor symmetry in the quark sector and a $(U(1)_{l+e})^3$ in the lepton sector.
$U(2)^3$ breaking terms are also available and they are implemented explicitly, i.e. without promoting the quark Yukawas to spurions of the flavor symmetry. Contractions inducing both flavor-conserving and violating neutral currents are included.   

All fermion masses and Yukawa couplings are neglected, except those of the top and bottom quarks of the tau lepton. The CKM is taken to be the unit matrix.

\item[Operators implemented.]
\dimsixtop\ contains only operators that modify the interactions of the top quark, and CP violating terms are included. Most operator definitions are identical to those in \smeftsim\ \top, \topsl. In a few cases, the invariants implemented differ by a Fierz rotation, as detailed in Ref.~\cite{AguilarSaavedra:2018nen}.

\item[Input parameters.]
Both input schemes \ascheme\ and \mwscheme\ are supported in \dimsixtop. Since purely bosonic and leptonic operators are omitted, this only affects the numerical values assigned to the SM parameters and not the dependence on the Wilson coefficients.

\item[SM loop-generated Higgs couplings.] Not implemented.
\end{description}

\dimsixtop\ matches very closely the \top\ and \topsl\ versions of \smeftsim\ and it can also be mapped to the \general\ one. A correspondence with other flavor versions of \smeftsim\ can only be established partially, due to incompatibilities in the assumed flavor structure. 

The mapping between Wilson coefficients defined in \dimsixtop\ and in the \top, \topsl\  versions of \smeftsim\ is provided in Tables~\ref{tab.cfr_dim6top},~\ref{tab.cfr_dim6top_rotation}. The mapping to the \general\ version of \smeftsim\ is provided in Tables~\ref{tab.cfr_dim6top_general},~\ref{tab.cfr_dim6top_general_rotation}.
In both cases, the first table contains parameters with a one-to-one correspondence, while the second contains parameters that require a basis rotation. 
For example, the point  {\tt cQlM1=1, cQl31=3} in \dimsixtop\ corresponds to {\tt cQl111=4, cQl311=3} (or {\tt clq1Re1133=4, clq3Re1133=3}) in \smeftsim.

Overall minus signs in the mapping are due to the fact that \dimsixtop\ and \smeftsim\ use opposite sign conventions for the definition of covariant derivatives. Although the operator definitions are identical, the \emph{relative} sign between the $\Lag_6$ contribution and the corresponding SM coupling is flipped in a few cases. The physics results are identical in both models once this is accounted for. The presence of explicit Yukawa couplings in the Tables is due to the different treatment of flavor symmetry breaking terms.

Wilson coefficients inducing flavor-changing neutral currents can be mapped to parameters in \smeftsim\ \general, and the corresponding tables are available upon request.

\subsection{\smeftatnlo}
\smeftatnlo~\cite{smeftatnlo,Degrande:2020evl} is equipped for NLO QCD calculations in \madgraph. Here we compare specifically to {\tt SMEFTatNLO v1.0} published in September 2020.
\begin{description}
\item[Flavor structure.] \smeftatnlo\ assumes a flavor symmetry $U(2)_q\times U(3)_d\times U(2)_u$ in the quark sector and $U(1)^3$ in the lepton sector, which is the same as in \smeftsim\ \top\ and in \dimsixtop, except for the treatment of down quarks.

All fermion masses and Yukawa couplings are neglected, except those of the top quark.

\item[Operators implemented.] \smeftatnlo\ contains all the operators in classes (1)-(7) and those in class (8) that contain a top quark. Terms that violate the flavor symmetry have been consistently dropped. CP violating terms are omitted.

\item[Input parameters.] \smeftatnlo\ implements the \mwscheme\ input scheme.

\item[SM loop-generated Higgs couplings.] Higgs couplings in the $m_t\to\infty$ limit are not implemented, but Higgs-gluon interactions can be fully reproduced at 1-loop in QCD.
\end{description}

Given its flavor structure, \smeftatnlo\ can be directly mapped to \smeftsim\ in the \top, \topsl\ and \general\ versions.
The mapping of Wilson coefficients between \smeftatnlo\ and the \top, \topsl\ versions of \smeftsim\ is provided in Tables~\ref{tab.cfr_SMEFTatNLO},~\ref{tab.cfr_SMEFTatNLO_rotation}.
The mapping to the \general\ version is provided in Tables~\ref{tab.cfr_SMEFTatNLO_general},~\ref{tab.cfr_SMEFTatNLO_general_rotation}. In both cases, the first table contains the mapping of parameters with a one-to-one correspondence, while the second contains parameters that require a basis rotation. 

As for \dimsixtop, the sign convention used in \smeftatnlo\ is the opposite compared to \smeftsim, which leads to some minus signs in the conversion.

\begin{table}[h!]\centering 
\renewcommand{\arraystretch}{1.2}
\hspace*{-1.5cm}
\scalebox{.97}{
\begin{tabular}{|c|*3{>{\tt}l>{\tt}l|}}
\toprule\rowcolor{tablesColor}
\multicolumn{7}{|c|}{\smeftsim\ \top\ vs \dimsixtop}\\
\toprule
class&
\multicolumn{2}{c|}{\smeftsim\ $\leftrightarrow$ \dimsixtop}&
\multicolumn{2}{c|}{\smeftsim\ $\leftrightarrow$ \dimsixtop}&
\multicolumn{2}{c|}{\smeftsim\ $\leftrightarrow$ \dimsixtop}\\

\midrule

$\Lag_6^{(5)}$&
ctHRe & ctp&
ctHIm & ctpI& 
&
\\
\midrule\multirow{2}{*}{
$\Lag_6^{(6)}$}&
ctGRe & - ctG &
ctGIm & - ctGI&
& 
\\
&
cbWRe & - cbW&
cbWIm & - cbWI&
&
 \\
\midrule\multirow{2}{*}{
$\Lag_6^{(7)}$}& 
cHt& cpt
&cHbq & cpb
& &
\\
&
cHtbRe & cptb & 
cHtbIm & cptbI&
& 
\\
\midrule\multirow{2}{*}{
$\Lag_6^{(8a)}$}&
cQQ1 & $\frac12$ cQQ1 &
cQj11 & cQq11&
cQj31 & cQq13
\\
&
cQQ8 & $\frac12$ cQQ8 &
cQj18 & cQq81&
cQj38 & cQq83
\\
\midrule\multirow{3}{*}{
$\Lag_6^{(8b)}$}&
ctt & ctt1&
ctu1&  ctu1&
ctu8 & ctu8\\
&
ctd1 & ctd1&
ctb8 & ctb8&
ctd8 & ctd8
\\
&
ctb1& ctb1 &
cte[pp] & cte[p]&
  &
\\
\midrule\multirow{4}{*}{
$\Lag_6^{(8c)}$}&
cQu1 & cQu1&
ctj1& ctq1&
cQt1& cQt1
\\
&
cQu8 & cQu8&
ctj8& ctq8&
cQt8& cQt8
\\
&
cQd1 & cQd1&
cQb1& cQb1  &
cQb8& cQb8
\\
&
cQd8& cQd8&
cQe[pp]& cQe[p]&
ctl[pp]& ctl[p]
\\
\midrule\multirow{6}{*}{
$\Lag_6^{(8d)}$}&
clebQRe[pp]& cblS[p]& 
cleQt1Re[pp]& ctlS[p]&
cleQt3Re[pp]& ctlT[p]
\\
&
clebQIm[pp]& cblSI[p]&
cleQt1Im[pp]& ctlSI[p]&
cleQt3Im[pp]& ctlTI[p]
\\
&
cQtQb1Re& cQtQb1&
cQtQb8Re& cQtQb8&
&
\\
&
cQtQb1Im& cQtQb1I&
cQtQb8Im& cQtQb8I&
&
\\
\bottomrule

\toprule\rowcolor{tablesColor}
\multicolumn{7}{|c|}{\smeftsim\ \topsl\ vs \dimsixtop}\\
\toprule
$\Lag_6^{(8b)}$&
cte & cte[p]&
&&&
\\
\midrule
$\Lag_6^{(8c)}$&
cQe& cQe[p]&
ctl& ctl[p]
&&
\\
\midrule\multirow{2}{*}{
$\Lag_6^{(8d)}$}&
clebQRe&  yl[p] cblS[p]& 
cleQt1Re& yl[p] ctlS[p]&
cleQt3Re& yl[p] ctlT[p]
\\
&
clebQIm]&  yl[p] cblSI[p]&
cleQt1Im& yl[p] ctlSI[p]&
cleQt3Im&  yl[p] ctlTI[p]
\\
\bottomrule
\end{tabular}}
\caption{Upper panel: conversion table between the SMEFT parameters defined in the \top\ version of \smeftsim\ and in \dimsixtop: parameters with a one-to-one translation. Lower panel: conversion table between  the \topsl\ version of \smeftsim\ and \dimsixtop. Only leptonic coefficients are reported, as the other parameters behave identically to the \top\ case. Lepton flavor indices {\tt p} take values {\tt p = $\{$1,2,3$\}$}, light quark indices {\tt r} take values {\tt r = $\{1,2\}$}. 
}\label{tab.cfr_dim6top}
\end{table}

\newpage\thispagestyle{empty}
\begin{table}[h!]\centering 
\renewcommand{\arraystretch}{1.2}
\vspace*{-1cm}
\hspace*{-1cm}\scalebox{.8}{
\begin{tabular}{|c|*2{>{\tt}l>{\tt}l|}}
\toprule\rowcolor{tablesColor}
\multicolumn{5}{|c|}{\smeftsim\ \top\ vs \dimsixtop}\\
\toprule
class&
\multicolumn{2}{c|}{\smeftsim $\to$ \dimsixtop}&
\multicolumn{2}{c|}{\dimsixtop $\to$ \smeftsim}
\\
\midrule
\multirow{4}{*}{$\Lag_6^{(6)}$}
&- ctWRe &  ctW
&- ctW & ctWRe
\\
&- ctWIm &  ctWI
&- ctWI & ctWIm
\\
& -ctWRe $\cw$ + ctBRe $\sw$  & ctZ
& ctZ/$\sw$ - ctW/$t_\theta$& ctBRe
\\
& -ctWIm $\cw$ + ctBIm $\sw$  & ctZI
& ctZI/$\sw$ - ctWI/$t_\theta$& ctBIm
\\
\midrule
\multirow{2}{*}{$\Lag_6^{(7)}$}
& cHQ3 &       cpQ3
& cpQ3 + cpQM& cHQ1
\\
& cHQ1 - cHQ3 &   cpQM
& cpQ3 & cHQ3
\\
\midrule
\multirow{2}{*}{$\Lag_6^{(8a)}$}
& cQl1[pp] - cQl3[pp]& cQlM[p]
& cQl3[p] + cQlM[p] & cQl1[pp]
\\
& cQl3[pp]& cQl3[p]
& cQl3[p]& cQl3[pp]
\\
\midrule\multirow{4}{*}{
$\Lag_6^{(8b)}$}
& $\frac13$ cutbd1Re + $\frac49$ cutbd8Re& yu[r] yd[s] cbtud1
& $\frac13$ cbtud1 + $\frac49$ cbtud8& $\frac{1}{\text{\tt yu[r] yd[s]}}$  cutbd1Re
\\
&$\frac13$ cutbd1Im + $\frac49$ cutbd8Im& yu[r] yd[s] cbtud1I
& $\frac13$ cbtud1I + $\frac49$ cbtud8I& $\frac{1}{\text{\tt yu[r] yd[s]}}$  cutbd1Im
\\
&$2$ cutbd1Re - $\frac13$ cutbd8Re& yu[r] yd[s] cbtud8
& $2$ cbtud1 - $\frac13$ cbtud8& $\frac{1}{\text{\tt yu[r] yd[s]}}$  cutbd8Re
\\
&$2$ cutbd1Im - $\frac13$ cutbd8Im& yu[r] yd[s] cbtud8I
& $2$ cbtud1I - $\frac13$ cbtud8I& $\frac{1}{\text{\tt yu[r] yd[s]}}$  cutbd8Im
\\
\midrule\multirow{8}{*}{
$\Lag_6^{(8c)}$}
& - $\frac23$ cjQtu1Re - $\frac89$ cjQtu8Re& yu[r] ctQqu1 
& - $\frac16$ ctQqu1 - $\frac29$ ctQqu8 & $\frac{1}{\text{\tt yu[r]}}$ cjQtu1Re
\\
&- $\frac23$ cjQtu1Im - $\frac89$ cjQtu8Im& yu[r] ctQqu1I
&- $\frac16$ ctQqu1I - $\frac29$ ctQqu8I & $\frac{1}{\text{\tt yu[r]}}$ cjQtu1Im
\\
& - $4$ cjQtu1Re + $\frac23$ cjQtu8Re& yu[r] ctQqu8
&- ctQqu1 + $\frac16$ ctQqu8 & $\frac{1}{\text{\tt yu[r]}}$ cjQtu8Re
\\
&- $4$ cjQtu1Im + $\frac23$ cjQtu8Im& yu[r] ctQqu8I
&- ctQqu1I + $\frac16$ ctQqu8I & $\frac{1}{\text{\tt yu[r]}}$ cjQtu8Im
\\
&- $\frac23$ cjQbd1Re - $\frac89$ cjQbd8Re& yd[r] cbQqd1
&- $\frac16$ cbQqd1 - $\frac29$ cbQqd8 & $\frac{1}{\text{\tt yd[r]}}$ cjQbd1Re
\\
&- $\frac23$ cjQbd1Im - $\frac89$ cjQbd8Im& yd[r] cbQqd1I
&- $\frac16$ cbQqd1I - $\frac29$ cbQqd8I & $\frac{1}{\text{\tt yd[r]}}$ cjQbd1Im
\\
&- $4$ cjQbd1Re + $\frac23$ cjQbd8Re& yd[r] cbQqd8
&- cbQqd1 + $\frac16$ cbQqd8 & $\frac{1}{\text{\tt yd[r]}}$ cjQbd8Re
\\
& - $4$ cjQbd1Im + $\frac23$ cjQbd8Im& yd[r] cbQqd8I 
&- cbQqd1I + $\frac16$ cbQqd8I& $\frac{1}{\text{\tt yd[r]}}$ cjQbd8Im
\\
\midrule\multirow{16}{*}{
$\Lag_6^{(8d)}$}
&cQtjd1Re - $\frac16$ cjtQd1Re - $\frac29$ cjtQd8Re & yd[r] cQtqd1&
cQtqd1 - $4$ cQtqd1T & $\frac{1}{\text{\tt yd[r]}}$ cQtjd1Re
\\
&cQtjd1Im - $\frac16$ cjtQd1Im - $\frac29$ cjtQd8Im & yd[r] cQtqd1I&
cQtqd1I - $4$ cQtqd1TI & $\frac{1}{\text{\tt yd[r]}}$ cQtjd1Im
\\
&cQtjd8Re - cjtQd1Re + $\frac16$ cjtQd8Re & yd[r] cQtqd8&
cQtqd8 - $4$ cQtqd8T& $\frac{1}{\text{\tt yd[r]}}$ cQtjd8Re
\\
&cQtjd8Im - cjtQd1Im + $\frac16$ cjtQd8Im & yd[r] cQtqd8I&
cQtqd8I - $4$ cQtqd8TI& $\frac{1}{\text{\tt yd[r]}}$ cQtjd8Im
\\

& - $\frac{1}{24}$ cjtQd1Re - $\frac{1}{18}$ cjtQd8Re & yd[r] cQtqd1T &
- $\frac83$ cQtqd1T - $\frac{32}{9}$ cQtqd8T& $\frac{1}{\text{\tt yd[r]}}$ cjtQd1Re
\\
& - $\frac{1}{24}$ cjtQd1Im - $\frac{1}{18}$ cjtQd8Im & yd[r] cQtqd1TI &
- $\frac83$ cQtqd1TI - $\frac{32}{9}$ cQtqd8TI& $\frac{1}{\text{\tt yd[r]}}$ cjtQd1Im  
\\
& - $\frac{1}{4}$ cjtQd1Re + $\frac{1}{24}$ cjtQd8Re & yd[r] cQtqd8T &
- $16$ cQtqd1T + $\frac83$ cQtqd8T& $\frac{1}{\text{\tt yd[r]}}$ cjtQd8Re  
\\
& - $\frac{1}{4}$ cjtQd1Im + $\frac{1}{24}$ cjtQd8Im & yd[r] cQtqd8TI &
- $16$ cQtqd1TI + $\frac83$ cQtqd8TI& $\frac{1}{\text{\tt yd[r]}}$ cjtQd8Im  
\\

&cjuQb1Re - $\frac16$ cQujb1Re - $\frac29$ cQujb8Re  & yu[r] cQbqu1&
cQbqu1 - $4$ cQbqu1T & $\frac{1}{\text{\tt yu[r]}}$ cjuQb1Re
\\
&cjuQb1Im - $\frac16$ cQujb1Im - $\frac29$ cQujb8Im & yu[r] cQbqu8&
cQbqu1I - $4$ cQbqu1TI & $\frac{1}{\text{\tt yu[r]}}$ cjuQb1Im
\\
&cjuQb8Re - cQujb1Re + $\frac16$ cQujb8Re  & yu[r] cQbqu1I&
cQbqu8 - $4$ cQbqu8T & $\frac{1}{\text{\tt yu[r]}}$ cjuQb8Re
\\
&cjuQb8Im - cQujb1Im + $\frac16$ cQujb8Im  & yu[r] cQbqu8I&
cQbqu8I - $4$ cQbqu8TI & $\frac{1}{\text{\tt yu[r]}}$ cjuQb8Im
\\

& - $\frac{1}{24}$ cQujb1Re - $\frac{1}{18}$ cQujb8Re  & yu[r] cQbqu1T &
- $\frac83$ cQbqu1T - $\frac{32}{9}$ cQbqu8T& $\frac{1}{\text{\tt yu[r]}}$ cQujb1Re 
\\
& - $\frac{1}{24}$ cQujb1Im - $\frac{1}{18}$ cQujb8Im  & yu[r] cQbqu1TI &
- $\frac83$ cQbqu1TI - $\frac{32}{9}$ cQbqu8TI& $\frac{1}{\text{\tt yu[r]}}$ cQujb1Im  
\\
& - $\frac{1}{4}$ cQujb1Re + $\frac{1}{24}$ cQujb8Re & yu[r] cQbqu8T &
- $16$ cQbqu1T + $\frac83$ cQbqu8T & $\frac{1}{\text{\tt yu[r]}}$ cQujb8Re  
\\
& - $\frac{1}{4}$ cQujb1Im + $\frac{1}{24}$ cQujb8Im & yu[r] cQbqu8TI &
- $16$ cQbqu1TI + $\frac83$ cQbqu8TI& $\frac{1}{\text{\tt yu[r]}}$ cQujb8Im
\\
\bottomrule

\toprule\rowcolor{tablesColor}
\multicolumn{5}{|c|}{\smeftsim\ \topsl\ vs \dimsixtop}\\
\toprule
\multirow{2}{*}{$\Lag_6^{(8a)}$}
& cQl1 - cQl3& cQlM[p]
& cQl3[p] + cQlM[p] & cQl1
\\
& cQl3& cQl3[p]
& cQl3[p]& cQl3
\\
\bottomrule
\end{tabular}}
\caption{Upper panel: conversion table between the SMEFT parameters defined in the \top\ version of \smeftsim\ and in \dimsixtop: parameters that require a basis rotation.  Lower panel: conversion table between  the \topsl\ version of \smeftsim\ and \dimsixtop. Only leptonic coefficients are reported here, as the other parameters behave identically to the \top\ case. Lepton flavor indices take values {\tt p = $\{$1,2,3$\}$}. In the notation {\tt yd[r], yu[r]} etc, the flavor index is that of the associated righthanded light field ($d$ for {\tt yd} and $u$ for {\tt yu}).  }\label{tab.cfr_dim6top_rotation}
\end{table}

\begin{table}[h!]\centering 
\hspace*{-5mm}
\renewcommand{\arraystretch}{1.3}
\begin{tabular}{|c|*3{>{\tt}l>{\tt}l|}}
\toprule\rowcolor{tablesColor}
\multicolumn{7}{|c|}{\smeftsim\ \general\ vs \dimsixtop}\\
\toprule
class&
\multicolumn{2}{c|}{\smeftsim\ $\leftrightarrow$ \dimsixtop}&
\multicolumn{2}{c|}{\smeftsim\ $\leftrightarrow$ \dimsixtop}&
\multicolumn{2}{c|}{\smeftsim\ $\leftrightarrow$ \dimsixtop}\\

\midrule
$\Lag_6^{(5)}$&
cuHRe33 & ctp&
cuHIm33 & ctpI& 
&
\\
\midrule\multirow{2}{*}{
$\Lag_6^{(6)}$}&
cuGRe33 & - ctG &
cuGIm33 & - ctGI&
& 
\\
&
cdWRe33 & - cbW&
cdWIm33 & - cbWI&
&
 \\
\midrule\multirow{2}{*}{
$\Lag_6^{(7)}$}& 
cHuRe33& cpt
& cHdRe33 & cpb
& &
\\
&
cHudRe33 & cptb & 
cHudIm33 & cptbI&
& 
\\
\midrule\multirow{2}{*}{
$\Lag_6^{(8b)}$}&
ceu[pp]33 & cte[p]&
cuuRe3333 & ctt1
& cud1Re33[rr] & ctd1
\\
& cud1Re3333 & ctb1
& cud8Re3333 & ctb8
& cud8Re33[rr] & ctd8
\\
\midrule\multirow{4}{*}{
$\Lag_6^{(8c)}$}&
cqu1Re33[rr] & cQu1&
cqu1Re[rr]33& ctq1&
cqu1Re3333& cQt1
\\
&
cqu8Re33[rr] & cQu8&
cqu8Re[rr]33& ctq8&
cqu8Re3333& cQt8
\\
&
cqd1Re33[rr] & cQd1
&cqd1Re3333& cQb1 &
cqd8Re3333& cQb8
\\
&
cqd8Re33[rr]& cQd8 &
cqeRe33[pp]& cQe[p]&
cluRe[pp]33& ctl[p]
\\
\midrule\multirow{6}{*}{
$\Lag_6^{(8d)}$}&
cledqRe[pp]33& cblS[p]& 
clequ1Re[pp]33& ctlS[p]&
clequ3Re[pp]33& ctlT[p]
\\
&
cledqIm[pp]33& cblSI[p]&
clequ1Im[pp]33& ctlSI[p]&
clequ3Im[pp]33& ctlTI[p]
\\
&
cquqd1Re3333& cQtQb1&
cquqd1Im3333& cQtQb1I&
&
\\
&
cquqd8Re3333& cQtQb8&
cquqd8Im3333& cQtQb8I&
&
\\
%
\bottomrule
\end{tabular}
\caption{Conversion table between the SMEFT parameters defined in the \general\ version of \smeftsim\ and in \dimsixtop: parameters with one-to-one conversion. Lepton flavor indices {\tt p} take values in {\tt $\{$1,2,3$\}$}. Quark flavor indices {\tt r} take values in {\tt $\{$1,2$\}$}.}\label{tab.cfr_dim6top_general}
\end{table}

\clearpage
\thispagestyle{empty}
\begin{table}[h!] 
\renewcommand{\arraystretch}{1.15}
\vspace*{-1cm}
\hspace*{-2cm}\scalebox{.8}{\begin{tabular}{|c|*2{>{\tt}l>{\tt}l|}}
\toprule\rowcolor{tablesColor}
\multicolumn{5}{|c|}{\smeftsim\ \general\ vs \dimsixtop}\\
\toprule
class&
\multicolumn{2}{c|}{\smeftsim\ $\to$ \dimsixtop}&
\multicolumn{2}{c|}{\dimsixtop\ $\to$ \smeftsim}
\\
\midrule
\multirow{4}{*}{$\Lag_6^{(6)}$}
& - cuWRe33 & ctW
& - ctW & cuWRe33
\\
& - cuWIm33 & ctWI
& - ctWI & cuWIm33
\\
& - cuWRe33 $\cw$ + cuBRe33 $\sw$  & ctZ
& ctZ$/\sw$ - ctW$/t_\theta$& cuBRe33
\\
& - cuWIm33 $\cw$ + cuBIm33 $\sw$  & ctZI
& ctZI$/\sw$ - ctWI$/t_\theta$& cuBIm33
\\                      
\midrule
\multirow{2}{*}{$\Lag_6^{(7)}$}
& cHq3Re33 &       cpQ3
& cpQ3 + cpQM& cHq1Re33
\\
& cHq1Re33 - cHq3Re33 &   cpQM
& cpQ3 & cHq3Re33
\\
\midrule
\multirow{8}{*}{$\Lag_6^{(8a)}$}
& clq1Re[pp]33 - clq3Re[pp]33& cQlM[p]
& cQl3[p] + cQlM[p] & clq1Re[pp]33
\\
& clq3[pp]33& cQl3[p]
& cQl3[p]& clq3Re[pp]33
\\
& $2$ cqq1Re3333 - $\frac23$ cqq3Re3333& cQQ1
& $\frac12$ cQQ1 + $\frac{1}{24}$ cQQ8 & cqq1Re3333
\\
& $8$ cqq3Re3333 & cQQ8
& $\frac18$ cQQ8   & cqq3Re3333
\\
& $\frac13$ cqq1Re[r]33[r] + $2$ cqq1Re[rr]33 + cqq3Re[r]33[r]& cQq11
& $\frac12$ cQq11 - $\frac{1}{12}$ cQq81  & cqq1Re[rr]33
\\
& $\frac13$ (cqq1Re[r]33[r] - cqq3Re[r]33[r]) + $2$ cqq3Re[rr]33& cQq13
& $\frac18$ cQq81 + $\frac38$ cQq83  & cqq1Re[r]33[r]
\\
& $2$ cqq1Re[r]33[r] + $6$ cqq3Re[r]33[r]& cQq81
& $\frac12$ cQq13 - $\frac{1}{12}$ cQq83 & cqq3Re[rr]33
\\
& $2$ (cqq1Re[r]33[r] - cqq3Re[r]33[r])& cQq83
& $\frac18$ (cQq81 - cQq83) & cqq3Re[r]33[r]
\\
\midrule
& $\frac23$ cuuRe[r]33[r] + $2$ cuuRe[rr]33 & ctu1
& $\frac12$ ctu1 - $\frac{1}{12}$ ctu8 & cuuRe[rr]33
\\
& $4$ cuuRe[r]33[r]& ctu8
& $\frac14$ ctu8 & cuuRe[r]33[r]
\\
& $\frac13$ cud1Re[r]33[r] + $\frac49$ cud8Re[r]33[r]   &cbtud1  
& $\frac13$ cbtud1 + $\frac49$ cbtud8& cud1Re[r]33[r]
\\
& $\frac13$ cud1Im[r]33[r] + $\frac49$ cud8Im[r]33[r]   &cbtud1I 
& $\frac13$ cbtud1I + $\frac49$ cbtud8I& cud1Im[r]33[r]
\\
& $2$ cud1Re[r]33[r] - $\frac13$ cud8Re[r]33[r]  &cbtud8  
& $2$ cbtud1 - $\frac13$ cbtud8& cud8Re[r]33[r]
\\
& $2$ cud1Im[r]33[r] - $\frac13$ cud8Im[r]33[r]  &cbtud8I 
& $2$ cbtud1I - $\frac13$ cbtud8I& cud8Im[r]33[r]
\\
\midrule\multirow{8}{*}{
$\Lag_6^{(8c)}$}
& - $\frac23$ cqu1Re[r]33[r] - $\frac89$ cqu8Re[r]33[r]    &ctQqu1  
& - $\frac16$ ctQqu1 - $\frac29$ ctQqu8& cqu1Re[r]33[r]
\\
& - $\frac23$ cqu1Im[r]33[r] - $\frac89$ cqu8Im[r]33[r]    &ctQqu1I 
& - $\frac16$ ctQqu1I - $\frac29$ ctQqu8I& cqu1Im[r]33[r]
\\
& - $4$ cqu1Re[r]33[r] + $\frac23$ cqu8Re[r]33[r]  &ctQqu8  
& - ctQqu1 + $\frac16$ ctQqu8 & cqu8Re[r]33[r]
\\        
& - $4$ cqu1Im[r]33[r] + $\frac23$ cqu8Im[r]33[r]   &ctQqu8I 
& - ctQqu1I + $\frac16$ ctQqu8I & cqu8Im[r]33[r]
\\  
& - $\frac23$ cqd1Re[r]33[r] - $\frac89$ cqd8Re[r]33[r]    &cbQqd1  
& - $\frac16$ cbQqd1 - $\frac29$ cbQqd8& cqd1Re[r]33[r]
\\
& - $\frac23$ cqd1Im[r]33[r] - $\frac89$ cqd8Im[r]33[r]    &cbQqd1I 
& - $\frac16$ cbQqd1I - $\frac29$ cbQqd8I& cqd1Im[r]33[r]
\\
& - $4$ cqd1Re[r]33[r] + $\frac23$ cqd8Re[r]33[r]    &cbQqd8  
& - cbQqd1 + $\frac16$ cbQqd8 & cqd8Re[r]33[r]
\\  
& - $4$ cqd1Im[r]33[r] + $\frac23$ cqd8Im[r]33[r]    &cbQqd8I 
& - cbQqd1I + $\frac16$ cbQqd8I & cqd8Im[r]33[r]
\\  
\midrule\multirow{16}{*}{
$\Lag_6^{(8d)}$}
&cquqd1Re33[rr] - $\frac16$ cquqd1Re[r]33[r] - $\frac29$ cquqd8Re[r]33[r] & cQtqd1&
cQtqd1 - $4$ cQtqd1T &  cquqd1Re33[rr]
\\
&cquqd1Im33[rr] - $\frac16$ cquqd1Im[r]33[r] - $\frac29$ cquqd8Im[r]33[r] & cQtqd1I&
cQtqd1I - $4$ cQtqd1TI &  cquqd1Im33[rr]
\\
&cquqd8Re33[rr] - cquqd1Re[r]33[r] + $\frac16$ cquqd8Re[r]33[r] & cQtqd8&
cQtqd8 - $4$ cQtqd8T&  cquqd8Re33[rr]
\\
&cquqd8Im33[rr] - cquqd1Im[r]33[r] + $\frac16$ cquqd8Im[r]33[r] & cQtqd8I&
cQtqd8I - $4$ cQtqd8TI&  cquqd8Im33[rr]
\\
& - $\frac{1}{24}$ cquqd1Re[r]33[r] - $\frac{1}{18}$ cquqd8Re[r]33[r] & cQtqd1T &
- $\frac83$ cQtqd1T - $\frac{32}{9}$ cQtqd8T& cquqd1Re[r]33[r]
\\
& - $\frac{1}{24}$ cquqd1Im[r]33[r] - $\frac{1}{18}$ cquqd8Im[r]33[r] &  cQtqd1TI &
- $\frac83$ cQtqd1TI - $\frac{32}{9}$ cQtqd8TI&  cquqd1Im[r]33[r]  
\\
& - $\frac{1}{4}$ cquqd1Re[r]33[r] + $\frac{1}{24}$ cquqd8Re[r]33[r] & cQtqd8T &
- $16$ cQtqd1T + $\frac83$ cQtqd8T& cquqd8Re[r]33[r]  
\\
& - $\frac{1}{4}$ cquqd1Im[r]33[r] + $\frac{1}{24}$ cquqd8Im[r]33[r] & cQtqd8TI &
- $16$ cQtqd1TI + $\frac83$ cQtqd8TI&  cquqd8Im[r]33[r]  
\\

&cquqd1Re[rr]33 - $\frac16$ cquqd1Re3[rr]3 - $\frac29$ cquqd8Re3[rr]3  & cQbqu1&
cQbqu1 - $4$ cQbqu1T &  cquqd1Re[rr]33
\\
&cquqd1Im[rr]33 - $\frac16$ cquqd1Im3[rr]3 - $\frac29$ cquqd8Im3[rr]3 & cQbqu8&
cQbqu1I - $4$ cQbqu1TI &  cquqd1Im[rr]33
\\
&cquqd8Re[rr]33 - cquqd1Re3[rr]3 + $\frac16$ cquqd8Re3[rr]3  & cQbqu1I&
cQbqu8 - $4$ cQbqu8T &  cquqd8Re[rr]33
\\
&cquqd8Im[rr]33 - cquqd1Im3[rr]3 + $\frac16$ cquqd8Im3[rr]3  &  cQbqu8I&
cQbqu8I - $4$ cQbqu8TI & cquqd8Im[rr]33
\\

& - $\frac{1}{24}$ cquqd1Re3[rr]3 - $\frac{1}{18}$ cquqd8Re3[rr]3  &  cQbqu1T &
- $\frac83$ cQbqu1T - $\frac{32}{9}$ cQbqu8T&  cquqd1Re3[rr]3 
\\
& - $\frac{1}{24}$ cquqd1Im3[rr]3 - $\frac{1}{18}$ cquqd8Im3[rr]3  &  cQbqu1TI &
- $\frac83$ cQbqu1TI - $\frac{32}{9}$ cQbqu8TI&  cquqd1Im3[rr]3  
\\
& - $\frac{1}{4}$ cquqd1Re3[rr]3 + $\frac{1}{24}$ cquqd8Re3[rr]3 & cQbqu8T &
- $16$ cQbqu1T + $\frac83$ cQbqu8T &  cquqd8Re3[rr]3  
\\
& - $\frac{1}{4}$ cquqd1Im3[rr]3 + $\frac{1}{24}$ cquqd8Im3[rr]3 & cQbqu8TI &
- $16$ cQbqu1TI + $\frac83$ cQbqu8TI&  cquqd8Im3[rr]3
\\
\bottomrule
\end{tabular}}
\caption{Conversion table between the SMEFT parameters defined in the \general\ version of \smeftsim\ and in \dimsixtop: parameters that require a basis rotation. Lepton flavor indices {\tt p} take values in {\tt $\{$1,2,3$\}$}. Quark flavor indices {\tt r} take values in {\tt $\{$1,2$\}$}.  }\label{tab.cfr_dim6top_general_rotation}
\end{table}

\begin{table}\centering 
\renewcommand{\arraystretch}{1.2}
\hspace*{-1cm}~\scalebox{.96}{\begin{tabular}{|p{1cm}|*3{>{\tt}p{2.5cm}>{\tt}p{2.25cm}|}}
\toprule\rowcolor{tablesColor}
\multicolumn{7}{|c|}{\smeftsim\ \top\ vs \smeftatnlo}\\
\toprule
class&
\multicolumn{2}{c|}{\smeftsim\ $\leftrightarrow$ \smeftatnlo}&
\multicolumn{2}{c|}{\smeftsim\ $\leftrightarrow$ \smeftatnlo}&
\multicolumn{2}{c|}{\smeftsim\ $\leftrightarrow$ \smeftatnlo}\\
\midrule\multirow{1}{*}{
$\Lag_6^{(1)}$}
& cG & - gs cG
& cW & - cWWW
& &
\\
\midrule\multirow{1}{*}{
$\Lag_6^{(2,3)}$}
& cH&      cp
& cHbox&  cdp
& cHDD &  cpDC
\\
\midrule\multirow{2}{*}{
$\Lag_6^{(4)}$}
& cHG & cpG
& cHW & cpW
& cHB & cpBB
\\
& cHWB & cpWB
& & 
& & 
\\
\midrule\multirow{1}{*}{
$\Lag_6^{(5)}$}
& ctHRe & ctp
& & 
& &
\\
\midrule\multirow{1}{*}{
$\Lag_6^{(6)}$}
& ctGRe & -  gs ctG
&  & 
&  & 
\\
\midrule\multirow{4}{*}{
$\Lag_6^{(7)}$}
& cHl1[pp]& cpl[p]
& cHl3[pp]& c3pl[p]
& cHd = cHbq& cpd 
\\
& cHe11 &  cpe
& cHe22 & cpmu
& cHe33 & cpta
\\
& cHu& cpu
& cHt& cpt
& &
\\
\midrule\multirow{3}{*}{
$\Lag_6^{(8a)}$}
& cQj11& cQq11
& cQj18& cQq81
& cQj31& cQq13
\\       
& cQj38& cQq83
& cQQ1&  $\frac12$ cQQ1
& cQQ8&  $\frac12$ cQQ8
\\
& cll[pppp] & cll[pppp]
& cll[pprr] & $2$ cll[pprr]
& cll[prrp] & $2$ cll[prrp]
\\
\midrule\multirow{3}{*}{
$\Lag_6^{(8b)}$}
& cte[pp]& cte[p]
& ctu1& ctu1
& ctu8& ctu8
\\
& ctt& ctt1 
& ctd1 = ctb1& ctd1
& ctd8 = ctb8& ctd8
\\
\midrule\multirow{4}{*}{
$\Lag_6^{(8c)}$}
& ctl[pp]& ctl[p]
& cQe[pp]& cQe[p]
& &
\\
& cQt1 & cQt1
& cQt8 & cQt8
& & 
\\
& cQu1& cQu1
& cQd1 = cQb1& cQd1
& ctj1& ctq1
\\
& cQu8& cQu8
& cQd8 = cQb8& cQd8
& ctj8& ctq8
\\
\midrule\multirow{1}{*}{
$\Lag_6^{(8d)}$}
& cleQt1Re33& ctlS3
& cleQt3Re33& ctlT3
& clebQRe33& cblS3
\\
\bottomrule
\end{tabular}}
\\
\hspace*{-1cm}\scalebox{.94}{
\begin{tabular}{|p{1cm}|>{\tt}p{2cm}>{\tt}p{4.2cm}|>{\tt}p{2cm}>{\tt}p{2.2cm}|>{\tt}p{2cm}>{\tt}p{2.2cm}|}
\toprule\rowcolor{tablesColor}
\multicolumn{7}{|c|}{\smeftsim\ \topsl\ vs \smeftatnlo}\\
\toprule
class&
\multicolumn{2}{c|}{\smeftsim\ $\leftrightarrow$ \smeftatnlo}&
\multicolumn{2}{c|}{\smeftsim\ $\leftrightarrow$ \smeftatnlo}&
\multicolumn{2}{c|}{\smeftsim\ $\leftrightarrow$ \smeftatnlo}\\
\midrule\multirow{1}{*}{
$\Lag_6^{(7)}$}
& cHe &  cpe = cpmu = cpta
& cHl1& cpl[p]
& cHl3& c3pl[p]
\\
\midrule\multirow{2}{*}{
$\Lag_6^{(8a)}$}
& cll & cll[pppp] = cll[pprr]
&&
&&
\\
& cll1 & cll[pppp] = cll[prrp]
&&
&&
\\
\midrule\multirow{1}{*}{
$\Lag_6^{(8b)}$}
& cte& cte[p]
&&
&&
\\
\midrule\multirow{1}{*}{
$\Lag_6^{(8c)}$}
& cQe& cQe[p]
& ctl& ctl[p]
&&
\\
\midrule\multirow{1}{*}{
$\Lag_6^{(8d)}$}
& cleQt1Re33& yl[3] ctlS3
& cleQt3Re33& yl[3] ctlT3
& clebQRe33& yl[3] cblS3
\\
\bottomrule
\end{tabular}}
\caption{
Upper panel: conversion table between the SMEFT parameters defined in the \top\ version of \smeftsim\ and in \smeftatnlo\ {\tt v1.0}, for parameters that have a one-to-one translation.  Lower panel: conversion table between  the \topsl\ version of \smeftsim\ and \smeftatnlo. Only leptonic coefficients are reported, as the other parameters behave identically to the \top\ case.
The {\tt =} sign indicates that all parameters need to be fixed to the same value. Lepton flavor indices {\tt p} take values {\tt p,r = $\{$1,2,3$\}$}. 
}
\label{tab.cfr_SMEFTatNLO}
\end{table}

\begin{table}[h!]\centering 
\renewcommand{\arraystretch}{1.3}
\begin{tabular}{|c|*2{>{\tt}l>{\tt}l|}}
\toprule\rowcolor{tablesColor}
\multicolumn{5}{|c|}{\smeftsim\ \top\ vs \smeftatnlo}\\
\toprule
class&
\multicolumn{2}{c|}{\smeftsim $\to$ \smeftatnlo}&
\multicolumn{2}{c|}{\smeftatnlo $\to$ \smeftsim}
\\
\midrule\multirow{2}{*}{
$\Lag_6^{(6)}$}
&- ctWRe &  ctW
&- ctW & ctWRe
\\
& -ctWRe $\cw$ + ctBRe $\sw$  & ctZ
& ctZ/$\sw$ - ctW/$t_\theta$& ctBRe
\\
\midrule
\multirow{4}{*}{$\Lag_6^{(7)}$}
& cHQ3 &       cpQ3
& cpQ3 + cpQM& cHQ1
\\
& cHQ1 - cHQ3 &   cpQM
& cpQ3 & cHQ3
\\
& cHj3 &       cpq3i
& cpq3i + cpqMi& cHj1
\\
& cHj1 - cHj3 &   cpqMi
& cpq3i & cHj3
\\
\midrule
\multirow{2}{*}{
$\Lag_6^{(8a)}$}
& cQl1[pp] - cQl3[pp]& cQlM[p]
& cQl3[p] + cQlM[p] & cQl1[pp]
\\
& cQl3[pp]& cQl3[p]
& cQl3[p]& cQl3[pp]
\\
\bottomrule

\toprule\rowcolor{tablesColor}
\multicolumn{5}{|c|}{\smeftsim\ \topsl\ vs \smeftatnlo}\\
\toprule
\multirow{2}{*}{
$\Lag_6^{(8a)}$}
& cQl1 - cQl3& cQlM[p]
& cQl3[p] + cQlM[p] & cQl1
\\
& cQl3& cQl3[p]
& cQl3[p]& cQl3
\\
\bottomrule
\end{tabular}
\caption{Upper panel: conversion table between the SMEFT parameters defined in the \top\ version of \smeftsim\ and in \smeftatnlo\ {\tt v1.0}: parameters that require a basis rotation.  Lower panel: conversion table between  the \topsl\ version of \smeftsim\ and \smeftatnlo. Only leptonic coefficients are reported here, as the other parameters behave identically to the \top\ case. Lepton flavor indices take values {\tt p = $\{$1,2,3$\}$}.   }\label{tab.cfr_SMEFTatNLO_rotation}
\end{table}

\begin{table}\centering 
\renewcommand{\arraystretch}{1.3}
\hspace*{-1.5cm}
\scalebox{.9}{
\begin{tabular}{|c|>{\tt}p{3cm}>{\tt}p{2cm}|>{\tt}p{3cm}>{\tt}p{2cm}|>{\tt}p{4.9cm}>{\tt}p{1.5cm}|}
\toprule\rowcolor{tablesColor}
\multicolumn{7}{|c|}{\smeftsim\ \general\ vs \smeftatnlo}\\\toprule
class&
\multicolumn{2}{c|}{\smeftsim\ $\leftrightarrow$ \smeftatnlo}&
\multicolumn{2}{c|}{\smeftsim\ $\leftrightarrow$ \smeftatnlo}&
\multicolumn{2}{c|}{\smeftsim\ $\leftrightarrow$ \smeftatnlo}\\
\midrule
$\Lag_6^{(1)}$ 
& cG & - gs cG
& cW & - cWWW
& &
\\
\midrule
$\Lag_6^{(2,3)}$
& cH&      cp
& cHbox&  cdp
& cHDD &  cpDC
\\
\midrule\multirow{2}{*}{
$\Lag_6^{(4)}$}
& cHG & cpG
& cHW & cpW
& cHB & cpBB
\\
& cHWB & cpWB
& & 
& & 
\\
\midrule
$\Lag_6^{(5)}$
& cuHRe33 & ctp
& & 
& &
\\
\midrule
$\Lag_6^{(6)}$
& cuGRe33 & - gs ctG
&& 
&&
\\
\midrule\multirow{4}{*}{
$\Lag_6^{(7)}$}
& cHl1Re[pp]& cpl[p]
& cHl3Re[pp]& c3pl[p]
& cHdRe[rr]=cHdRe33& cpd 
\\
& cHeRe11 &  cpe
& cHe2Re2 & cpmu
& cHeRe33 & cpta
\\
& cHuRe[rr]& cpu
& cHuRe33& cpt
&&
\\
\midrule
$\Lag_6^{(8a)}$
& cllRe[prst] & cll[prst]
& & 
& &
\\
\midrule\multirow{2}{*}{
$\Lag_6^{(8b)}$}
& ceuRe[pp]33& cte[p]
& cuuRe3333& ctt1
& cud1Re33[rr]=cud1Re3333& ctd1
\\
&& 
&&
& cud8Re33[rr]=cud8Re3333& ctd8
\\
\midrule\multirow{4}{*}{
$\Lag_6^{(8c)}$}
& cluRe[pp]33& ctl[p]
& cqeRe33[pp]& cQe[p]
&&
\\
& cqu1Re3333 & cQt1
& cqu8Re3333 & cQt8
&&
\\
& cqu1Re[rr]33& ctq1
& cqu1Re33[rr]& cQu1
& cqd1Re33[rr]=cqd1Re3333& cQd1
\\
& cqu8Re[rr]33& ctq8
& cqu8Re33[rr]& cQu8
& cqd8Re33[rr]=cqd8Re3333& cQd8
\\
\midrule $\Lag_6^{(8d)}$
&clequ1Re3333& ctlS3
&clequ3Re3333& ctlT3
&cledqRe3333& cblS3
\\
\bottomrule
\end{tabular}}
\caption{Conversion table between the SMEFT parameters defined in the \general\ version of \smeftsim\ and in \smeftatnlo\ {\tt v1.0}: set of parameters with one-to-one conversion. Lepton flavor indices {\tt p} take values in {\tt $\{$1,2,3$\}$}. Quark indices {\tt r} take values in {\tt $\{$1,2$\}$}. }
\label{tab.cfr_SMEFTatNLO_general}
\end{table}

\begin{table}[h!] 
\renewcommand{\arraystretch}{1.3}
\hspace*{-1.5cm}\scalebox{.9}{\begin{tabular}{|c|*2{>{\tt}l>{\tt}l|}}
\toprule\rowcolor{tablesColor}
\multicolumn{5}{|c|}{\smeftsim\ \general\ vs \smeftatnlo}\\
\toprule
class&
\multicolumn{2}{c|}{\smeftsim\ $\to$ \smeftatnlo}&
\multicolumn{2}{c|}{\smeftatnlo $\to$ \smeftsim}
\\
\midrule
\multirow{2}{*}{
$\Lag_6^{(6)}$}
& - cuWRe33 & ctW
& - ctW & cuWRe33
\\
& - cuWRe33 $\cw$ + cuBRe33 $\sw$  & ctZ
& ctZ$/\sw$ - ctW$/t_\theta$& cuBRe33
\\                      
\midrule
\multirow{4}{*}{$\Lag_6^{(7)}$}
& cHq3Re33 &       cpQ3
& cpQ3 + cpQM& cHq1Re33
\\
& cHq1Re33 - cHq3Re33 &   cpQM
& cpQ3 & cHq3Re33
\\
& cHq3Re[rr] &       cpq3i
& cpq3i + cpqMi& cHq1Re[rr]
\\
& cHq1Re[rr] - cHq3Re[rr] &   cpqMi
& cpq3i & cHq3Re[rr]
\\
\midrule
\multirow{8}{*}{$\Lag_6^{(8a)}$}
& clq1Re[pp]33 - clq3Re[pp]33& cQlM[p]
& cQl3[p] + cQlM[p] & clq1Re[pp]33
\\
& clq3[pp]33& cQl3[p]
& cQl3[p]& clq3Re[pp]33
\\
& $2$ cqq1Re3333 - $\frac23$ cqq3Re3333& cQQ1
& $\frac12$ cQQ1 + $\frac{1}{24}$ cQQ8 & cqq1Re3333
\\
& $8$ cqq3Re3333 & cQQ8
& $\frac18$ cQQ8   & cqq3Re3333
\\
& $\frac13$ cqq1Re[r]33[r] + $2$ cqq1Re[rr]33 + cqq3Re[r]33[r]& cQq11
& $\frac12$ cQq11 - $\frac{1}{12}$ cQq81  & cqq1Re[rr]33
\\
& $\frac13$ (cqq1Re[r]33[r] - cqq3Re[r]33[r]) + $2$ cqq3Re[rr]33& cQq13
& $\frac18$ cQq81 + $\frac38$ cQq83  & cqq1Re[r]33[r]
\\
& $2$ cqq1Re[r]33[r] + $6$ cqq3Re[r]33[r]& cQq81
& $\frac12$ cQq13 - $\frac{1}{12}$ cQq83 & cqq3Re[rr]33
\\
& $2$ (cqq1Re[r]33[r] - cqq3Re[r]33[r])& cQq83
& $\frac18$ (cQq81 - cQq83) & cqq3Re[r]33[r]
\\
\midrule
\multirow{2}{*}{$\Lag_6^{(8b)}$}
& $\frac23$ cuuRe[r]33[r] + $2$ cuuRe[rr]33 & ctu1
& $\frac12$ ctu1 - $\frac{1}{12}$ ctu8 & cuuRe[rr]33
\\
& $4$ cuuRe[r]33[r]& ctu8
& $\frac14$ ctu8 & cuuRe[r]33[r]
\\
\bottomrule
\end{tabular}}
\caption{Conversion table between the SMEFT parameters defined in the \general\ version of \smeftsim\ and in \smeftatnlo\ {\tt v1.0}: parameters that require a basis rotation. Lepton flavor indices {\tt p} take values in {\tt $\{$1,2,3$\}$}. Quark flavor indices {\tt r} take values in {\tt $\{$1,2$\}$}.}\label{tab.cfr_SMEFTatNLO_general_rotation}
\end{table}

\clearpage
\section{Validation of the UFO models}
\label{app.validation}
\begin{figure}[t]\centering
\includegraphics[width=12cm]{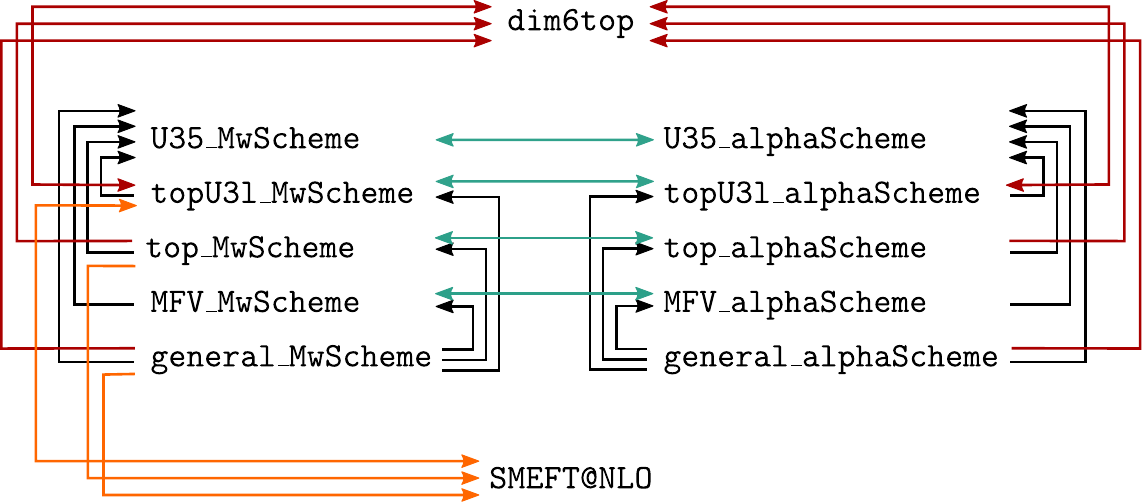}
\caption{Diagram of the pairwise comparisons performed to validate the \ufo\ models. The arrow indicates the direction in which the SMEFT parameters are mapped.}\label{fig.validation_scheme}
\end{figure}

The 10 \ufo\ models contained in the \smeftsim\ package have been validated following the recommendations of Ref.~\cite{Durieux:2019lnv}: the procedure relies on pairwise comparisons between models, based on the values returned for a set of squared amplitudes. Each comparison is performed with the dedicated \madgraph\ plugin~\cite{MG-plugin}: given a list of $2\to n$ processes and of points in parameter space, 
the SM squared amplitude $|A_{\rm SM}|^2$, the pure SM-$\Lag_6$ interference $2\re A_{\rm SM}A_6^*$ and the quadratic $\Lag_6$ contribution $|A_6|^2$ are calculated at one random phase-space point for each process and parameter point.
The validation is considered successful if the squared amplitudes evaluated with each model pair agree within a permille. Larger discrepancies are ignored if they do not show a consistent pattern across different processes and the squared amplitude is $<10^{-16}$ for both models. 

Figure~\ref{fig.validation_scheme} illustrates diagrammatically the set of comparisons performed: the \top, \topsl\ and \general\ versions of \smeftsim\ have been compared to \dimsixtop\ (version of May 2020) and \smeftatnlo\ (both versions of August 2019 and September 2020, only for models with \mwscheme\ scheme). An internal validation was also carried out, comparing models with different flavor assumptions and same input scheme, and vice versa. 
The arrows in the figure indicate that, in the comparison, the parameters of the first model were mapped onto those of the latter: the flow generally goes towards more restrictive flavor assumptions.

The validation was performed on the processes listed in Tab.~\ref{tab.validation_processes}, that were chosen so as to probe most effective operators independently. All Wilson coefficients have been included in the comparison, with the exception of those inducing flavor-changing neutral currents.

\begin{table}[t]\centering 
\hspace*{-1.2cm}
\scalebox{.9}{
\begin{tabular}{|*4{>{\tt}p{3cm}|}>{\tt}p{3.4cm}|>{\tt}p{3cm}|}
\toprule
g g > t t~
&
g g > u u~
&
g g > g g
&
w+ w- > w+ w-
&
a w+ > z w+
&
z z > w+ w-
\\
w+ w- > a a
&
z z > z z
&
h h > w+ w-
&
h h > z z
&
h h > h h
&
h w- > t~ b
\\
h w+ > u d~
&
h w- > mu- vm~
&
h z > b b~
&
h z > s s~
&
h z > t t~
&
h z > u u~
\\
h a > b b~
&
h a > t t~
&
h a > d d~
&
h a > c c~
&
h z > mu+ mu-
&
h z > vt vt~
\\
h a > e+ e-
&
b b~ > t t~
&
b b~ > b b~
&
b b~ > u u~
&
b b~ > d d~
&
b b~ > a a
\\
t t~ > t t~
&
t t~ > c c~
&
t t~ > s s~
&
t t~ > d d~
&
b~ t > d~ u
&
u u~ > u u~
\\
u u~ > c c~
&
s s~ > s s~
&
d d~ > s s~
&
d d~ > c c~
&
u d~ > u d~
&
u d~ > c s~
\\
e+ e- > mu+ mu-
&
ve ve~ > vm vm~
&
vt vt~ > vt vt~
&
e+ ta- > e+ ta-
&
mu+ mu- > ta+ ta-
&
e+ e- > e+ e-
\\
e+ ve > e+ ve
&
e+ vm > e+ vm
&
e+ ve > mu+ vm
&
b t~ > e- ve~
&
u d~ > mu+ vm
&
c s~ > ta+ vt
\\
d d~ > e+ e-
&
c c~ > mu+ mu-
&
b b~ > mu+ mu-
&
t t~ > ta+ ta-
&
b b~ > ve ve~
&
s s~ > vt vt~
\\
t t~ > vm vm~
&
u u~ > ve ve~
&
w+ w- > h d d~
&
w+ w- > h c c~
&
w+ w- > h t t~
&
w+ w- > h b b~
\\
w+ w- > a a a
&
h w+ > a u d~
&
h w- > a e- ve~
&
h w+ > a t b~
&
h h > h u u~
&
h h > h ta+ ta-
\\
h h > h s s~
&
h h > h t t~
&
h h > h b b~
&
h h > z ve ve~
&
h h > z c c~
&
h h > z t t~
\\
h h > z e+ e-
&
h h > z d d~
&
h h > z b~ b
&
h h > w+ e- ve~
&
h h > w- u d~
&
h h > w+ t~ b
\\
h h > w+ w- a
&
g g > g g g
&
h h > h h h
&
h h > h z z
&&
\\\bottomrule
\end{tabular}}
\caption{Set of processes that have been used to validate the \ufo\ models.}\label{tab.validation_processes}
\end{table}

\begin{table}[t]\centering 
\hspace*{-1cm}
\scalebox{.9}{
\begin{tabular}{|*6{>{\tt}p{3cm}|}}
\toprule
w+ w- > w+ w-
&
z z > w+ w-
&
w+ w- > a a
&
h h > w+ w-
&
h h > z z
&
h h > h h
\\
h w+ > u d~
&
h w- > mu- vm~
&
h z > s s~
&
h z > t t~
&
h z > mu+ mu-
&
h z > vt vt~
\\
b b~ > t t~
&
b b~ > b b~
&
t t~ > t t~
&
b~ t > d~ u
&
u u~ > u u~
&
s s~ > s s~
\\
d d~ > c c~
&
u d~ > c s~
&
e+ e- > mu+ mu-
&
ve ve~ > vm vm~
&
e+ ta- > e+ ta-
&
e+ e- > e+ e-
\\
e+ ve > e+ ve
&
e+ vm > e+ vm
&
e+ ve > mu+ vm
&
b t~ > e- ve~
&
c s~ > ta+ vt
&
c c~ > mu+ mu-
\\
b b~ > mu+ mu-
&
b b~ > ve ve~
&
s s~ > vt vt~
&
u u~ > ve ve~
&
t g > b w+
&
w+ w- > h d d~
\\
w+ w- > h c c~
&
w+ w- > h b b~
&
w+ w- > a a a
&
h w+ > a u d~
&
h w- > a e- ve~
&
h w+ > a t b~
\\
h h > h ta+ ta-
&
h h > h b b~
&
h h > z c c~
&
h h > z t t~
&
h h > z b~ b
&
h h > w+ e- ve~
\\
h h > w- u d~
&
h h > w+ w- a
&&&&
\\\bottomrule
\end{tabular}}
\caption{Set of processes that have been used to validate linearized propagator corrections implemented in the \ufo\ models.}\label{tab.validation_processes_prop}
\end{table}

\begin{table}[t]\centering 
\hspace*{-1cm}
\scalebox{.9}{
\begin{tabular}{|*6{>{\tt}p{3cm}|}}
\toprule
g g > t t~
&
g g > u u~
&
g g > g g
&
a w+ > z w+
&
z z > w+ w-
&
w+ w- > a a
\\
h a > b b~
&
h a > d d~
&
h a > c c~
&
h a > e+ e-
&
b b~ > a a
&
t t~ > z a
\\
w+ w- > h d d~
&
w+ w- > h t t~
&
w+ w- > h b b~
&
w+ w- > a a a
&
h h > z c c~
&
h h > z e+ e-
\\
h h > z d d~
&
h h > z b~ b
&
g g > g g g
&
h h > h h h
&&
\\\bottomrule
\end{tabular}}
\caption{Set of processes that have been used to validate loop-induced SM Higgs couplings implemented in the \ufo\ models.}\label{tab.validation_processes_smhloop}
\end{table}

All fermion masses and Yukawa couplings were retained for internal validation, while only those implemented in \dimsixtop\ or \smeftatnlo\ were included when comparing to these models. CKM mixing has been neglected in all cases.  

Linearized propagator corrections have been validated with an analogous procedure, using the processes listed in Tab.~\ref{tab.validation_processes_prop} and carrying out internal comparisons across models with same inputs and different flavor assumption.
Loop-induced SM Higgs couplings have been validated comparing the SM squared amplitudes at one phase-space point for each of the processes in Tab.~\ref{tab.validation_processes_smhloop} with all models.

The output files generated  by the \madgraph\ plugin are available at the github repository.
No significant residual differences are present between models. Only one exception was observed: potentially large discrepancies are present between \smeftsim\ (both \top\ and \general) and \dimsixtop, for the Wilson coefficients {\tt cQbqu1T, cQbqu8T, cQtqd1T, cQtqd8T} and the associated imaginary parts. These differences have been already noted in the past and are currently not fully understood.

\bibliographystyle{JHEP}
\bibliography{bibliography}

\end{document}